\documentclass[12pt]{aastex}\usepackage{emulateapj5} %AASPP

\righthead{Ion Abundances in HVCs and IVCs}
\def\to{$-$}
\def\dash{$-$}
\def\tdex#1{$\times$10$^{#1}$}
\def\dex#1{10$^{#1}$}
\def\cmm#1{cm$^{-#1}$}
\def\kms{km\,s$^{-1}$}
\def\Msun{M$_{\odot}$}
\def\AA{\leavevmode\setbox0=\hbox{\rm A}\raise\ht0\rlap{$\scriptstyle\circ$}\hskip-0.15em {\rm A}}
\def\vlsr{$v_{\rm LSR}$}
\let\deg=\arcdeg

\let\r=\relax
\def\ref{\par\noindent}

\def\Ha{H$\alpha$}
\def\HI{\protect\ion{H}{1}}

\def\CI{\protect\ion{C}{1}}
\def\CII{\protect\ion{C}{2}}
\def\CIII{\protect\ion{C}{3}}
\def\CIV{\protect\ion{C}{4}}
\def\NI{\protect\ion{N}{1}}
\def\NII{\protect\ion{N}{2}}

\def\NV{\protect\ion{N}{5}}
\def\OI{\protect\ion{O}{1}}

\def\OVI{\protect\ion{O}{6}}
\def\NaI{\protect\ion{Na}{1}}

\def\MgI{\protect\ion{Mg}{1}}
\def\MgII{\protect\ion{Mg}{2}}

\def\AlII{\protect\ion{Al}{2}}
\def\AlIII{\protect\ion{Al}{3}}

\def\SiII{\protect\ion{Si}{2}}
\def\SiIII{\protect\ion{Si}{3}}
\def\SiIV{\protect\ion{Si}{4}}

\def\PII{\protect\ion{P}{2}}

\def\SII{\protect\ion{S}{2}}
\def\SIII{\protect\ion{S}{3}}

\def\ClI{\protect\ion{Cl}{1}}
\def\ClII{\protect\ion{Cl}{2}}
\def\ArI{\protect\ion{Ar}{1}}

\def\KI{\protect\ion{K}{1}}
\def\CaI{\protect\ion{Ca}{1}}
\def\CaII{\protect\ion{Ca}{2}}

\def\TiII{\protect\ion{Ti}{2}}
\def\CrII{\protect\ion{Cr}{2}}
\def\MnII{\protect\ion{Mn}{2}}

\def\FeII{\protect\ion{Fe}{2}}
\def\FeIII{\protect\ion{Fe}{3}}
\def\NiII{\protect\ion{Ni}{2}}
\def\ZnII{\protect\ion{Zn}{2}}

\def\Sformat{2}
\def\Sion{3}         \def\Fmultiple{1}

\def\SionNa{\Sion.5} \def\FCaNa{2}
\def\SionCa{\Sion.6}

\def\Scloud{4}       \def\Fpattern{3}
\def\SSA{1}          \def\FgA{4}
\def\SSM{2}          \def\FgM{5}
\def\SSC{3}          \def\FgC{6}

          \def\FgH{7}
        \def\FgACS{8}

\def\SSACot{8}       \def\FgACc{9}
\def\SSCHS{9}

\def\SSGP{11}

\def\SSMS{14}        \def\FgMS{10}

\def\SSWE{19}        \def\FgWE{11}

\def\SSnoHI{22}
\def\SSLMC{23}
\def\SSIVa{24}       \def\FgIVupp{12} \def\FgIVlow{13} \def\Fivdist{14}
      \def\Fgspur{15}
\def\SSLLIV{26}      \def\FgLLIV{16}
\def\SSK{27}         \def\FgK{17}
\def\SSSouth{28}     \def\Fgsouth{18}
\def\SSgp{29}        \def\Fggp{19}

\def\Ssumm{5}

\def\Tindex{1}
\def\Tmain{2}
\def\Trefs{3}
\def\Tsumm{4}

\def\Cname{1}
\def\Cglon{2}
\def\Cglat{3}
\def\Cdist{4}
\def\Cheight{5}
\def\Ctype{6}
\def\Ccloud{7}
\def\Cvlsr{8}
\def\CNHI{9}
\def\Ctel{10}
\def\Cion{11}
\def\Cvion{12}
\def\CNion{13}
\def\Cionflag{14}
\def\CAion{15}
\def\CAref{16}
\def\CAsolar{17}
\def\Cdflag{18}
\def\Cref{19}
\def\Cnote{20}

\def\NoteCsouth{30}

\def\PGname{22}
\def\NEffNew{113}
\def\NEffOld{9}
\def\NLDSHI{67}

\def\NAr{20}
\def\NJB{6}
\def\NPNarrow{10}
\def\NHIPASS{4}

\def\NFOS{21}
\def\NPKSLMC{112}
\def\NWHI{4}
\def\NWVT{8}
\def\NWVTM15{21}

\begin{document}

\submitted{\null} %AASPP
\title{Distances and Metallicities of High- and Intermediate-Velocity Clouds}
\author{B.P. Wakker}
\affil{Department of Astronomy, University of Wisconsin \\
475 N Charter St, Madison, WI\,53706, USA \\
wakker@astro.wisc.edu}

%VALUES: table 4
\begin{abstract}
A table is presented that summarizes published absorption line measurements for
the high- and intermediate velocity clouds (HVCs and IVCs). New values are
derived for N(\HI) in the direction of observed probes, in order to arrive at
reliable abundances and abundance limits (the \HI\ data are described in
Paper~II). Distances to stellar probes are revisited and calculated
consistently, in order to derive distance brackets or limits for many of the
clouds, taking care to properly interpret non-detections. The main conclusions
are the following. 1) Absolute abundances have been measured using lines of
\SII, \NI\ and \OI, with the following resulting values: $\sim$0.1 solar for one
HVC (complex~C), $\sim$0.3 solar for the Magellanic Stream, $\sim$0.5 solar for
a southern IVC, and $\sim$ solar for two northern IVCs (the IV Arch and LLIV
Arch). Finally, approximate values in the range 0.5\to2 solar are found for
three more IVCs. 2) Depletion patterns in IVCs are like those in warm disk or
halo gas. 3) Most distance limits are based on strong UV lines of \CII, \SiII\
and \MgII, a few on \CaII. Distance limits for major HVCs are $>$5\,kpc, while
distance brackets for several IVCs are in the range 0.5\to2\,kpc. 4) Mass limits
for major IVCs are 0.5\to8\tdex{5}\,\Msun, but for major HVCs they are
$>$\dex{6}\,\Msun. 5) The \CaII/\,\HI\ ratio varies by up to a factor 2\to5
within a single cloud, somewhat more between clouds. 6) The NaI/\,\HI\ ratio
varies by a factor $>$10 within a cloud, and even more between clouds. Thus,
\CaII\ can be useful for determining both lower and upper distance limits, but
\NaI\ only yields upper limits.
\end{abstract}


\newpage

%%%%%%%%%%%%%%%%%%%%%%%%%%%%%%%%%%%%%%%%%%%%%%%%%%%%%%%%%%%%%%%%%%%%%%%%%%%%%%%%
%%%%%%%%%%%%%%%%%%%%%%%%%%%%%%%%%%%%%%%%%%%%%%%%%%%%%%%%%%%%%%%%%%%%%%%%%%%%%%%%
%%%%%%%%%%%%%%%%%%%%%%%%%%%%%%%%%%%%%%%%%%%%%%%%%%%%%%%%%%%%%%%%%%%%%%%%%%%%%%%%

\section{Introduction}
The high- and intermediate-velocity clouds (HVCs and IVCs) consist of gas moving
at velocities incompatible with a simple model of differential galactic rotation
(Wakker 1991). An operational definition has been that HVCs have velocities
larger than $\sim$90\,\kms\ (positive or negative) relative to the LSR. The
definition of IVCs has been less strict. In this paper they are defined as
clouds with velocities relative to the LSR between $\sim$40 and 90\,\kms. In a
few directions slightly lower velocities are included, if there is a clear
connection with gas at higher velocities. This definition excludes many clouds
with $\vert$\vlsr$\vert$$<$40\,\kms\ that were considered an IVC by other
authors, but in most cases these have ill-defined borders and often such
velocities can be understood within the framework of differential galactic
rotation, after allowing for turbulent velocities of up to 30\,\kms.
\par The HVCs and IVCs are still poorly understood, although much progress has
been made in the most recent decade. They appear to serve as tracers of
energetic processes in the Galactic Disk and Halo (as part of a ``Galactic
Fountain''), but also as an ingredient in the continuing formation of the Galaxy
(some are examples of accreting gas). Finally, some may be tidal remnants (most
prominently the Magellanic Stream), or isolated clouds in the Local Group. A
review of our understanding as of a few years ago was presented by Wakker \& van
Woerden (1997); an update has already proven necessary (Wakker et al.\ 1999a).
Ever since the discovery of the HVCs (Muller et al.\ 1963), it has been clear
that the key to a proper understanding lies in using interstellar absorption
lines to determine distances (and metallicities) for this class of clouds. Thus,
some of the recent progress has come from new models and improved mapping, but
most has come from new observations of interstellar absorption lines.
\par Mapping \dash\ The published HVC all-sky maps (Hulsbosch \& Wakker 1988,
Bajaja et al.\ 1985) have proven very useful to understand the properties and
statistics of the HVCs. However, both these surveys suffer from low velocity
resolution (16\,\kms) and incomplete mapping. The Hulsbosch \& Wakker (1988)
survey covered the sky north of declination $-$18\deg\ on a 1\deg\ grid with a
0\fdg5 beam, but the southern survey of Bajaja et al.\ (1985) suffered from lack
of coverage (2\deg\ grid with a 0\fdg5 beam) (although Bajaja et al.\ (1989)
mapped selected areas on a 0\fdg5 grid). IVC maps have only been presented for
the ``Intermediate-Velocity Arch'', a large structure crossing the northern
Galactic sky at latitudes $>$30\deg\ (Kuntz \& Danly 1996). This paper was based
on the data in the ``Bell Labs Survey'' (Stark et al.\ 1992), which has a 3\deg\
beam and 1 \,\kms\ velocity resolution).
\par New datasets allow great improvements in mapping, especially for IVCs and
southern HVCs. a) The Leiden-Dwingeloo Survey (LDS) (Hartmann \& Burton 1997),
covers the sky north of declination $-$35\deg\ on a 0\fdg5 grid with 1\,\kms\
velocity resolution. b) The HVC survey made at the ``Instituto Argentina de
Radioastronomia'' (IAR) (Morras et al.\ 2000) provides a list of \HI\ components
with $\vert$\vlsr$\vert$$>$80\,\kms\ for declinations south of $-$23\deg\ on a
0\fdg5 grid, extracted from spectra with 16\,\kms\ velocity resolution. c) The
\HI\ Parkes All-Sky Survey ({\it HIPASS}) (Staveley-Smith 1997) covers the sky
south of declination 0\deg\ on a 14\farcm4 grid with 26\,\kms\ velocity
resolution. d) The Parkes Narrow Band Survey (Haynes et al.\ 1999, Br\"uns et
al.\ 2001) covers the Magellanic Stream (both trailing and leading parts) on a
14\farcm4 grid with 1\,\kms\ velocity resolution. This paper presents new IVC
maps based on the LDS, and new maps for southern HVCs based on the IAR list. The
LDS and both Parkes surveys are further used to construct \HI\ spectra toward
HVC and IVC probes.
\par Models \dash\ Gardiner \& Noguchi (1996) presented a modern version of the
model in which the Magellanic Stream is formed by tidal stripping. Combined with
the observational identification of the predicted leading arm by Lu et al.\
(1998) and Putman et al.\ (1998), this has led to a better understanding of the
Stream and of which other HVCs could be part of the same tidal feature. Blitz et
al.\ (1999) suggested that the majority of the HVCs are remnants of the
formation of the Local Group, and are similar to the original building blocks of
the Milky Way and the Andromeda Nebula. Braun \& Burton (1999) presented a
variant of this interpretation, in which only some small HVCs are Local Group
clouds. These models contrast with previous ones in which the HVCs/IVCs are
generated in a Galactic Fountain (Bregman 1980) or are remnants of the formation
of the Milky Way (Oort 1970). It now appears that examples of at least three
(and possibly all four) of the proposed origins can be found (see e.g.\ Wakker
et al.\ 1999a).
\par Absorption-line studies \dash\ Metallicities and distances are best
determined using absorption-line studies, both in the optical and the
ultra-violet. In the future, the optical emission lines of [\SII] and \Ha\ may
prove useful (Tufte et al.\ 1998; Bland-Hawthorn \& Maloney 1999), but their
potential has not yet been realized. Most HVC/IVC metallicity and distance
estimates are fairly recent, with half of the relevant papers published since
1994. This is partly due to the availability of the ``Goddard High Resolution
Spectrograph'' ({\it GHRS}), the ``Space Telescope Imaging Spectrograph'' ({\it
STIS}), and the ``Far Ultra-Violet Spectroscopic Explorer'' ({\it FUSE}), and
partly to an increase in sensitivity of ground-based telescopes. A table of
detections of absorptions associated with HVCs was presented by Wakker \& van
Woerden (1997). Several major discoveries have been made since. This paper aims
at summarizing all relevant literature pertaining to deriving distances and
metallicities for HVCs and IVCs.
\par Many of the papers in the literature discuss the implications of
absorption-line detections and non-detections for deriving distance brackets or
limits. Here, a consistent set of criteria was applied to (re)derive these
distance brackets and limits. In general, the conclusions agree with the
original papers, but some new limits are found, and some are shown to have been
in error. For this re-analysis, consistent distances were determined for the
stellar probes (see description of Col.~\Cheight\ in Appendix~A), and all
published equivalent widths and logarithmic column densities were converted to
column densities (see description of Col.~\CNion). Further, improved \HI\ data
were obtained for almost all sightlines, superseding the published value for
directions to probes (see description of Cols.~\Cvlsr\to\Ctel). For about 50\%
of the probes new Effelsberg \HI\ data were obtained, which are presented in a
companion paper (Wakker et al.\ 2001, Paper~II, this issue). For about 25\% of
the probes, N(\HI) is based on the LDS. The remainder are based on (refitted)
published spectra, either of the two new Parkes surveys or (in a small number of
cases) on published numbers. These \HI\ column densities were used to rederive
absolute and relative (to solar) abundances (see description of
Cols.~\CAion\to\CAsolar). Finally, a consistent set of criteria was applied to
determine the significance of non-detections used to derive lower distance
limits (see description of Col.~\Cdflag).
\par This paper is organized as follows: first, a general overview is given of
the format of the main Table~\Tmain\ (Sect.~\Sformat). In Sect.~\Sion\ a short
discussion is given of the abundance results for the individual ions.
Section~\Scloud\ summarizes the derived distance limits, metallicities and
abundance patterns for each of the clouds for which relevant information is
known, and Sect.~\Ssumm\ presents a final analysis. The values in each column of
the main table are described in detail in Appendix~A.

%%%%%%%%%%%%%%%%%%%%%%%%%%%%%%%%%%%%%%%%%%%%%%%%%%%%%%%%%%%%%%%%%%%%%%%%%%%%%%%%
%%%%%%%%%%%%%%%%%%%%%%%%%%%%%%%%%%%%%%%%%%%%%%%%%%%%%%%%%%%%%%%%%%%%%%%%%%%%%%%%
%%%%%%%%%%%%%%%%%%%%%%%%%%%%%%%%%%%%%%%%%%%%%%%%%%%%%%%%%%%%%%%%%%%%%%%%%%%%%%%%

\section{General description of the table}
Table~\Tindex\ provides an index to the main table (Table~\Tmain), to help find
particular clouds or probes. The first part of this table lists the HVC and IVC
names and the table page on which results for each cloud can be found. For a
number of clouds an abbreviation is given in parentheses (e.g.\ ``(=CHS)'').
This is used in the second part of the index, which lists the clouds seen toward
each individual stellar or extra-galactic target.
%A                    M                C
%AI AIII AIV AVI WW81 MI IV4 MIII WW63 CIA CIB CIIIA CIIIC CD CeI CeIII CeIV c C/K
%H           ACH                      ACVH                  MS
%HI HII HIII AC0 ACII CHS WW363 WW468 ACI WW507 WW363 WW525 WW485 WW509 WW532 WW524
%EN         EP                      GN
%WW84 WW274 WW211 WW487 WW187 WW219 WW419 WW302 WW345
%OA G GP L WB    WE    wb                   Other
%          WW225 WW364 WW137 WW29 WW34 WW62 100-7
%ACS IVa                                            LLIV                    Sp K PP gp
%    IV7 IV6 IV17 IV11 IV9 IV15 IV24 IV26 IV21 IV19 LLIV1 LLIV3 LLIV4 LLIVe S1      g1 g2
%11 ON  10 OP
%source count:
%anomabs
%cat ABSLINES.tex|egrep '&[patsrg]&|\\D\ '|egrep -v '^%|&X&' | awk -F\& '{print $1}'|sort|uniq | wc
\par Table~\Tmain\ lists the published results for all probes of each HVC/IVC,
sorted by cloud. Results are given for 7 HVC complexes, 7 IVC complexes, and 21
smaller HVCs/IVCs. In the complexes a total of 47 cores have been observed. In
addition there are 21 unnamed clouds, which often are only seen in absorption.
Data are given for 326 different targets (stars and AGNs), with 1078 entries,
one for each observation of an ion. Stars less distant than $\sim$0.2\,kpc were
excluded to avoid including many nearby stars that provide little distance
information (i.e.\ neither upper nor lower limits).
\par The classical HVCs (A, M, C, H, Anti-Center \dash\ see Wakker \& van
Woerden 1991) are listed first, followed by the GCN and GCP complexes, the Outer
Arm, the Magellanic Stream and the smaller HVC complexes. A new HVC complex
(complex~WE) is introduced here (see Sect.~\Scloud.\SSWE). Next follow results
for the +165, +120 and +65\,\kms\ clouds projected onto the LMC. Then the
classical IVCs (IV Arch, LLIV Arch \dash\ see Kuntz \& Danly 1996) are listed,
followed by other IVC complexes, including three named ones that are introduced
in this paper: ``K'', the ``PP Arch'' and ``gp'', see Sects.~\Scloud.\SSK,
\Scloud.\SSSouth\ and \Scloud.\SSgp.
\par For most clouds the table first gives a summary of measured and expected
abundances for each of the ions observed in the cloud. Also, on the first two
lines for each cloud, the derived metallicity and upper/lower distance limit are
summarized, if known. Details of the method used to derive these limits can be
found in Appendix~A under the description of Cols.~\Cdist\ and \Cdflag.
\par If an abundance was actually measured for a particular ion, that value is
listed in Col.~\CAion\ in the cloud summary lines. For the few cases where
multiple determinations for one ion were made in the same cloud, the average
value is listed (using only the higher quality measurements). A discussion of
these abundances is provided in Sects.~\Sion\ and \Scloud.
%VALUES: absolute levels
\par Column~\CAref\ lists the expected ion abundances in square brackets, but
only for ions that are the dominant ionization stage in the diffuse ISM. These
values are used to determine whether non-detections are significant and can be
used to set a lower distance limit. Expected abundances are derived by assuming
an overall abundance level combined with a halo depletion pattern, as given by
Savage \& Sembach (1996a); see Sect.~\Sion\ for a more detailed description. For
the highly-ionized ions (\CIV, \NV, \OVI, \SiIV) a ``typical'' value is shown
within square brackets in Col.~\CNion; this serves as a point of comparison for
the value observed in the HVC.
\par The overall abundance is assumed to be near solar, unless shown otherwise
in the notes column (Col.~\Cnote), on the first line pertaining to the cloud. A
number followed by Z$_\odot$ in the notes column indicates that the abundance
has actually been measured. This is the case for complex~MI (0.8 solar),
complex~CI (0.1 solar), the Magellanic Stream (0.25 solar), the PP Arch (0.5
solar), IV6, IV9, IV19 (1 solar), and the LLIV Arch (1 solar). Parentheses
around the abundance indicates clouds where an abundance different from solar is
suspected, but not directly proven.
\par After the cloud summary, the table lists the individual observations
relevant for that cloud. In many cases the sightline to a probe intersects more
than one cloud, so that one probe may be listed under two, three or even four
different clouds (see Table~\Tindex).

%%%%%%%%%%%%%%%%%%%%%%%%%%%%%%%%%%%%%%%%%%%%%%%%%%%%%%%%%%%%%%%%%%%%%%%%%%%%%%%%
%%%%%%%%%%%%%%%%%%%%%%%%%%%%%%%%%%%%%%%%%%%%%%%%%%%%%%%%%%%%%%%%%%%%%%%%%%%%%%%%
%%%%%%%%%%%%%%%%%%%%%%%%%%%%%%%%%%%%%%%%%%%%%%%%%%%%%%%%%%%%%%%%%%%%%%%%%%%%%%%%

\section{Notes on ion abundances}
\subsection{General remarks}
\subsubsection{Organization}
In this section, some general remarks are given to accompany the discussion of
results for most of the observed ions below. The symbol $A$ refers to the
abundance of the element, while $\delta$ is used to refer the ratio (observed
abundance in the gas)/(solar abundance). Savage \& Sembach (1996a, Sect.~7)
define $\delta$ in this way, but call it ``depletion'', implying that the
gaseous abundances appear lower than solar because most of the element's atoms
sit in dust grains. However, with this definition $\delta$ really refers to the
combination of depletion and ionization, as both depletion onto dust grains and
the presence of different ionization stages can make the elemental abundance in
the gas appear lower than the intrinsic abundance. Here, we will use $\delta$ to
stand for the observed relative abundance in the gas, i.e.\ the product of
depletion onto dust and ionization.
\par Below, some general remarks are given concerning reference abundances,
oscillator strengths, complications due to ionization, and comparisons of
different measurements of the same ion in the same cloud. A correlation between
ion abundances and \HI\ column density was found, which is summarized. Then, a
summary is given of absorption by \HI, which yields the kinetic (spin)
temperature. Early results on molecular hydrogen are listed. Next, dominant ions
of undepleted ions are discussed, which yield intrinsic abundances. This is
followed by a discussion of dominant ions of depleted elements, which yield
depletion patterns and insight into the presence (and composition) of the dust,
Then, results for non-dominant ions are summarized. Finally, the highly-ionized
ions are described. A more detailed discussion of the numerical results for each
cloud is presented in Sect.~\Scloud.

\subsubsection{Reference abundances}
Reference abundances are summarized in Table~\Trefs, following Savage \& Sembach
(1996a). This table lists the Solar System (meteoritic) abundances of the
element, as given by Anders \& Grevesse (1989) (with photospheric updates for C,
N and O from Grevesse \& Noels 1993), the depletion $\times$ ionization
(assuming a halo-like pattern, as given by Savage \& Sembach 1996a), the
resulting expected halo abundance (in parts per billion, ppb), and the
ionization potential of the previous and next ionization stage (i.e.\ the energy
required to produce and destroy the ion).
\par Solar System abundances are used as a reference because these are
comparatively well-determined. However, as Savage \& Sembach (1996a) point out,
abundances in nearby B stars tend to be 0.15\to0.25 dex lower. Since B stars
have formed recently, they may be a better reference for the local ISM. ISM
abundances may also show inhomogeneities between individual open clusters. Such
differences mostly influence the interpretation of depletion patterns and the
composition of dust.
\par Meyer et al.\ (1998) determined from a set of high-quality measurements
that the typical gaseous abundance of oxygen in nearby low-velocity gas is about
320000 ppb, or 0.4 solar. They also argued that dust contains at most 180000 ppb
of oxygen, so that the total oxygen abundance in the nearby ISM is about 0.66
solar. This is similar to the value derived for nearby B-stars. The suggested
explanations are that a) the early solar system was enriched by a supernova, b)
the ISM has recently been diluted by metal-poor gas, or c) the Sun has moved
outward from the Galactic center since it formed. This abundance difference is
mostly important when using a depletion pattern to derive the composition of
dust particles. Further, if the local ISM indeed has intrinsic abundances below
solar, this has some bearing on understanding the origin of relatively nearby
intermediate-velocity halo gas, which appears to have solar abundance (as
derived from sulphur, see Sects.~\Scloud.\SSIVa, \Scloud.\SSLLIV, \Scloud.\SSK\
and \Scloud.\SSSouth).
\par For non-dominant ions, the results listed in Table~\Tmain\ were used to
derive fiducial values for HVCs/IVCs. This pertains to \CI, \NII, \NaI, \MgI,
\SIII, \KI, \CaI, \CaII, and \FeIII. Table~\Trefs\ shows the range of abundances
found in the HVCs and IVCs for these elements (in the column labeled $A$(halo)).
An average value is given for $\delta$. These abundances clearly can vary by a
large amount, which is not unexpected considering that they depend on the
detailed physical conditions in a cloud (temperature, density, radiation field).
\par Rarely are there complementary data for other ionization stages for clouds
where non-dominant ions were measured. And in cases where both kinds of ions
have been measured, usually no analysis of physical conditions has been done
\dash\ the exceptions being HD\,93521 (Spitzer \& Fitzpatrick 1993), HD\,215733
(Fitzpatrick \& Spitzer 1997), SN\,1987\,A (Welty et al.\ 1999) and PG\,0804+761
(Richter et al.\ 2001a).

\subsubsection{Oscillator strength issues}
To convert the observed amount of absorption into a column density, it is
necessary to know the oscillator strength or $f$-value of the line. For papers
published since 1990, most authors take $f$-values from the list of Morton
(1991). No attempt was made to correct column densities in older papers or
papers where a different source was used for the $f$-values. In general,
differences tend to be relatively small ($<$20\%), although there are
exceptions. Still, this is an extra source of systematic uncertainty in the
tabulated column densities. If a paper gave the column density, this was used in
Table~\Tmain, independent of the actual $f$-value and method used to derive this
column density.
\par If a paper gave an equivalent width, this was converted to a column density
as described in Appendix~A under Col.~\CNion. The $f$-value was then taken from
the compilation of Verner et al.\ (1994), which claims to be a digitized update
of the Morton (1991) list. For \SiII-1260, \SiII-1304, \SiII-1526, \ArI-1048,
1066, \CrII-2056, 2062, 2066 and \ZnII-2026, 2062 updated values were taken from
Savage \& Sembach (1996a), while for \MgII-1239, 1240 updated values were found
by Fitzpatrick \& Spitzer (1997). Further, for the \FeII\ lines between 1121 and
1144\,\AA, new $f$-values were determined experimentally from {\it FUSE} spectra
by Howk (priv.\ comm.).
\par However, for some lines the $f$-values that Verner et al.\ (1994) give are
rather different than those given by Morton (1991). In particular, the ratio
Verner/Morton is 0.95 for \SII-1250, 1253, 1259, 2.88 for \NI-1134.16, 1134.41,
1134.98 and 1.22 for \NI-1199.54, 1200.22, 1200.71. These differences are
unexplained. However, for the case of \NI\ the Morton values appear more
reliable. This conclusion is based on comparing the relative equivalents widths
predicted for the \NI-1134 and \NI-1200 triplets with the high-quality
measurements made by Howk et al.\ (1999) for the sightline to $\mu$\,Col.
\par Morton et al.\ (2001) present a new compilation of $f$-values. These may
differ from the previously published values, sometimes by as much as a factor 2.
However, such differences mostly pertain to far-UV lines in the wavelength range
observed by {\it FUSE}. In the publications using such data, the new $f$-values
have been used.

\subsubsection{Ionization issues}
%VALUES: PG0804+761 - H+
Note that the numbers in the table give the ratio N(ion)/N(\HI). That is, they
do {\it not} include a correction for hydrogen ionization. In fact, only for a
few sightlines was a study of ionization possible. Toward HD\,93521 (Spitzer \&
Fitzpatrick 1993) and HD\,215733 (Fitzpatrick \& Spitzer 1997) N(H$^+$)/N(\HI)
does not seem to be high in the IVCs studied. However, toward Mrk\,509 Sembach
et al.\ (1999) find that N(\CIV)$>$N(\CII), which suggests almost completely
ionized gas. Also, toward Mrk\,876 (Murphy et al.\ 2000) N(H$^+$) appears to
dominate in the complex~C component. In contrast, toward Mrk\,290 (complex~C,
Wakker et al.\ 1999) and PG\,0804+761 (LLIV Arch, Richter et al.\ 2001a)
N(H$^+$)/N(\HI)$\sim$0.2, while toward HD\,215733 hydrogen ionization appears
negligible (Fitzpatrick \& Spitzer 1997). Clearly, ionization issues are
important in general, but usually they cannot be addressed with the existing
data.

\subsubsection{Multiple measurements in one cloud}
For a few ions multiple determinations were made within the same cloud. This is
the case for both \NaI\ and \CaII\ in the IV Arch (Sect.~\Scloud.\SSIVa),
complex~gp (Sect.~\Scloud.\SSgp) and the +65\,kms\ IVC toward the LMC
(Sect.~\Scloud.\SSLMC). Further, \CaII\ is seen in several sightlines on
complex~A (Sect.~\Scloud.\SSA), complex~C (Sect.~\Scloud.\SSC), the LLIV Arch
(Sect.~\Scloud.\SSLLIV) and the HVCs toward the LMC.
\par These values are compared later in this section (Sect.~\SionNa\ and
\SionCa). There it is shown that in general \NaI\ can vary by a factor $>$10
within a single cloud and by a factor up to 100 for any given \HI\ column
density. \CaII\ is more constant, varying by a factor 2\to5 within a single
cloud, and a factor $<$10 at any given N(\HI).
\par Multiple measurements were also made for \SII\ in complex~C, for \MgII\ in
the Magellanic Stream, and for \OI, \MgII, \SII\ and \FeII\ in the +120\,\kms\
HVC toward the LMC. These are discussed under the subsection pertaining to each
cloud (Sects.~\Scloud.\SSC, \Scloud.\SSMS\ and \Scloud.\SSLMC).
\par Figure~\Fmultiple\ presents scatter plots of the \HI\ column density vs the
ion column density for the clouds with multiple determinations.

\subsection{A correlation between ion abundances and N(\HI)}
Previous studies of the depletion in the ISM have found that the abundance of
elements such as \MgII, \PII, \ClII, \MnII, \FeII, \CaII\ and \TiII\ correlates
with the average density of \HI\ in the sightline (calculated as N(\HI) divided
by the distance to the background star) (e.g.\ Jenkins et al.\ 1986, Crinklaw et
al.\ 1994).
\par In these studies of nearby, low-velocity gas the lowest observed \HI\
column density is $\sim$\dex{20}\,\cmm2. As HVC and IVC components stand out in
velocity, it is possible to measure abundances in clouds with much lower column
density, down to \dex{18}\,\cmm2. However, it is not possible to derive an
average volume density since the cloud's distances and depths are not known.
Instead, the gaseous abundances of \MgII, \MnII, \CaII, \TiII\ and \FeII\ in
HVCs/IVCs (from this paper) and low-velocity gas (from Jenkins et al.\ 1986 and
Crinklaw et al.\ 1994) were plotted against \HI\ column density (rather than
against average \HI\ volume density).
\par Unexpectedly, significant correlations were found. The high column density
low-velocity gas and the low column density high-velocity gas lie on the same
curves. Least-squares fits yield a rms scatter of 0.3\to0.4 dex around the mean,
for N(\HI) between \dex{18} and \dex{22}\,\cmm2. These results are presented in
detail in a separate paper (Wakker \& Mathis 2000) and have implications for the
density structure of the ISM and for understanding the formation and destruction
of dust. Here, the fit results are used to predict abundances at a given N(\HI);
these predictions are compared to the actual observed values to help understand
the observed abundance patterns.

%%%%%%%%%%%%%%%%%%%%%%%%%%%%%%%%%%%%%%%%%%%%%%%%%%%%%%%%%%%%%%%%%%%%%%%%%%%%%%%%

\subsection{\HI\ 21-cm absorption}
21-cm \HI\ absorption is seen in only four HVCs/IVCs. The radio continuum source
showing absorption in complex~H is unidentified, but is probably extra-galactic
(Wakker et al.\ 1991). The derived spin temperature of 50\,K is typical for cold
gas. As N(\HI,absorption)$\sim$N(\HI,emission), there appears to be no warm \HI\
in the very core of complex~H.
\par Three \HI\ absorption components associated with the Anti-Center shell are
seen in the spectrum of 3C\,123 (Payne et al.\ 1980, Kulkarni et al.\ 1985),
although only one emission component can be discerned. The column density of the
broad (30\,\kms) absorption component is similar to that of the emission column
density, suggesting the presence of mostly warm (T$>$200\,K) gas. One might
argue that the feature sampled by 3C\,123 is unconnected with the main AC Shell
feature that runs at $b$$\sim$7\deg\ from $l$=170\deg\ to $l$=200\deg\ (see
Fig.~\FgACS), so a separate measurement of T$_s$ in the main feature would be
useful.
\par Relatively many radio continuum sources lie projected onto the Outer Arm,
and \HI\ absorption is seen toward two of those (3C\,395 and QSO\,2005+403
\dash\ Payne et al.\ 1980, Akeson \& Blitz 1999), yielding a spin temperature of
$\sim$50\,K and $\sim$150\,K, respectively, i.e.\ typical values for Galactic
\HI. In several other background sources the lack of \HI\ absorption sets lower
limits of 300\to1000\,K, although a few of these may be Galactic sources in
front of the Outer Arm. This deserves a more thorough study.
\par Payne et al.\ (1980) reported \HI\ absorption associated with cloud~R in
the spectrum of 4C\,33.48, which is probably an extra-galactic radio source. The
derived spin temperature is 69\,K. The ratio of absorption to emission \HI\
column density is only $\sim$0.2, suggesting that most of the \HI\ is warm.
\par In all other cases where observations have been made, 21-cm \HI\ absorption
was not found. This often implies lower limits to the temperature of $>$20\,K,
which is not a very interesting limit. However, lower limits of 70\,K or more
have been found for complex~C (Akeson \& Blitz 1999), the Cohen Stream (Colgan
et al.\ 1990), the Magellanic Stream (Payne et al.\ 1980, Colgan et al.\ 1990,
Mebold et al.\ 1991, Akeson \& Blitz 1999) and complex~K (Colgan et al.\ 1990).

%%%%%%%%%%%%%%%%%%%%%%%%%%%%%%%%%%%%%%%%%%%%%%%%%%%%%%%%%%%%%%%%%%%%%%%%%%%%%%%%

\subsection{Molecular hydrogen}
The study of H$_2$ in HVCs and IVCs is in its infancy. Attempts to detect
molecular gas in HVCs using CO emission have always been unsuccessful (see
Wakker et al.\ 1997). However, H$_2$ is easy to detect in ultra-violet
absorption, as is shown by the recent {\it ORFEUS} (Orbiting and Retrievable Far
and Extreme Ultraviolet Spectrometer) and {\it FUSE} instruments. H$_2$ has now
been found in two HVCs and three IVCs.
\par The first detection was that in the {\it ORFEUS} spectrum of Sk$-$68\,82
for the +120\,\kms\ cloud projected on the LMC (Richter et al.\ 1999). This was
also measured by Bluhm et al.\ (2001) and yields a ratio for N(H$_2$)/N(\HI) of
3.0$\pm$0.6\tdex{-5}, with N(\HI)=12\tdex{18}\,\cmm2. However, toward the star
Sk$-$68\,80 (less than an arcmin away) Richter (priv.\ comm.) finds no H$_2$,
showing large variations on small scales in this cloud. The +60\,\kms\ cloud
toward the LMC also contains H$_2$, with a fraction of 5\tdex{-3}, although
N(\HI) is only $\sim$\dex{18}\,\cmm2.
\par Murphy et al.\ (2000) reported a limit of N(H$_2$)/N(\HI) $<$\dex{-5}, for
complex~C, where N(\HI)=19\tdex{18}\,\cmm2. H$_2$ in the IV arch was found in
the {\it ORFEUS} spectrum of HD\,93521 (Gringel et al.\ 2000), where
N(\HI)=37\tdex{18}\,\cmm2\ and N(H$_2$)/N(\HI)= 0.8$\pm$0.5\tdex{-5}. H$_2$ in
the LLIV arch was detected by Richter et al.\ (2001a), in the {\it FUSE}
spectrum of PG\,0804+761. The N(H$_2$)/N(\HI) ratio there is 1.5\tdex{-5}, with
N(\HI)=35\tdex{18}\,\cmm2. H$_2$ is also found in the +240\,\kms\ HVC seen in
the spectrum of NGC\,3783 (cloud WW187), where N(\HI)=83\tdex{18}\,\cmm2\ and
N(H$_2$)/N(\HI)=0.8$\pm$0.2\tdex{-3} (Sembach et al.\ 2001).
\par Thus, high- and intermediate-velocity molecular hydrogen has been found,
with molecular fractions that are not atypical for the amount of hydrogen in the
sight line. This requires the presence of dust in the detected IVCs/HVCs, which
is also suggested by the depletion pattern of these particular clouds. The
only non-detection is toward complex~C, where a smaller absolute amount of dust
is expected because this cloud has low metallicity. However, it is unclear
whether the sensitivity was sufficiently high to detect H$_2$ at the low \HI\
column density.

%%%%%%%%%%%%%%%%%%%%%%%%%%%%%%%%%%%%%%%%%%%%%%%%%%%%%%%%%%%%%%%%%%%%%%%%%%%%%%%%

\subsection{Dominant ions of lightly depleted elements}
Intrinsic abundances can be determined or estimated using lines of \NI, \OI,
\PII, \SII\ and \ZnII. That these elements are (almost) completely in the gas
phase is an expectation based on measurements of low-velocity gas (Savage \&
Sembach 1996a). Results have been reported for seven clouds, as summarized
below.
%VALUES: IZw18 - OI
\par First, for HVC complex~A an abundance of about 0.1 solar is suggested by
the \OI-1302 absorption seen toward I\,Zw\,18 (Kunth et al.\ 1994). However,
this measurement is based on a strong line, the error in the equivalent width is
large and the component structure complicates matters further (see also
Sect.~\Scloud.\SSA).
%VALUES: Mrk290 - SII  PG1259 - NI, OI, SiII, FeII  Mrk876 - NI ArI FeII
\par For complex~C abundances were measured in two sightlines with high N(\HI)
($\sim$90\tdex{18}\,\cmm2 \dash\ Mrk\,290 and PG\,1259+593), two with
intermediate N(\HI) ($\sim$30\tdex{18}\,\cmm2 \dash\ Mrk\,817 and Mrk\,279) and
one with relatively low N(\HI) ($\sim$20\tdex{19}\,\cmm2 \dash\ Mrk\,876).
\par A value of 0.09$\pm$0.02 solar was measured for sulphur toward Mrk\,290
(Wakker et al.\ 1999), which was corrected for 20\% H ionization (based on a
measurement of \Ha\ emission) and for \HI\ small-scale structure. Richter et
al.\ (2001b) find \OI/\HI=0.11$^{+0.13}_{-0.08}$ solar toward PG\,1259+593. \OI\
is a particularly useful ion as it has an ionization potential similar to that
of \HI\ and its ionization is strongly coupled to that of hydrogen through a
charge-exchange reaction (Sofia \& Jenkins 1998).
\par Toward Mrk\,876 Murphy et al.\ (2000) found N(\NI)/N(\HI)$\sim$0.08 solar
and N(\ArI)/N(\HI)$<$0.1 solar. \NI\ and \HI\ have similar ionization potential
and the nitrogen ionization also tends to couple to that of hydrogen, although
not as strongly as that of oxygen. Argon is a noble gas, and therefore probably
undepleted. However, \ArI\ has a larger photo-ionization cross section than \HI\
and is not coupled to \HI. It is therefore more easily ionized than \HI\ and in
a situation where neutral and (photo-)ionized gas are mixed, \ArI\ can appear
deficient. In the sightline toward Mrk\,876 Murphy et al.\ (2000) further find
that N(\FeII)/N(\HI)$\sim$0.5 solar. To reconcile this with the \NI\ abundance
requires a large H$^+$/\HI\ ratio.
%VALUES: Mrk817 Mrk279 - SII
\par For two other complex~C probes (Mrk\,817, Mrk\,279) Gibson et al.\ (2001)
found N(\SII)/N(\HI)=0.3\to0.4 solar. However, in neither of these directions
has \HI\ small-scale structure or ionization been taken into account. Ionization
probably is important, since if photo-ionization is responsible for the H$^+$,
N(H$^+$) should not vary by more than a factor $\sim$2 across the face of the
cloud. Since, in the sightline toward Mrk\,290 N(H$^+$) is
$\sim$2\tdex{19}\,\cmm2, and toward Mrk\,817 and Mrk\,279 N(\HI) is
$\sim$3\tdex{19}\,\cmm2, the ionization correction could easily be a factor 2,
bringing the S$^+$ abundances in line with those found toward Mrk\,290.
%VALUES: Fairall9 NGC3783 - SII
\par Two sulphur abundance measurements exist for the Magellanic Stream: 0.33
solar (Fairall~9, Gibson et al.\ 2000) and 0.25 solar (NGC\,3783, Lu et al.\
1998). Ionization corrections have not yet been addressed, but they are expected
to be minimal as N(\HI) is high and the gas is far from both the Galaxy and the
LMC. Note that toward NGC\,3783 a 1\arcmin\ resolution map of the \HI\
small-scale structure resulted in a 50\% correction of N(\HI) (and thus the
S$^+$ abundance) relative to a measurement with a large beam (Lu et al.\ 1998).
%VALUES: HD121800 HD93521 PG0953+414 - SII
\par For three sightlines through the IV Arch \SII/\HI\ ratios have been
derived, resulting in values of 0.78 solar for IV9/IV19 (HD\,121800, Howk et
al.\ 2001), 0.8 and 1.2 solar for off-core components at $-$58 and $-$51\,\kms\
(HD\,93521, Spitzer \& Fitzpatrick 1993), and 1.1 solar toward PG\,0953+414
(Fabian et al.\ 2001). Further, \OI/\HI\ is found to be 0.9$\pm$0.7 solar toward
PG\,1259+593.
%VALUES: PG0804+761 - OI NI PII ArI  SN1993J - ZnII
\par PG\,0804+761 gives the best estimate of abundances in the LLIV Arch
(Richter et al.\ 2001a). N(\OI)/N(\HI) is $\sim$1 solar, while N(\NI)/N(\HI) is
$\sim$0.6 solar. The fact that N(\PII)/N(\HI)$\sim$1.3 solar is interpreted as
evidence that 20\% of the H is in the form of H$^+$, as P$^+$ can co-exist with
H$^+$ as well as with \HI, whereas \OI\ becomes ionized when \HI\ gets ionized.
This is supported by the value of 1.6$\pm$0.4 solar for Zn$^+$ (which behaves
like P$^+$) found from SN\,1993\,J. The lower \NI/\HI\ ratio may be interpreted
as evidence for partial ionization (following Sofia \& Jenkins 1998), which is
further supported by the low ratio N(\ArI)/N(\HI)$\sim$0.3.
%VALUES: SII ZnII - HD215733
\par Toward HD\,215733 Fitzpatrick \& Spitzer (1997) decomposed the \HI\
spectrum based on the absorption components. This results in components at
$-$92\,\kms\ with N(\SII)/N(\HI)=0.17 solar, at $-$56\,\kms\ with
N(\SII)/N(\HI)=0.32 solar and at $-$43\,\kms\ with N(\SII)/N(\HI)=1.2 solar. The
latter two are part of what is defined in this paper as the ``PP Arch'' (see
Sect.~\Scloud.\SSSouth). The combined abundance is 0.5 solar. In both components
\ZnII\ absorption is also seen, yielding abundances of 0.23 and 0.95 solar, 0.37
solar when combined. Ionization appears to be unimportant as Fitzpatrick \&
Spitzer (1997) derive a low electron column density, but the complex component
structure hampers the interpretation. Still, the measurements are consistent
with an intrinsic abundance of $\sim$0.5 solar for the PP Arch.
%VALUES: SII - Mrk509
\par Penton et al.\ (2000) fitted three components to the \SII-1250, 1253 and
1259 lines in a {\it GHRS} spectrum of Mrk\,509. One of these is at +60\,\kms\
and is associated with complex~gp. This spectrum is also shown (but not
measured) by Sembach et al.\ (1999). Penton et al.\ (2000) give equivalent
widths of 58, 85 and 30\,m\AA\ for the three sulphur lines. The \SII-1259 line
is strongly blended with \SiII-1260, and thus unreliable. The better view of the
spectrum presented by Sembach et al.\ (1999) shows that the equivalent width
must be about half the value given by Penton et al.\ (2000). Therefore, to
calculate the \SII\ column density, values of 29 and 42\,m\AA\ were used. Using
the \HI\ linewidth of 29\,\kms\ seen in this direction, these equivalent widths
imply an average \SII/\HI\ ratio of 0.8 solar. Since N(\HI) is small
(24.5\tdex{18}\,\cmm2), a large correction for H$^+$ is quite possible, and the
intrinsic abundance of complex~gp remains unknown. However, this result strongly
suggests that the abundance of complex~gp is within a factor 2 of solar.

%%%%%%%%%%%%%%%%%%%%%%%%%%%%%%%%%%%%%%%%%%%%%%%%%%%%%%%%%%%%%%%%%%%%%%%%%%%%%%%%

\subsection{Dominant ions of depleted elements}
\medskip\par\noindent
\CII, \MgII, \SiII\ \dash\ Several depleted ions have very strong lines,
reaching an optical depth of 3 for clouds with standard depletion patterns and
solar abundances at quite low column densities ($\sim$2\tdex{18}\,\cmm2\ for
\CII-1334, $\sim$\dex{18}\,\cmm2\ for \MgII-2796 and \SiII-1260). These lines
are therefore very useful for determining distance limits, but in general they
are less useful for determining depletion patterns.
\par \CII-1334 is clearly detected in many clouds, and has been used to derive a
lower distance limit for complex~H. Unresolved \MgII-2796, 2803 absorption has
been seen in many AGNs observed with the FOS, as described by Savage et al.\
(2000a). \MgII-2796 is the main line used to derive lower distance limits for
complexes~A (Wakker et al.\ 1996b), C (de Boer et al.\ 1994), H (Wakker et al.\
1998), the Cohen Stream and WW507 (Kemp et al.\ 1994), and the LLIV Arch (Wakker
et al.\ 1996b), while \SiII\ is the main ion used to derive limits for IV4, IV6,
IV9, IV11, IV17, IV19, IV24, IV26 and the IV spur (Kuntz \& Danly 1996)
\par For a few sightlines with low N(\HI), \MgII-2796 has been used to derive
the \MgII\ abundance: for complex~C and cloud WW84 using Mrk\,205 (Bowen et al.\
1995b), and for 11 sightlines through the Magellanic Stream (see
Sect.~\Scloud.\SSMS). \MgII-1239, 1240 has been measured toward SN\,1987\,A
(Welty et al.\ 1999), HD\,93521 (Spitzer \& Fitzpatrick 1993) and HD\,215733
(Fitzpatrick \& Spitzer 1997).

\medskip\par\noindent
\AlII, \AlIII, \ClI\ \dash\ These ions are difficult to observe and thus data
exist only for three sightlines: 4\,Lac (Bates et al.\ 1990), SN\,1993\,J (de
Boer et al.\ 1993) and SN\,1987\,A (Welty et al.\ 1999).

\medskip\par\noindent
\TiII, \CrII, \MnII, \NiII\ \dash\ These elements have weak lines, and are
rarely seen. In fact, \CrII\ and \NiII\ have only been found toward the IVC
toward the LMC and in the PP Arch (Welty et al.\ 1999, Fitzpatrick \& Spitzer
1997), while \TiII\ and \MnII\ are also seen in the IVC toward the LMC (Caulet
\& Newell 1996) and several IV Arch cores (Albert 1983, Albert et al.\ 1993,
Spitzer \& Fitzpatrick 1993, Lipman \& Pettini 1995, Fitzpatrick \& Spitzer
1997, Lehner et al.\ 1999a, Bowen et al.\ 2000).

\medskip\par\noindent
\FeII\ \dash\ This is the most useful ion for obtaining an indication of the
presence and amount of dust in HVCs/IVCs, as it has many strong lines in the UV,
with a large range of oscillator strengths. Sembach \& Savage (1996) found that
the depletion of Fe is maximal in cold disk gas (typically $\delta$$\sim$0.01,
less in warm disk gas ($\delta$$\sim$0.1), and least in halo gas
($\delta$$\sim$0.2). Higher gaseous abundances were not observed in the
sightlines studied, which was interpreted by Sembach \& Savage (1996) as
evidence for a hard-to-destroy iron core of the dust particles.
\par High- and intermediate velocity \FeII\ absorption has been seen in many
clouds: complex~C (Murphy et al.\ 2000, Richter et al.\ 2001b), the Magellanic
Stream (Jannuzi et al.\ 1998, Savage et al.\ 2000a), WW187 (Lu et al.\ 1998),
HVC\,100$-$7+100 (Bates et al.\ 1990), the HVCs/IVCs toward the LMC (Savage \&
de Boer 1979, 1981, Welty et al.\ 1999, Richter et al.\ 1999), IV4 (Bowen et
al.\ 2000), IV6 (Spitzer \& Fitzpatrick 1993), the LLIV Arch (de Boer et al.\
1993, Richter et al.\ 2001a), and the PP Arch (Fitzpatrick \& Spitzer 1997).
%VALUES: Mrk876     - Fe/N 0.50/0.08+
%VALUES: PG1259+593 - Fe/O 0.05 Fe/N<1
%VALUES: PKS0637-75+Fairall9 - Fe/S 0.062/0.33
%VALUES: NGC3783    - Fe/S 0.033/0.25
%VALUES: PG1259+593   Fe/O 0.41 Fe/N 0.86
%VALUES: HD93521    - Fe/S 0.42/2.1 0.36/0.78 0.18/1.2
%VALUES: PG0804+761 - Fe/O 0.27/1   Fe/P 0.27/1.3
%VALUES: HD215733   - Fe/S 0.057/0.32 0.28/1.2
\par For clouds where an undepleted element was also measured the ratio with
\FeII\ can be derived and the depletion of Fe can be derived. This results in
(Fe/N)$>$5 solar for complex~C (Mrk\,876 \dash\ Murphy et al.\ 2000; ionization
corrections are likely to be large in this sightline, however); (Fe/N)$<$1 solar
and (Fe/O)=0.37$\pm$0.26 solar in complex~C (PG\,1259+593 \dash\ Richter et al.\
2001b); (Fe/S)=0.19$\pm$0.07 solar in the Magellanic Stream proper
(PKS\,0637$-$75 combined with Fairall\,9 \dash\ Jannuzi et al.\ 1998, Gibson et
al.\ 2000). (Fe/S)=0.13$\pm$0.05 solar in the leading arm (NGC\,3783 \dash\ Lu
et al.\ 1998); (Fe/O)$\sim$0.4 solar in the IV Arch (PG\,1259+593 \dash Richter
et al.\ 2001b); (Fe/S)$\sim$0.2 solar in the IV Arch (HD\,93521 \dash\ Spitzer
\& Fitzpatrick 1993); (Fe/O)$\sim$0.27 solar and (Fe/P)$\sim$0.2 solar in the
LLIV Arch (PG\,0804+762 \dash\ Richter et al.\ 2001a), and (Fe/S)$\sim$0.2 solar
in the PP Arch (HD\,215733 \dash\ Fitzpatrick \& Spitzer 1997).
\par Thus, the Magellanic Stream and three of the major IVCs appear to have Fe
depletions of about a factor 5, which is similar to the typical halo value
derived by Savage \& Sembach (1996a). Note, however, that the IVC results for
HD\,93521 and HD\,215733 were used to arrive at this typical value. Note also
that Wakker \& Mathis (2000) show that the apparent depletion of \FeII\ depends
on N(\HI). It is thus unclear whether the high Fe/S ratios in HVCs/IVCs are due
to environmental conditions or to the fact that the HVC/IVC \HI\ column
densities are relatively low.

%%%%%%%%%%%%%%%%%%%%%%%%%%%%%%%%%%%%%%%%%%%%%%%%%%%%%%%%%%%%%%%%%%%%%%%%%%%%%%%%

\subsection{Non-dominant ions}
Several non-dominant ions have been seen. These are discussed individually
below, in order to derive a reference value for incorporation in Table~\Trefs.

\medskip\par\noindent
%VALUES: CI - NGC3783 HD215733 SN1987A
\CI\ \dash\ For this ion an abundance has been found for clouds WW187
(NGC\,3783, 280$\pm$180 ppb \dash\ Lu et al.\ 1994a) and the PP Arch
(HD\,215733, 240$\pm$30 ppb \dash\ Fitzpatrick \& Spitzer 1997). Both values are
about 0.2\% of the reference \CII\ abundance. \CI\ was also seen in the
+65\,\kms\ IVC in the SN\,1987\,A sightline (Welty et al.\ 1999), with an
abundance of 340$\pm$130 ppb (however, N(\HI) is comparatively uncertain).

\medskip\par\noindent
%VALUES: NII - PG0804+761
\NII\ \dash\ The {\it FUSE} bandpass contains the \NII-1083 line, which is a
good complement to the \NI\ triplets at 1134 and 1199\,\AA. An analysis of the
PG\,0804+761 sightline yields an \NII/\,\NI\ ratio $\sim$0.15 (Richter et al.\
2001a). Only a lower limit to \NII\ is found for complex~C toward Mrk\,876
(Murphy et al.\ 2000).

\medskip\par\noindent
%VALUES: MgI - 4Lac SN1998S SN1993J HD215733 SN1987A
\MgI\ \dash\ This ion has been measured toward HVC\,100$-$7+110 (4\,Lac,
100$\pm$20 ppb \dash\ Bates et al.\ 1990), IV4 (SN\,1998\,S, 6.6$\pm$1.4 ppb
\dash\ Bowen et al.\ 2000), the +120 and +60\,\kms\ clouds toward the LMC (Bluhm
et al.\ 2001, with large errors but showing large variations), IV17
(SN\,1998\,S, 36 ppb \dash\ Bowen et al.\ 2000), LLIV1 (SN\,1993\,J, 44 ppb
\dash\ de Boer et al.\ 1993), and the PP Arch (HD\,215733, 30$\pm$14 ppb \dash\ 
Fitzpatrick \& Spitzer 1997). These values are 0.04\to0.7\% of the reference for
$A$(\MgII). The approximate values for the clouds in the SN\,1987\,A sightline
are slightly higher (110, 110 and 390 ppb, Welty et al.\ 1999). Since each of
the listed clouds has near-solar abundance (see Sects.~\Scloud.\SSLMC\ and
\Scloud.\SSIVa), there clearly are large variations in the relative \MgI\
abundance.
%VALUES: MgI/MgII SN1987A+165 - 6/1300   SN1987A+125 - 8.9/1700  HD215733 - 5.9/1900
\par Only for the +165 and +122\,\kms\ LMC clouds and the PP Arch has an \MgII\
abundance also been measured directly. The resulting \MgI/\,\MgII\ ratios are
0.5\%, 0.5\% and 0.3\%, respectively.

\medskip\par\noindent
\SiIII\ \dash\ The 1206\,\AA\ line of \SiIII\ is very strong, so that all six
detections only give lower limits to the Si$^{+2}$ abundance: in complex~C
(Mrk\,876 \dash\ Gibson et al.\ 2001), WW487 (NGC\,1705 \dash\ Sahu \& Blades
1997), complex GCN (Mrk\,509 \dash\ Sembach et al.\ 1995, 1999), complex~WE
(HD\,156359 \dash\ Sembach et al.\ 1991), the $-$150\,\kms\ and +130\,\kms\
clouds against PG\,0953+414 (Fabian et al.\ 2001), and the +65\,\kms\ IVC toward
the LMC (Sk$-$67\,104 \dash\ Savage \& Jeske 1981).

\medskip\par\noindent
%VALUES: SIII - HD215733
\SIII\ \dash\ This ion has only been seen in the three IVCs toward HD\,93521,
where the \SIII/\,\SII\ ratios are 0.11, 0.04 and 0.15 (Spitzer \& Fitzpatrick
1993).

\medskip\par\noindent
%VALUES: KI - SN1993J M15 K144 M15 IV-38
\KI\ \dash\ This ion has weak optical lines. It was detected toward SN\,1993\,J
at +122 and +140\,\kms\ with abundances $>$9.2 and $>$5.8 ppb (Vladilo et al.\
1993, 1994). It was also detected in complex~gp in the spectrum of M 15 K\,144,
with an abundance of 4.9 ppb (Kennedy et al.\ 1998), which is about 5\% of the
solar K abundance. However, Kennedy et al.\ (1998) did not detect \KI\ toward
the star M 15 IV-38, with a limit of 0.82 ppb, suggesting large variations in
the \KI\ abundance on small scales.

\medskip\par\noindent
%VALUES: CaI - SN1993J SN1987A
\CaI\ \dash\ Like \KI, \CaI\ has weak optical lines, and was found toward
SN\,1993\,J at +122 and +140\,\kms\ (Vladilo et al.\ 1993, 1994). The abundance
is $>$1.7 ppb for both clouds. For the +165 and +65\,\kms\ clouds toward
SN\,1987\,A approximate abundances of 0.87 and 0.31 ppb were found, whereas for
the +120\,\kms\ cloud an upper limit of 0.05 ppb was set (Magain 1987,
Vidal-Madjar 1987, Welty et al.\ 1999).

\medskip\par\noindent
%VALUES: FeIII/FeII - PG0804+761
\FeIII\ \dash\ The {\it FUSE} bandpass contains the \FeIII-1122 line, which is a
good complement to the many \FeII\ lines. The analysis of the PG\,0804+761
sightline gives an \FeIII/\,\FeII\ ratio $\sim$0.03 for the LLIV Arch.

%%%%%%%%%%%%%%%%%%%%%%%%%%%%%%%%%%%%%%%%%%%%%%%%%%%%%%%%%%%%%%%%%%%%%%%%%%%%%%%%

\subsection{\NaI}
\subsubsection{\NaI\ detections}
%VALUES: NaI - SN1986G BD+38.2182 HD83206 Mrk595 Mrk205 Fairall9 NGC3783
In low-velocity gas, an average \NaI\ abundance of $\sim$5 ppb is seen, although
the range is large (more than a factor 10 either way). Only one catalogued HVC
has been detected in \NaI\ absorption: cloud WW219 toward SN\,1986\,G (d'Odorico
et al.\ 1989) yields $A$(\NaI)=50$\pm$10 ppb. High-velocity \NaI\ has been
searched for in only two stars known to be sufficiently distant. Yet, it was not
detected toward BD+38\,2182 (MIII \dash\ Keenan et al.\ 1995) and HD\,83206
(WW63 \dash\ Lehner et al.\ 1999a). Four extra-galactic probes give upper limits:
$<$4 ppb (Mrk\,595, Cohen Stream \dash\ Kemp \& Bates 1998), $<$55 ppb
(Mrk\,205, WW84 \dash\ Bowen et al.\ 1995a), $<$2.5 ppb (Fairall\,9, Magellanic
Stream \dash\ Songaila \& York 1981) and $<$3.7 ppb (NGC\,3783, cloud WW187
\dash\ West et al.\ 1985). Considering the large observed range for N(\NaI) in
low-velocity gas, these limits clearly are not very significant.
%VALUES: NaI - SN1993J SN1994D SN1994I
\par Detections of \NaI\ at high velocity have been reported for some
extreme-positive velocity gas, in all but one case for directions where no \HI\
was detected: SN\,1993\,J (Vladilo et al.\ 1993, 1994), SN\,1994\,D and
SN\,1994\,I (Ho \& Filippenko 1995, 1996) give $A$(\NaI)$>$20 to$>$800 ppb. In
all these cases the rather large abundances and low value of N(\HI) suggest that
the gas is mostly ionized.
%VALUES: NaI - LMCIVC
\par For the +65\,\kms\ IVC toward the LMC, \NaI\ abundances in directions with
N(\HI)$>$5\tdex{18}\,\cmm2\ vary by a factor $>$10 ($<$7 to 70 ppb; see
Fig.~\Fmultiple f). The highest abundances ($>$70 ppb) occur for two directions
with low N(\HI) (Sk$-$68\,82 and Sk$-$71\,03 \dash\ Songaila \& York 1981,
Songaila et al.\ 1981).
%VALUES: NaI - IVArch
\par In the IV Arch \NaI\ detections exist for cores IV6, IV21 and IV26, as well
as in some off-core probes (Benjamin et al.\ 1996, Ryans et al.\ 1997a, Lehner
et al.\ 1999a). Abundances range from 0.57 ppb to 6.1 ppb, i.e., inside the
range typical for low-velocity gas (Fig.~\Fmultiple h). One higher value ($>$50
ppb for BD+38\,2182, Ryans et al.\ 1997a) is associated with a low value for
N(\HI), in which case ionization issues may be important.
%VALUES: NaI - SN1993J HD77770
\par The two measurements in the LLIV Arch yield normal abundances of 2.4 and
4.2 ppb (SN\,1993\,J \dash\ Vladilo et al.\ 1993, 1994, HD\,77770 \dash Welsh et
al.\ 1996).
%VALUES: NaI - M13
\par Toward several stars in M\,13 that probe complex~K, Shaw et al.\ (1996)
determined N(\HI) at 1\arcmin\ resolution, and found that \NaI\ was not always
detected toward stars with similar N(\HI), giving limits $A$$<$5 ppb, while the
detections range from 13 ppb to 45 ppb.
%VALUES: NaI - M15* HD203664 Mrk509
\par Detections associated with complex~gp toward 8 stars in M\,15 yield
abundances between 10 and 40 ppb (Morton \& Blades 1986, Langer et al.\ 1990,
Kennedy et al.\ 1998; Fig.~\Fmultiple j). Meyer \& Lauroesch (1999) found a
change from 10 to $>$160 ppb over several arcminutes in M\,15. These values were
derived using N(\HI) as measured at 9\farcm1 resolution and interpolated between
9 beams placed at 5\arcmin\ intervals. When probed in the sightline toward
HD\,203664, this IVC shows three absorption components with an average abundance
of 77 ppb (Ryans et al.\ 1996). Finally, toward Mrk\,509 $A$(\NaI)=10 ppb (York
et al.\ 1982).
\par Thus, for the IV and LLIV Arch, $A$(\NaI) tends to be similar to that in
low-velocity gas, whereas for the IVC toward the LMC and complex~gp it is
higher. Further, within a single cloud the measured abundance can vary by a
factor $>$10.

\subsubsection{Correlation between $A$(\NaI) and N(\HI)}
Figure~\FCaNa\ a shows the correlation between $A$(\NaI) and N(\HI) for the
high- and intermediate-velocity detections. The mostly-horizontal straight line
is the relation claimed by Ferlet et al.\ (1985b) for low-velocity \HI\ (log
N(\NaI) = 1.04 log N(\HI) $-$ 9.09), while the dotted lines show the 1$\sigma$
spread in that relation (0.5 in the log). Note that this relation is supposed to
correlate N(\NaI) with the combined atomic and molecular column density, not
just the neutral hydrogen column density. For HVCs and IVCs, N(H$_2$) is
relatively low, so N(H)$\sim$N(\HI). This correlation is often used to infer
N(\HI) from a measurement of N(\NaI). For instance, Ho \& Filippenko (1995)
stated that the \NaI\ column density for the +243\,\kms\ component seen in the
spectrum of SN\,1994\,D implies that N(\HI) should be $\sim$6\tdex{19}\,\cmm2.
\par However, the N(\NaI) vs N(\HI) relation is not as well defined as it has
been made out to be, and here it is shown to be invalid for the HVC/IVC gas.
Also, Welty et al.\ (1994, Sect.~4.3) pointed out that below
N(\HI)=\dex{19}\,\cmm2\ the relation was defined from five points, three of
which they showed to be inaccurate. Welty et al.\ (1994) also point out that
Ferlet et al.\ 1985b) mixed \NaI/\HI\ ratios derived for individual components
with velocity-integrated values, which is invalid if the abundance vs N(H)
relation is not linear. Finally, the spread in the nominal relation is large:
for any given value of N(\NaI), the predicted value of N(\HI) has a range of at
least a factor 100.
%VALUES: HI - SN1994D
\par In the case of the HVCs/IVCs, a direct observation of N(\HI) toward
SN\,1994\,D shows that N(\HI)$<$2.5\tdex{18}\,\cmm2\ (Paper~II), a factor 25
lower than would be inferred from N(\NaI). For the HVC/IVC points in Fig.~\FCaNa
a the formal fit shows a non-linear correlation, but the 1$\sigma$ range of log
N(\NaI) at any given value of log N(\HI) is $\pm$1, even larger than the
$\pm$0.5 shown by Ferlet et al.\ (1985b) for low-velocity gas. Thus, for any
given N(\HI) there is a range of about a factor 100 in the \NaI\ column density,
rather than the factor 10 for low-velocity gas. In conclusion, for high- and
intermediate-velocity gas N(\NaI) is an even worse predictor of N(\HI) than it
is for low-velocity gas.

\subsection{\CaII}
\subsubsection{\CaII\ abundances in HVCs}
The most important non-dominant ion is \CaII. HVCs and IVCs have \HI\ column
densities in the range a few \dex{18} to \dex{20}\,\cmm2, and the \CaII\
abundance then tends to be 10\to100 ppb, which corresponds to a logarithmic
value for the product depletion $\times$ ionization (log $\delta$) of $-$1.3 to
$-$2.3. \CaII\ has been detected in many different clouds, HVCs as well as IVCs,
often in multiple sightlines through the same cloud. It is therefore possible to
check the constancy of its abundance across a cloud. So far, assuming constant
\CaII\ abundance appears to be reasonable, although it may be even better to
calculate a predicted value for $A$(\CaII) using the observed correlation
between $A$(\CaII) and N(\HI) (see Wakker \& Mathis 2000). This correlation can
also be seen in Fig.~\FCaNa b, which shows N(\CaII) vs N(\HI).
%VALUES: CaII - Mrk106 38E18/17ppb; ADUMa 80/21               (pred=16,9)
%VALUES: CaII - Mrk290 92/21 33/12; PG1351+640 7.2/110 56/18  (pred=8,18,57,12)
%VALUES: CaII - BD+38.2182 <0.5/>290; SN1998S 43/3.3          (pred=440,15)
%VALUES: CaII - Fairall9 89/27; NGC3783 80/69                 (pred=8,9)
%VALUES: CaII - PKS0837-12 7.9/280; SN1986G 2.8/150 4.0/150   (pred=53,118,90)
\par \CaII\ is detected in the following HVCs: complex~A (Mrk\,106 \dash\
Schwarz et al.\ 1995, AD\,UMa \dash\ van Woerden et al.\ 1999a, complex~C
(Mrk\,290 and PG\,1351+640 \dash\ Wakker et al.\ 1996a), cloud MIII (BD+38 2182
\dash\ Ryans et al.\ 1997a), cloud IV4 (SN\,1998\,S \dash\ Bowen et al.\ 2000),
complex~WB (PKS\,0837$-$12 \dash\ Robertson et al.\ 1991), the Magellanic Stream
(Fairall 9 \dash\ Songaila 1981, NGC\,3783 \dash\ West et al.\ 1985), and cloud
WW219 (SN\,1986\,G \dash\ d'Odorico et al.\ 1989). For 8 components with
N(\HI)=30\to90\tdex{18}\,\cmm2\ six \CaII\ abundances are in the range 12\to27
ppb, while a high value of 69 ppb is found toward NGC\,3783 and a low value of
3.3$\pm$0.9 ppb is found toward SN\,1998\,S in IV4. The latter may be low
because IV4 shows a disk-like (i.e.\ more depleted) pattern, rather than the
halo-like pattern seen in other HVCs (see Sect.~\Scloud.\SSM). The
$A$(\CaII)-N(\HI) correlation (Wakker \& Mathis 2000) predicts a range of 8\to18
ppb for the \HI\ column densities in these components.
\par In five lower column density components (N(\HI)$<$8\tdex{18}\,\cmm2) the
values are higher, 110\to$>$290 pbb, where the $A$(\CaII)-N(\HI) correlation
predicts a range of 53\to440 ppb for the observed \HI\ column densities.
\par Within a single complex multiple determinations are within a factor 2.5 of
each other (18/21 ppb in complex~A, 21/12/18 ppb in complex~C, 27/69 ppb in the
Magellanic Stream). However, the \HI\ column densities in these directions are
also within a factor 2 of each other, so that the abundances are expected to be
similar.
%VALUES: CaII - SN1983N <6/>27; SN1991T <0.3/>250 <0.3/>370  (pred 66,650)
%VALUES: CaII - SN1993J <6/>43; SN1994D <1.5/>140,500,140    (pred 66,190)
%VALUES: BD+38.2182 <0.5/>290; HD83206 <2/>35                (pred 440,152)
\par \CaII\ absorption without associated high-velocity \HI\ was seen toward
four extra-galactic supernovae (SN\,1983\,N \dash\ d'Odorico et al.\ 1985,
SN\,1991\,T \dash\ Meyer \& Roth 1991, SN\,1993\,J \dash\ Vladilo et al.\ 1993,
1994, and SN\,1994\,D \dash\ King et al.\ 1995) and two stars (BD+38 2182 \dash\
Keenan et al.\ 1995 and HD\,83206 \dash\ Lehner et al.\ 1999a). This sets lower
limits to $A$(\CaII) of $>$27 to $>$500 ppb. In all cases, the \HI\ column
density must be very low ($<$ a few \dex{18}\,\cmm2), so the limit on the
abundance is below (and thus consistent with) the value predicted from the
$A$(\CaII)-N(\HI) relation that is discussed by Wakker \& Mathis (2000).

\subsubsection{\CaII\ abundances in HVCs/IVCs toward the LMC}
%VALUES: LMCHVC LMCIVC
Many detections exist toward the HVCs/IVC toward the LMC. If the \HI\ column
densities toward the background stars were more reliable, this cloud would
provide an ideal testing ground to look for abundance variations. However, many
\HI\ column densities are interpolated between three directions (see item PI in
the description of Col.~\Ctel), or based on using a ruler on the plots of Wayte
(1990) (item PR). Still, some patterns do emerge, see Fig.~\Fmultiple g, h, i).
Excluding two discrepant values from Songaila et al.\ (1981), in the two HVCs
(at +165, and +120\,\kms) the abundance range is about a factor 5 (26\to110 ppb
and 20\to100 ppb, respectively), when N(\HI)$>$4.5\tdex{18}\,\cmm2, while for
sightlines with lower N(\HI) the typical abundances are higher (180\to370 ppb
and 160\to1300 ppb, respectively). The pattern in the IVC is similar, but with
higher values. $A$(\CaII) ranges from 31 to 230 ppb for
N(\HI)$>$4.5\tdex{18}\,\cmm2, with two outliers at 400 and 480 ppb (both have
interpolated N(\HI)), and from 80 to 430 ppb at lower N(\HI). This cloud shows
an unusual depletion pattern for all the other elements (see Welty et al.\
1999), which may indicate that unusual processes are going on.

\subsubsection{\CaII\ abundances in IVCs}
%VALUES: CaII - IVArch H.O.+41B BD+49.2137
Many detections exist in the IV Arch (Fig.~\Fmultiple c) (Wesselius \& Fejes
1973, Albert 1983, Songaila et al.\ 1985, 1988, Spitzer \& Fitzpatrick 1993,
Albert et al.\ 1993, Schwarz et al.\ 1995, Wakker et al.\ 1996a, Ryans et al.\
1997a, Lehner et al.\ 1999a, Ryans et al.\ 1999, van Woerden et al.\ 1999b,
Bowen et al.\ 2000).
%VALUES: CaII - IVArch IV17 102E18/26ppb 78/16 24/<5.2 28/4.3 18/4.3
%                                                           (pred 8,9,23,20,28)
%VALUES: CaII - IV19 105/37 10/13                           (pred 7,44)
%VALUES: CaII - IV6 <0.5/>250 3.2/30                        (pred 440,106)
%VALUES: CaII - IV7 8/23                                    (pred 52)
%VALUES: CaII - IV8 72/7.5                                  (pred 10)
%VALUES: CaII - IV26 166/4.6 154/11                         (pred 5,5)
%VALUES: CaII - IVupp <3/>32>25>68>120 5.2/150              (pred 112,73)
%VALUES: CaII - IVlow 15/36 9.5/55 34/55 22/16 22/7.9 27/18 (pred33,46,14,24,21)
%ratios 3 2 <0.25 0.2 0.14 5 0.3 >0.5 0.3 0.5 0.8 1 0.5 2 1.1 1.2 0.7 0.3 0.9
%ratios 0.14 0.2 <0.25 0.3 0.3 0.3 >0.5 0.5 0.5 0.7 0.8 0.9 1 1.1 1.2 2 2 3 5
\par Within a single core some variations exist: 16/26 ppb in IV17, 13/37 ppb in
IV19, 4.6/11 ppb in IV26. Between cores the variations are larger: 30 ppb in
IV6, 23 ppb in IV7, 7.5 ppb in IV9. Outside cores the range is 8\to55 ppb.
Clearly, variations are about a factor 5. However, this relatively small range
may be an artifact of the small range in N(\HI), as for 15 out of 19 cases the
observed value is within a factor 3 of the value predicted by the
$A$(\CaII)-N(\HI) relation discussed by Wakker \& Mathis (2000). One of the
exceptions is BD+49\,2137 where two \CaII\ components are seen, but only one
\HI\ component can be measured.
%VALUES: CaII LLIV 70/14 92/14 24/12 77/2.5 34/13 24/13 32/14 20/17 31/14 16/9.1
%             pred    10     8    23    9      17    23    18    26    19    31
%             ratio  1.4    1.6  0.5    0.28  0.8   0.6   0.8   0.7   0.7   0.3
\par Abundances in the LLIV Arch have been measured toward 8 different
background probes (Fig.~\Fmultiple b) (Vladilo et al.\ 1993, 1994, Welsh et al.\
1996, Ryans et al.\ 1997b, Lehner et al.\ 1999a), and tend to lie in the narrow
range of 12 to 17 ppb (with two outliers at 9 and 2.5 ppb; for the latter the
derived \HI\ column density is probably affected by the spectral decomposition).
Thus, variations appear to be very small, and on average the value is lower than
in other clouds. This is consistent with the warm-disk-like depletion pattern
for this object (Sect.~\Scloud.\SSLLIV). The values predicted from the
$A$(\CaII)-N(\HI) relation discussed in Sect.~\SionCa\ lie in the range 9\to26
ppb, i.e.\ the low \CaII\ abundance may also be due to the relatively high \HI\
column densities seen in the LLIV Arch.
%VALUES: CaII - Mrk509 25E18/74  (pred 22)
%VALUES: CaII - HD203664 2.2/450 or 1.1/0.5/0.5  (pred 22/141 or 240,440,440)
%VALUES: CaII - M15* 54/27 50/51 53/53 47/20 47/14 44/25 51/25 52/79
%                    48/43 50/29 50/19 46/44 50/29 (pred 12-14)
\par A cloud where future work will allow a study of small-scale variations in
$A$(\CaII) is complex~gp (see Sect.~\Scloud.\SSgp), which is probed by the
globular cluster M\,15 (Fig.~\Fmultiple e). Meyer \& Lauroesch (1999) found that
toward M\,15 the \NaI\ abundance varies from 10 to 160 ppb (i.e.\ a factor 16)
on a scale of a few arcsec. Lehner et al.\ (1999b) measured N(\CaII) toward 12
stars, but did not provide a detailed \HI\ map. N(\HI) was measured at 9
positions with Effelsberg, and interpolated to the stellar positions. This
results in abundances of 20\to55 ppb. The same IVC is also seen toward Mrk\,509
($A$(\CaII)=74 ppb, York et al.\ 1982) and toward HD\,203664, where N(\HI) is
very low (2.2\tdex{18}\,\cmm2), and the \CaII\ abundance is correspondingly high
(440 ppb, Ryans et al.\ 1996). The $A$(\CaII)-N(\HI) relation (Wakker \& Mathis
2000) predicts values that are a factor $\sim$3 lower, i.e.\ the \CaII\
abundance in complex~gp is relatively high compared to expectations.

\subsubsection{Correlations between N(\NaI), N(\CaII) and velocity}
For the IVCs and HVCs no evidence exists for a correlation between velocity and
\NaI\ (or \CaII) column density (Fig.~\FCaNa c, d). For low-velocity gas, Routly
\& Spitzer (1952) found that the column density ratio N(\CaII)/\,N(\NaI)
increases with LSR velocity. Vallerga et al.\ (1993) show that the effect occurs
when studying nearby ($<$100\,pc), low-velocity ($<$20\,\kms) gas, while Sembach
\& Danks (1994) find a correlation of this ratio with the deviation velocity
(the difference between the observed LSR velocity and the maximum velocity that
is expected from a simple model of galactic rotation), but only for gas with LSR
velocities below 50\,\kms. This effect is usually interpreted as showing that Ca
is less depleted at higher peculiar velocities. Fig.~\FCaNa e and f show that
the \CaII/\NaI\ column density ratio does not depend on velocity for sightlines
through IVCs and HVCs. This is not entirely unexpected since for the HVCs/IVCs
the LSR velocity is not a good measure of the peculiar velocity relative to
their surroundings. Further, it should be noted that components at higher
velocity tend to have lower column density. Wakker \& Mathis (2000) showed that
on average the abundance of \NaI\ is independent of N(\HI), whereas $A$(\CaII)
is larger at lower N(\HI). Thus the \CaII/\NaI\ ratio depends on N(\HI), which
tends to be lower at higher velocity.

\subsubsection{Implications of the \NaI\ and \CaII\ results}
\NaI\ varies by a factor $>$10 within clouds and by an even larger factor
between clouds. \CaII\ also shows a large range (a factor $>$100), although at
any given \HI\ column density the range is more like a factor 10), and within
any given cloud the range is a factor up to $\sim$5.
\par These results can be used to derive a safety factor for interpreting the
significance of a non-detection of \NaI\ or \CaII. This is discussed in more
detail in Appendix~A, under the description of Col.~\Cdflag. In summary, \HI\
small-scale structure requires a safety factor $\sim$2. For both \NaI\ and
\CaII\ the depletion is uncertain by a factor $\sim$2.5. The ionization is
uncertain by a factor $\sim$2 for \CaII, $\sim$5 for \NaI. Finally, if the ion
abundance in the cloud has to be assumed (rather than being measured toward
another probe), a final factor $\sim$2 is needed.
\par Thus, in practice non-detections of \NaI\ never result in a lower limit to
a cloud distance, as the combined safety factor needs to be 25\to50. Since at an
abundance of 4.6 ppb $\tau$(\NaI\ D2) is 0.022 (N(\HI)/\dex{19}), a significant
non-detection requires a spectrum with S/N$>$1000\to2000 (\dex{19}/\,N(\HI)).
Thus, although \NaI\ tends to be the easiest interstellar line to observe, it is
hard to determine upper limits to HVC/IVC distances and not useful for
determining lower limits. None of the published non-detections can be considered
to give a significant lower distance limit.
\par \CaII\ gives much stronger absorption than \NaI\ ($\tau$(\CaII\ K) is 0.13
(N(\HI)/\dex{19}) for a standard abundance of 22 ppb). Thus a spectrum with
S/N$>$75\to100 (\dex{19}/\,N(\HI)) can yield a lower distance limit. However, it
remains necessary to derive $A$(\CaII) for any given cloud. Also, such S/N
ratios are difficult to achieve for the more distant, fainter stars that will be
needed to derive distances to HVCs.

%%%%%%%%%%%%%%%%%%%%%%%%%%%%%%%%%%%%%%%%%%%%%%%%%%%%%%%%%%%%%%%%%%%%%%%%%%%%%%%%

\subsection{Highly-ionized species}
For the high ionization ions (\CIV, \NV, \SiIV), the results of Sembach \&
Savage (1992) were used to provide fiducial values for the column density. They
found that, on average, N(\CIV)=1.6\tdex{14}\,\cmm2\ toward the Galactic pole.
Averages for the other ions follow from the average ratios they give, yielding
3.5\tdex{13}\,\cmm2\ for \NV, and 4.4\tdex{13}\,\cmm2\ for \SiIV.
\par Preliminary results for halo \OVI\ were presented by Savage et al.\
(2000b), showing column densities in the range 1\to7\tdex{14}\,\cmm2. The polar
value (the average of N\,sin\,{\it b}) is about 2\tdex{14}\,\cmm2.

\medskip\par\noindent
%VALUES: H1821+643   -  610 600        120
%VALUES: Fairall9    -                <410
%VALUES: NGC3783     -                <330
%VALUES: Mrk509      - >480     >1600
%VALUES: PKS2155-304 -  320 320
%VALUES: SN1993J     -  400     >2000
%VALUES: SN1987A     -  270
%VALUES: Sk-66 118   -               >300 >1000
%VALUES: Sk-67 169   -                 70 600
%VALUES: Sk-67 104   -               1000
%VALUES: HD93521     -  240
%VALUES: HD121800    -  360
%VALUES: HD215733    -  370            14
\CIV\ \dash\ High-velocity \CIV\ is observed in the Outer Arm (H\,1821+643
\dash\ Savage et al.\ 1995), complex~GCN (Mrk\,509 and PKS\,2155$-$304 \dash\
Bruhweiler et al.\ 1993, Sembach et al.\ 1999), in the +65 and +120\,\kms\ IVCs
toward the LMC (Savage \& Jeske 1981, Savage et al.\ 1989, Bomans et al.\ 1996),
IV6 (HD\,93521 \dash\ Spitzer \& Fitzpatrick 1992), IV9/IV19 (HD\,121800 \dash\
Howk et al.\ 2001), LLIV1 (SN\,1993\,J \dash\ de Boer et al.\ 1993), and the PP
Arch (HD\,215733 \dash\ Fitzpatrick \& Spitzer 1997). The more reliable
determinations have column densities in the range 2.4\to6.0\tdex{13}\,\cmm2,
i.e.\ a factor 3\to6 below the typical value though the Galactic halo. Higher
values exist toward Mrk\,509 ($>$1.6\tdex{14}\,\cmm2) and SN\,1993\,J
($>$2\tdex{14}\,\cmm2).

\medskip\par\noindent
%VALUES: H1821+643   - <420 <350
%VALUES: Fairall9    - <510
%VALUES: NGC3783     - <330
%VALUES: Mrk509      - <150 <95 85
%VALUES: PKS2155-304 - <110 <140
%VALUES: HD156359    -   58
%VALUES: Sk-67 104   - <300
\NV\ \dash\ \NV\ was observed but not detected toward 5 AGNs with high-velocity
gas in the sightline (H\,1821+643 \dash\ Savage et al.\ 1995, Fairall\,9 \dash\
Lu et al.\ 1994b, NGC\,3783 \dash\ Lu et al.\ 1994a, Mrk\,509 and
PKS\,2155$-$304 \dash\ Sembach et al.\ 1999), although the upper limits are only
1\to5\tdex{13}\,\cmm2, i.e.\ comparable to the total amount through the Galactic
halo. A small amount of high-velocity \NV\ was found in complex~WE, toward
HD\,156359, with a column density of 5.8\tdex{12}\,\cmm2\ (Sembach et al.\
1995).

\medskip\par\noindent
\OVI\ \dash\ Sembach et al.\ (2000) found high-velocity \OVI\ absorption
associated with complex~C (Mrk\,876), complex~GCN (toward Mrk\,509 and
PKS\,2155$-$304), the Magellanic Stream (three sightlines) and the Outer Arm
(H\,1821+643), with column densities that are a significant fraction of the
total \OVI\ column density in these directions. High-velocity \OVI\ absorption
has not been found in other AGNs that are not projected on or close to a
previously-known \HI\ HVC.

\medskip\par\noindent
%VALUES: Mrk509    - >280 <85
%VALUES: SN1993J   -  250 630
%VALUES: SN1987A   -   59
%VALUES: Sk-67 104 -  300
%VALUES: HD93521   -  820
%VALUES: HD121800  -  140
%VALUES: HD215733  -   66
\SiIV\ \dash\ High-velocity \SiIV\ was found toward Mrk\,509 (Sembach et al.\
1999) and SN\,1993\,J (de Boer et al.\ 1993) with column densities of
$\sim$2.5\tdex{13}\,\cmm2. Intermediate-velocity \SiIV\ has been detected toward
IV6 (HD\,93521 \dash\ Spitzer \& Fitzpatrick 1992), LLIV1 (SN\,1993\,J \dash\ de
Boer et al.\ 1993), the PP Arch (HD\,215733 \dash\ Fitzpatrick \& Spitzer 1997),
and the +65 and +120\,\kms\ HVC toward the LMC (Savage \& Jeske 1981, Savage et
al.\ 1989). The measured column densities range from 5.9 to 82\tdex{12}\,\cmm2,
where for low-velocity gas values between 50 and 100\tdex{12}\,\cmm2\ are
typical for the galactic latitude range of the probes.

%%%%%%%%%%%%%%%%%%%%%%%%%%%%%%%%%%%%%%%%%%%%%%%%%%%%%%%%%%%%%%%%%%%%%%%%%%%%%%%%
%%%%%%%%%%%%%%%%%%%%%%%%%%%%%%%%%%%%%%%%%%%%%%%%%%%%%%%%%%%%%%%%%%%%%%%%%%%%%%%%
%%%%%%%%%%%%%%%%%%%%%%%%%%%%%%%%%%%%%%%%%%%%%%%%%%%%%%%%%%%%%%%%%%%%%%%%%%%%%%%%

\section{Notes on individual clouds}
In this section, some remarks are made concerning the metallicity and distance
determinations for each listed cloud. For about half the clouds a map is
presented that shows the positions of the probes relative to the \HI. Further,
for 18 clouds the pattern of abundances is plotted in Fig.~\Fpattern, where it
is compared to the standard patterns for cool disk, warm disk and halo gas.
\par An estimated mass range is given for most of the clouds, using the observed
distance range. For the HVCs the mass is based on the integrated flux measured
by Hulsbosch \& Wakker (1988). The flux, $S$, is converted to an \HI\ mass
(M$_{HI}$ = 0.236 ($S$/Jy\,\kms) (D/1 kpc)$^2$ \Msun. In the case of IVCs, a
column density map is made using the LDS survey, and this is summed over an
irregular region outlining the cloud. Since each beam represents an area
A=7.25\tdex{38}\,cm$^2$ at a distance of 1\,kpc, the sum is multiplied by
A*${{\rm m}_H\over{\rm M}_\odot}$ to obtain the cloud mass at a distance of
1\,kpc. In both cases, a (hopefully typical) factor 1.2 is included to account
for ionized hydrogen, and a factor 1.39 is included to account for helium.

%%%%%%%%%%%%%%%%%%%%%%%%%%%%%%%%%%%%%%%%%%%%%%%%%%%%%%%%%%%%%%%%%%%%%%%%%%%%%%%%

\subsection{Complex A}
%VALUES: D+M A
This is the only HVC for which a distance bracket is known: 4.0\to9.9\,kpc,
which implies a mass of 0.3\to2\tdex{6}\,\Msun. Figure~\FgA\ shows the velocity
field and probe positions. The bracket is based on the stars AD\,UMa (van
Woerden et al.\ 1999a) and PG\,0859+593 (Wakker et al.\ 1996b). The
non-detection of \CaII\ in PG\,0832+675 (d=8.1\,kpc) is not significant, the
limit being only a factor 2 below the value measured toward Mrk\,106 and
AD\,UMa.
\par Welsh et al.\ (1996) did not detect \CaII\ and \NaI\ toward three stars
(HD\,77770, HD\,75755 and HD\,68164), and claimed to derive a lower limit of
1\,kpc from this result. However, this provides a clear example of the
difficulties of interpreting non-detections, as all their limits are similar to
the values observed in more distant targets. They are thus not significant in
terms of setting a lower distance limit.
%VALUES: OI - IZw18
\par The metallicity of complex~A has not yet been determined, due to a lack of
known UV-bright background probes. \OI-1302 and \SiII-1304 associated with
complex~A were seen in the spectrum of I\,Zw\,18 by Kunth et al.\ (1994). The
low flux of that galaxy makes the measured equivalent widths of
310$\pm$100\,m\AA\ and 110$\pm$75\,m\AA\ rather uncertain. The derived
abundances of course depend on the assumed linewidth and component structure.
The most likely case is that the absorption width is comparable to the \HI\
linewidth of 53\,\kms. Then the \OI\ abundance is 0.06$_{-0.03}^{+0.07}$ times
solar. If a two-component structure is assumed, with two equal lines with the
same intrinsic width, the derived \OI\ abundance is 0.09$_{-0.06}^{+0.3}$ solar
(for widths of 20\,\kms) to 0.05$_{-0.02}^{+0.05}$ solar (for widths of
30\,\kms). Thus, for reasonable assumptions, the implied \OI\ abundance of
complex~A is most likely to be on the order of 0.05\to0.1 solar, though the
uncertainty is large.
%lines wrange=1300,1310 ion=OI NH=20e18 vturb=53 EW=OI-1302,210 => 0.028
%lines wrange=1300,1310 ion=OI NH=20e18 vturb=53 EW=OI-1302,310 => 0.057
%lines wrange=1300,1310 ion=OI NH=20e18 vturb=53 EW=OI-1302,410 => 0.130
%lines wrange=1300,1310 ion=OI NH=10e18 vturb=20 EW=OI-1302,105 => 0.034
%lines wrange=1300,1310 ion=OI NH=10e18 vturb=20 EW=OI-1302,155 => 0.093
%lines wrange=1300,1310 ion=OI NH=10e18 vturb=20 EW=OI-1302,205 => 0.417
%lines wrange=1300,1310 ion=OI NH=10e18 vturb=30 EW=OI-1302,105 => 0.025
%lines wrange=1300,1310 ion=OI NH=10e18 vturb=30 EW=OI-1302,155 => 0.049
%lines wrange=1300,1310 ion=OI NH=10e18 vturb=30 EW=OI-1302,205 => 0.095
\par Combined with the \MgII\ and \CaII\ abundances, a halo-like depletion
pattern is suggested for complex~A (Fig.~\Fpattern a). It is clear, however,
that better measurements at higher angular and spectral resolution are needed to
confirm this.

%%%%%%%%%%%%%%%%%%%%%%%%%%%%%%%%%%%%%%%%%%%%%%%%%%%%%%%%%%%%%%%%%%%%%%%%%%%%%%%%

\subsection{Complex M}
Complex~M consists of several clouds. These have similar velocities but they are
not connected. Whether these clouds are spatially close or only close in
position on the sky is an open question, though the former seems most likely.
The different clouds in the complex are discussed separately below. Figure~\FgM\
shows the positions of the probe stars.
%VALUES: A+M - MI
\par Cloud MI has the most negative velocities ($<$$-$100\,\kms), and lies at
longitudes $l$$>$150\deg; it was also classified as IV2 by Kuntz \& Danly
(1996). The \CaII\ non-detections in Schwarz et al.\ (1995) turn out to be
not-significant, so no distance limits are known. Its mass is
4\tdex{3}\,(D/1\,kpc)$^2$\,\Msun.
\par Tufte et al.\ (1998) observed \Ha\ and [\SII]$\lambda$6716 emission
associated with cloud MI (see Wakker \& van Woerden 1991), in the direction of
the brightest \HI\ emission in cloud MI. So far, this is the only observation of
[\SII] emission in a HVC or IVC. The observed [\SII]/\Ha\ ratio is
0.64$\pm$0.14. This can be used to constrain the metallicity of cloud MI. To do
this three assumptions are needed: a) what is the geometry? b) what is the
temperature? and c) what is the fraction of S$^{+2}$? It is most likely that
toward the bright \HI\ core the cloud has a neutral core surrounded by a fully
ionized envelope. A temperature in the range 6000\to10000\,K is suggested by the
systematic study of \Ha\ and [\SII] emission at heights up to 1 kpc above the
Perseus arm (Haffner et al.\ 1999). They also find that N(S$^{+2}$)/N(S$^+$)
lies in the range 0.3\to0.7. With this range of parameters, the S$^+$ abundance
of cloud MI has a most likely value of $\sim$0.8 solar, but is only restricted
to lie in the range 0.4\to1.8 times solar. That is, a metallicity of $\sim$0.1
solar such as that found for complex~C (see Sect.~\Scloud.\SSC) is excluded, and
the most likely value is consistent with the idea that complex M is part of the
IV Arch, for which near solar metallicity has been found (see
Sect.~\Scloud.\SSIVa).
%VALUES: SN1998S - CaII MnII FeII   CaII - BD+49.2137 PG1213+456
%VALUES: M - IV4
%VALUES: distance BD+49.2137
\par Toward lower longitudes, the velocity of MI changes to $\sim$$-$80\,\kms.
This part was classified as IV4 by Kuntz \& Danly (1996) and is probably part of
the IV Arch (see Sect.~\Scloud.\SSIVa). Behind IV4 lies SN\,1998\,S, which was
observed by Bowen et al.\ (2000). The \CaII, \MnII\ and \FeII\ abundances (3.3,
44 and 1600 ppb) are more like those in warm disk gas (6.9, 42 and 1800 ppb)
than like in halo gas (69, 87 and 7800 ppb) (see also Fig.~\Fpattern e). The
same can be said for the abundance ratios (e.g.\ \CaII/\,\MnII=0.08 in IV4, 0.16
in warm gas and 0.79 in halo gas). Such abundances are consistent with the small
distance of 0.6\to0.8\,kpc ($z$=0.5\to0.7\,kpc) that can be derived by combining
the detection of \SiII\ toward BD+49\,2137 with the non-detection of \SiII\
toward HD\,106420 (Kuntz \& Danly 1996). Ryans et al.\ (1997a) did not detect
\CaII\ toward BD+49\,2137 nor toward PG\,1213+456 (at 2.9\,kpc), but the limits
of $<$2.1 and $<$50 ppb are not inconsistent with the value of 3.3$\pm$0.9 ppb
found toward SN\,1998\,S. Note also that the distance of BD+49\,2137 was revised
from 1.8\,kpc (Kuntz \& Danly 1996) to 0.8\,kpc by Ryans et al.\ (1997a). It
remains to be seen whether this distance bracket applies to the higher-velocity
part of cloud MI. The mass of IV4 is constrained to the range
1.5\to2.5\tdex{3}\,\Msun.
\par In the gap between clouds MII and MIII lie BD+38\,2182 and HD\,93521, which
are just 3\arcmin\ apart, \CII, \OI, \SiII\ and \CaII\ absorption at the
velocity of MIII are clearly seen toward BD+38\,2182, and clearly absent toward
HD\,93521 (Danly et al.\ 1993). Taking the distance to BD+38\,2182 from Ryans et
al.\ (1997a), this sets an upper limit on the distance of MII/MIII of 4.0\,kpc
and an upper limit on the mass of 6\tdex{4}\,\Msun. However, no strong lower
limit can be derived from HD\,93521, as a 12\arcmin\ beam \HI\ spectrum shows no
emission at the velocity of the absorption toward either star, down to a
5-$\sigma$ limit of 0.5\tdex{18}\,\cmm2\ (Ryans et al.\ 1997a). If H$^+$ were
present, no \OI\ absorption is expected, but \CII\ and \SiII\ should still be
seen. It remains unclear whether the high-velocity absorption toward HD\,93521
is absent because the cloud is behind the star or because there is no
high-velocity material in the sightline. The position in the sky of MII/MIII and
their velocity strongly suggest that these clouds are part of the IV Arch. A
lower distance limit of 1.9\,kpc would then be at odds with the other distance
brackets derived for the IV Arch (see Sect.~\Scloud.\SSIVa).
\par Ryans et al.\ (1997b) and Lehner et al.\ (1999a) reported \CaII\ absorption
at a velocity of $-$108\,\kms\ in the spectrum of HD\,83206. No \HI\ is seen at
this velocity down to a limit of 2\tdex{18}\,\cmm2. The nearest \HI\ cloud with
similar velocity in the HVC survey of Hulsbosch \& Wakker (1988) is 1\fdg5 away
(cloud WW63; \vlsr=$-$112\,\kms; $l$=166\deg, $b$=46\deg), although this faint
cloud can not be discerned in the Leiden-Dwingeloo Survey. Just north of this
position are the southernmost clouds of complex~M (specifically WW51 and WW39,
as well as some uncatalogued clouds, all with velocities $\sim$$-$120 to
$\sim$$-$100\,\kms). Quite possibly the weak \CaII\ absorption is associated
with these faint \HI\ clouds that line the edge of complex~M. The implied high
\CaII\ abundance is consistent with the abundance vs \HI\ column density
relation found by Wakker \& Mathis (2000).

%%%%%%%%%%%%%%%%%%%%%%%%%%%%%%%%%%%%%%%%%%%%%%%%%%%%%%%%%%%%%%%%%%%%%%%%%%%%%%%%

\subsection{Complex~C}
%VALUES: note number for C-south=30
Since complex~C covers a large area of the sky (1600 square degrees), and since
it contains a number of well-defined cores, it was split into 5 parts in
Table~\Tmain. CI-A,B,C and CIII-A,B,C were defined by Giovanelli et al.\ (1973).
``C-south'' refers to the lower-latitude part of C at longitudes $>$80\deg\ (for
a precise specification see Note \NoteCsouth). The core at $l$=68\deg,
$b$=38\deg\ (``CeI'') was previously named ``C-extension'' by Giovanelli et al.\
(1973). Here, cores CeI through CeV are defined by analogy. Complex~D was
defined by Wakker \& van Woerden (1991); core ``CD'' within complex~C is defined
here as the core at $l$=90\deg, $b$=34\deg, \vlsr$<$$-$150\,\kms, which is close
to complex~D and has similarly high-negative velocities. Finally, core C/K is
defined as the core at $l$=75\deg, $b$=37\deg, \vlsr$\sim$$-$90\,\kms, which
lies close in angle and velocity to IVC complex~K, and may be part of that
complex, rather than a core of complex~C. A detailed analysis of the spectra in
this region is called for.
%VALUES: Mrk290 PG1351+640
\par The distance to HVC complex~C is not well known. \CaII\ absorption was
measured toward Mrk\,290 and PG\,1351+640, so that good abundance comparisons
can be made for distance-determination purposes. However, toward both of these
the component structure is slightly complicated. A best-guess fit yields two
components toward Mrk\,290 with \CaII\ abundances of 12$\pm$3 and 22$\pm$1 ppb,
while toward PG\,1351+640 abundances of 18$\pm$2 and 110$\pm$50 ppb are found.
The last value is for a low-column density component with uncertain N(\HI), and
thus the fact that it differs from the other three is probably not significant.
%VALUES: exp/obs(CaII) BS16034-0114: 2.3/22
%VALUES: M - C
\par Figure~\FgC\ shows the positions of the probes. Most observations of
stellar probes lead to a non-detection, but do not give significant distance
information, usually because N(\HI) is small. Most of the few significant
non-detections are toward rather nearby stars, giving a strong lower distance
limit of 1.2\,kpc. Limits were set on $A$(\CaII) for seven distant ($>$3\,kpc)
stars. However, only one of these is significant (BS\,16034-0114 at 6.1\,kpc),
although a better analysis of these spectra is still needed. For BS\,16034-0114
the ratio (expected)/(observed) is $\sim$10, which thus yields a (weak) lower
limit of 6.1\,kpc to the distance of core CIA. The strong lower distance limit
gives a mass for complex~C of $>$\dex{5}\,\Msun, but the probable lower limit of
6.1\,kpc yields M$>$3\tdex{6}\,\Msun.
\par Complex~C is the only cloud for which several different metallicity
measurements have been made, although in all cases there are some special
considerations or special problems to be solved. \SII-1250 and/or 1253 and/or
1259 have been measured toward 3 AGNs projected onto complex~C: Mrk\,290 (Wakker
et al.\ 1999b) and Mrk\,817, Mrk\,279 (Gibson et al.\ 2001). \NI\ was measured
toward Mrk\,876 (Gibson et al.\ 2001, Murphy et al.\ 2000) and PG\,1259+593
(Richter et al.\ 2001b), while \OI/\HI\ has been measured toward PG\,1259+593
(Richter et al.\ 2001b).
%VALUES: Mrk290 - S/H   Mrk817/Mrk279 - SII/HI
\par The best determination is the one using Mrk\,290, where N(\HI) is measured
at 2\arcmin\ resolution and \Ha\ was detected. The result is
N(S)/N(H)=0.09$\pm$0.02 times solar, with a small dependence on the assumed
distance and geometry. The measurement toward Mrk\,817 yields
N(\SII)/N(\HI)=0.33$\pm$0.02 solar, while that toward Mrk\,279 gives
0.43$\pm$0.10 solar. In both cases the (unknown) ionization correction may be
high, as N(\HI) in these directions is much lower than toward Mrk\,290, whereas
N(H$^+$) is expected to be more or less constant if the hydrogen is
photoionized. High-resolution \HI\ maps are also imperative to understand the
difference between the Mrk\,290 and Mrk\,817/279 sightlines.
%VALUES: PG1259+593 - NI OI
\par One other sightline with high N(\HI) has been studied: PG\,1259+593
(Richter et al.\ 2001b). Here \OI/\HI=0.11$^{+0.13}_{-0.08}$ solar. Richter et
al.\ (2001b) also find that \NI/HI=$<$0.046 solar, as well as \SiII/\OI$\sim$1,
but \FeII/\OI$\sim$0.4. No H$_2$ was detected (fraction $<$\dex{-5}), but this
is not unexpected considering the relatively low total hydrogen column density
and the relatively small amount of dust expected to go with the low metallicity.
If there is no or little dust, the \NI, \OI, \SiII\ and \FeII\ abundances are
consistent with an enhancement in the $\alpha$ elements (Si/O$>$1), combined
with a relatively low amount in the iron peak elements (Fe/O$<$1) and no
secondary nitrogen (N/O$<$0.5). This is consistent with the idea that that the
heavy elements in complex~C were created in type II supernovae, and that there
has been no subsequent star formation.
%VALUES: Mrk876 - NI, NII, PII, ArI FeII
\par The sightline toward Mrk\,876 was analyzed by Murphy et al.\ (2000). It
shows unusual abundances. Nominally undepleted elements have low abundances
relative to \HI\ ($A$(\NI)$\sim$0.08 solar, $A$(\PII)$<$0.8, $A$(\ArI)$<$0.1
solar), consistent with the values found using the other probes. However, the
usually depleted \FeII\ has a high abundance ($A$(\FeII)=0.5). The probable
interpretation in this case is that in the region around the sightline toward
Mrk\,876 the gas is partially photo-ionized throughout by a soft radiation
field, so that \HI\ is a $\sim$25\% contaminant. Then Ar and N can become
overionized relative to H, unlike Fe, which remains in the form of \FeII\ (Sofia
\& Jenkins 1998).
\par Nevertheless, the maximum ionization correction consistent with the
absorption line and \Ha\ data in the direction of Mrk\,876 seems to be on the
order of a factor 3\to5. Since S/H=0.1, this implies Fe/S$\sim$1\to2. Such a
ratio is very unlike the typical ratio of 0.2 seen in halo gas (Sembach \&
Savage 1996) or in other HVCs and IVCs (this paper). The implication is that
complex~C appears to have low dust content. This is further supported by the
\SiII/\OI\ and \FeII/\OI\ ratio toward PG\,1259+593 (Richter et al.\ 2001b).
\par From the results described above, it has become clear that in order to
derive reliable abundances and abundance ratios, it will be necessary to observe
\Ha\ emission for every probe, to pay close attention to ionization corrections
and unusual circumstances, and to combine the absorption and emission data with
modeling.

%%%%%%%%%%%%%%%%%%%%%%%%%%%%%%%%%%%%%%%%%%%%%%%%%%%%%%%%%%%%%%%%%%%%%%%%%%%%%%%%

\subsection{Complex G}
%VALUES: D+M - G
This HVC has velocities that deviate only slightly from those allowed by
differential galactic rotation, especially at the lowest latitudes. A better
definition is needed. The non-detections of several elements with strong lines
in the spectrum of 4\,Lac are not commented upon by Bates et al.\ (1990), but
they do set a (not unsurprising) lower limit of 1.3\,kpc, and thus a mass limit
$>$5\tdex4\,\Msun.

%%%%%%%%%%%%%%%%%%%%%%%%%%%%%%%%%%%%%%%%%%%%%%%%%%%%%%%%%%%%%%%%%%%%%%%%%%%%%%%%

\subsection{Complex H}
%VALUES: D+M - H
This HVC complex lies in the Galactic plane, which allows the use of luminous O
and B stars as probes. So far, only a lower distance limit has been set, using
the non-detection of \MgII, \CII\ and \OI\ in IUE spectra (Wakker et al.\ 1998).
Centuri\'on et al.\ (1994) did not see \CaII\ absorption in seven stars
projected onto complex~H, but only one of the non-detections can be considered
significant (assuming complex~H has a \CaII\ abundance similar to that seen in
other HVCs).
\par Figure~\FgH\ shows the positions of the probes. The HVC can be subdivided
into several parts, although these are probably all spatially connected, as
there are no clear boundaries. The distance limit for the brightest, central,
core is d$>$5\,kpc, which implies a mass for the whole complex of
$>$\dex{6}\,\Msun.

%%%%%%%%%%%%%%%%%%%%%%%%%%%%%%%%%%%%%%%%%%%%%%%%%%%%%%%%%%%%%%%%%%%%%%%%%%%%%%%%

\subsection{Anti-Center Shell}
The Anti-Center Shell (Fig.~\FgACS) was first delineated by Heiles (1984).
Tamanaha (1997) made a detailed study of the region and argued that the Shell as
such does not exist, but is an artifact of the data display being based on
channels at constant \vlsr. He rather sees it as the point of impact of a stream
of HVCs falling toward the plane \dash\ the Anti-Center HVCs
(Sects.~\Scloud.\SSACot, \Scloud.\SSCHS) being the rest of that stream.
%VALUES: HDE248894 - dist CaII:3.5/22=0.16
\par Kulkarni \& Mathieu (1986) failed to find \CaII\ absorption toward 6 OB
stars in the direction of the Shell. They put the most distant star
(HDE\,248894) at 2.7\,kpc (recalculated here as 3.0\,kpc). The abundance of
\CaII\ has not yet been measured, so although $A$(\CaII) is $<$15\% of the
``normal'' value, the non-detection is not considered significant.
%VALUES: HDE256725 CaII 4.3/22=0.19
\par Kulkarni \& Mathieu (1986) also observed HDE\,256725, classified as an
O-star at d$<$2.5\,kpc. The non-detection of \CaII\ is not significant (a factor
5 below the expected value, rather than a factor $>$20). According to SIMBAD,
this is a B-star, but Garmany et al.\ (1987) classified it as O5V and gave a
distance of 8.0\,kpc. If confirmed and reobserved with higher S/N, this star
might set an upper/lower distance limit of 8.0\,kpc.

%%%%%%%%%%%%%%%%%%%%%%%%%%%%%%%%%%%%%%%%%%%%%%%%%%%%%%%%%%%%%%%%%%%%%%%%%%%%%%%%

\subsection{Cloud AC0}
Cloud AC0 was defined by Tamanaha (1997), by analogy with the chain of cores
ACIII, ACII, ACI (see Fig.~\FgACc). It is embedded in the Anti-Center Shell, but
stands out in velocity space. Tamanaha (1996) failed to find \CaII\ and \NaI\
absorption toward 4 stars. None of the \NaI\ non-detections are significant,
however (the best being only a factor 9 below the average expected value), but
the \CaII\ non-detection sets a weak lower limit of just 0.3\,kpc. Note that
Tamanaha (1996) includes LS V+30 31 as a probe of AC0, but the Effelsberg \HI\
spectrum shows that this star lies just off the core.

%%%%%%%%%%%%%%%%%%%%%%%%%%%%%%%%%%%%%%%%%%%%%%%%%%%%%%%%%%%%%%%%%%%%%%%%%%%%%%%%

\subsection{Clouds ACI and ACII}
Two studies of \NaI\ and \CaII\ absorption were made for these two clouds, using
bright (V$<$10) B and A stars (Songaila et al.\ 1988, Tamanaha 1996). See
Fig.~\FgACc\ for the positions of the probes. Again, none of the \NaI\
non-detections contains distance information, whereas the \CaII\ non-detections
toward ACI are mostly significant, setting a lower limit of 0.4\,kpc to the
distance. This is a not very interesting limit, as the expected distance is on
the order of several to tens of kpc, if indeed the AC clouds are infalling
intergalactic clouds, as proposed by Mirabel (1982) and Tamanaha (1995, 1997).
\par The same set of stars probes \HI\ components assigned to ACII, which partly
overlaps with ACI. However, since the column densities associated with ACII are
lower, most non-detections are not considered significant.

%%%%%%%%%%%%%%%%%%%%%%%%%%%%%%%%%%%%%%%%%%%%%%%%%%%%%%%%%%%%%%%%%%%%%%%%%%%%%%%%

\subsection{Cohen Stream and HVC\,168$-$43+280 (cloud WW507)}
Two studies of absorption were made for these two clouds, using bright (V$<$10)
B and A stars (Kemp et al.\ 1994, Tamanaha 1996, see Fig.~\FgACc). None of the
\NaI\ non-detections and only one of the \CaII\ non-detections are significant.
However, from \MgII\ non-detections a lower distance limit of 0.3\,kpc is set.
Weiner et al.\ (2001) detect \Ha\ emission from three directions in cloud WW507,
with intensities of 50\to250\,mR (1 Rayleigh is \dex/4$\pi$
photons\,\cmm2\,s$^{-1}$\,sr$^{-1}$). Applying a model for the distribution of
ionizing radiation (Bland-Hawthorn \& Maloney 1999) then suggests distances on
the order of 10\to20\,kpc.
%VALUES: Mrk595 - NaI
\par Kemp \& Bates (1998) searched for, but did not detect \NaI\ absorption
toward the Seyfert galaxy Mrk\,595. They calculated an expected value for
EW(\NaI) of $\sim$30\,m\AA\ (1.3\tdex{11}\,\cmm2), assuming that the \NaI\
abundance is similar to that seen in low-velocity gas. Since their detection
limit is 21\,m\AA, and considering the large observed variations in $A$(\NaI)
(see Sect.~\Sion), this non-detection contains no information about the
conditions in the Anti-Center clouds.

%%%%%%%%%%%%%%%%%%%%%%%%%%%%%%%%%%%%%%%%%%%%%%%%%%%%%%%%%%%%%%%%%%%%%%%%%%%%%%%%

\subsection{Complex GCN}
These widely scattered, faint clouds have high negative velocities.
High-ionization \CIV\ and \SiIV\ absorption was seen in the spectra of Mrk\,509
and PKS\,2155$-$304 by Sembach et al.\ (1995, 1999), as was \OVI\ (Sembach et
al.\ 2001). Sembach et al.\ (1999) proposed that a) the gas is photo-ionized by
the extra-galactic background radiation and b) the thermal pressure in the
sightlines is $\sim$1\to5\,K\,\cmm3, c) the distance of the gas is 10\to100\,kpc
(depending on the metallicity, which was assumed to lie between 0.1 and 1 times
solar). If it is assumed that the \HI\ clouds in complex~GCN are the denser and
cooler tips of the iceberg of a large, coherent, mostly ionized cloud, then the
implied cloud size is 8\to80\,kpc. These parameters are consistent with this
(group of) clouds being a low-metallicity, tenuous, extra-galactic cloud.
\par If the metallicity is eventually found to be $\sim$0.1 solar, complex~GCN
may represent an early stage in the process proposed by Oort (1970), in which
hot gas at the outskirts of the sphere of influence of the Milky Way condenses
and then accretes. Complex~C may be a late stage.

%%%%%%%%%%%%%%%%%%%%%%%%%%%%%%%%%%%%%%%%%%%%%%%%%%%%%%%%%%%%%%%%%%%%%%%%%%%%%%%%

\subsection{Complex GCP}
Although there are many bright stars as well as distant RR\,Lyrae stars
projected on this cloud, the only published data are from the first-ever paper
concerning HVC absorption lines (Prata \& Wallerstein 1967). The non-detection
toward HD\,187350 sets a lower distance limit of 0.3\,kpc (and thus a lower mass
limit of \dex3\,\Msun). Bland-Hawthorn et al.\ (1998) detected \Ha\ emission
associated with complex~GCP, and argued that it is related to the Sagittarius
dwarf, at a distance of $\sim$25\,kpc. However, if this cloud is related to
lower velocity gas that seems connected to it, its distance can be no more than
a few kpc (see Sect.~\Scloud.\SSgp\ for further discussion).

%%%%%%%%%%%%%%%%%%%%%%%%%%%%%%%%%%%%%%%%%%%%%%%%%%%%%%%%%%%%%%%%%%%%%%%%%%%%%%%%

\subsection{Outer Arm}
The Outer Arm was first recognized as a separate, distant spiral arm by Habing
(1966). It was further analyzed by Kepner (1970) and Haud (1992). The velocities
are only 20\to30\,kms\ higher than expected from a flat rotation curve for gas
at galactocentric radii of 15\to25\,kpc. Toward $l$$\sim$90\deg\ a large
kinematical leverage is provided, allowing one to study the outer Galaxy. In the
direction to H\,1821+643, high-ionization \CIV\ absorption was found by Savage
et al.\ (1995), while \OVI\ was reported by Oegerle et al.\ (2000) and Sembach
et al.\ (2000).

%%%%%%%%%%%%%%%%%%%%%%%%%%%%%%%%%%%%%%%%%%%%%%%%%%%%%%%%%%%%%%%%%%%%%%%%%%%%%%%%

\subsection{Cloud R}
This feature was defined by Kepner (1970), and may be associated with the Outer
Arm, or it may be a separate, distant HVC.

%%%%%%%%%%%%%%%%%%%%%%%%%%%%%%%%%%%%%%%%%%%%%%%%%%%%%%%%%%%%%%%%%%%%%%%%%%%%%%%%

\subsection{Magellanic Stream}
The Magellanic Stream is most likely a tidal tail torn out of the SMC during the
tidal interaction between the Milky Way, LMC and SMC one orbit (2\,Gyr) ago
(Gardiner \& Noguchi 1996). Some of the gas has been decelerated, forming the
well-known Stream, and some of it has been accelerated ahead of the SMC system.
Most of the latter fell on the LMC, but some has gotten past, forming a
scattered leading arm. Lu et al.\ (1998) identified the
extreme-positive-velocity clouds of Wakker \& van Woerden (1991) as part of this
leading arm. Fig.~\FgMS\ shows the Stream and these clouds in a projection that
has galactic longitude 270\deg\ along the equator. Overlaid is the model of
Gardiner \& Noguchi (1996).
\par This projection was chosen as the simplest way to make the Magellanic
Stream lie as near the equator as possible. This is basically done by turning
the great circle $l$=90\deg\ and $l$=270\deg\ into the equator. Formally, the
pole lies at galactic longitude $l$=180\deg, galactic latitude $b$=0\deg, and it
is further rotated by 33.42779 degrees so that the current position of the LMC
has new longitude L=0.
% tab mode=catalogue catpar=cld cmparrange=l,0,180,b,b0,b1
%                        HW      MP
% l=0,180    b0,b1=-48,-32 = MSVI  2.6E6  1.3E6
%            b0,b1=-60,-48 = MSV   8.4E6  3.9E6
%            b0,b1=-72,-60 = MSIV  1.0E7  6.7E6
%            b0,b1=-85,-72 = MSIII 1.3E7  1.4E7
% l=180,270  b0,b1=-90,-75   MSII  3.5E6  3.2E7
%            b0,b1=-75,-60 = MSI   1.8E7  4.0E7
%            b0,b1=-60,-45 = Head  6.9E7  1.1E8
% l=0,180                          3.3E7
% l=180,360  b0,b1=-90,-45         9.0E7
%                            Total 123E6  206E6
\par The distances to the trailing and leading parts of the stream are not well
known. If 50\,kpc is assumed, the mass of the trailing part of the stream is
about 1.5\dex8\,\Msun\ (not counting the gas in the Magellanic Bridge, at
galactic latitudes $>$$-$45\deg). For the leading part, all clouds in complex~EP
of Wakker \& van Woerden (1991) add up to about 5\tdex7\,\Msun\ at an assumed
distance of 50\,kpc. The model predicts distances ranging from 50 to 100\,kpc
for this gas.
\par Also added in Fig.~\FgMS\ are some small positive-velocity clouds in the
northern Galactic hemisphere that were not considered part of HVC complexes WA
and WB by Wakker \& van Woerden (1991). These appear to line up ahead of the
curve defined by the SMC orbit and the EP clouds near $l$=290, $b$=+20 (60,$-$20
in Fig.~\FgMS), although they are not in the orbital plane of the Magellanic
Clouds. For further discussion of these small clouds, see Sect.~\Scloud.\SSnoHI.
%VALUES: SII - Fairall9 NGC3783
\par The tidal model predicts a metallicity similar to that of the SMC. For
sulphur that means S/H$\sim$0.2\to0.3 solar (Russel \& Dopita 1992). Indeed,
such values are found toward Fairall\,9 in the tail (0.33$\pm$0.05 solar, Gibson
et al.\ 2000) and NGC\,3783 in the leading arm (0.25$\pm$0.08 solar, Lu et al.\
1998). For NGC\,3783 N(\HI) was measured at 1\arcmin\ resolution by combining
ATCA and Parkes data. No high-resolution \HI\ map has yet been made for the
field around Fairall\,9; a 25\% correction is easily possible. In both
directions a measurement of \Ha\ will be needed to account for H$^+$, although
N(\HI) is high enough and the radiation field is expected to be low enough that
only a small correction is expected.
\par Although the distance to different parts of the Stream is expected to be
30\to80\,kpc, no formal limit can be set, as only non-significant non-detections
have been found toward a few nearby stars. Considering the large sky area
covered by the Magellanic Stream, it is possible that more non-detections lie
hidden in the literature, in papers not aimed at studying the ISM, and therefore
not noted.
%VALUES: MgII -avg
\par Several AGNs behind the Magellanic Stream were observed with the FOS.
Savage et al.\ (2000a) report equivalent widths for detections of \MgII; Jannuzi
et al.\ (1998) give the wavelength offsets, from which velocities are
calculated. Although the velocity resolution is only 220\,\kms, this is
sufficient to separate absorption by the Magellanic Stream from that by disk
gas. As the \HI\ profiles are broad (30\to50\,\kms; see Paper~II), the estimated
\MgII\ optical depths are 0.2\to3, allowing to convert the equivalent widths to
column densities. Figure \Fmultiple a shows the resulting correlation between
N(\HI) and $A$(\MgII). For N(\HI)$>$2\tdex{18}\,\cmm2, $A$(\MgII) is nearly
constant (average 3600 ppb, or 0.10 times solar). At lower N(\HI) much higher
values and lower limits are found. The most likely explanation for this change
is that the hydrogen becomes mostly ionized, as the ionization potential of
\MgII\ is higher than that of hydrogen, while its depletion seems to be
independent of N(\HI) (Wakker \& Mathis 2000). The average abundance of 0.10
times solar implies an \MgII\ depletion of 0.3\to0.4 for the Magellanic Stream,
i.e., similar to the value in halo gas.
%VALUES: FeII  PKS0637-12 NGC3783
\par Abundances in the Stream can also be derived for \CaII, \FeII\ and \SiII,
see Fig.~\Fpattern c. The table in Jannuzi et al.\ (1998) gives \FeII\ toward
PKS\,0637$-$75 near the SMC, which yields $A$(\FeII)=0.062$\pm$0.025 times
solar. In the leading arm, NGC\,3783 gives $A$(\FeII)=0.033$\pm$0.006 times
solar. The implied ratios of Fe/S=0.18$\pm$0.06 and 0.13$\pm$0.05 times solar
are similar to the typical ratio in halo-like gas (0.23). The \MgII\ and \FeII\
abundances both imply that there is dust in the Magellanic Stream, and that it
has properties similar to the dust in the Galactic Halo.

%%%%%%%%%%%%%%%%%%%%%%%%%%%%%%%%%%%%%%%%%%%%%%%%%%%%%%%%%%%%%%%%%%%%%%%%%%%%%%%%

\subsection{Population EP}
Several of the extreme-positive velocity clouds have been studied. Cloud WW187
was discussed above, in the context of it being part of the leading arm of the
Magellanic Stream.
\par Sahu \& Blades (1997) and Sahu (1998) observed NGC\,1705 with a velocity
resolution of 140\,\kms\ and found absorption at \vlsr$\sim$+260\,\kms\ in
several \SiII\ lines. This was interpreted as being associated with HVC WW487,
which has \vlsr=+240\,\kms\ and is 2 degrees away from NGC\,1705; it is probably
a shred of the Magellanic Stream (see Fig.~\FgMS). A map of this area was made
by Putman (priv.\ comm.) using {\it HIPASS} data, after reprocessing in order to
extract extended structure. This clearly shows WW487, but no other HVCs in the
area. Faint (1.7\tdex{18}\,\cmm2) high-velocity \HI\ is detected directly in the
sightline toward NGC\,1705 (Paper~II). \SiII-1190, \SiII-1304 and \SiII-1526
absorption are clearly seen, but the listed equivalent widths are inconsistent.
The derived abundances are 0.48$\pm$0.14, 2.9$\pm$1.5 and 2.9$\pm$1.5,
respectively. This poses an unsolved problem. However, the result does suggest
that most of the gas in this direction is in the form of H$^+$.
%lines wrange=1300,1600 ion=SiII NH=2e18 vturb=24 EW=SiII-1190,88  => 0.504
%lines wrange=1300,1600 ion=SiII NH=2e18 vturb=24 EW=SiII-1193,88  => 0.252
%lines wrange=1300,1600 ion=SiII NH=2e18 vturb=24 EW=SiII-1304,157 => 4.086
%lines wrange=1300,1600 ion=SiII NH=2e18 vturb=24 EW=SiII-1526,239 => 7.117
\par Sahu (1998) also lists an equivalent width for \NI-1199, but at 140\,\kms\
resolution this absorption is a hopeless blend of $\lambda$1199.55 and
$\lambda$1200.22 absorption due to the Galaxy, the HVC and NGC\,1705.
\par The only cloud in population EP for which a formal distance limit exists is
cloud WW187. HD\,101274 is a few arcmin from NGC\,3783 and sets a lower limit of
0.4\,kpc. Wakker \& van Woerden (1997) listed a lower limit of 6.2\,kpc to the
distance of the extreme-positive velocity cloud WW211, based on the
non-detection of \SiII\ toward HD\,86248 by Danly et al.\ (1993). The revised
distance of this star is 7.6$\pm$3.0\,kpc. Since WW211 is part of the leading
arm of the Magellanic Stream, the expected \SiII\ abundance is a factor 4 below
the standard halo value of 19000 ppb, i.e.\ $\sim$5000 ppb. The observed limit
of 3300 ppb may therefore not be significant, and the non-detection does not set
a lower limit to the distance of cloud WW211 after all.

%%%%%%%%%%%%%%%%%%%%%%%%%%%%%%%%%%%%%%%%%%%%%%%%%%%%%%%%%%%%%%%%%%%%%%%%%%%%%%%%

\subsection{Very-high-velocity clouds}
%VALUES: Mrk205 - MgII
The sightline to Mrk\,205 passes through HVC WW84, which has \vlsr=$-$202\,\kms.
Bowen et al.\ (1991a, 1991b) detected weak \MgII\ absorption from this cloud.
The \HI\ column density has not yet been properly measured toward this object.
In the 9\farcm1 Effelsberg beam it appears to be 15\tdex{18}\,\cmm2. However, as
the high-resolution (1\arcmin) WSRT map presented by Braun \& Burton (2000)
shows, such a beam picks up some of the very bright small-scale structure that
lies nearby. A proper correction requires to combine the WSRT map with a grid of
single-dish data to produce a fully-sampled interferometer map. This has not yet
been done. In the meantime, the best guess for N(\HI) is
8$\pm$5\tdex{18}\,\cmm2. This yields an \MgII\ abundance of 0.020$\pm$0.014
times solar.
\par This value is much lower than usual in halo gas (0.4 solar is typical for
gas with intrinsically solar abundance), and even lower than is found in cool
disk gas with large amounts of dust (0.03 times solar). A value near 0.02 solar
can also be found in a low-metallicity cloud. If there is some dust, the
intrinsic metallicity could be $\sim$0.05\to0.1 solar. Confirmation using other
absorption lines is needed, as well as an improved value for the \HI\ column
density and an assessment of the ionization correction.
\par Combes \& Charmandaris (2000) report a possible detection of HCO$^+$ at
$-$198\,\kms\ in the spectrum of the radio continuum source 1923+210 in a
direction lying between the VHVCs WW274 (\vlsr=$-$200\,\kms) and WW283
(\vlsr=$-$198\,\kms). If the ratio of HCO$^+$ to H$_2$ were similar to that
found in low-velocity gas (6\tdex{-9}, Lucas \& Liszt 1996), this would imply
N(H$_2$)=7\tdex{19}\,\cmm2, whereas the observed limit to N(\HI) is
$<$2\tdex{18}\,\cmm2. Clearly, this cloud is unusual in that it either has large
molecular content, or relatively high HCO$^+$ abundance. These VHVCs may be
outliers of the GCN complex, outliers of the Magellanic Stream, or genuine
isolated VHVCs; their position in the sky does not allow an unambiguous
identification.

%%%%%%%%%%%%%%%%%%%%%%%%%%%%%%%%%%%%%%%%%%%%%%%%%%%%%%%%%%%%%%%%%%%%%%%%%%%%%%%%

\subsection{Complex L}
Albert et al.\ (1993) reported \CaII\ absorption components at $-$98 and
$-$127\,\kms\ in the spectrum of HD\,135485, thought to be a B5IIp star at a
distance of 2.5\,kpc. Van Woerden (1993) suggested that this absorption is due
to complex~L, but Danly et al.\ (1995) showed that interstellar \CII, \OI\ and
\SiII\ absorption are absent at these velocities, implying that the \CaII\
components are circumstellar. They also reclassified HD\,135485 and revised the
distance down to 0.8\,kpc. According to Hipparcos, the parallax is 5.52$\pm$1.13
mas, which gives an even lower distance of 0.18$\pm$0.04\,kpc. Thus, in spite of
earlier suggestions, all that is known for complex~L through HD\,135485 is a
lower limit to its distance of 0.2\,kpc.
\par Weiner et al.\ (2001) find that complex~L shines brightly in \Ha\ emission,
with detections of 0.3\to1.7\,Rayleigh (=\dex6/4$\pi$
photons\,\cmm2\,s$^{-1}$\,sr$^{-1}$). [\NII] $\lambda$6583 emission is similarly
strong, with [\NII]/\,\Ha = 1.1. Applying a model for the distribution of
ionizing radiation (Bland-Hawthorn \& Maloney 1999) then suggests that complex~L
lies at a distance of 8\to15\,kpc, and 2\to10\,kpc above the galactic plane,
either near the Galactic Center or at the other side of the Center. Of course,
because of individual features such as spiral arms, the radiation model is least
accurate close to the Galactic plane and close to the Galactic Center.
Nevertheless, the bright \Ha\ emission associated with complex~L shows that this
HVC must lie in the lower Galactic Halo.

%%%%%%%%%%%%%%%%%%%%%%%%%%%%%%%%%%%%%%%%%%%%%%%%%%%%%%%%%%%%%%%%%%%%%%%%%%%%%%%%

\subsection{Complex~WB}
Only one resolved detection is known for a previously catalogued HVC complex
with positive velocities: \CaII\ absorption associated with cloud WW225 is seen
in the spectrum of PKS\,0837$-$12 (Robertson et al.\ 1991). This gives the
highest measured \CaII\ abundance for any HVC (280 ppb). The published value of
160 ppb was based on N(\HI)=14\tdex{18}\,\cmm2\ as measured with the Parkes
telescope (15\arcmin\ beam) whereas with the 9\farcm1 Effelsberg beam
N(\HI)=7.9\tdex{18}\,\cmm2. Since N(\HI) is low, and no other measurements exist
for cloud WW225, it is unclear whether the high $A$(\CaII) is due to hydrogen
ionization, to anomalously low calcium depletion or to low \HI\ column density
(the latter effect is discussed by Wakker \& Mathis 2000).

%%%%%%%%%%%%%%%%%%%%%%%%%%%%%%%%%%%%%%%%%%%%%%%%%%%%%%%%%%%%%%%%%%%%%%%%%%%%%%%%

\subsection{Complex~WE}
In the sightline to HD\,156359 Sembach et al.\ (1991) found \CII, \MgII, \SiII\
and \FeII\ absorption at a velocity of +110\,\kms, while Sembach \& Savage
(1995) reported \NV\ at +128\,\kms. Differential galactic rotation predicts
velocities between 0 and $-$100\,\kms\ in this direction.
\par At the position of HD\,156359 no \HI\ is found in the list of Morras et
al.\ (2000), but many small, faint HVCs with similar velocities exist nearby
(Fig.~\FgWE). These include some clouds previously catalogued on a
2\deg$\times$2\deg\ grid (WW356, WW364, WW373, WW412). With the better view
provided by the Morras et al.\ (2000) list, these clouds were swept together
into ``complex~WE'', by analogy with the positive-velocity complexes WA through
WD defined by Wakker \& van Woerden (1991).
\par It is now clear that HD\,156359 lies less than 1\deg\ away from one of the
brighter cores of this complex, which has a velocity of +110\,\kms. The star
thus sets an upper limit to the distance of the HVC of 12.8\,kpc
($z$$<$3.2\,kpc). Most likely, the star samples the faint outer envelope of this
cloud. Assuming solar abundance and halo-like depletion, the implied value of
N(H) is about \dex{17}\,\cmm2. The mass of the collection of clouds forming
complex~WE is limited to be $<$2.5\tdex5\,\Msun.

%%%%%%%%%%%%%%%%%%%%%%%%%%%%%%%%%%%%%%%%%%%%%%%%%%%%%%%%%%%%%%%%%%%%%%%%%%%%%%%%

\subsection{Small positive-velocity clouds}
The sightline to the star BD+10\,2179 crosses a small (1 square degree) cloud,
WW29. Danly et al.\ (1993) give a distance for this star of 4.0\,kpc, with
spectral type ``Bp''. SIMBAD does not provide a better type, so this distance is
rather uncertain. The non-detection of \SiII\ is only a factor three below the
expected value, so it may not be significant.
\par The presence of heavy elements in some of the small positive-velocity
clouds is suggested by wide \MgII\ lines toward 4C\,06.41 and PKS\,1136$-$13
(Savage et al.\ 2000a).

%%%%%%%%%%%%%%%%%%%%%%%%%%%%%%%%%%%%%%%%%%%%%%%%%%%%%%%%%%%%%%%%%%%%%%%%%%%%%%%%

\subsection{HVC\,100$-$7+110}
In HVC\,100$-$7+110 (probed by 4\,Lac) several elements were detected, giving an
abundance pattern. Its velocity of +106\,\kms\ is opposite to that expected from
differential galactic rotation at a longitude of 100\deg. The cloud appears to
be small, and is within 200\,pc of the Galactic plane. It is not seen in the
Leiden-Dwingeloo Survey, but it is clearly detected in spectra taken with the
Jodrell Bank telescope (12\arcmin\ beam) and in a high-angular resolution map
made at Westerbork (Stoppelenburg et al.\ 1998). The upper limit on its mass is
1\,\Msun, making it a very unusual low-z solar-metallicity blob moving away from
the plane.
\par Bates et al.\ (1990) measured equivalent widths for several elements and
converted these to column densities. For this step they quoted a $b$-value of
6.5$\pm$0.5\,\kms, as derived from the \FeII\ lines. This $b$-value is then used
to convert the equivalent widths of the strongly saturated \OI\, \MgII\ and
\AlII\ lines to a column density. For a range of 1\,\kms\ in $b$, and using the
listed equivalent width errors, this implies optical depths $>$10 for the \OI\
and \MgII\ lines and $>$3 for \AlII. Thus, all one can say is that $A$(\OI) lies
in the range 0.2\to2.2 times solar, $A$(\MgII) in the range 0.6\to8.4 times
solar, and $A$(\AlII) in the range 0.3\to0.8 times solar, with a most likely
value for \AlII\ of 0.42 solar (1300 ppb). The \FeII\ abundance is slightly more
reliable and is 0.25$\pm$0.06 times solar (8000 ppb). The \AlII\ and \FeII\
abundances are comparable to the expected values for a cloud with intrinsically
solar abundance and halo-like depletion (1600 and 7800 ppb, respectively).

%%%%%%%%%%%%%%%%%%%%%%%%%%%%%%%%%%%%%%%%%%%%%%%%%%%%%%%%%%%%%%%%%%%%%%%%%%%%%%%%

\subsection{HVCs with weak \HI}
In eight extra-galactic probes absorption is detected at high velocities but
without corresponding \HI\ emission, even though the upper limits on N(\HI) can
be quite good (3\tdex{17}\,\cmm2\ in the case of SN\,1991\,T). Only toward
SN\,1986\,G is such high-velocity absorption accompanied by a weak \HI\
component. Below, possible explanations are provided for all but two or three of
these absorption components.
\par An absorption that remains unexplained is the +250\,\kms\ \NaI\ component
toward SN\,1994\,I (Ho \& Filippenko 1995). No other positive-velocity IVCs are
known in the neighbourhood ($l$=104\deg, $b$=68\deg). Also mysterious are the
components at +125, +140 and +230\,\kms\ seen toward SN\,1993\,J ($l$=142\deg,
$b$=41\deg), if they are indeed unrelated to the galaxy M\,81, as de Boer et
al.\ (1993) argued. Finally, the +275\,\kms\ \MgII\ absorption seen toward
PKS\,0232$-$04 (Savage et al.\ 2000a, $l$=174\deg, $b$=$-$56\deg) is strange. No
high-positive velocity gas is known in this part of the sky (a few degrees from
the edge of the Anti-Center HVCs, which have high-negative velocity). As
PKS\,0232$-$94 has a redshift of 1.434, it is possible that two Lyman-$\alpha$
absorbers at z=1.3 mimic the \MgII\ doublet.
%VALUES: PG0953+414   CII=13.55 SiII=12.51 SiIII=12.55 AlII=11.95
%                  NH=   1E17          1.9E17             3E17   (solar abd)
\par The \CII, \SiII, \SiIII\ and \AlII\ absorption at $-$150\,\kms\ seen toward
PG\,0953+414 (Fabian et al.\ 2001) may be related to an extended low-column
density tail of either HVC complex~A or IV1. Both of these have velocities of
$-$120 to $-$150\,\kms. However, IV1 is about 8\deg\ distant, while the \HI\
edge of complex~A is about 10\deg\ away. The ionic column densities imply
N(H)$\sim$\dex{18}\,\cmm2\ for gas with 0.1 solar abundance. However,
considering that N(\SiII)$\sim$N(\SiIII), most of the hydrogen should be
ionized, so that N(\HI) is much lower. A possible counterargument is that toward
PG\,0804+761 absorption associated with complex~A is not seen in any line
(Richter et al.\ 2001a), even though that probe lies just 0\fdg5 off the \HI\
edge of complex~A. That would suggest a rather sharp edge. However, PG\,0804+761
lies near the more constrained low-latitude part of complex~A, rather than near
the flared-out high-latitude end. This problem requires further observations of
AGNs in the region between PG\,0953+414 and the edge of complex~A.
%VALUES: NaI CaII - SN1994D SN1991T SN1986G SN1983N
\par Three of the remaining \HI-less absorptions are seen at a velocity
$>$+200\,\kms\ in \NaI\ and \CaII\ (SN\,1994\,D, King et al.\ 1995, Ho \&
Filippenko 1995) SN\,1991\,T (Meyer \& Roth 1991) and SN\,1983\,N (d'Odorico
1985). \HI-less absorption is also seen at \vlsr=+130\,\kms\ toward PG\,0953+414
(Fabian et al.\ 2001). Weak \HI\ is detected at \vlsr$>$+200\,\kms\ toward
SN\,1986\,G (d'Odorico et al.\ 1989). Finally, +200\,\kms\ \CII\ and \SiII\
absorption is seen toward PG\,1116+215 (Tripp et al.\ 1998). In these directions
the \NaI/\HI\ ratios are $>$57, $>$59, $>$20, 46$\pm$20 and 60$\pm$18 ppb. All
of these are much higher than the average value in neutral gas (4.6 ppb).
\CaII/\HI\ ratios are $>$140, $>$500, $>$140, $>$250, $>$370, 150$\pm$70,
150$\pm$50 and $>$27 ppb, again much higher than the reference value of 22 ppb.
These high values are consistent with the relation found between $A$(\CaII) and
N(\HI), which is discussed by Wakker \& Mathis (2000), who show that high
apparent abundances are correlated with low \HI\ column density.
\par These six sightlines all lie in the region $l$=180\deg\to320\deg,
$b$=20\deg\to70\deg. Many small positive-velocity clouds are also present in
this region. Figure~\FgMS\ shows that this region further lies in the extension
of the curve defined by the SMC orbit and the EP clouds near $l$=290, $b$=+20
(at 60,$-$20 in Fig.~\FgMS). Thus, it is suggested that these high-positive
velocity absorptions and small positive-velocity clouds are associated with the
tenuous leading edge of the leading arm of the Magellanic Stream.
\par A possible problem with this model is that it requires that the leading-arm
gas no longer follows the SMC's orbit, deviating more as it gets farther ahead.
However, since the gas had to pass the LMC first, it seems not unreasonable to
suggest that it was given an additional nudge at that time. A more complete
model is required to test this hypothesis.
%VALUES:
%SN1994D: CaII 1.7  3.4   13  0.5; SN1991T: 0.74  1.1; SN1986G: 4.3 6.2; SN1983N: 1.6
%           HI <2.5                         <0.3                2.8 4.0          <6.0
%            A 100 >140 >500 >140           >250 >370           150 150           >27
% HIth = HIobs * A/69 /1E18
%         HIth 3.6  5.1   18  5.1            1.1  1.6                             2.3
%
% n = N / L = N / (3.1E18 * 10E3)
\par In the Gardiner \& Noguchi (1996) model, the tip of the leading arm gas is
supposed to be at a distance of 30\to80\,kpc. As the clouds in the leading arm's
tip cover about 20\deg\ on the sky, this corresponds to pathlengths of
$>$10\,kpc. Then column densities of $\sim$\dex{18}\,\cmm2\ correspond to a
volume density of $\sim$5\tdex{-5}\,\cmm3, which implies \Ha\ emission
intensities $<$0.01\,R, which is below the current detection limit (Reynolds et
al.\ 1998). Not detecting \Ha\ emission from the small positive-velocity HVCs
would therefore not conclusively favor a Local Group over a Magellanic origin.
\par A further check would be to determine whether high-velocity absorption is
not seen toward extra-galactic supernovae that lie away from the orbits of the
Magellanic Clouds.

%%%%%%%%%%%%%%%%%%%%%%%%%%%%%%%%%%%%%%%%%%%%%%%%%%%%%%%%%%%%%%%%%%%%%%%%%%%%%%%%

\subsection{HVCs/IVC toward the LMC}
In the spectrum of many stars in the LMC absorption is seen at velocities of
+165, +120 and +65\,\kms. \CaII\ and \NaI\ are the most-observed elements
(Blades 1980, Songaila \& York 1981, Songaila et al.\ 1981, Blades et al.\ 1982,
Songaila et al.\ 1986, Magain 1987, Vidal-Madjar et al.\ 1987, Molaro et al.\
1989, Wayte 1990, Molaro et al.\ 1993, Caulet \& Newell 1996, Welty et al.\
1999). Other studies concentrate on dominant elements (Welty et al.\ 1999, Bluhm
et al.\ 2001). In principle these results could be used to study abundance
variations in the HVCs and IVC. However, in practice there are many problems: a)
the earlier spectra suffer from low signal-to-noise ratios and the measured
equivalent widths are not very reliable; b) in almost all cases only equivalent
widths were published and linewidths need to be assumed to convert to column
densities; c) toward all but a few stars N(\HI) must be based on an
interpolation between nearby observations, often with barely sufficient
sensitivity; d) the clouds have not been mapped in \HI, as they are have rather
low column density (the Parkes Narrow Band Survey should make this possible,
eventually). Section~\SionCa\ presents a summary of the conclusions achievable
with these caveats in mind.
\par From the spectrum of SN\,1987\,A in the LMC, a large set of accurate ion
column densities was derived by Welty et al.\ (1999) for all three \HI\
components. However, a good, directly measured, value for N(\HI) was not
obtained, and instead N(\HI) was inferred from the abundance patterns (see item
``AB'' in description of Col.~\Ctel). The absolute abundance therefore remains
uncertain. However, for the two HVCs the abundance pattern is similar to the
halo pattern (Sembach \& Savage 1995). The pattern for the IVC would suggest
zero depletion. These matters are discussed in more detail by Welty et al.\
(1999).
\par For the +120\,\kms\ HVC a notable detection is that of H$_2$ toward
Sk$-$68\,82 by Richter et al.\ (1999), showing the presence of molecular
hydrogen. Bluhm et al.\ (2001) also find H$_2$ toward Sk$-$68\,82 in the IVC
with \vlsr=+60\,\kms.
%VALUES: LMCm FeII
\par \FeII\ absorption in the +120\,\kms\ HVC has been seen in five probes, but
with uncertain results. $A$(\FeII)=0.28$\pm$0.07 solar toward SN\,1987\,A (Welty
et al.\ 1999), 0.8$\pm$0.6 solar toward Sk$-$67\,104 (Bluhm et al.\ 2001),
1.3$\pm$0.9 solar toward Sk$-$67\,166 (Bluhm et al.\ 2001), 0.6$\pm$0.4 solar
Sk$-$68\,82 (Bluhm et al.\ 2001) and 0.16 solar toward Sk$-$69\,246 (Savage \&
de Boer 1981). N(\HI) is uncertain toward SN\,1987\,A (based on the depletion
pattern), Sk$-$67\,104, Sk$-$67\,166 and Sk$-$69\,246 (based on an
interpolation) and better measurements are needed to reconcile the factor 8
range. On the other hand, Bluhm et al.\ (2001) argue that the high \SII, \SiII\
and \FeII\ abundances in this cloud combined with the low \OI\ abundance
($\sim$0.5 solar) argues in favor of substantial ionization, possibly as high as
90\%. In this case variations in N(\FeII)/N(\HI) are due to variations in the
ionized fraction.

%%%%%%%%%%%%%%%%%%%%%%%%%%%%%%%%%%%%%%%%%%%%%%%%%%%%%%%%%%%%%%%%%%%%%%%%%%%%%%%%

\subsection{Intermediate-Velocity Arch}
This structure was first studied by Wesselius \& Fejes (1973). Kuntz \& Danly
(1996) presented a catalogue of cores. These designations were used here to sort
the many observed probes. At many positions, a higher- ($<$$-$60\,\kms) and a
lower- ($>$$-$60\,\kms) velocity component overlap, so the IV Arch is further
divided into two parts, one consisting of cores IV5 through IV17, the other
containing cores IV18 through IV26 (IV1\to4 are identical with HVC complex~M).
Within each part the cores are sorted along the Arch by decreasing galactic
longitude. Figs.~\FgIVupp\ and \FgIVlow\ show the structure, with the probe
positions overlaid.
\par Kuntz \& Danly (1996) derived a z-height of 0.8\to1.5\,kpc for the IV Arch,
and brackets for IV17 and IV26. The distance limits derived here are consistent
with their results, and they are graphically shown in Fig.~\Fivdist. The
horizontal axis in this figures shows an approximate ``angle along the Arch'',
with 0 degrees at the highest-longitude core. As the IV Arch lies at high
galactic latitudes, it is more useful to discuss z-heights than distances. A
fuller discussion now follows.
%VALUES: d - IV Arch
\par For the {\it higher-velocity part of the IV Arch}, z-height brackets can be
set for 3 cores: z=0.4\to1.7\,kpc for IV6, z=0.5\to0.7\,kpc for IV17 and
z=0.2\to1.6\,kpc for IV9. These are consistent with the upper limit of 3.9\,kpc
for IV7 and the lower limit of 0.7\,kpc for IV11. The upper limit for IV17 is
based on the HB star BD+49\,2137, for which Kuntz \& Danly (1996) gave a
z-height of 1.7\,kpc, which is correct for the spectral type given in SIMBAD
(B7V). Ryans et al.\ (1997a) did a more detailed spectroscopic analysis and
argue that it is a HB star at $z$=0.7\,kpc, although the final number is
somewhat model-dependent. The distance bracket for IV17 is set by the same stars
as those that bracket IV4, suggesting that these two clouds are close together
in space, even though they differ in velocity by 30\,\kms.
%VALUES: d - IV Arch
\par For the {\it lower velocity part of the IV Arch, at angles along the Arch
$<$50\deg\ ($l$$>$150\deg)}, a z-height bracket of 0.4\to2.6\,kpc can be set for
IV26, as well as a lower limit for IV24 of 0.4\,kpc. Off cores HD\,93521 is the
only star with z$<$2.6\,kpc, and it reduces the upper limit to 1.7\,kpc.
%VALUES: d(HDE233791: M(B9V)=0.5 d=10^(5+10.4-0.5)/5
%VALUES: d - IV Arch
\par For the {\it lower velocity part of the IV Arch, at angles along the Arch
$>$50\deg} IV19 sets an upper limit to the z-height of 1.6\,kpc. The limits
toward HDE\,233791 and PG\,1255+546 are in apparent conflict. However, the
classification of HDE\,233791 is uncertain (Ryans et al.\ 1997a) and instead of
an HB star at 0.5\,kpc distance it could be a B9V star at 0.9\,kpc distance
(z=0.8\,kpc). For PG\,1255+546 the 1-$\sigma$ uncertainty in the distance is
0.3\,kpc, so that this star could be at z=0.4\,kpc, rather than z=0.7\,kpc. If
these uncertainties are taken into account, combining the lower limit for
PG\,1255+456 with the upper limit for HDE\,233791 yields a possible z-height
bracket for this part of the Arch of 0.4\to0.8\,kpc. HDE\,233791 may even reduce
the upper limit to 0.4\,kpc.
%VALUES: d - IV Arch
\par A final complication is the low upper limit of 0.3\,kpc derived from {\it
core IV21}. This core is slightly off the main axis of the IV Arch, and is
further unusual in that CO and 100\,$\mu$m emission have been detected in it
(Weiss et al.\ 1999). These authors study the cloud as an example of a high-z
molecular cloud. Benjamin et al.\ (1996) claimed a distance bracket of
0.3\to0.4\,kpc. However, the lower limit is based on \NaI\ non-detections that
are not significant, as they are only a factor $\sim$6 below the measured \NaI\
abundance. Therefore, the distance may be less than 0.3\,kpc ($z$$<$0.2\,kpc).
More observations, preferably using \CaII\, are required to settle this
question.
%VALUES: d - IV Arch
\par In summary, for the higher-velocity part of the IV Arch a strong bracket of
z=0.7\to1.7\,kpc can be derived. The upper limit depends mostly on the z-height
of BD+49\,2137, which may be as low as 0.7\,kpc. For the lower-velocity IV Arch
the strong bracket is z=0.4\to1.6\,kpc for the part with $l$$>$150\deg. At lower
longitudes the situation is slightly confusing, but a possible z-height is
0.4\,kpc, although the strong bracket is 0.4\to1.6\,kpc.
%VALUES: M - IV Arch
\par Using these distance brackets, the mass of the IV Arch was estimated from
the Leiden-Dwingeloo Survey to be 1\to4.5\tdex5\,\Msun\ for velocities
$<$$-$60\,\kms, and 1\to8\tdex5\,\Msun\ for lower velocities. If the lower
distances are used for BD+49\,2137 and HDE\,233791, the implied total mass of
the IV Arch is about 2\tdex5\,\Msun.
%VALUES: A - IV Arch
\par The metallicity of the IV Arch has not yet been well-determined. Sulphur
was observed to have near solar abundance toward HD\,93521 (Fitzpatrick \&
Spitzer 1993); PG\,0953+414 (Fabian et al.\ 2001) and HD\,121800 (IV9/IV19
\dash\ Howk et al.\ 2001). Similarly, oxygen is found to have an abundance of
$\sim$1 solar toward PG\,1259+593 (Richter et al.\ 2001b).
%VALUES: SII - HD93521
%VALUES: SII - HD121800 PG0953+414
\par Toward HD\,93521 absorption components are seen at several velocities
($-$65, $-$57, $-$51 and $-$36\,\kms), even though the \HI\ profile just shows a
20.8\,\kms\ wide component centered at $-$56\,\kms. The component at
$-$65\,\kms\ is probably related to IV6. Applying a component analysis to the
\HI\ spectrum observed at Green Bank (21\arcmin\ beam), Fitzpatrick \& Spitzer
(1993) derived \SII/\HI\ ratios of 2.1$\pm$2, 0.78$\pm$0.12, 1.2$\pm$0.25 and
0.74$\pm$0.31 times solar (0.97$\pm$0.06 when summing all components). They
further determine $n_e$ and use the result to argue that hydrogen ionization is
unimportant. Toward HD\,121800 two \HI\ and two \SII\ components occur, which
are not well-separated. The average \SII/\HI\ ratio is 0.8 solar. In the
sightline to PG\,0953+414 Fabian et al.\ (2001) find \SII/\,\HI=1.1$\pm$0.2
times solar, but N(\HI) is low (23\tdex{18}\,\cmm2), so the ionization
correction is potentially large. In general, these results suggest near-solar
abundance for the IV Arch although more work is needed on component structure,
\HI\ small-scale structure and ionization corrections.
\par The abundance patterns toward HD\,93521 were used by Savage \& Sembach
(1996a) to {\it define} the reference for halo gas. These patterns are shown in
Fig.~\Fpattern f and k.

%%%%%%%%%%%%%%%%%%%%%%%%%%%%%%%%%%%%%%%%%%%%%%%%%%%%%%%%%%%%%%%%%%%%%%%%%%%%%%%%

\subsection{Intermediate-Velocity Spur}
These cores form an extension to the IV Arch, at somewhat lower velocities and
higher longitudes and latitudes, and were defined by Kuntz \& Danly (1996).
These authors also derived the distance bracket of 0.3\to2.1\,kpc based on the
\SiII\ lines in IUE spectra. The implied mass bracket is 0.2\to8\tdex5\,\Msun.
The positions of the probes are shown in Fig.~\Fgspur.

%%%%%%%%%%%%%%%%%%%%%%%%%%%%%%%%%%%%%%%%%%%%%%%%%%%%%%%%%%%%%%%%%%%%%%%%%%%%%%%%

\subsection{Low-Latitude Intermediate-Velocity Arch}
%VALUES: d,M LLIV
This structure was named by Kuntz \& Danly (1996). It crosses over complex~A,
resulting in a relatively large number of observed probes (see Fig.~\FgLLIV).
The distance can be constrained to lie in the range 0.9 to 1.8\,kpc
($z$=0.6\to1.2\,kpc). The implied mass is 1.5\to6\tdex{5}\,\Msun. In the
Galactic plane at these longitudes lies the Perseus Arm, a spiral arm that is
2.5\,kpc distant (Reynolds et al.\ 1995). The LLIV Arch thus appears to be
high-z interarm gas.
%VALUES: PG0804+761 - OI, NI, PII
%VALUES: SN1993J - ZnII
\par Several probes lie off the main structure, and are collected under the
heading ``LLIV Arch extension''. This gas has velocities similar to that in the
LLIV Arch proper, and most likely is spatially close to it. If so, an upper
distance limit of 0.9\,kpc is set by HD\,83206, equal to the lower distance
limit for the main part of the LLIV Arch.
\par The metallicity and depletion pattern can be determined from the spectra of
SN\,1993\,J (de Boer et al.\ 1993), PG\,0804+761 (Richter et al.\ 2001a) and
HDE\,233622 (Ryans et al.\ 1997b). The metallicity follows from
N(\OI)/N(\HI)=1.0$\pm$0.5 solar and N(\NI)/N(\HI)=0.55$\pm$0.14 solar, as O and
N are undepleted, and (especially \OI) tied to \HI. If hydrogen ionization were
ignored, this would be inconsistent with the apparently supersolar abundance of
\PII\ (1.3$\pm$0.6 solar) and \ZnII\ (1.6$\pm$0.4 solar). However, both these
elements can co-exist with both \HI\ and H$^+$. The degree of ionization can be
estimated by assuming that \OI\ and \HI\ go together. Then N(\PII)/N(\OI) $\sim$
(N(\HI)+N(H$^+$))/N(\HI). This yields an ionization fraction of $\sim$20\%.
%VALUES: PG0804+761 - SiII FeII
%VALUES: SN1993J    - SiII AlII FeII
%VALUES: HDE233622  - TiII
\par The depletions of some refractory elements are typical for warm disk gas:
$\delta$(\SiII)$\sim$0.3 (typical is 0.15), $\delta$(\AlII)=0.09 (typical is
0.15), $\delta$(\TiII)=0.11 (typical is 0.05), and $\delta$(\FeII)=0.1\to0.3
(typical is 0.1). See Fig.~\Fpattern d for a graphical representation. %VALUES:
CaII in LLIV probes (see ~ line 950)
\par \CaII\ abundances in the LLIV Arch are remarkable for their constancy. Out
of 10 determinations 7 are in the range 12\to14 ppb, while ``outliers'' are 17
and 9.1 ppb. The only really deviating value is 2.5 ppb for the weaker component
in the spectrum of PG\,0833+698. However, for this probe the decomposition of
the \HI\ spectrum is somewhat suspect.
\par Intermediate-velocity \OVI\ and \CIV\ are also found toward PG\,0804+762
and SN\,1993\,J, respectively (Richter et al.\ 2001a, de Boer et al.\ 1993).
Thus, the LLIV Arch appears to be a near-solar metallicity cloud, with a
substantial H$^+$ fraction, disk-like dust, embedded in hot gas, and located in
the interarm region between the Local and Perseus spiral arms. Its velocity is
20\to30\,\kms\ more negative than expected from differential galactic rotation.
\par All these characteristics are typically those expected for a cloud that is
part of the return flow of the Galactic Fountain. That is, gas that was ejected
into the Galactic Halo from inside the solar radius is expected to have a
metallicity slightly above that in the local ISM, the dust is expected to
survive in the hot phase, the rotational velocity is expected to decrease as the
gas rises and moves outward (because of conservation of angular momentum,
Bregman 1980), and after condensations grow they will remain embedded in the
as-yet uncooled part of the gas.

%%%%%%%%%%%%%%%%%%%%%%%%%%%%%%%%%%%%%%%%%%%%%%%%%%%%%%%%%%%%%%%%%%%%%%%%%%%%%%%%

\subsection{Complex K}
This object was first seen in 21-cm emission by Kerr \& Knapp (1972) in the
direction of M\,13. De Boer \& Savage (1983) detected \MgII\ toward the star
Barnard\,29 in M\,13. Figure~\FgK\ shows the gas between velocities of $-$95 and
$-$60\,\kms\ in this region of the sky, based on the Leiden-Dwingeloo Survey
(Hartmann \& Burton 1997). Below latitudes of 45\deg, the velocity of the peak
is $<$$-$70\,\kms, and those components were included in the survey of Hulsbosch
\& Wakker (1988), and considered part of complex~C. However, they stand out from
the main body of C, although near $l$$\sim$80\deg\ the components merge and it
is difficult to assign them to either C or K. Here, complex~K is defined as the
intermediate-velocity gas with $-$95$<$\vlsr$<$$-$60\,\kms\ in this region of
the sky. A better definition requires detailed component fitting.
\par From differential galactic rotation, velocities in the range 0 to
+30\,\kms\ are expected, whereas complex~K has \vlsr$\sim$$-$80\,\kms. Thus, it
rotates too slowly by about 100\,\kms. A metallicity measurement is required to
determine whether this is an infalling cloud, such as complex~C, or whether it
is a Galactic Fountain-type cloud.
\par Haffner (2000) detects faint \Ha\ emission (0.1\to0.2\,R) associated with
complex~K. The \Ha\ emission map correlates very well with the \HI\ column
density map for the fainter extended part and for the minor cores. However, in
the brightest core (near $l$=55\deg, $b$=38\deg) the \Ha\ peaks east of the \HI.
\par Shaw et al.\ (1996) mapped the \HI\ at high angular resolution
(3\arcmin$\times$2\arcmin) in the direction of the globular cluster M\,13, using
a combination of Jodrell Bank and DRAO data. They found variations of a factor 2
on arcminute scales. They also observed \NaI\ toward several stars, giving 3
detections (at 45, 14 and 13 ppb) and 6 upper limits ($<$5 ppb). Clearly, there
are large variations in the \NaI\ abundance on arcminute scales.
\par The \MgII\ and \NaI, detections toward stars in M\,13 set an upper limit to
the distance of K of 6.8\,kpc, and a mass limit of $<$7.5\tdex5\,\Msun.
\par Toward Mrk\,501 the abundance of \CaII\ is found to be 89 ppb, although
with large errors. There are 31 PG stars with a range of distances that lie
projected onto the main core of complex~K at $l$=55\deg, $b$=35\deg, but for
none of these has a \CaII\ spectrum yet been taken. Clearly, a good distance
determination is possible. However, since N(\HI) is low, spectra with high
signal-to-noise ratio will be necessary.
\par For the star M\,3\,vz1128 ($l$=42\deg, $b$=79\deg, shown on Fig.~\FgK) de
Boer \& Savage (1984) claimed strong \CII\ absorption at a velocity of
$-$70\,\kms, although no associated \HI\ is seen. Recent data from FUSE do not
show this absorption in the \CII-1036 and other strong lines (Howk, priv.\
comm.). Thus, the claim by de Boer \& Savage (1984) turns out to have been
spurious.

%%%%%%%%%%%%%%%%%%%%%%%%%%%%%%%%%%%%%%%%%%%%%%%%%%%%%%%%%%%%%%%%%%%%%%%%%%%%%%%%

%\subsection{Draco Nebula}
%Ref to Herbstmeier / Gladders => nearby tiny IVC.

%%%%%%%%%%%%%%%%%%%%%%%%%%%%%%%%%%%%%%%%%%%%%%%%%%%%%%%%%%%%%%%%%%%%%%%%%%%%%%%%

\subsection{Southern Intermediate-Velocity Clouds}
At velocities between $-$85 and $-$45\,\kms\ the southern Galactic sky contains
a number of IVCs in the region between galactic longitudes 60\deg\ and 150\deg,
as shown in Fig.~\Fgsouth. Note that by placing the gas in the outer Galaxy,
velocities up to $-$40\,\kms\ can still be understood within the framework of
differential galactic rotation for latitudes $>$$-$40\deg, and longitudes
90\to150\deg. High-velocity gas with \vlsr$<$$-$100\,\kms\ also occurs in this
area, as shown by the thick solid outlines (see figure caption for details).
\par These southern IVCs have not been studied, with the exception of the
HD\,215733 sightline, which was analyzed in great detail by Fitzpatrick \&
Spitzer (1997). This sightline shows intermediate-velocity absorption at $-$92,
$-$56 and $-$43\,\kms. The first of these is very weak in \HI. The \HI\ spectrum
shows a single component centered at $-$44\,\kms. Fitzpatrick \& Spitzer (1997)
decomposed it based on the UV absorption lines. The components can be associated
with an Arch running through the constellations Pegasus and Pisces ($l$=90\deg,
$b$=$-$40\deg\ to $l$=130\deg, $b$=$-$60\deg). Here, and elsewhere in the paper
this structure is referred to as the Pegasus-Pisces Arch, or the PP Arch.
Although much weaker, this is the closest southern counterpart to the IV Arch in
the north.
%VALUES: HD215733 - SII
\par Individually, the absorption components toward HD\,215733 have \SII/\HI\
ratios of 0.17, 0.32 and 1.2 times solar, but when combined this is
0.54$\pm$0.04 times solar. As Fitzpatrick \& Spitzer exclude a large ionization
correction, this is clearly subsolar. Together with HD\,93521, the depletion
pattern in these two components (see Fig.~\Fpattern n, p, r) {\it defines} the
halo pattern.
%VALUES: d - HD215733 PG0039+048
\par From HD\,215733 an upper distance limit of 2.7\,kpc can be set to the
northern knot of the Pegasus-Pisces Arch, or $z$$<$1.6\,kpc. For the southern
part, PG\,0039+048 seems to set an upper limit of 1.1\,kpc, as \NaI\ absorption
is seen in its spectrum (Centuri\'on et al.\ 1994). However, this component is
only seen in the line wing and a more accurate measurement is needed.
%VALUES: M - PPArch
\par Using the distance limit implied by PG\,0039+048 and integrating the LDS
data in the velocity range between $-$85 and $-$45\,\kms\ sets an upper mass
limit of about 0.5\tdex{5}\,\Msun.

%%%%%%%%%%%%%%%%%%%%%%%%%%%%%%%%%%%%%%%%%%%%%%%%%%%%%%%%%%%%%%%%%%%%%%%%%%%%%%%%

\subsection{Complex~gp}
The IVC centered on $l$=65\deg, $b$=$-$27\deg\ was the first IVC detected in
absorption, against the globular cluster M\,15 (Cohen 1979). A Leiden-Dwingeloo
Survey map of the intermediate-velocity gas in this area shows a large number of
scattered, faint IVCs (see Fig.~\Fggp). These seem to be an extension toward
lower velocities of HVC\,40$-$15+100, which is also known as the Smith cloud
(Smith 1963), or as the main cloud in complex~GP (Wakker \& van Woerden 1991).
This shows up prominently at $l$=35\to50\deg, $b$$>$$-$25\deg. The IVCs between
+60 and +90\,\kms\ seem to extend complex~GP toward more negative latitudes and
velocities. Hence, these IVCs will be collectively referred to as
``complex~gp''. Here it is worthwhile to note that at velocities between +30 and
+60\,\kms\ there are no coherent structures in this region of the sky.
Differential galactic rotation can account for about 20\to40\,\kms\ of the
observed radial velocity.
%VALUES: d,M - gp
\par The detection of the IVC toward HD\,203664 sets an upper distance limit of
4.3\,kpc ($z$$<$2.0\,kpc), while the non-detection of the IVC toward HD\,203699
sets a lower limit of 0.8\,kpc ($z$$>$0.3\,kpc) (Albert et al.\ 1993, Little et
al.\ 1994, Ryans et al.\ 1996, Kennedy et al.\ 1998). A rough integration of the
LDS data in the velocity range between +55 and +100\,\kms\ then gives a mass
range of 0.1\to3\tdex{5}\,\Msun\ for the intermediate-velocity gas. Compare this
to the mass of the higher-velocity gas (i.e.\ complex~GP, Sect.~\Scloud.\SSGP),
which for the same distance limits would have a mass of 0.3\to8\tdex{5}\,\Msun.
Thus, if they are related, the high-velocity gas has about twice the mass of the
intermediate-velocity gas.
%VALUES: A - gp
\par An detailed study of the metallicity of this cloud has not yet been made.
However, Penton et al.\ (2000) list equivalent widths for \SII-1250, 1253 in the
direction toward Mrk\,509. These imply an \SII/\HI\ ratio of 2.0 solar. Sembach
et al.\ (1999) show the spectrum in more detail, and from this it is clear that
the numbers given by Penton et al.\ (2000) are a factor 2 too high, and that a
better abundance estimate is 0.8 solar. Since N(\HI) for the +60\,\kms\
component toward Mrk\,509 is only about 24.5\tdex{18}\,\cmm2, it seems likely
that a substantial ionization correction is needed. However, even if that is a
factor two, a near-solar metallicity is implied. As Mrk\,509 is one of the
brightest AGNs in the UV, a good study of abundances and ionization will be
possible with {\it FUSE}. \NaI\, \SiII\ and \CaII\ toward Mrk\,509 were already
measured by York et al.\ (1982), Blades \& Morton (1983) and Morton \& Blades
(1986).
\par Bland-Hawthorn et al.\ (1998) detected \Ha\ emission from two positions
within the Smith cloud, and they proposed that complex~GP is a tidal stream
related to the Sagittarius dwarf galaxy, at a distance on the order of 25\,kpc.
This is difficult to reconcile with the upper distance limit of 4.3\,kpc for the
IVC, unless the apparent spatial relation between complex~GP and the IVCs is
accidental.
%VALUES: M15* - CaII
\par The intermediate-velocity gas is seen in the spectrum of many stars in the
globular cluster M\,15, the sightline to which intersects one of the four cores
of complex~gp. Lehner et al.\ (1999b) measure \CaII\ toward 12 stars (see
Sect.~\SionCa), finding values in the range 14\to79 ppb. \NaI\ was measured by
Langer et al.\ (1990) toward 7 stars, and found to be in the range 6\to21 ppb.
However, Meyer \& Lauroesch (1999) find that \NaI\ varies by a factor 15 across
the face of M\,15. Toward HD\,203664 (a few degrees from M\,15), the \CaII/\HI\
ratio is much higher (440 ppb). However, at the very low N(\HI) seen in that
direction (2.2\tdex{18}\,\cmm2) high \CaII/\HI\ ratios are to be expected (see
Wakker \& Mathis 2000).

%%%%%%%%%%%%%%%%%%%%%%%%%%%%%%%%%%%%%%%%%%%%%%%%%%%%%%%%%%%%%%%%%%%%%%%%%%%%%%%%

\subsection{Other IVCs}
In 23 sightlines, observations of a probe of an identified HVC or IVC also show
absorption and/or \HI\ emission associated with another IVC, usually small
and/or faint clouds. These are collected under the heading ``Other Negative-
(Positive-) velocity IVCs''. The data are usually too sparse to learn useful
things about these IVCs, except that the abundances tend to be normal.

%%%%%%%%%%%%%%%%%%%%%%%%%%%%%%%%%%%%%%%%%%%%%%%%%%%%%%%%%%%%%%%%%%%%%%%%%%%%%%%%
%%%%%%%%%%%%%%%%%%%%%%%%%%%%%%%%%%%%%%%%%%%%%%%%%%%%%%%%%%%%%%%%%%%%%%%%%%%%%%%%
%%%%%%%%%%%%%%%%%%%%%%%%%%%%%%%%%%%%%%%%%%%%%%%%%%%%%%%%%%%%%%%%%%%%%%%%%%%%%%%%

\section{Summary}
Table~\Tsumm\ summarizes the distances, masses, metallicities and depletion
patterns discussed in the previous section. The main new conclusions that can be
drawn from this table are:
%VALUES: d,A - A
\par\noindent
1) HVC complex~A is the only HVC for which a distance bracket is known, which is
4.0\to9.9\,kpc ($z$=2.6\to6.8\,kpc). It may have a metallicity of about 0.1 solar.
%VALUES: A,d - M
\par\noindent
2) Complex~M is a grab bag collection of clouds, which may or may not be
physically related. The most likely metallicity of cloud MI is near-solar. Cloud
IV4 (or MI-extension) lies at $z$=0.5\to0.7\,kpc and has a warm-disk-like
depletion pattern. Cloud MII/MIII lies at $z$$<$3.5\,kpc; a possible lower limit
of 1.7\,kpc remains controversial.
%VALUES: A,d - C  SII - Mrk290 Mrk817 Mrk279  NI - Mrk876
\par\noindent
3) Complex~C appears to have a metallicity of $\sim$0.1 solar, based on \SII\
absorption toward Mrk\,290 and \NI\ toward Mrk\,876. Richter et al.\ (2001b)
report \OI/\HI=0.1 solar toward PG\,1259+593. \SII/\HI\ ratios of 0.3\to0.4
solar that have been reported toward two other probes are uncorrected for
ionization and \HI\ small-scale structure. The Mrk\,876 sightline gives some
anomalous abundances, which are most likely due to high partial ionization.
However, \NI/\HI\ is consistent with \SII/\HI. There seems to be little dust in
complex~C. In that case, the \NI, \OI, \SiII\ and \FeII\ abundance pattern
toward PG\,1259+593 is consistent with the idea that the heavy elements were
produced in type II supernovae, with no subsequent star formation. Finally,
complex~C shows the importance and necessity of corrections for ionization and
\HI\ fine structure.
\par\noindent
%VALUES: d - C
4) The strong limit to the distance to complex~C is $>$1.2\,kpc, but the
distance is probably $>$6\,kpc.
%VALUES: S - MS
5) For the trailing and leading part of the Magellanic Stream \SII\ abundances
of 0.33$\pm$0.05 and 0.25$\pm$0.08 solar are found toward Fairall\,9 (Gibson et
al.\ 2000) and NGC\,3783 (Lu et al.\ 1998), consistent with the idea that the
Stream is a tidal feature extracted from the SMC (Gardiner \& Noguchi 1996).
\par\noindent
%VALUES: Mg/H - MS
6) Using the FOS, Savage et al.\ (2000a) found \MgII\ absorptions associated
with the Magellanic Stream, showing a nearly constant abundance of 0.1 solar
when N(\HI)$>$2\tdex{18}\,\cmm2, but showing larger \MgII/\HI\ ratios at lower
N(\HI), suggesting increasing ionization.
%VALUES: Fe/S - MS
\par\noindent
7) Fe/S was determined for the trailing and leading parts of the Stream, giving
values of 0.18$\pm$0.06 and 0.13$\pm$0.05 times solar. This is comparable to the
value in SMC gas, and indicates the presence of dust in the Stream.
%VALUES: Mrk205 - MgII
\par\noindent
8) \MgII\ absorption due to cloud WW84 is seen in the spectrum of Mrk\,205. The
\MgII\ abundance remains uncertain, however, due to \HI\ small-scale structure.
$A$(\MgII) could possibly be as low as 0.02 solar.
%VALUES: d - WE
\par\noindent
9) A second upper limit was set to the distance of a HVC: complex~WE (which is
identified and named in this paper) has $d$$<$12.8\,kpc, $z$$<$3.2\,kpc, based
on the detection of absorption in the star HD\,156359.
\par\noindent
10) It is suggested that many small positive-velocity HVCs and the high-positive
velocity absorptions without associated \HI\ in the region
$l$=180\deg\to320\deg, $b$=20\deg\to70\deg\ are due to a spread-out leading edge
to the leading arm of the Magellanic Stream. These occur in the extension of the
curve defined by the SMC orbit and the EP clouds (Fig.~\FgMS).
%VALUES: d - IVArch
\par\noindent
11) For the higher-velocity part of the IV Arch a strong bracket of
$z$=0.7\to1.7\,kpc can be derived. The upper limit depends mostly on the
distance to BD+49\,2137, which may be as low as 0.7\,kpc. For the lower-velocity
IV Arch the strong bracket is $z$=0.4\to1.7\,kpc for the part with
$l$$>$150\deg. At lower longitudes the situation is confusing, but a possible
z-height is 0.4\,kpc.
\par\noindent
%VALUES: A - IV Arch  SII - HD93521 HD121800 PG0953+414
12) The IV Arch appears to have near solar abundances, based on \SII/\HI\ ratios
of 0.97$\pm$0.06, 0.8$\pm$0.1 and 1.1$\pm$0.2 found toward HD\,93521, HD\,121800
and PG\,0953+414, respectively.
%VALUES: d,A - LLIV
\par\noindent
13) A metallicity of $\sim$0.7 solar and a distance of 0.9\,kpc ($z$=0.6\,kpc)
are known for the LLIV Arch. This object appears to be a prime example of the
return flow of a Galactic Fountain: it lies at high-z in an interarm region, it
is rotating somewhat slower than expected from differential galactic rotation,
it has a metallicity similar to that in the local ISM, the depletion pattern is
similar to that in warm disk gas, and it is embedded in hot gas.
%VALUES: d - K
\par\noindent
14) IVC complex~K is newly defined in this paper, as IVC components with
\vlsr$\sim$$-$80\,\kms\ near $l$=50\deg. It has a distance $<$6.8\,kpc
($z$$<$4.5\,kpc), and many potentially useful targets exist. It is notable for a
strong correlation between \Ha\ emission and \HI\ column density (Haffner et
al.\ 2000).
%VALUES: d,A - PP Arch
\par\noindent
15) The PP Arch is the best southern counterpart to the IV Arch in the northern
Galactic hemisphere. An upper distance limit of 2.7\,kpc ($z$$<$1.6\,kpc) can be
set for the northern part, and of 1.1\,kpc ($z$$<$0.9\,kpc) for the southern
part. A metallicity of 0.5 solar is derived. This object deserves further study.
%VALUES: d - gp
\par\noindent
16) Positive-velocity IVCs near $l$=50\deg, $b$=$-$25\deg\ may be related to HVC
complex~GP (also known as HVC\,40$-$15+100 or as the ``Smith cloud''). Some of
the IVCs lie in the distance range 0.8\to4.3\,kpc ($z$=0.3\to2.0\,kpc) If the
IVCs and complex~GP are related, this excludes the suggestion by Bland-Hawthorn
et al.\ (1998) that complex~GP is a tidal stream connected to the Sagittarius
dwarf. Complex~gp is probed by the bright UV probe Mrk\,509, towards which \SII\
absorption suggests an abundance of 1\to2 times solar.
%VALUES: A - C
\par\noindent
17) The metallicity of HVCs and IVCs ranges from 0.1 to 1 solar. The most
accurate value (0.089 times solar) is for complex~C. For complex~A $\sim$0.1
solar is suggested. The Magellanic Stream (both the trailing and the leading
arms) has Magellanic-like abundances (0.25 solar for sulphur). IVCs tends to
have higher metallicities: the PP Arch has Z$\sim$0.5 solar, the LLIV Arch has
Z$\sim$0.8 solar, while the IV Arch has Z$\sim$1 solar. Thus, the classical HVCs
A and C most likely represent material that has never before been part of the
Milky Way and is now being accreted (see Wakker et al.\ 1999 for more
discussion). The IVCs, on the other hand consist of Galactic gas. It is an open
question which clouds are previously-hot halo gas compressed by infalling
material, and which are part of the return flow of a Galactic Fountain. For the
LLIV Arch the second possibility is preferred.
%VALUES: FeII - Mrk876
\par\noindent
18) Very little information exists on dust in HVCs. The Magellanic Stream shows
the pattern typical for halo gas. In complex~C there are indications for a lack
of dust, based on the ratio \FeII/\HI=0.5 solar toward Mrk\,876. Since S/H=0.1,
a halo-like ratio of Fe/S=0.25 would require N(H$^+$)/N(\HI)$>$20, which is
incompatible with the other absorption lines and the non-detection of \Ha\
emission (Murphy et al.\ 2000).
\par\noindent
19) The IVC depletion pattern varies from that typical for warm disk gas (IV4,
LLIV Arch) to that typical of halo gas (IV Arch, PP Arch).
%VALUES: d - A WE C H IV..
\par\noindent
20) Only one HVC distance bracket is known (4\to10\,kpc for complex~A). An upper
limit of $<$12.8\,kpc was set for complex~WE, and lower limits of $>$6 and
$>$5\,kpc exist for complexes~C and H. In contrast, several IVC distance
brackets were set. Expressed as z-heights, these tend to be about
0.5\to1.5\,kpc. Thus, the major IVCs appear to be a rather local phenomenon,
whereas the major HVCs appear to be large clouds away from the Galactic plane.
%VALUES M - IV HV
\par\noindent
21) For the large IVCs (IV Arch, LLIV Arch, PP Arch), mass limits typically are
0.1\to8\tdex{5}\,\Msun, whereas for the larger HVCs mass limits typically are
$>$\dex{6}\,\Msun\ (with an upper limit of 2\tdex{6}\,\Msun\ for complex~A). The
Magellanic Stream is the most massive HVC, having M$>$\dex{8}\,\Msun. Thus, the
major HVCs (A, C, H, MS) appear to be more massive than the major IVCs.
\par\noindent
22) The abundance of \CaII\ tends to vary by a factor 2\to5 within clouds and by
a factor 5\to10 between clouds. At any given value of N(\HI), N(\CaII) can vary
by a factor 10. Nevertheless, a strong correlation is found between the \CaII\
abundance and N(\HI), which is discussed in more detail by Wakker \& Mathis
(2000).
\par\noindent
23) The abundance of \NaI\ varies by a factor 100 for a given N(\HI), and by a
factor $>$10 within a single cloud. No correlation is seen between log N(\NaI)
and log N(\HI), showing that for HVCs and IVCs N(\NaI) is an even worse
predictor of N(\HI) than it is for low-velocity gas.
\par\noindent
24) It is possible to use \CaII\ to set lower limits to cloud distances, but
this requires a) knowledge of the \CaII\ abundance in the cloud and b)
sightlines with sufficiently high N(\HI) ($>$5\tdex{18}\,\cmm2) and c)
sufficiently sensitive observations of the \CaII\ lines. For \NaI\ the safety
factor required to interpret a non-detection as a lower limit is prohibitively
large and in practice non-detection of \NaI\ should not be considered
significant.
\acknowledgements
This research has made extensive use of the SIMBAD database, operated at CDS,
Strasbourg, France.

%%%%%%%%%%%%%%%%%%%%%%%%%%%%%%%%%%%%%%%%%%%%%%%%%%%%%%%%%%%%%%%%%%%%%%%%%%%%%%%%

%%%%%%%%%%%%%%%%%%%%%%%%%%%%%%%%%%%%%%%%%%%%%%%%%%%%%%%%%%%%%%%%%%%%%%%%%%%%%%%%

\newpage
\appendix
\section{Description of the columns}

\par\noindent {\it Column \Cname}
\par The name of the stellar or extra-galactic probe. In most cases this is the
name given in the publication (see Col.~\Cref). In a few cases the HD, BD, or PG
number is substituted, most notably for the SAO stars in Lilienthal et al.\
(1990) and Tamanaha (1997), for HZ\,22 (=PG\,1212+369), HZ\,25 (=BD\,+36~2268)
and AG+53~783 (=HDE\,233791) in Ryans et al.\ (1997a), and for H.O.+23B
(=PG\,1205+228) in Kuntz \& Danly (1996) [a revised distance was also determined
for the latter star by Quinn et al.\ (1991)]. In the case of extra-galactic
objects, an effort was made to use the names in the V\'eron-Cetty \& V\'eron
(1996) catalogue of QSOs and Seyfert galaxies, even if a different name was used
in the absorption-line publication.

\par\noindent {\it Column \Cglon, \Cglat}
\par The galactic longitude and latitude of the probe, rounded to 1/100th of a
degree. For more accurate values, see SIMBAD or the original publication.

\par\noindent {\it Column \Cdist, \Cheight}
\par Column~\Cdist\ gives the distance to the star and the distance error, in
kpc. This is followed by a flag that indicates the method used to make the
estimate. Column~\Cheight\ shows the z-height in kpc, derived as $d$\,sin({\it
b}). The following flags are used.
%VALUES d - HD135485
\par\noindent ``p'' \dash\ Distance is 1/parallax, as measured by Hipparcos, and
as given in SIMBAD. This is used if the parallax is known to better than about
30\%. The distance error is calculated from the given parallax error, using
$\Delta$d/d=$\Delta$p/p. Usually, the resulting distance is consistent with the
spectroscopic distance (see below, under flag ``t'') to within 0.1\,kpc. The
most notable exception is HD\,135485 (type B5IIp), which was estimated to be at
2.5\,kpc by Albert et al.\ (1993) and at 0.8\,kpc by Danly et al.\ (1995), but
whose parallax puts it at 0.18\,kpc.
\par\noindent ``a'' \dash\ Distance determined from a detailed atmospheric
analysis based on intermediate-resolution spectroscopy. For \PGname\ PG stars,
the distance values are taken from Moehler et al.\ (1990), Theissen et al.\
(1995), Wakker et al.\ (1996b), and Ryans et al.\ (1997a, 1997b). In the latter
paper 3 non-PG stars were also analyzed (BD\,+38~2182, BD\,+49~2137 and
HDE\,233781). The nominal errors (as given in the publications above) are
typically 0.3\to0.4\,kpc. In some cases the spectral analysis is difficult
because the star may be either a Horizontal-Branch or a post-AGB star. In this
case, the actual uncertainty is much larger than the formal uncertainty. In
particular, PG\,0832+675 was at some point thought to be a main-sequence B1V
star at 31\,kpc (Brown et al.\ 1989), but Hambly et al.\ (1996) found it to be
an evolved star at 8.1\,kpc, which is what is used here.
\par\noindent ``g'' \dash\ For stars in globular clusters and in the LMC, the
distance is the distance to the cluster or to the LMC. Globular cluster
distances are taken from Harris (1996). The LMC is assumed to be at a distance
of 50\,kpc.
\par\noindent ``s'' \dash\ Distance determined from the spectral type, which
yields the absolute magnitude of the star. An extinction correction is applied,
using the map of Lucke (1978), which gives the average A$_V$ out to 2\,kpc in
mag/kpc for 5\deg$\times$5\deg\ regions on the sky. Except for stars on
complex~H and clouds AC0 and ACI, the extinction estimates always are below 0.2
magnitudes. A distance calculation is done even if the original reference gives
a spectroscopic distance, so that all distances are calculated on the same
absolute-magnitude system. The differences between published and re-calculated
values are usually less than 10\%.
\par For RR\,Lyrae stars, M$_V$=0.58 is assumed, based on recent calibrations of
their absolute magnitude (Fernley et al.\ 1998). A 10\% error is assumed, which
corresponds to 0.2 magnitudes.
\par For confirmed Horizontal Branch stars that have not been subjected to a
detailed analysis (PG\,1126+468, PG\,1510+635, PG\,1008+689 and PG\,1343+577) it
is assumed that M$_V$ is in the range 0.55 to 1.15 (Preston et al.\ 1991). The
average of these two values gives the most likely distance. The distance error
is found as half of the range implied by the range in M$_V$. Five probes of
complex~C are probably also Horizontal Branch stars and the M$_V$ values above
are also used; better classifications are needed to confirm their status.
\par For stars with an MK classification the table of Straizys \& Kurilene
(1981) is used to convert from spectral type to M$_V$. The resulting distances
are normally within 0.2\,kpc of estimates made for the same star by other
authors. The (1$\sigma$) distance error is estimated by also calculating the
distance using a classification that differs by one subtype and one luminosity
class. This gives a maximum range for the distance, which is assumed to be
equivalent to $\pm$3$\sigma$. For main-sequence stars, as well as for all O-type
stars, the typical error is less than 10\%. Large relative errors (10\to100\%)
occur for (super)giant A and B stars, where a difference of one luminosity class
can make a difference of 2 magnitudes in the estimated absolute magnitude.
\par\noindent ``t'' \dash\ The spectroscopic distance is given, even though the
parallax was measured by Hipparcos. However, the measured parallax is more
uncertain than the spectroscopic distance.
\par\noindent ``r'' \dash\ Distance given is value quoted in reference, as the
spectral type is not well-specified in SIMBAD or the reference, and thus the
distance cannot be recalculated on the same system as for other stars. This is
the case for BD\,+10~2179 and H.O.+41B.

\par\noindent {\it Column \Ctype}
\par The classification of the probe. This can be an MK spectral type, ``sd''
for subdwarfs (sometimes with subtype), ``HB'' for confirmed Horizontal Branch
stars, ``(HB)'' for suspected Horizontal Branch stars, ``PAGB'' for Post
Asymptotic Giant Branch,  ``SN'' for extra-galactic supernovae, as well as the
obvious ``RR Lyr'', ``QSO'', `BL Lac'', ``Sey'' or ``Gal''. ``radio'' is used
for 21-cm or 3-mm continuum sources where high-velocity \HI\ absorption has been
searched for.

\par\noindent {\it Column \Ccloud}
\par The name or number of the HVC/IVC on which the probe is projected. For IVCs
in the IV and LLIV Arch, the core numbers in the catalogue of Kuntz \& Danly
(1996) are used. For stars that probe the IV Arch away from its cores ``IVa'' is
given. The HVC names and numbers are from the catalogue of Wakker \& Woerden
(1991). For clouds outside complexes the catalogue number is preceded by the
acronym ``WW''. One new HVC complex and three new IVC complexes are introduced
in this paper: ``WE'', ``K'', the ``PP Arch'' and ``gp'', see
Sects.~\Scloud.\SSWE, \Scloud.\SSK, \Scloud.\SSSouth\ and \Scloud.\SSgp.

\par\noindent {\it Column \Cvlsr, \CNHI}
\par The best value for the \HI\ velocity and column density in the direction of
the probe. The velocity is given in \kms, relative to the LSR. The column
density and its error are given in units of \dex{18}\,\cmm2. Upper limits are 5
$\sigma$. If there is absorption, but no corresponding \HI\ component was
detected, the value within parentheses refers to the velocity at which an upper
limit to the \HI\ column density is set. For all probes, the \HI\ spectrum and
gaussian fits to the components are shown in Paper~II.
%VALUES: split of \HI\ based on absorption
\par For a few probes the \CaII\ and/or \NaI\ absorption spectra show multiple
components, although only one \HI\ component can be discerned, When these are
closer together than half the FWHM of the \HI\ spectrum, the \HI\ column density
is split in two, assuming that the ion abundance in both components is the same.
This results in two \HI\ components listed at the same velocity, suffixed by
``a'' and ``b'' for BD+49\,2137 (IV17), BD+38\,2182 and BD+36\,2268 (Lower IV
Arch), PG\,1008+689 (LLIV Arch), HD\,203664 (complex~gp), as well as for seven
stars in the LMC.

\par\noindent {\it Column \Ctel}
\par A code that shows the telescope used to measure the \HI\ column density. In
general, the value determined at the highest-available resolution is used, as
(especially in cores) \HI\ fine structure can produce variations of a factor of
a few at arcminute scales (Wakker \& Schwarz 1991, Wakker et al.\ 1996b). The
following codes are used:
\par\noindent We \dash\ N(\HI) from a combination of Westerbork and Effelsberg
data (1 or 2 arcmin beam), derived as described by Wakker et al.\ (1996b). This
was done for 5 probes (Mrk\,106, PG\,0832+675, PG\,0859+593, PG\,0906+597,
Mrk\,290).
\par\noindent Ws \dash\ N(\HI) from a Westerbork map only (0159+625 in
complex~H).
\par\noindent AP \dash\ N(\HI) from a combination of ATCA and Parkes data (1
arcmin beam). This was done for NGC\,3783 and HD\,101274 (Lu et al.\ 1998).
\par\noindent WJ \dash\ N(\HI) from a combination of Westerbork and Jodrell Bank
data (2 arcmin beam). This was done for HD\,135485 and 4\,Lac by Stoppelenburg
et al.\ (1998).
\par\noindent JD \dash\ N(\HI) from a combination of Jodrell Bank and DRAO (2
arcmin) data. This was done for the stars in M\,13 (Shaw et al.\ 1996).
\par\noindent Ar \dash\ N(\HI) from a spectrum taken at Arecibo (3 arcmin beam).
This is the case for \NAr\ probes in the \HI\ absorption studies of Payne et
al.\ (1980), Colgan et al.\ (1990) and Akeson \& Blitz (1999).
\par\noindent Ef, ef \dash\ N(\HI) derived from a spectrum taken at Effelsberg
(9\farcm1 beam). For \NEffOld\ probes in complexes~A, C, H, and K this is a
published value (code ``ef'') (Lilienthal et al.\ 1990, de Boer et al.\ 1994 and
Centuri\'on et al.\ 1994). For \NEffNew\ other probes new spectra were taken
with Effelsberg (code ``Ef''), see Paper~II).
\par\noindent JB \dash\ N(\HI) derived from a published spectrum taken with the
Jodrell Bank Mark\,III telescope (12 arcmin beam). New gaussian fits were made
to \NJB\ of the probes in Ryans et al.\ (1997a, b), for which no Effelsberg
spectrum was obtained (see Paper~II). These usually agree with the published
values, except when the intermediate-velocity \HI\ component is in the line
wing.
\par\noindent Pk \dash\ N(\HI) based on an observation at Parkes (15\arcmin\
beam). This is the case for \NPKSLMC\ stars.
\par\noindent PN \dash\ N(\HI) from a fit to a spectrum extracted from the
Parkes Narrow Band Survey (Haynes et al.\ 1999, \NPNarrow\ sightlines)). These
spectra are presented in Paper~II.
\par\noindent PK \dash\ N(\HI) from a fit to a spectrum extracted from the
Parkes Multi-Beam Survey ({\it HIPASS)} (Staveley-Smith 1997). These \NHIPASS\
spectra are presented in Paper~II.
\par\noindent GB \dash\ N(\HI) derived from an observation with the Green Bank
140-ft telescope (21 arcmin beam). For 14 extra-galactic sources this is a deep
observation done by Murphy (detection limit $\sim$3\tdex{17}\,\cmm2), see
summary in Savage et al.\ (2000a). For three stars (HD\,86248, HD\,137569 and
HD\,100340) values were taken from Danly et al.\ (1992) and Albert et al.\
(1993). Finally for HD\,93521 and HD\,215733 the component fits of Spitzer \&
Fitzpatrick (1993) and Fitzpatrick \& Spitzer (1997) were used.
\par\noindent PI \dash\ For some LMC stars, N(\HI) is found from an
interpolation between the value toward SN\,1987\,A (see code ``AB'' below) and
Parkes observations toward Sk$-$69\,246 and position JM43 in McGee \& Newton
(1986).
\par\noindent PR \dash\ For some LMC stars, N(\HI) in the +165, +120 and
+65\,\kms\ clouds was found by using a ruler on the plot of Wayte (1990). These
numbers are therefore not very reliable.
\par\noindent HC \dash\ For SN\,1991\,T a special 24-hour integration was done
at Hat Creek (see Paper~II), in order to achieve a 5 $\sigma$ upper limit of
3\tdex{17}\,\cmm2.
\par\noindent Dw \dash\ For \NLDSHI\ sightlines, N(\HI) is based on an
interpolation between gridpoints in the LDS (Hartmann \& Burton 1997) (36 arcmin
beam). That atlas gives the spectrum on every half degree in galactic longitude
and latitude. Therefore, for each velocity channel a weighted average was
constructed using the four gridpoints surrounding the direction to the probe;
the weights are given by max(0,0.5\to$R$), where $R$ is the distance to the
nearest gridpoint in degrees.
\par\noindent VE \dash\ For HD\,156359 the limit of 2\tdex{18}\,\cmm2\ is based
on the fact that Morras et al.\ (2000) do not list a component at this position,
as determined from observations with the IAR telescope at Villa Elisa.
\par\noindent AB \dash\ No \HI\ observation directly centered on SN\,1987\,A has
been published. Welty et al.\ (1999) derived the value of N(\HI) for each
spectral component from the abundance pattern of absorption, combined with the
likely abundance of undepleted elements. In particular, if solar abundance (from
Anders \& Grevesse 1989) and standard halo depletion (from Savage \& Sembach
1996a) are assumed, an implied N(\HI) can be derived from the column densities
of \MgII, \AlII, \SiII, \MnII\ and \FeII. This yields column densities of
8.0$\pm$2.5 and 5.4$\pm$1.9\tdex{18}\,\cmm2\ for the components at +120 and
+165\,\kms, respectively, which are consistent with the N(\HI) measured half a
degree away by McGee \& Newton (1986) in the directions toward Sk$-$69\,246 and
their position ``JM43''. Using this method for the IVC yields
N(\HI)=31$\pm$6\tdex{18}\,\cmm2, a factor 5 higher than the \HI\ spectrum would
suggest, indicating either no depletion, low abundance or a high ionization
fraction in this cloud. The analysis of Welty et al.\ (1999) does not
definitively allow a choice between these three possibilities. Here, a column
density of 6.5\tdex{18}\,\cmm2\ will be used, which is the average of the 6.1
and 7.1\tdex{18}\,\cmm2\ observed half a degree away.
%VALUES:
% ion  &   Asol &  depl  &  N65  &   NH65       & N120  & NH120 & N165  & NH165
%\MgII & --4.42 & --0.42 &       &              & 14.22 & 19.06 & 14.10 & 18.92
%\AlII & --5.52 & --0.27 &       &              & 12.97 & 18.76 & 12.71 & 18.50
%\SiII & --4.45 & --0.27 & 14.68 & 19.40(19.13) & 14.16 & 18.88 & 13.98 & 18.70
%\CrII & --6.32 & --0.50 & 12.78 & 19.60(19.10)
%\MnII & --6.47 & --0.59 & 12.40 & 19.46(18.87) & 11.70 & 18.76 & 11.60 & 18.66
%\FeII & --4.49 & --0.62 & 14.41 & 19.52(18.90) & 13.86 & 18.97 & 13.67 & 18.78
%avg   &        &        &       & 31.7   10.0  &       &  8.0  &       &  5.4
%rms                                6.3    3.1             2.5             1.9
%avg of log                        19.5   19.0            18.89           18.71
%rms of log avg                     0.85   0.14            0.13            0.15
%convert avg of log                31.6   10.0             7.8             5.1
\par\noindent ab \dash\ In the \HI\ spectrum of 3C\,123 three absorption
components are seen, but only one emission component. Except for the main
component, the absorption column density is therefore given in Col.~\CNion.
\par\noindent md \dash\ For Mrk\,509 and PKS\,2155$-$304, Sembach et al.\ (1999)
use the Green Bank 140-ft and find that N(\HI) is $<$6\tdex{17}\,\cmm2. A
photo-ionization model is used to convert the observed column densities of
several ions (\CIV, \CII, \SiIII) to a most likely value for N(H) = N(\HI) +
N(H$^+$). This depends on the assumed metallicity, which may be in the range
0.1\to1 times solar. For these two exceptional sightlines, the value in
Col.~\CNHI\ therefore is the total hydrogen column density, with an ``error''
expressing the possible range.

\par\noindent {\it Column \Cion}
\par The ion observed in absorption. Some entries between ``('' (in Col.~\Cion)
and '')'' (in Col.~\Cref) refer to measurements superseded by a more recent
reference.

\par\noindent {\it Column \Cvion}
\par The velocity (relative to the LSR) of the measured absorption line.

\par\noindent {\it Column \CNion}
This column gives the column density (and error or upper/lower limit) for the
observed ion, in units of \dex{11}\,\cmm2.

\par\noindent {\it Column \Cionflag}
A code showing how the column density was derived.
\par\noindent ``N'' \dash\ Column density quoted in reference. This is sometimes
derived from assuming a linear relation between equivalent width and column
density, but in many cases from a fit to the apparent optical depth profile or a
curve-of-growth.
\par\noindent ``L'' \dash\ The logarithm of the column density was given in the
reference. In this case the errors are usually given as logarithmic errors.
These are converted to a linear error by calculating
(dexp($L$+$\delta$)$-$dexp($L$$-$$\delta$))/2, where $L$ is the logarithmic
column density and $\delta$ the logarithmic error. E.g.\ log(N)=11.25$\pm$0.15
indicates a range of 1.3\to2.5\tdex{11}, and becomes N=1.8$\pm$0.6\tdex{11}.
\par\noindent ``W'' \dash\ The equivalent width is given in the reference. This
is converted to a column density by assuming that the gas temperature is 7000\,K
and using this to calculate the turbulent width from the observed \HI\ linewidth
($W$(\HI)): v$_{\rm turb}$ = sqrt($W$(HI)$^2$$-$$W$(T)$^2$), where $W$(T) is
given by sqrt(8\,ln2 kT/m(H)). The width of the absorption line for the ion is
then found as $W$(ion) = sqrt($W$(T,m(ion))$^2$+v$_{\rm turb}^2$). Then the
formula for the equivalent width (EW = $\int$ 1$-$exp($-\tau(v)$)) is inverted,
assuming $\tau(v)$ is gaussian. If the \HI\ linewidth is less than
$W$(7000)=17.9\,\kms, v$_{\rm turb}$=0 and a lower temperature are assumed and a
'V' is added to the notes column (this happens in \NWVT\ directions). If the
\HI\ linewidth was not measured (\NWHI\ directions as well as all LMC
sightlines), a value of 30\,\kms\ is assumed, and a 'W' is added to the notes
column. Note that the derived column density is not very sensitive to the
assumed temperature for weak lines.
\par\noindent ``E'' \dash\ Detection inferred from large equivalent width. This
is the case for \NFOS\ extra-galactic sources observed with the FOS (Burks et
al.\ 1994, Savage et al.\ 2000a). I.e., when expressed as a velocity width, the
equivalent width of the \MgII\ or \CII\ absorption line is as large or larger
than the total width of the \HI\ 21-cm line, including the HVC.
\par\noindent ``S'' \dash\ Upper limit to column density derived from a quoted
signal-to-noise (S/N) level, which is converted to an equivalent error using the
error formula for fitting a gaussian given by Kaper et al.\ (1963): $\sigma(W) =
{\lambda\over c}\,\sqrt{3\sqrt{\pi\over8\ln2}}\ \sqrt{W\,h}\ {1\over SN}$, where
the observed linewidth, $W$, and a grid spacing h=8\,\kms\ are used.
\par\noindent ``$\tau$'' \dash\ This happens for observations of 21-cm
absorption, and the optical depth of the absorption is given. The ion column
density column then contains the column density of the absorbing cool \HI\ (but
in units of \dex{18}\,\cmm2\ rather than \dex{11}\,\cmm2).
%\par\noindent ``$\alpha$'' \dash\ The abundance was derived from the ratio
%between the [\SII]$\lambda$6716 and \Ha\ intensities. This is the case for one
%direction in cloud MI.
\par Note that for rows giving a summary of cloud parameters, this column may
give, within square brackets a reference value for the column density of
highly-ionized atoms (\CIV, \NV, \OVI\ and \SiIV).

\par\noindent {\it Column \CAion}
\par The ratio N(ion)/N(\HI), in units of \dex{-9} (i.e., parts per billion or
ppb). In the case of \HI\ absorption the ratio N(\HI,absorption)/N(\HI,emission)
is given. These values should be compared to the expected abundance of the ion,
which is given in Col.~\CAref\ in the cloud overview (see Sect.~\Sion).

\par\noindent {\it Column \CAref}
\par For rows giving an overview of abundances for a particular cloud,
Col.~\CAref\ shows the expected abundance values (see Sect.~\Sion), in square
brackets. In the case of complexes~A and C all expected abundances are scaled
down by a factor 10, as an abundance of 0.089 times solar was found for sulphur
in complex~C (Wakker et al.\ 1999), and an oxygen abundance of $\sim$0.06 times
solar was suggested for complex~A (Kunth et al.\ 1994, Sect.~\Scloud.\SSA. Both
sulphur and oxygen are not or lightly depleted onto dust. For directions toward
the Magellanic Stream (both the trailing and leading arms), the expected
abundances are scaled down by a factor 4, as indicated by the sulphur abundances
found toward NGC\,3783 and Fairall\,9.

\par\noindent {\it Column \CAsolar}
\par The derived abundance relative to the solar abundance, given by Anders \&
Grevesse (1989), with updates for C, N and O from Grevesse \& Noels (1993).

\par\noindent {\it Column \Cdflag}
\par A flag indicating whether the result gives an upper or lower distance
limit. Detections toward stellar probes yield an upper limit (flag ``U'').
Non-detections may set a lower limit (flags ``l'' and ``L'', see below), but
only if it can be shown that the line should have been seen if the probe were
behind the cloud. To get to that point, there are several effects to consider.
\par Fine structure in the \HI\ distribution can lead to variations of up to a
factor 5 on arcminute scales in HVC cores (Wakker \& Schwarz 1991, Wakker et
al.\ 2001), although the variations may be less outside cores (Wakker et al.\
1996b). For eight probes N(\HI) has been observed at 35\arcmin, 9\arcmin\ and
2\arcmin\ resolution (see Paper~II). As the covering factor of the very bright
spots is low ($\sim$10\% of the surface area of the cloud), an observation with
a 10\to20\arcmin\ beam will usually give a column density accurate to within
50\%, i.e.\ a factor 2. This factor $<$2 is supported by the results of Savage
et al.\ (2000a). They compared N(\HI) measured using Ly$\alpha$ with the value
measured in 21-cm emission toward 12 extra-galactic probes. The Ly$\alpha$
values are 60\to90\% of those measured at 21-cm. A more detailed discussion of
the comparison between values of N(\HI) derived with different beams is
presented in Paper~II.
\par The gaseous abundance of an ion may also vary within a cloud, as both the
depletion and ionization can vary. For dominant ions, variations are expected to
be small if the \HI\ column density is high, say above \dex{19}\,\cmm2, while
for non-dominant ions (e.g. \CaII\ and especially \NaI), ionization variations
are expected to be large. As discussed in Sect.~\Sion, multiple measurements in
the same cloud exist for several ions (mainly \NaI, \MgII, \SII, \CaII, \FeII).
As the quality of these measurements is not always the same, and the \HI\ values
are usually derived using different telescopes, no systematic study can be made.
Yet, it is clear that for N(\HI)$>$5\tdex{18}\,\cmm2\ most ions show variations
of at most a factor 3\to5 within a single cloud, whereas \NaI\ can vary by a
factor 20 for different sightlines through the same cloud (Meyer \& Lauroesch
1999; Sect.~\SionNa).
\par Thus, to determine whether a non-detection is significant and sets a lower
limit to the distance of a cloud, the observed limit is compared with an
expected value. This is done in terms of optical depths, since the absorption
has to be deep enough to be recognizable. The expected optical depth is
calculated by combining the expected column density for the ion in question with
the measured \HI\ line width. The expected column density is found as the
product of the \HI\ column density and either the measured or the reference
abundance (see Sect.~\Sion\ and Col.~\CAref).
\par The expected optical depth is then divided by a safety factor, to account
for possible \HI\ small-scale structure and depletion and/or ionization
variations. \HI\ small-scale structure introduces a safety factor of 2. For
\NaI\ and for strongly depleted elements (atomic numbers $>$18, except Zn), the
depletion safety factor is 2.5; for other elements it is 1.5. The ionization
safety factor is 5 for \NaI, 2 for other elements, except for \CII, \MgII\ and
\OI, where it is 1. For \CII\ this is because the ionization potential of the
next ionization stage is high, and \CIII\ will only occur where H is (almost)
fully ionized. In the case of \OI, ionization is not a problem, as \OI\ and \HI\
are coupled through a strong charge-exchange reaction (Osterbrock 1989). A final
safety factor of 2 is introduced if the ion abundance had to be assumed. The
combined safety factor ranges from 3 for \OI, \CII, and \MgII\ for clouds where
their abundance has been measured, to typical factors of 10 for \CaII\ in clouds
with known abundance, 12 for \SiII\ and 20 for \FeII\ for clouds with assumed
abundance and a factor of 50 for \NaI\ if its abundance has not been measured in
the cloud.
\par If the safety-factor-multiplied expected optical depth is larger than the
observed lower limit, the non-detection is considered significant enough to set
a lower limit to the cloud's distance (flag ``l''). If the expected value is
more than 3 times larger than the observed limit, the distance limit is
considered strong (flag ``L'').
\par For HVC complex~H and the Anti-Center clouds, no direct abundance
measurement exists, but in the framework of the ``Local Group model'' (Blitz et
al.\ 1999) an abundance of 0.1 times solar is predicted. Therefore, the safety
factor needs to be calculated as if the abundance were 1/10th solar.

\par\noindent {\it Column \Cref\ \dash\ References}
1: Prata \& Wallerstein 1967;\ 
2: Wesselius \& Fejes 1973;\ 
3: Savage \& de Boer 1979;\ 
4: Cohen \hbox {} 1979;\ 
5: Payne et al.\ 1978, 1980;\ 
6: Blades 1980, Blades \& Meaburn 1980;\ 
7: Meaburn \hbox {} 1980;\ 
8: Songaila \& York 1981;\ 
9: Songaila \hbox {} 1981;\ 
10: Savage \& de Boer 1981;\ 
11: Savage \& Jeske 1981;\ 
12: Songaila et al. 1981;\ 
13: York et al. 1982;\ 
14: Blades et al. 1982;\ 
15: de Boer \& Savage 1983;\ 
16: Albert \hbox {} 1983;\ 
17: de Boer \& Savage 1984;\ 
18: Ferlet et al. 1985a;\ 
19: Songaila et al. 1985;\ 
20: d'Odorico et al. 1985;\ 
21: Kulkarni et al. 1985;\ 
22: West et al. 1985;\ 
23: Songaila et al. 1986;\ 
24: Kulkarni \& Mathieu 1986;\ 
25: Morton \& Blades 1986;\ 
26: Magain \hbox {} 1987;\ 
27: Vidal-Madjar et al. 1987, Andreani et al. 1987;\ 
28: de Boer et al. 1987;\ 
29: Dupree et al. 1987;\ 
30: Keenan et al. 1988;\ 
31: d'Odorico et al. 1989;\ 
32: Conlon et al. 1988;\ 
33: Songaila et al. 1988;\ 
34: Blades et al. 1988;\ 
35: Meaburn et al. 1989;\ 
36: Molaro et al. 1989;\ 
37: Savage et al. 1989;\ 
38: Lilienthal et al. 1990;\ 
39: Colgan et al. 1990;\ 
40: Wayte \hbox {} 1990;\ 
41: Langer et al. 1990;\ 
42: Bates et al.\ 1990, 1991;\ 
43: Wakker et al. 1991;\ 
44: Mebold et al. 1991;\ 
45: Sembach et al. 1991;\ 
46: Quin et al. 1991;\ 
47: Meyer \& Roth 1991;\ 
48: Robertson et al. 1991;\ 
49: Bowen et al.\ 1991a, 1991b;\ 
50: Danly et al. 1992;\ 
51: Spitzer \& Fitzpatrick 1992;\ 
52: Molaro et al. 1993;\ 
53: de Boer et al. 1993;\ 
54: Sembach et al. 1993;\ 
55: Bruhweiler et al. 1993;\ 
56: Spitzer \& Fitzpatrick 1993;\ 
57: Danly et al. 1993;\ 
58: Albert et al. 1993;\ 
59: Kunth et al. 1994;\ 
60: de Boer et al. 1994;\ 
61: Vladilo et al.\ 1993, 1994;\ 
62: Centuri{\accent 19 o}n et al. 1994;\ 
63: Bowen et al. 1994;\ 
64: Lu et al. 1994a;\ 
65: Little et al. 1994;\ 
66: Lu et al. 1994b;\ 
67: Burks et al. 1994;\ 
68: Kemp et al. 1994;\ 
69: King et al. 1995;\ 
70: Schwarz et al. 1995;\ 
71: Sembach et al. 1995;\ 
72: Lipman \& Pettini 1995;\ 
73: Ho \& Filippenko 1995, 1996;\ 
74: Bowen et al. 1995a;\ 
75: Bowen et al. 1995b;\ 
76: Savage et al. 1995;\ 
77: Keenan et al. 1995;\ 
78: Danly et al. 1995;\ 
79: Wakker et al. 1996a;\ 
80: Welsh et al. 1996;\ 
81: Bomans et al. 1996;\ 
82: Ryans et al. 1996;\ 
83: Kuntz \& Danly 1996;\ 
84: Benjamin et al. 1996;\ 
85: Caulet \& Newell 1996;\ 
86: Savage \& Sembach 1996b;\ 
87: Wakker et al. 1996b;\ 
88: Shaw et al. 1996;\ 
89: Lockman \& Savage 1995;\ 
90: Fitzpatrick \& Spitzer 1997;\ 
91: Sahu \& Blades 1997;\ 
92: Tamanaha \hbox {} 1997;\ 
93: Ryans et al. 1997a;\ 
94: Ryans et al. 1997b;\ 
95: Stoppelenburg et al. 1998;\ 
96: Wakker et al. 1998;\ 
97: Tufte et al. 1998;\ 
98: Tripp et al. 1998;\ 
99: Jannuzi et al. 1998;\ 
100: Lu et al. 1998;\ 
101: Sahu \hbox {} 1998;\ 
102: Kennedy et al. 1998;\ 
103: Kemp \& Bates 1998;\ 
104: van Woerden et al. 1999a;\ 
105: Richter et al. 1999;\ 
106: Wakker et al. 1999b;\ 
107: Lehner et al. 1999a;\ 
108: Lehner et al. 1999b;\ 
109: Welty et al. 1999;\ 
110: Sembach et al.\ 1995, 1999;\ 
111: Meyer \& Lauroesch 1999;\ 
112: Akeson \& Blitz 1999;\ 
113: Ryans et al. 1999;\ 
114: van Woerden et al. 1999b;\ 
115: Combes \& Charmandaris 2000;\ 
116: Gringel et al. 2000;\ 
117: Bowen et al. 2000;\ 
118: Sembach et al. 2000;\ 
119: Murphy et al. 2000;\ 
120: Penton et al. 2000;\ 
121: Savage et al.\ 1993, 2000a;\ 
122: Gibson et al. 2000;\ 
123: Richter et al. 2001a;\ 
124: Bluhm et al. 2001;\ 
125: Gibson et al. 2001;\ 
126: Richter et al. 2001b;\ 
127: Sembach et al. 2001;\ 
128: Fabian et al. 2001;\ 
129: Howk et al. 2001;\ 

\par\noindent {\it Column \Cnote\ \dash\ Notes}
\let \eightrm =\rm \let \eightit =\it 
\par \noindent {$\tau $:} With the measured FWHM of the {\fam 0 \eightrm H}$\mskip \thickmuskip ${{\textfont 0=\scriptfont 1I}}\r \penalty 10000\ profile, the column density or equivalent width implies $\tau $$>$10.
\par \noindent {W:} For the conversion of equivalent width to column density, the {\fam 0 \eightrm H}$\mskip \thickmuskip ${{\textfont 0=\scriptfont 1I}}\r \penalty 10000\ linewidth is assumed to be 30\r \kern .16667em \r \hbox {km\r \kern .16667em s$^{-1}$}\r \penalty 10000\kern 0.05em\penalty 10000.
\par \noindent {V:} {\fam 0 \eightrm H}$\mskip \thickmuskip ${{\textfont 0=\scriptfont 1I}}\r \penalty 10000\ linewidth is $<$17.9\r \kern .16667em \r \hbox {km\r \kern .16667em s$^{-1}$}\r \penalty 10000\kern 0.05em\penalty 10000, implying T$<$7000\r \kern .16667em K; v$_{\fam 0 \eightrm turb}$=0 is assumed.
\par \noindent {T:} Last two data columns give spin temperature and width of absorption feature; the N(ion) column gives N({\fam 0 \eightrm H}$\mskip \thickmuskip ${{\textfont 0=\scriptfont 1I}}\r \penalty 10000\kern 0.1em\penalty 10000) of the cold, absorbing gas, in units of \r \hbox {10$^{18}$}\r \kern .16667em \r \hbox {cm$^{-2}$}\r \penalty 10000\r \kern .16667em \penalty 10000; the abundance column is N(abs)/N(em).
\par \noindent {s:} Approximate limit for N({\fam 0 \eightrm Si}$\mskip \thickmuskip ${{\textfont 0=\scriptfont 1II}}\r \penalty 10000\kern 0.1em\penalty 10000), assuming S/N=10 in the line; also: N({\fam 0 \eightrm C}$\mskip \thickmuskip ${{\textfont 0=\scriptfont 1II}}\r \penalty 10000\kern 0.1em\penalty 10000)$<$\r \hbox {10$^{13}$}\r \kern .16667em \r \hbox {cm$^{-2}$}\r \penalty 10000\kern 0.05em\penalty 10000.
\par \noindent {S:} Based on IUE spectra in which {\fam 0 \eightrm Si}$\mskip \thickmuskip ${{\textfont 0=\scriptfont 1II}}\r \penalty 10000\ is the strongest uncontaminated line; spectra for HD\r \kern .16667em 100600, HD\r \kern .16667em 97991 and HD\r \kern .16667em 121800 in Danly et al.\ (1992), others unpublished.
\par \noindent {f:} Low-velocity resolution (220\r \kern .16667em \r \hbox {km\r \kern .16667em s$^{-1}$}\r \penalty 10000\kern 0.1em\penalty 10000) FOS spectrum, HVC absorption implied by large equivalent width; {\fam 0 \eightrm H}$\mskip \thickmuskip ${{\textfont 0=\scriptfont 1I}}\r \penalty 10000\ spectrum in Lockman \& Savage (1995).
\par \noindent {1:} Also known as Mrk\r \kern .16667em 116; the measurement has a very large error.
\par \noindent {2:} Distance estimate of star has changed a lot. Brown et al.\ (1989) found it had type B1V, and z=18\r \kern .16667em kpc. Hambly et al.\ (1996) find z=4.6\r \kern .16667em kpc.
\par \noindent {3:} Error visually estimated from spectrum and quoted typical S/N.
\par \noindent {4:} Parallax is 2.19$\pm $0.92 mas, or $d$=0.45\r \kern .16667em kpc; 1.2\r \kern .16667em kpc is 1.5 $\sigma $ off; $d$=0.45\r \kern .16667em kpc would imply type B6IV or B5V.
\par \noindent {5:} Na limit not given, but estimated from S/N=50.
\par \noindent {6:} Is SAO\r \kern .16667em 6225 in Lilienthal et al.\ (1990).
\par \noindent {7:} Is SAO\r \kern .16667em 6253 in Lilienthal et al.\ (1990).
\par \noindent {8:} Is SAO\r \kern .16667em 14733 in Songaila et al.\ (1988).
\par \noindent {9:} Definitively shown to be stellar by Lilienthal et al.\ (1990).
\par \noindent {10:} In NGC\r \kern .16667em 3877, which has \r \hbox {{\fam 0 \eightrm v}$_{\fam 0 \eightrm lsr}$}\r \penalty 10000\penalty 10000=+910\r \kern .16667em \r \hbox {km\r \kern .16667em s$^{-1}$}\r \penalty 10000\kern 0.05em\penalty 10000.
\par \noindent {11:} Kuntz \& Danly (1996) give the distance to this star as 1.8\r \kern .16667em kpc.
\par \noindent {12:} The Hipparcos parallax is 3.86$\pm $1.17 mas, which gives $d$=250\r \kern .16667em pc, with 1.6\r \kern .16667em kpc at 2.5 $\sigma $; Lehner et al.\ (1999a) argue for a spectroscopic distance of 1.2\r \kern .16667em kpc.
\par \noindent {13:} Cloud 63 is $\sim $1\r \r $^{\circ }$\penalty 10000\ away, while cloud 51, at the southern edge of complex M, is $\sim $2\r \r $^{\circ }$\penalty 10000\ away; Ryans et al.\ (1997a) do not list this component in {\fam 0 \eightrm H}$\mskip \thickmuskip ${{\textfont 0=\scriptfont 1I}}\r \penalty 10000\r \kern .16667em \penalty 10000; it could be in the IV Arch.
\par \noindent {14:} 5 $\sigma $ upper limits for N({\fam 0 \eightrm H}$\mskip \thickmuskip ${{\textfont 0=\scriptfont 1I}}\r \penalty 10000\kern 0.1em\penalty 10000) with 12\r \r $^{\prime }$\penalty 10000\ beam, (Ryans et al.\ 1997); with a 20\r \r $^{\prime }$\penalty 10000\ beam N({\fam 0 \eightrm H}$\mskip \thickmuskip ${{\textfont 0=\scriptfont 1I}}\r \penalty 10000\kern 0.1em\penalty 10000)=3.5\r \hbox {$\cdot $10$^{18}$}\r \kern .16667em \r \hbox {cm$^{-2}$}\r \penalty 10000\ for BD\r \kern .16667em +38 2182 and 4.1\r \hbox {$\cdot $10$^{18}$}\r \kern .16667em \r \hbox {cm$^{-2}$}\r \penalty 10000\ for HD\r \kern .16667em 93521 (Danly et al.\ 1993).
\par \noindent {15:} Limits for N({\fam 0 \eightrm Na}$\mskip \thickmuskip ${{\textfont 0=\scriptfont 1I}}\r \penalty 10000\kern 0.1em\penalty 10000) and for N({\fam 0 \eightrm Ca}$\mskip \thickmuskip ${{\textfont 0=\scriptfont 1II}}\r \penalty 10000\kern 0.1em\penalty 10000) toward HD\r \kern .16667em 93521 are implied but not actually given, the errors for BD\r \kern .16667em +38 2182 are used as estimate.
\par \noindent {16:} Limit to N({\fam 0 \eightrm C}$\mskip \thickmuskip ${{\textfont 0=\scriptfont 1II}}\r \penalty 10000\kern 0.1em\penalty 10000), N({\fam 0 \eightrm O}$\mskip \thickmuskip ${{\textfont 0=\scriptfont 1I}}\r \penalty 10000\kern 0.1em\penalty 10000) and N({\fam 0 \eightrm Si}$\mskip \thickmuskip ${{\textfont 0=\scriptfont 1II}}\r \penalty 10000\kern 0.1em\penalty 10000) from assuming S/N=20 in the continuum.
\par \noindent {17:} Complex C in region $\ell $=80\r \r $^{\circ }$\penalty 10000\ $-$90\r \r $^{\circ }$\penalty 10000\kern 0.05em\penalty 10000, $b$=37\r \r $^{\circ }$\penalty 10000\ $-$43\r \r $^{\circ }$\penalty 10000\kern 0.05em\penalty 10000, and \r \hbox {{\fam 0 \eightrm v}$_{\fam 0 \eightrm lsr}$}\r \penalty 10000\ $>$$-$140; distance limit assumes HB classifications are correct for the BS stars.
\par \noindent {18:} Corrected for S$^{+2}$ and H$^+$ (i.e.\ (N(S$^+$)+N(S$^{+2}$))/(N(H$^0$)+N(H$^+$))), the S abundance is 0.089 solar.
\par \noindent {19:} The {\fam 0 \eightrm N}$\mskip \thickmuskip ${{\textfont 0=\scriptfont 1I}}\r \penalty 10000\ spectrum is noisy, and was fit by a single component; half of this was assigned to each {\fam 0 \eightrm H}$\mskip \thickmuskip ${{\textfont 0=\scriptfont 1I}}\r \penalty 10000\ component.
\par \noindent {20:} Total {\fam 0 \eightrm O}$\mskip \thickmuskip ${{\textfont 0=\scriptfont 1VI}}\r \penalty 10000\ column density split evenly between the two {\fam 0 \eightrm H}$\mskip \thickmuskip ${{\textfont 0=\scriptfont 1I}}\r \penalty 10000\ components.
\par \noindent {21:} Limit to N(CO) from non-detection of absorption (3-sigma significance), assuming excitation temperature of 10\r \kern .16667em K.
\par \noindent {22:} Danly et al.\ (1992) classify HD\r \kern .16667em 146813 as B1.5 at $d$=2.4\r \kern .16667em kpc, Diplas \& Savage (1994) give 2.6\r \kern .16667em kpc, I would find 3.3\r \kern .16667em kpc; the SIMBAD type B8V is consistent with the Hipparcos parallax of 2.41$\pm $0.78\r \kern .16667em mas, implying $d$=0.4\r \kern .16667em kpc.
\par \noindent {23:} A non-existent {\fam 0 \eightrm Na}$\mskip \thickmuskip ${{\textfont 0=\scriptfont 1I}}\r \penalty 10000\ detection at $-$133\r \kern .16667em \r \hbox {km\r \kern .16667em s$^{-1}$}\r \penalty 10000\ was also claimed.
\par \noindent {24:} Disputed, very high probability that the line is stellar; moreover, there is no {\fam 0 \eightrm H}$\mskip \thickmuskip ${{\textfont 0=\scriptfont 1I}}\r \penalty 10000\ at $-$136\r \kern .16667em \r \hbox {km\r \kern .16667em s$^{-1}$}\r \penalty 10000\kern 0.05em\penalty 10000, the velocity of the claimed absorption.
\par \noindent {25:} Complex C at $\ell $$>$109\r \r $^{\circ }$\penalty 10000\ and $b$$>$49\r \r $^{\circ }$\penalty 10000\kern 0.05em\penalty 10000.
\par \noindent {26:} Detection not mentioned in paper (Burks et al.\ 1994), but equivalent widths are 280\r \kern .16667em \r \hbox {km\r \kern .16667em s$^{-1}$}\r \penalty 10000\ for {\fam 0 \eightrm C}$\mskip \thickmuskip ${{\textfont 0=\scriptfont 1II}}\r \penalty 10000\ and 160\r \kern .16667em \r \hbox {km\r \kern .16667em s$^{-1}$}\r \penalty 10000\ for {\fam 0 \eightrm Si}$\mskip \thickmuskip ${{\textfont 0=\scriptfont 1II}}\r \penalty 10000\kern 0.05em\penalty 10000.
\par \noindent {27:} {\fam 0 \eightrm H}$\mskip \thickmuskip ${{\textfont 0=\scriptfont 1I}}\r \penalty 10000\ also has component at $-$118\r \kern .16667em \r \hbox {km\r \kern .16667em s$^{-1}$}\r \penalty 10000\ (2.7$\pm $0.3\r \hbox {$\times $10$^{18}$}\r \kern .16667em \r \hbox {cm$^{-2}$}\r \penalty 10000\kern 0.1em\penalty 10000).
\par \noindent {28:} Is HD\r \kern .16667em 100971.
\par \noindent {29:} Is Feige\r \kern .16667em 87 in Schwarz et al.\ (1995).
\par \noindent {30:} Complex C at $\ell $$>$101\r \r $^{\circ }$\penalty 10000\kern 0.05em\penalty 10000, $b$$<$48\r \r $^{\circ }$\penalty 10000\kern 0.05em\penalty 10000, and $\ell $=80\r \r $^{\circ }$\penalty 10000\ $-$100\r \r $^{\circ }$\penalty 10000\kern 0.05em\penalty 10000, $b$$<$43\r \r $^{\circ }$\penalty 10000\kern 0.05em\penalty 10000, plus \r \hbox {{\fam 0 \eightrm v}$_{\fam 0 \eightrm lsr}$}\r \penalty 10000\ $<$$-$140 in region of overlap with C\r \kern .16667em I.
\par \noindent {31:} This is a very doubtful {\fam 0 \eightrm S}$\mskip \thickmuskip ${{\textfont 0=\scriptfont 1II}}\r \penalty 10000\ component; {\fam 0 \eightrm H}$\mskip \thickmuskip ${{\textfont 0=\scriptfont 1I}}\r \penalty 10000\ has two components of 19.1 and 11.6\r \hbox {$\times $10$^{18}$}\r \kern .16667em \r \hbox {cm$^{-2}$}\r \penalty 10000\ at $-$137 and $-$102\r \kern .16667em \r \hbox {km\r \kern .16667em s$^{-1}$}\r \penalty 10000\kern 0.05em\penalty 10000.
\par \noindent {32:} Revised equivalent widths from Bowen et al.\ 1995b.
\par \noindent {33:} Source name is 0959+68W1 in FOS Key Project; {\fam 0 \eightrm H}$\mskip \thickmuskip ${{\textfont 0=\scriptfont 1I}}\r \penalty 10000\ has tail to $-$175\r \kern .16667em \r \hbox {km\r \kern .16667em s$^{-1}$}\r \penalty 10000\kern 0.05em\penalty 10000.
\par \noindent {34:} Complex C at $\ell $$<$80\r \r $^{\circ }$\penalty 10000\kern 0.05em\penalty 10000.
\par \noindent {35:} Is ``1749+096'' in Akeson \& Blitz (1999).
\par \noindent {36:} 4\r \kern .16667em Lac sets a distance limit on complex G, although this is not commented upon by Bates et al.\ (1991).
\par \noindent {37:} Components with \r \hbox {T$_{\fam 0 \eightrm B}$}\r \penalty 10000\ $>$0.5\r \kern .16667em K in the region $\ell $=123\r \r $^{\circ }$\penalty 10000\ to 133\r \r $^{\circ }$\penalty 10000\kern 0.05em\penalty 10000, $b$=$-$3\r \r $^{\circ }$\penalty 10000\ to +5\r \r $^{\circ }$\penalty 10000\kern 0.05em\penalty 10000.
\par \noindent {38:} Continuum source in Westerbork map of core of complex H.
\par \noindent {39:} Centuri{\accent 19 o}n et al.\ (1994) give a distance of 1.1 kpc for HD\r \kern .16667em 10125 and 4.1\r \kern .16667em kpc for HD\r \kern .16667em 13256.
\par \noindent {40:} Is Hiltner 198 in Centuri{\accent 19 o}n et al.\ (1994).
\par \noindent {41:} Is Hiltner 190 in Centuri{\accent 19 o}n et al.\ (1994).
\par \noindent {42:} Components in the region $\ell $=110\r \r $^{\circ }$\penalty 10000\ to 120\r \r $^{\circ }$\penalty 10000\kern 0.05em\penalty 10000.
\par \noindent {43:} Components within a few degrees of the core at $\ell $=132\r \r $^{\circ }$\penalty 10000\kern 0.05em\penalty 10000, $b$=$-$5\r \r $^{\circ }$\penalty 10000\kern 0.05em\penalty 10000.
\par \noindent {44:} The spectral type would suggest a distance of 3.4\r \kern .16667em kpc, if the extinction is assumed to be 1.6 mag, it may be 5 mag.
\par \noindent {45:} Complex H components outside the three cores.
\par \noindent {46:} Is ``0224+671'' in Akeson \& Blitz (1999).
\par \noindent {47:} Is ``0300+470'' in Akeson \& Blitz (1999).
\par \noindent {48:} The spectral type would suggest a distance of 2.1\r \kern .16667em kpc, if the extinction is indeed 2.4 mag, it may be 5 mag.
\par \noindent {49:} This star was erroneously called BD\r \kern .16667em +61 2619 by Centuri{\accent 19 o}n et al.
\par \noindent {50:} In the {\fam 0 \eightrm H}$\mskip \thickmuskip ${{\textfont 0=\scriptfont 1I}}\r \penalty 10000\ emission profile only a single, broad component at $\sim $$-$60\r \kern .16667em \r \hbox {km\r \kern .16667em s$^{-1}$}\r \penalty 10000\ can be discerned, listed on the first line. The $-$73 and $-$58\r \kern .16667em \r \hbox {km\r \kern .16667em s$^{-1}$}\r \penalty 10000\ components are only clearly seen in absorption; the listed {\fam 0 \eightrm H}$\mskip \thickmuskip ${{\textfont 0=\scriptfont 1I}}\r \penalty 10000\ column densities are those of the absorption.
\par \noindent {51:} Kulkarni \& Mathieu (1986) classify this as an O star at 2.5\r \kern .16667em kpc. According to SIMBAD it is a B star. According to Garmany et al.\ (1987) it is O5V at 8.0\r \kern .16667em kpc.
\par \noindent {52:} Error assumed from quoted S/N of 100, width 15\r \kern .16667em \r \hbox {km\r \kern .16667em s$^{-1}$}\r \penalty 10000\ and grid 6\r \kern .16667em \r \hbox {km\r \kern .16667em s$^{-1}$}\r \penalty 10000\kern 0.05em\penalty 10000.
\par \noindent {53:} Is SAO\r \kern .16667em 76016 in Songaila et al.\ (1988).
\par \noindent {54:} Is SAO\r \kern .16667em 76994 in Songaila et al.\ (1988).
\par \noindent {55:} Is SAO\r \kern .16667em 76980 in Songaila et al.\ (1988).
\par \noindent {56:} Is SAO\r \kern .16667em 76954 in Songaila et al.\ (1988).
\par \noindent {57:} Estimated S/N=20 for the IUE spectra.
\par \noindent {58:} A general 5\r \kern .16667em m\AA \relax \ detection limit is given.
\par \noindent {59:} Is ``0239+108'' in Akeson \& Blitz (1999).
\par \noindent {60:} Cloud 363 is the ``Giovanelli Stream'', is ``0428+20'' in Akeson \& Blitz (1999).
\par \noindent {61:} Cloud 525 (v=$-$231) is 1 degree away in the Dwingeloo survey.
\par \noindent {62:} Cloud 419 in pop.\ GCN $\sim $1 degree away.
\par \noindent {63:} N(H,total) based on photo-ionization models; N({\fam 0 \eightrm H}$\mskip \thickmuskip ${{\textfont 0=\scriptfont 1I}}\r \penalty 10000\kern 0.1em\penalty 10000)$<$0.6\r \hbox {$\times $10$^{18}$}\r \kern .16667em cm$^{-2}$ from Green Bank data.
\par \noindent {64:} Total {\fam 0 \eightrm O}$\mskip \thickmuskip ${{\textfont 0=\scriptfont 1VI}}\r \penalty 10000\ column density split in ratio 1:2 between the two components.
\par \noindent {65:} About 10\r \r $^{\circ }$\penalty 10000\ away from nearest GCN cloud; an unclassified {\fam 0 \eightrm H}$\mskip \thickmuskip ${{\textfont 0=\scriptfont 1I}}\r \penalty 10000\ HVC at $-$133\r \kern .16667em \r \hbox {km\r \kern .16667em s$^{-1}$}\r \penalty 10000\ is detected 15 arcmin away.
\par \noindent {66:} Is ``1829+29'' in Akeson \& Blitz (1999).
\par \noindent {67:} Is ``1901+319'' in Akeson \& Blitz (1999).
\par \noindent {68:} Is ``1828+487'' in Akeson \& Blitz (1999).
\par \noindent {69:} Is ``2037+511'' in Akeson \& Blitz (1999).
\par \noindent {70:} A deeper STIS spectrum suggests that the CIV component at $-$213 km/s may not be real.
\par \noindent {71:} Lockman \& Savage (1995) give a 72\r \kern .16667em \r \hbox {km\r \kern .16667em s$^{-1}$}\r \penalty 10000\ wide component from $-$150\r \kern .16667em \r \hbox {km\r \kern .16667em s$^{-1}$}\r \penalty 10000\ to $-$78\r \kern .16667em \r \hbox {km\r \kern .16667em s$^{-1}$}\r \penalty 10000\r \kern .16667em \penalty 10000; this was split in two.
\par \noindent {72:} Is ``1928+738'' in Akeson \& Blitz (1999); {\fam 0 \eightrm H}$\mskip \thickmuskip ${{\textfont 0=\scriptfont 1I}}\r \penalty 10000\ has tail to $-$161\r \kern .16667em \r \hbox {km\r \kern .16667em s$^{-1}$}\r \penalty 10000\kern 0.05em\penalty 10000.
\par \noindent {73:} {\fam 0 \eightrm H}$\mskip \thickmuskip ${{\textfont 0=\scriptfont 1I}}\r \penalty 10000\ has tail to $-$128\r \kern .16667em \r \hbox {km\r \kern .16667em s$^{-1}$}\r \penalty 10000\kern 0.05em\penalty 10000.
\par \noindent {74:} Is ``0538+498'' in Akeson \& Blitz (1999).
\par \noindent {75:} Average of values toward 5 probes with N({\fam 0 \eightrm H}$\mskip \thickmuskip ${{\textfont 0=\scriptfont 1I}}\r \penalty 10000\kern 0.1em\penalty 10000)$>$2.5\r \hbox {$\times $10$^{18}$}\r \kern .16667em \r \hbox {cm$^{-2}$}\r \penalty 10000\kern 0.05em\penalty 10000.
\par \noindent {76:} Low-velocity resolution (220\r \kern .16667em \r \hbox {km\r \kern .16667em s$^{-1}$}\r \penalty 10000\kern 0.1em\penalty 10000) FOS spectrum, but HVC component is separately seen; the actual EW values and velocities are from Jannuzi et al.\ (1998).
\par \noindent {77:} Is PG\r \kern .16667em 0007+106.
\par \noindent {78:} N({\fam 0 \eightrm H}$\mskip \thickmuskip ${{\textfont 0=\scriptfont 1I}}\r \penalty 10000\kern 0.1em\penalty 10000) from single-component fit, although there are two {\fam 0 \eightrm H}$\mskip \thickmuskip ${{\textfont 0=\scriptfont 1I}}\r \penalty 10000\ components: at +194\r \kern .16667em \r \hbox {km\r \kern .16667em s$^{-1}$}\r \penalty 10000\kern 0.05em\penalty 10000, N({\fam 0 \eightrm H}$\mskip \thickmuskip ${{\textfont 0=\scriptfont 1I}}\r \penalty 10000\kern 0.1em\penalty 10000)=75.3$\pm $0.8\r \hbox {$\times $10$^{18}$}, width 45.0\r \kern .16667em \r \hbox {km\r \kern .16667em s$^{-1}$}\r \penalty 10000\kern 0.05em\penalty 10000, and +140\r \kern .16667em \r \hbox {km\r \kern .16667em s$^{-1}$}\r \penalty 10000\kern 0.05em\penalty 10000, N({\fam 0 \eightrm H}$\mskip \thickmuskip ${{\textfont 0=\scriptfont 1I}}\r \penalty 10000\kern 0.1em\penalty 10000)=13.7$\pm $0.6\r \kern .16667em \r \hbox {km\r \kern .16667em s$^{-1}$}\r \penalty 10000\kern 0.05em\penalty 10000, width 27.

7\r \kern .16667em \r \hbox {km\r \kern .16667em s$^{-1}$}\r \penalty 10000\r \kern .16667em \penalty 10000; in absorption usually only one component can be discerned.
\par \noindent {79:} The {\fam 0 \eightrm Na}$\mskip \thickmuskip ${{\textfont 0=\scriptfont 1I}}\r \penalty 10000\ \ and {\fam 0 \eightrm Ca}$\mskip \thickmuskip ${{\textfont 0=\scriptfont 1II}}\r \penalty 10000\ \ absorptions toward Fairall 9 found by Songaila \& York (1981) were published again by Songaila (1981).
\par \noindent {80:} Lu et al.\ (1994) give EW({\fam 0 \eightrm Si}$\mskip \thickmuskip ${{\textfont 0=\scriptfont 1II}}\r \penalty 10000\penalty 10000-1260)=368\r \kern .16667em m\AA \relax , but this is for absorption from +160 to +300\r \kern .16667em \r \hbox {km\r \kern .16667em s$^{-1}$}\r \penalty 10000\kern 0.05em\penalty 10000, and is larger than possible for the observed {\fam 0 \eightrm H}$\mskip \thickmuskip ${{\textfont 0=\scriptfont 1I}}\r \penalty 10000\ linewidth of 37\r \kern .16667em \r \hbox {km\r \kern .16667em s$^{-1}$}\r \penalty 10000\kern 0.05em\penalty 10000. Possibly there is an extra component at v$\sim $+200\r \kern .16667em \r \hbox {km\r \kern .16667em s$^{-1}$}\r \penalty 10000\kern 0.05em\penalty 10000. The EW listed here is the maximum possible given the observed $W$({\fam 0 \eightrm H}$\mskip \thickmuskip ${{\textfont 0=\scriptfont 1I}}\r \penalty 10000\kern 0.1em\penalty 10000).
\par \noindent {81:} Lu et al.\ (1994) use a linear conversion from equivalent width.
\par \noindent {82:} Parallax measured as 1.84$\pm $1.22\r \kern .16667em mas; distance given as 6.2\r \kern .16667em kpc by Danly et al.\ (1992), as 6.6\r \kern .16667em kpc by Diplas \& Savage (1994).
\par \noindent {83:} Cloud WW84 is very centrally condensated, and a 10\r \r $^{\prime }$\penalty 10000\ Effelsberg beam centered on Mrk\r \kern .16667em 205 picks up some of the very bright emission from the cloud center. WSRT observations have not yet been properly combined with single-dish data. The tabulated value is the current best guess.
\par \noindent {84:} Parallax measured as 5.52$\pm $1.13\r \kern .16667em mas, revising distance down from 0.8\r \kern .16667em kpc in Danly et al.\ (1995), or 2.5\r \kern .16667em kpc in Albert et al.\ (1993); N({\fam 0 \eightrm H}$\mskip \thickmuskip ${{\textfont 0=\scriptfont 1I}}\r \penalty 10000\kern 0.1em\penalty 10000) from Stoppelenburg et al.\ (1998).
\par \noindent {85:} Probably circumstellar (Danly et al.\ 1996). Albert et al.\ (1993) do not actually claim this as a detection.
\par \noindent {86:} Is 3C\r \kern .16667em 206.
\par \noindent {87:} Nearest HVC (WW364, v=+120) is 6\r \r $^{\circ }$\penalty 10000\ away, but the region was only sampled on a 2$\times $2\r \r $^{\circ }$\penalty 10000\ grid.
\par \noindent {88:} In NGC\r \kern .16667em 4527, which has \r \hbox {{\fam 0 \eightrm v}$_{\fam 0 \eightrm lsr}$}\r \penalty 10000\penalty 10000=+1733\r \kern .16667em \r \hbox {km\r \kern .16667em s$^{-1}$}\r \penalty 10000\kern 0.05em\penalty 10000.
\par \noindent {89:} HVC\r \kern .16667em 100$-$7+100; N({\fam 0 \eightrm H}$\mskip \thickmuskip ${{\textfont 0=\scriptfont 1I}}\r \penalty 10000\kern 0.1em\penalty 10000) from Stoppelenburg et al.\ (1998); parallax is 1.54$\pm $0.52\r \kern .16667em mas, giving d$<$1.0\r \kern .16667em kpc (1 $\sigma $).
\par \noindent {90:} In NGC\r \kern .16667em 5194, which has \r \hbox {{\fam 0 \eightrm v}$_{\fam 0 \eightrm lsr}$}\r \penalty 10000\penalty 10000=+463\r \kern .16667em \r \hbox {km\r \kern .16667em s$^{-1}$}\r \penalty 10000\r \kern .16667em \penalty 10000; paper assumes a ratio N({\fam 0 \eightrm Na}$\mskip \thickmuskip ${{\textfont 0=\scriptfont 1I}}\r \penalty 10000\kern 0.1em\penalty 10000)/N({\fam 0 \eightrm H}$\mskip \thickmuskip ${{\textfont 0=\scriptfont 1I}}\r \penalty 10000\kern 0.1em\penalty 10000), but the implied N({\fam 0 \eightrm H}$\mskip \thickmuskip ${{\textfont 0=\scriptfont 1I}}\r \penalty 10000\kern 0.1em\penalty 10000) then is much larger than limit from the Dwingeloo profile.
\par \noindent {91:} In M\r \kern .16667em 81, which has \r \hbox {{\fam 0 \eightrm v}$_{\fam 0 \eightrm lsr}$}\r \penalty 10000\penalty 10000=$-$130\r \kern .16667em \r \hbox {km\r \kern .16667em s$^{-1}$}\r \penalty 10000\kern 0.05em\penalty 10000, $-$40\r \kern .16667em \r \hbox {km\r \kern .16667em s$^{-1}$}\r \penalty 10000\ \ at position of SN\r \kern .16667em 1993\r \kern .16667em J
\par \noindent {92:} In NGC\r \kern .16667em 4526, which has \r \hbox {{\fam 0 \eightrm v}$_{\fam 0 \eightrm lsr}$}\r \penalty 10000\penalty 10000=+625\r \kern .16667em \r \hbox {km\r \kern .16667em s$^{-1}$}\r \penalty 10000\kern 0.05em\penalty 10000, 880\r \kern .16667em \r \hbox {km\r \kern .16667em s$^{-1}$}\r \penalty 10000\ \ at position of SN\r \kern .16667em 1994\r \kern .16667em D; faint {\fam 0 \eightrm H}$\mskip \thickmuskip ${{\textfont 0=\scriptfont 1I}}\r \penalty 10000\ found at Arecibo; the erratum corrects the misconception that this originates inside NGC\r \kern .16667em 4526.
\par \noindent {93:} In NGC\r \kern .16667em 5128, which has \r \hbox {{\fam 0 \eightrm v}$_{\fam 0 \eightrm lsr}$}\r \penalty 10000\ $\sim $+400\r \kern .16667em \r \hbox {km\r \kern .16667em s$^{-1}$}\r \penalty 10000\r \kern .16667em \penalty 10000; HVCs 219 and 208 are 1\r \r $^{\circ }$\penalty 10000\ and 3\r \r $^{\circ }$\penalty 10000\ away.
\par \noindent {94:} In M\r \kern .16667em 83, which has \r \hbox {{\fam 0 \eightrm v}$_{\fam 0 \eightrm lsr}$}\r \penalty 10000\penalty 10000=320$-$620\r \kern .16667em \r \hbox {km\r \kern .16667em s$^{-1}$}\r \penalty 10000\kern 0.05em\penalty 10000.
\par \noindent {95:} Average of values toward SN\r \kern .16667em 1987\r \kern .16667em A and Sk$-$69 243. These are consistent with the values for other probes, for which N({\fam 0 \eightrm H}$\mskip \thickmuskip ${{\textfont 0=\scriptfont 1I}}\r \penalty 10000\kern 0.1em\penalty 10000) is uncertain and low.
\par \noindent {96:} IUE spectra without measurements are presented by de Boer et al.\ (1987) and Dupree et al.\ (1987).
\par \noindent {97:} {\fam 0 \eightrm Ca}$\mskip \thickmuskip ${{\textfont 0=\scriptfont 1II}}\r \penalty 10000\ profile was published, but not the measurements; estimated equivalent widths.
\par \noindent {98:} N({\fam 0 \eightrm H}$\mskip \thickmuskip ${{\textfont 0=\scriptfont 1I}}\r \penalty 10000\kern 0.1em\penalty 10000) from McGee \& Newton (1986).
\par \noindent {99:} Average of the values found toward Sk$-$67 104, Sk$-$67 166, and Sk$-$68 82.
\par \noindent {100:} Average of the values found toward SN\r \kern .16667em 1987\r \kern .16667em A, Sk$-$67 104, Sk$-$67 166, and Sk$-$68 82.
\par \noindent {101:} Average of the values found toward Sk$-$67 104, Sk$-$67 166, and Sk$-$68 82.
\par \noindent {102:} Average of the best determinations, toward SN\r \kern .16667em 1987\r \kern .16667em A, Sk$-$69 243, Sk$-$69 246, Sk$-$69 247, Sk$-$71 41, Sk$-$71 42 and Sk$-$71 45.
\par \noindent {103:} Average of the values found toward SN\r \kern .16667em 1987\r \kern .16667em A, Sk$-$67 104, Sk$-$67 166, Sk$-$68 82 and Sk$-$69 246.
\par \noindent {104:} Absorption occurs at all velocities between 0 and +300\r \kern .16667em \r \hbox {km\r \kern .16667em s$^{-1}$}\r \penalty 10000\r \kern .16667em \penalty 10000; results given for velocity range 120 to 190\r \kern .16667em \r \hbox {km\r \kern .16667em s$^{-1}$}\r \penalty 10000\r \kern .16667em \penalty 10000; for AlIII the 1854 line gives logN=11.64, the 1862 line has logN=12.08.
\par \noindent {105:} HDE\r \kern .16667em 268605.
\par \noindent {106:} N({\fam 0 \eightrm H}$\mskip \thickmuskip ${{\textfont 0=\scriptfont 1I}}\r \penalty 10000\kern 0.1em\penalty 10000) from McGee \& Newton (1986), position JM43; also has component at +182 (5.9\r \hbox {$\times $10$^{18}$}).
\par \noindent {107:} Value toward SN\r \kern .16667em 1987\r \kern .16667em A, which is the only reliable one; equivalent widths toward Sk$-$67 05, Sk$-$68 82 and Sk$-$71 3 are only rough values; Molaro et al.\ (1993) find an upper limit of 5.2 ppb toward Sk$-$69 246.
\par \noindent {108:} Average of the values toward Sk$-$67 05 and Sk$-$67 104; Blades et al.\ (1988) imply 2.6 solar toward SN\r \kern .16667em 1987\r \kern .16667em A, but that column density is very uncertain and was not measured by Welty et al.\ (1999).
\par \noindent {109:} The three measurements are very discrepant: 0.08 solar toward Sk$-$69 246 (Savage \& de Boer 1981), 0.85 solar toward Sk$-$67 104 (Savage \& Jeske 1981) and 2.1 solar toward SN\r \kern .16667em 1987\r \kern .16667em A (Welty et al.\ 1999).
\par \noindent {110:} Average of values toward SN\r \kern .16667em 1987\r \kern .16667em A, R\r \kern .16667em 139, R\r \kern .16667em 140, Sk$-$69 243, Sk$-$69 246, Sk$-$69 248, Sk$-$69 255, Sk$-$71 41, Sk$-$71 42 and Sk$-$71 45. For other measurements N({\fam 0 \eightrm H}$\mskip \thickmuskip ${{\textfont 0=\scriptfont 1I}}\r \penalty 10000\kern 0.1em\penalty 10000) is based on interpolation or reading off plots.
\par \noindent {111:} Value toward SN\r \kern .16667em 1987\r \kern .16667em A; toward Sk$-$69 246 Savage \& de Boer (1981) found 0.2 solar, but this is uncertain; the values toward Sk$-$67 05 and Sk$-$67 104 are for the combined +65 and +120\r \kern .16667em \r \hbox {km\r \kern .16667em s$^{-1}$}\r \penalty 10000\ components.
\par \noindent {112:} In absorption components are seen at $-$64, $-$45 and $-$30 \r \hbox {km\r \kern .16667em s$^{-1}$}\r \penalty 10000\kern 0.05em\penalty 10000, but in the {\fam 0 \eightrm H}$\mskip \thickmuskip ${{\textfont 0=\scriptfont 1I}}\r \penalty 10000\ spectrum these are blended. The fit was forced to have components at $-$65 and $-$45 \r \hbox {km\r \kern .16667em s$^{-1}$}\r \penalty 10000\kern 0.05em\penalty 10000.
\par \noindent {113:} At the edge of IV6, which is not separately visible in the {\fam 0 \eightrm H}$\mskip \thickmuskip ${{\textfont 0=\scriptfont 1I}}\r \penalty 10000\ emission spectrum.
\par \noindent {114:} {\fam 0 \eightrm Mg}$\mskip \thickmuskip ${{\textfont 0=\scriptfont 1II}}\r \penalty 10000\ column densities determined from the 1239/1240 lines are corrected downward by 0.67 dex, see Savage \& Sembach (1996a).
\par \noindent {115:} In the paper, the fits to these components are shown, but no numbers are given $-$ they were provided by Bowen.
\par \noindent {116:} Mixture of IV17 \& IV26; IVC absorption not listed in Schwarz et al.\ (1995).
\par \noindent {117:} Ca absorption does not resolve IV9 (v=$-$73, T$_B$=0.56 K) and IV19 (v=$-$48, T$_B$=2.36 K); column densities are combined values.
\par \noindent {118:} In the GHRS absorption spectrum IV9 and IV19 are hard to separate, the combined {\fam 0 \eightrm S}$\mskip \thickmuskip ${{\textfont 0=\scriptfont 1II}}\r \penalty 10000\ column density was proportionally divided between the two {\fam 0 \eightrm H}$\mskip \thickmuskip ${{\textfont 0=\scriptfont 1I}}\r \penalty 10000\ components.
\par \noindent {119:} Spectrum analyzed by Conlon et al.\ (1988); coordinates differ from those quoted by Kuntz \& Danly (1996), who also confuse z and d.
\par \noindent {120:} Is HZ\r \kern .16667em 22 in Ryans et al.\ (1997a).
\par \noindent {121:} The discrepancy between the ``normal'' value for the {\fam 0 \eightrm Ca}$\mskip \thickmuskip ${{\textfont 0=\scriptfont 1II}}\r \penalty 10000\ of IV15 abundance measured toward Mrk\r \kern .16667em 290 and the unusually high (by a factor 10) value toward BT\r \kern .16667em Dra suggests that the $-$83\r \kern .16667em \r \hbox {km\r \kern .16667em s$^{-1}$}\r \penalty 10000\ absorption toward BT\r \kern .16667em Dra may be stellar (its b-value is also unusually low).
\par \noindent {122:} Is HZ\r \kern .16667em 25 in ref.
\par \noindent {123:} Ryans et al.\ (1997a) call this star AG+53 783 and give $\ell $=154\r \r \hbox {$.\mkern -4mu^{\circ }$}\r \hskip -0.3333em\r \penalty 10000\ 4, $b$=56\r \r \hbox {$.\mkern -4mu^{\circ }$}\r \hskip -0.3333em\r \penalty 10000\ 6; is this the same star? Possibly it is a B9V star at 1.1\r \kern .16667em kpc; the identification of AG+53 783 with HDE\r \kern .16667em 233791 is based on SIMBAD.
\par \noindent {124:} {\fam 0 \eightrm O}$\mskip \thickmuskip ${{\textfont 0=\scriptfont 1I}}\r \penalty 10000\penalty 10000-1304, {\fam 0 \eightrm C}$\mskip \thickmuskip ${{\textfont 0=\scriptfont 1II}}\r \penalty 10000\penalty 10000-1334, {\fam 0 \eightrm Si}$\mskip \thickmuskip ${{\textfont 0=\scriptfont 1II}}\r \penalty 10000\ are seen in the line wing.
\par \noindent {125:} Is HIP55461.
\par \noindent {126:} In reference as H.O.+23B, to which SIMBAD has 11 references with coordinates differing by up to 5\r \r $^{\prime \prime }$\penalty 10000\r \kern .16667em \penalty 10000; z determined by Quin et al.\ (1991).
\par \noindent {127:} N({\fam 0 \eightrm H}$\mskip \thickmuskip ${{\textfont 0=\scriptfont 1I}}\r \penalty 10000\kern 0.1em\penalty 10000) fitted to a spectrum in which M\r \kern .16667em 81 was taken out by means of a third order polynomial fit; the uncertainty reflects the uncertainty in the zero level at v=$-$50\r \kern .16667em \r \hbox {km\r \kern .16667em s$^{-1}$}\r \penalty 10000\penalty 10000\r 
\par \noindent {128:} Assuming S/N=30.
\par \noindent {129:} BD names given in paper: HDE\r \kern .16667em 237844 = BD\r \kern .16667em +56 1411, HDE\r \kern .16667em 233622 = BD\r \kern .16667em +50 1631.
\par \noindent {130:} Combined component groups 1$-$2, 3$-$5 and 6$-$9 to form $-$92, $-$56 and $-$43\r \kern .16667em \r \hbox {km\r \kern .16667em s$^{-1}$}\r \penalty 10000\ components.
\par \noindent {131:} For the $-$56 and $-$43\r \kern .16667em \r \hbox {km\r \kern .16667em s$^{-1}$}\r \penalty 10000\ components, the (logarithmic) ionization fraction has been determined to be $-$3.6 to $-$1.3 and $-$3.6 to $-$1.9, respectively.
\par \noindent {132:} Penton et al.\ (2000) fit a component at +60\r \kern .16667em \r \hbox {km\r \kern .16667em s$^{-1}$}\r \penalty 10000\ to the {\fam 0 \eightrm S}$\mskip \thickmuskip ${{\textfont 0=\scriptfont 1II}}\r \penalty 10000\penalty 10000-1250, 1253 and 1259 lines, with EWs of 58, 85 and 30 \r \kern .16667em m\AA \relax ; {\fam 0 \eightrm S}$\mskip \thickmuskip ${{\textfont 0=\scriptfont 1II}}\r \penalty 10000\penalty 10000-1259 is blended with {\fam 0 \eightrm Si}$\mskip \thickmuskip ${{\textfont 0=\scriptfont 1II}}\r \penalty 10000\penalty 10000-1260. The weaker lines are shown by Sembach et al.\ (1999), and this shows that they half as strong. The listed column density is based on the lower EWs.
\par \noindent {133:} Originally found by York et al.\ (1982), analyzed by Blades \& Morton (1983), and improved by Morton \& Blades (1986).
\par \noindent {134:} N({\fam 0 \eightrm H}$\mskip \thickmuskip ${{\textfont 0=\scriptfont 1I}}\r \penalty 10000\kern 0.1em\penalty 10000) determined by interpolating between the values on a 3$\times $3 grid with 9\r \r $^{\prime }$\penalty 10000\ spacing.
\par \noindent {135:} M 15 NW/SE core refer to the positions with the highest and lowest {\fam 0 \eightrm Na}$\mskip \thickmuskip ${{\textfont 0=\scriptfont 1I}}\r \penalty 10000\ abundance measured using a 7x13 (27\r \r $^{\prime \prime }$\penalty 10000\ $\times $43\r \r $^{\prime \prime }$\penalty 10000\kern 0.1em\penalty 10000) array of 3 arcsec fibers.
\par \noindent {136:} In NGC\r \kern .16667em 1316, which has \r \hbox {{\fam 0 \eightrm v}$_{\fam 0 \eightrm lsr}$}\r \penalty 10000\penalty 10000=+1521\r \kern .16667em \r \hbox {km\r \kern .16667em s$^{-1}$}\r \penalty 10000\kern 0.05em\penalty 10000.
\par \noindent {137:} de Boer \& Savage (1984) claimed to have found this {\fam 0 \eightrm C}$\mskip \thickmuskip ${{\textfont 0=\scriptfont 1II}}\r \penalty 10000\ \ absorption line; however recent FUSE data (Howk priv.\ comm.) do not show the absorption in the {\fam 0 \eightrm C}$\mskip \thickmuskip ${{\textfont 0=\scriptfont 1II}}\r \penalty 10000\penalty 10000-1036 and other strong lines.

%%%%%%%%%%%%%%%%%%%%%%%%%%%%%%%%%%%%%%%%%%%%%%%%%%%%%%%%%%%%%%%%%%%%%%%%%%%%%%%%
%%%%%%%%%%%%%%%%%%%%%%%%%%%%%%%%%%%%%%%%%%%%%%%%%%%%%%%%%%%%%%%%%%%%%%%%%%%%%%%%
%%%%%%%%%%%%%%%%%%%%%%%%%%%%%%%%%%%%%%%%%%%%%%%%%%%%%%%%%%%%%%%%%%%%%%%%%%%%%%%%
%%%%%%%%%%%%%%%%%%%%%%%%%%%%%%%%%%%%%%%%%%%%%%%%%%%%%%%%%%%%%%%%%%%%%%%%%%%%%%%%

\def\fgnumber#1{\noindent Figure #1.\ } %AASPP

\def\InsertPage#1#2#3#4{\newpage\vbox to \vsize{\vss{\small#4}}%AASPP
                        \includegraphics{#1}} %AASPP

%%%%%%%%%%%%%%%%%%%%%%%%%%%%%%%%%%%%%%%%%%%%%%%%%%%%%%%%%%%%%%%%%%%%%%%%%%%%%%%%

\InsertPage{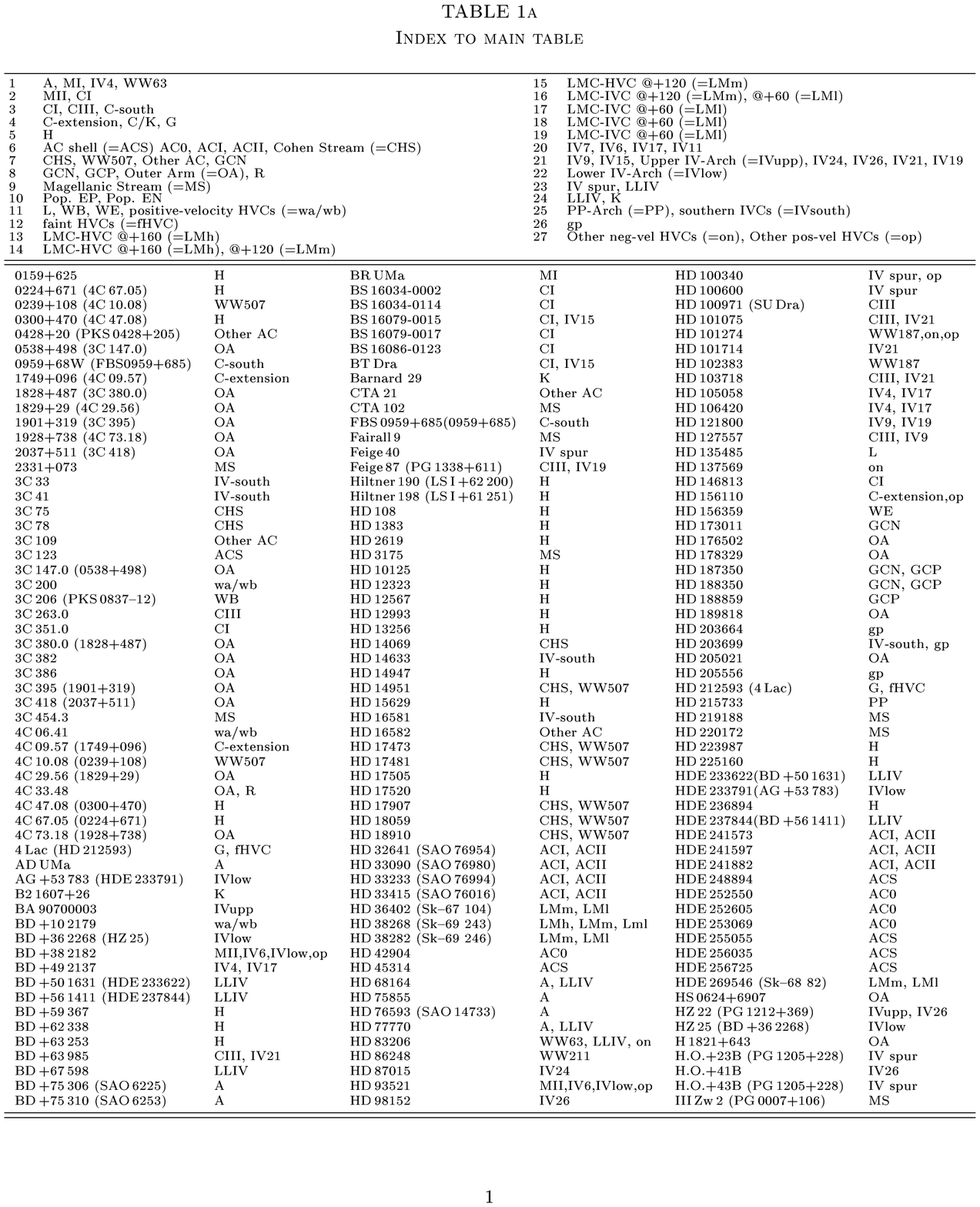}{-20}{-60}{\null} %AASPP
\InsertPage{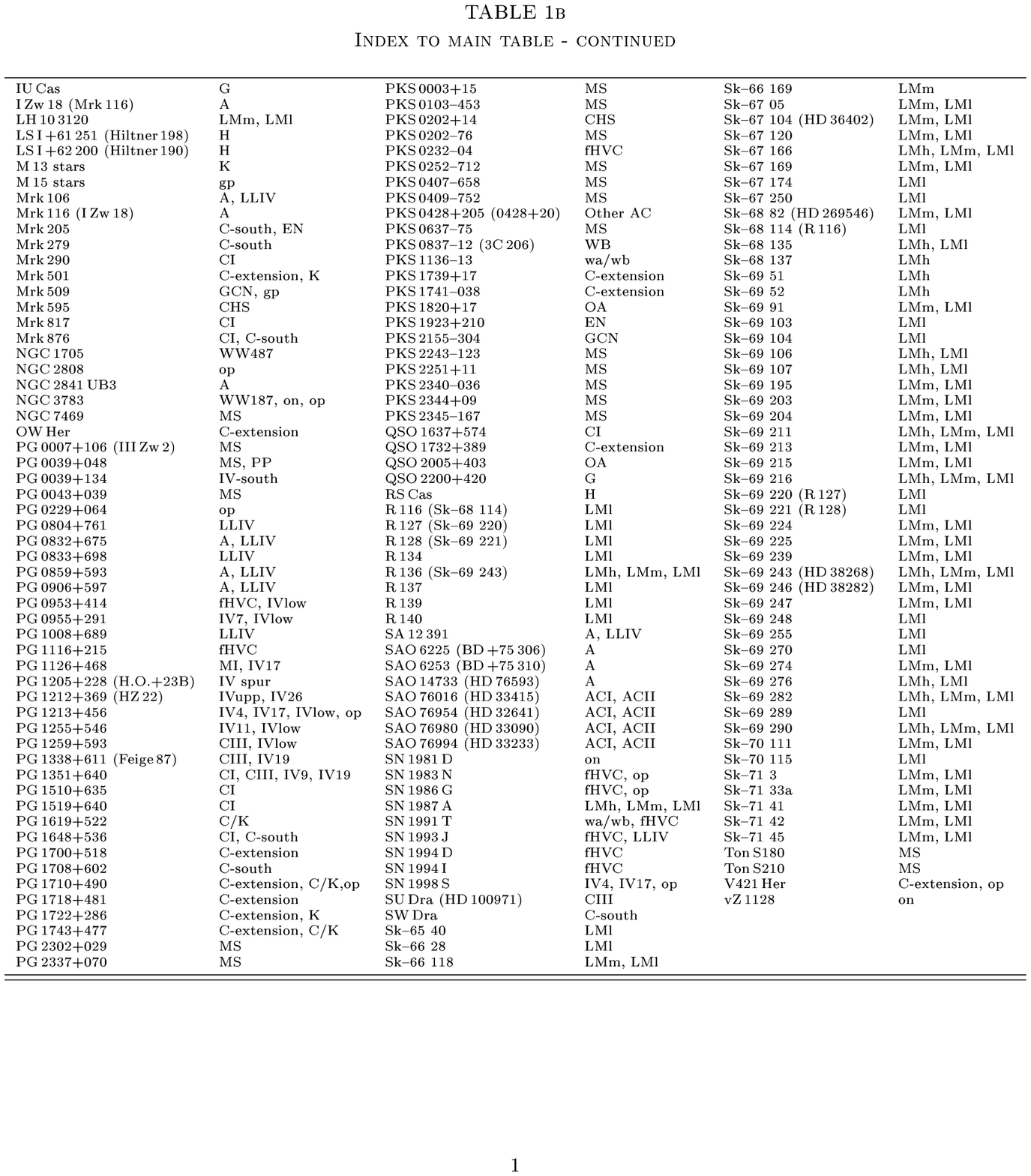}{-20}{-60}{\null} %AASPP
\InsertPage{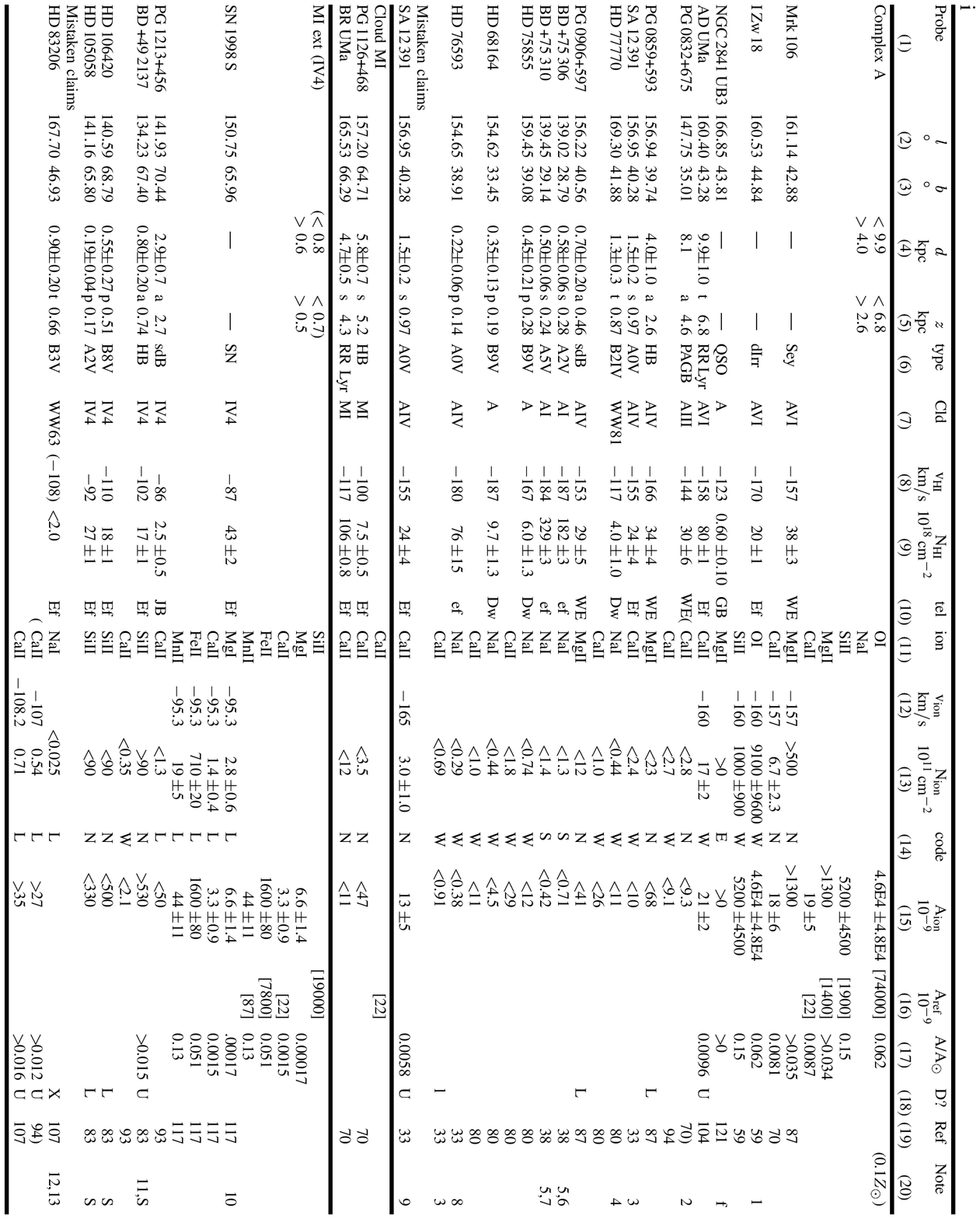}{-25}{-60}{\null} %AASPP
\InsertPage{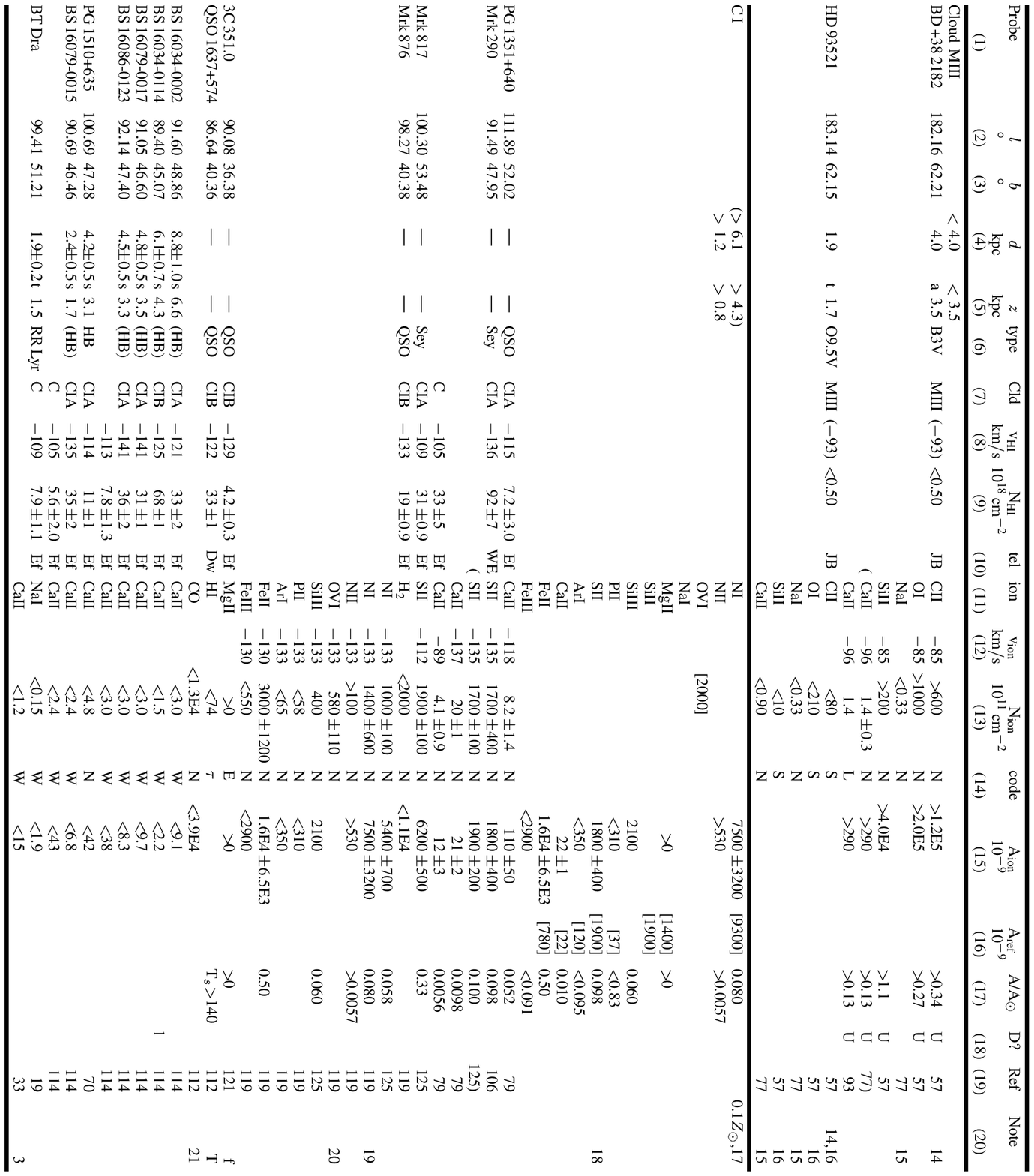}{-25}{-60}{\null} %AASPP
\InsertPage{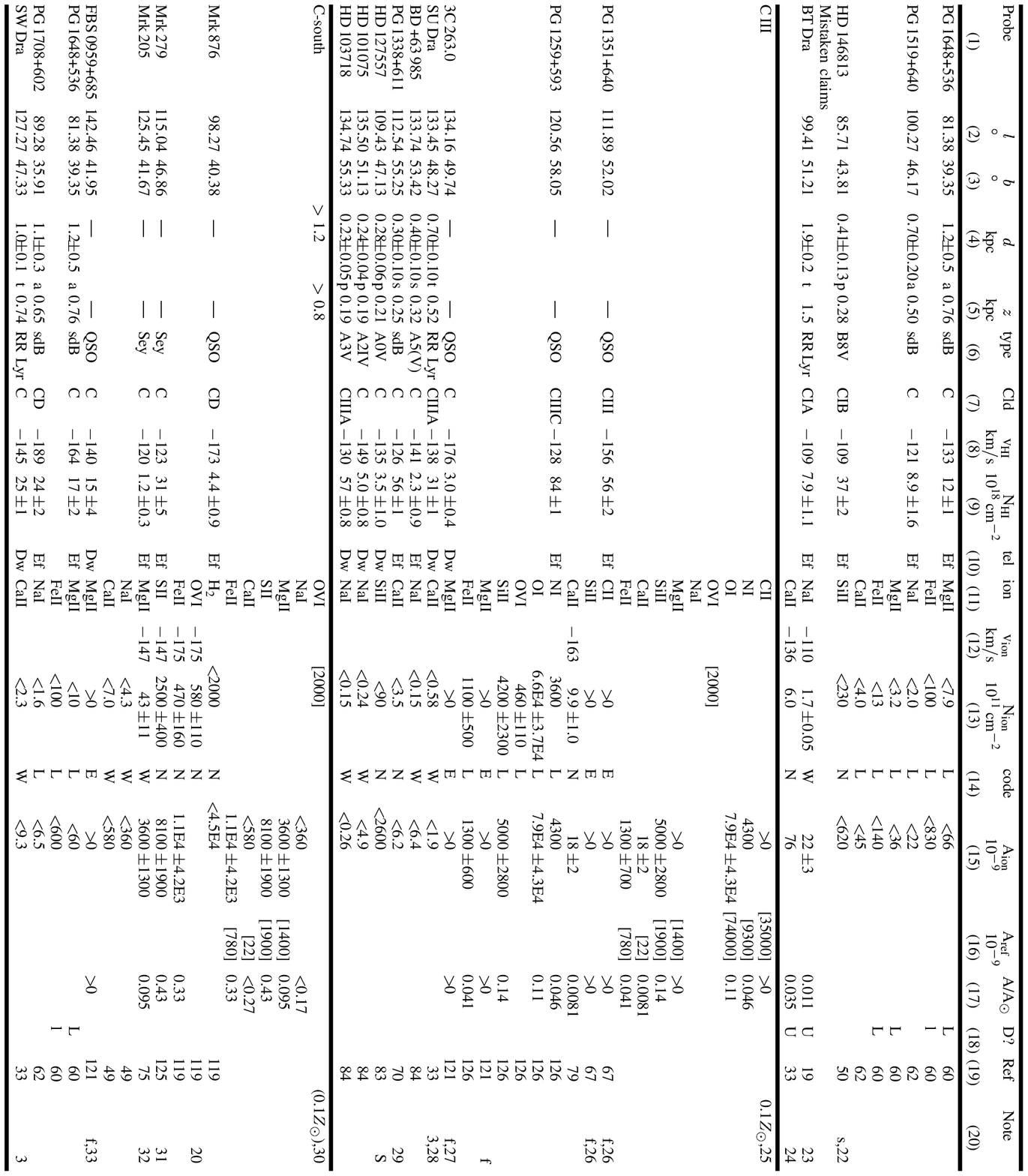}{-25}{-60}{\null} %AASPP
\InsertPage{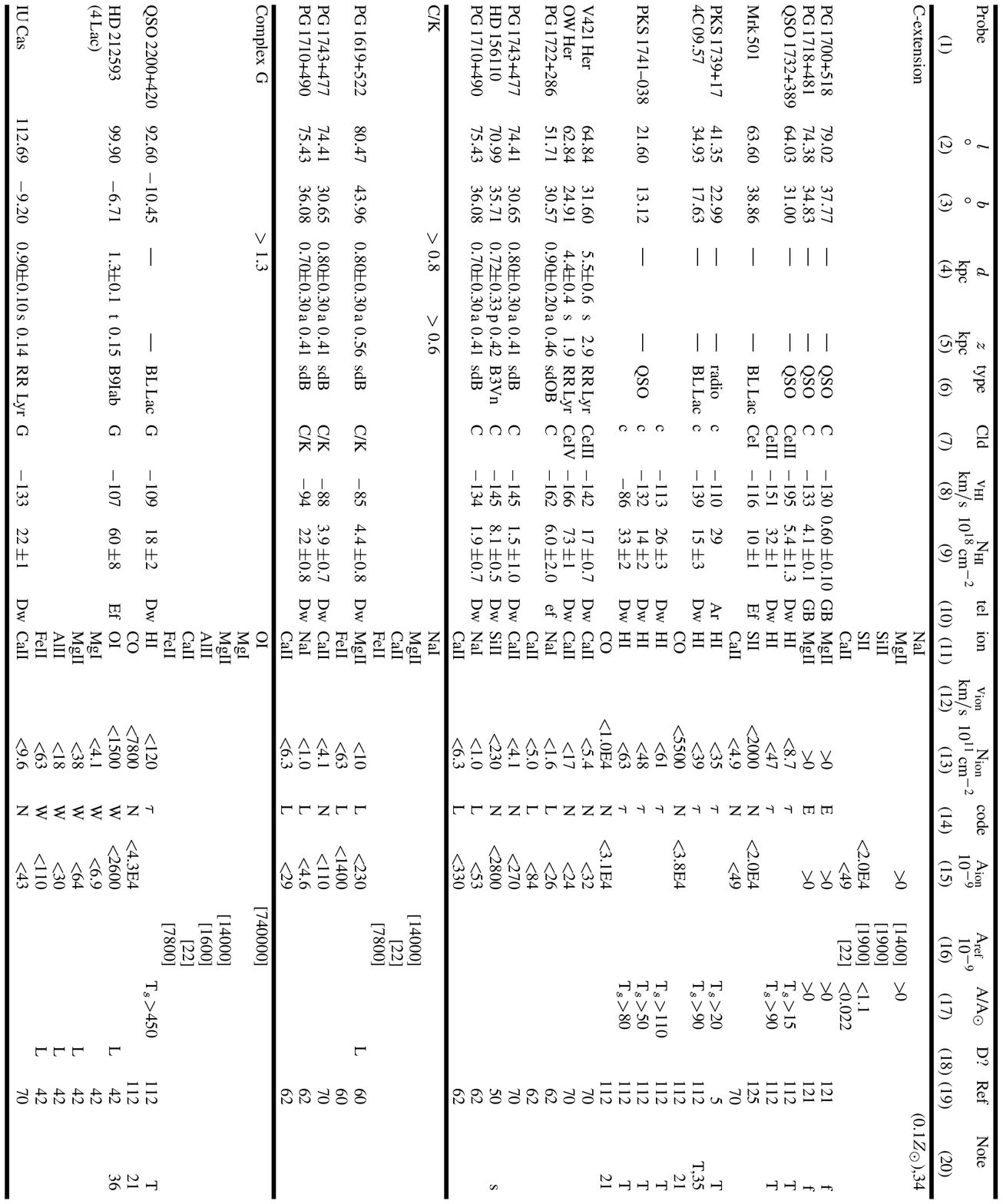}{-25}{-60}{\null} %AASPP
\InsertPage{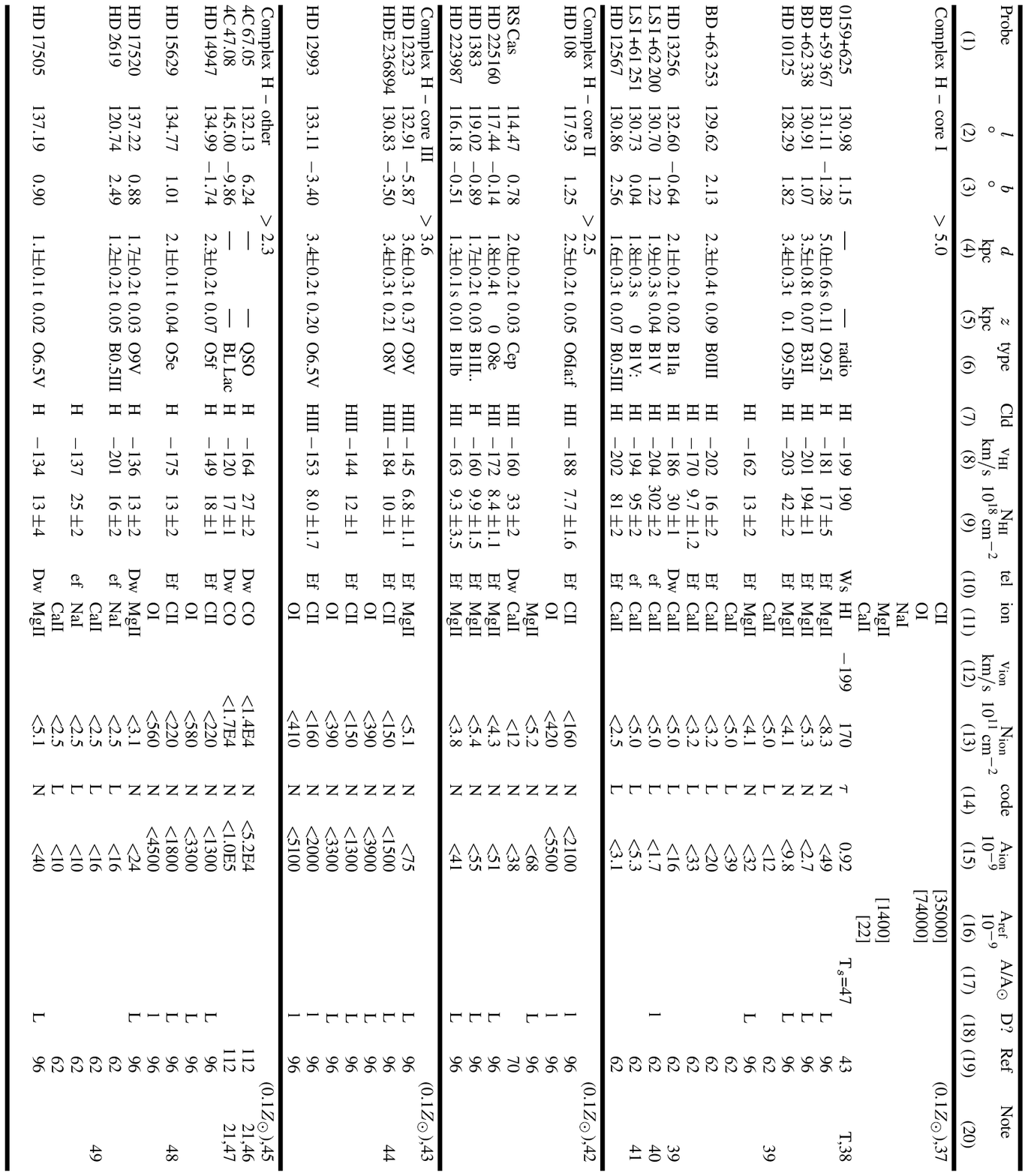}{-25}{-60}{\null} %AASPP
\InsertPage{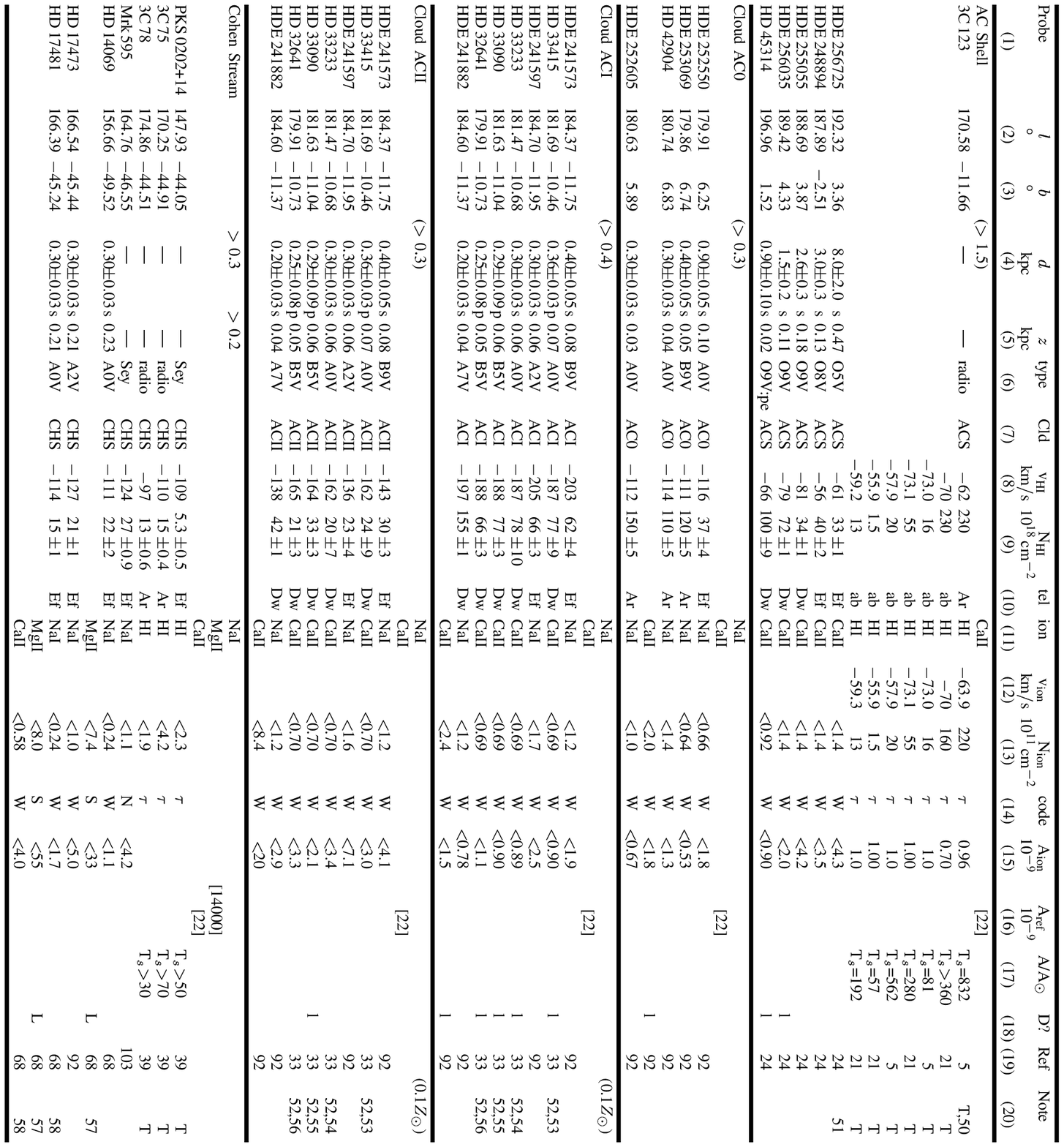}{-25}{-60}{\null} %AASPP
\InsertPage{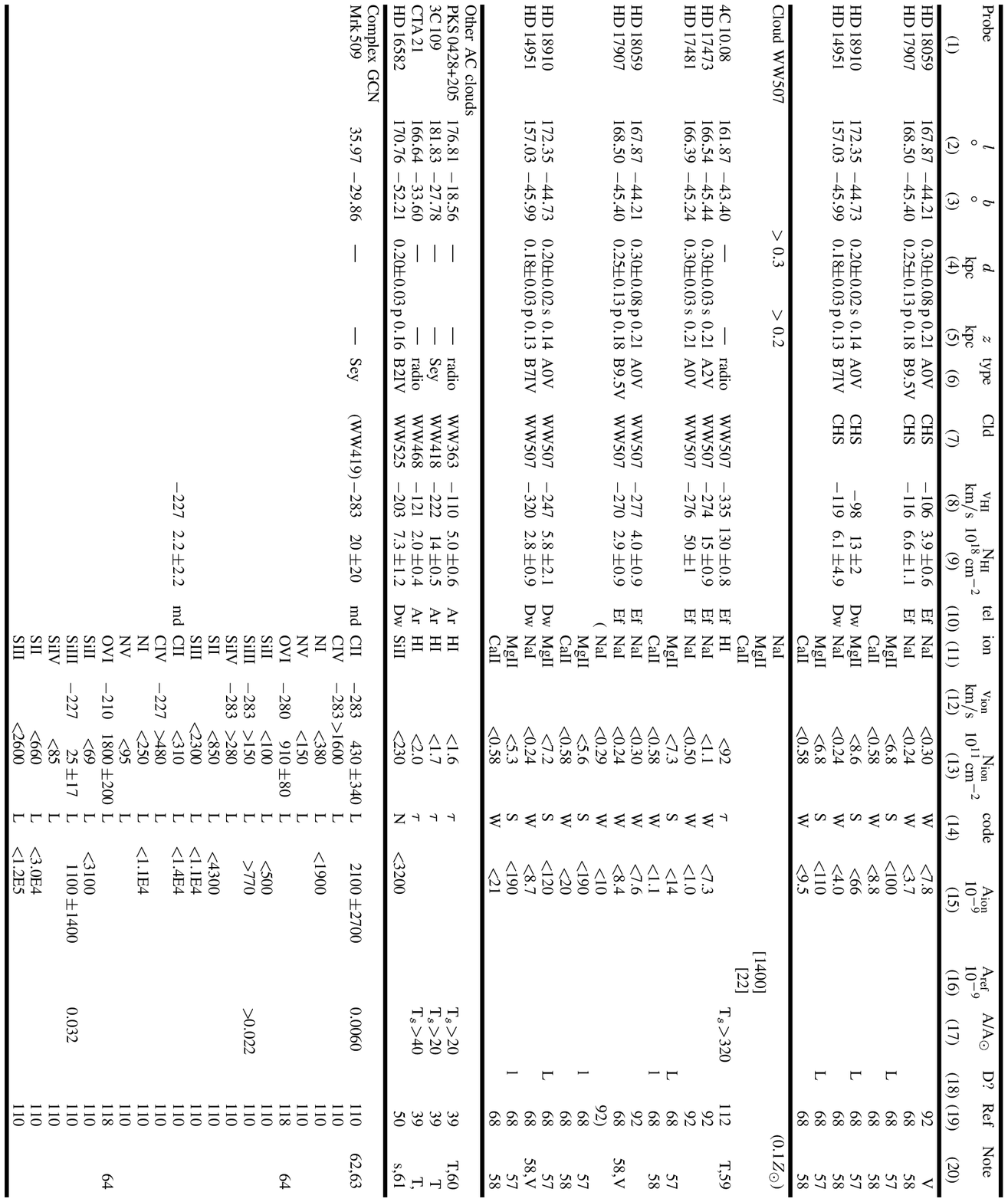}{-25}{-60}{\null} %AASPP
\InsertPage{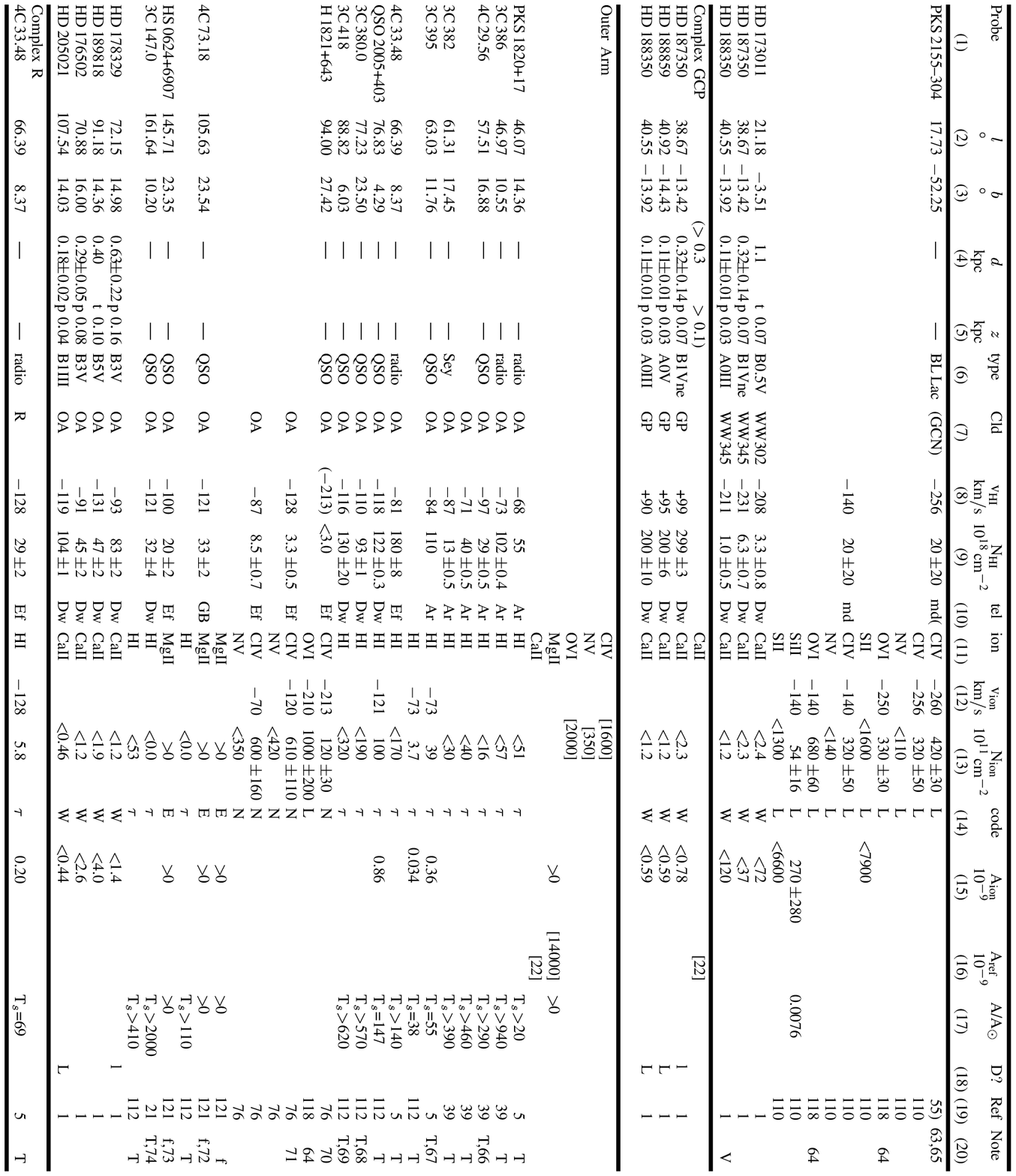}{-25}{-60}{\null} %AASPP
\InsertPage{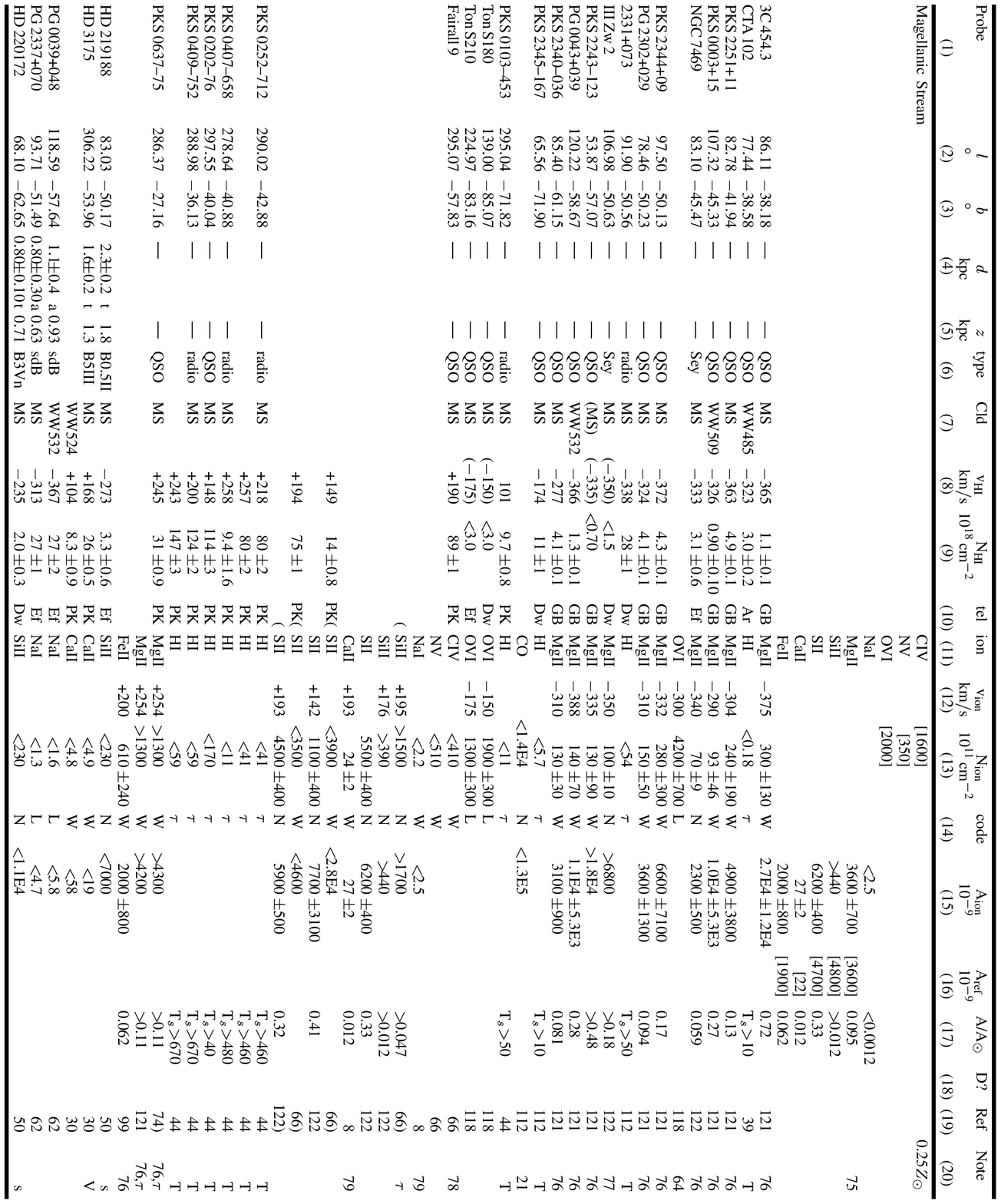}{-25}{-60}{\null} %AASPP
\InsertPage{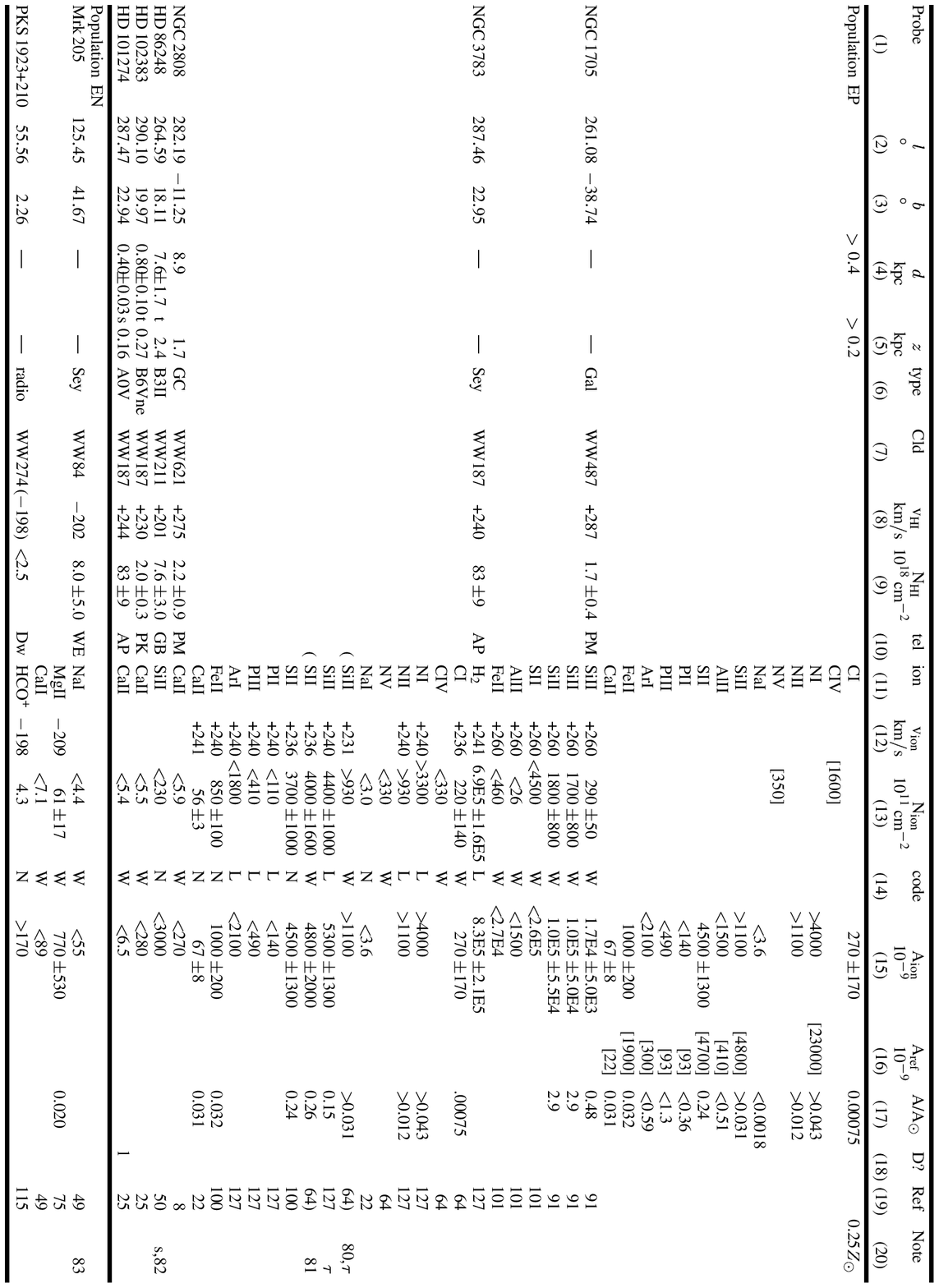}{-25}{-60}{\null} %AASPP
\InsertPage{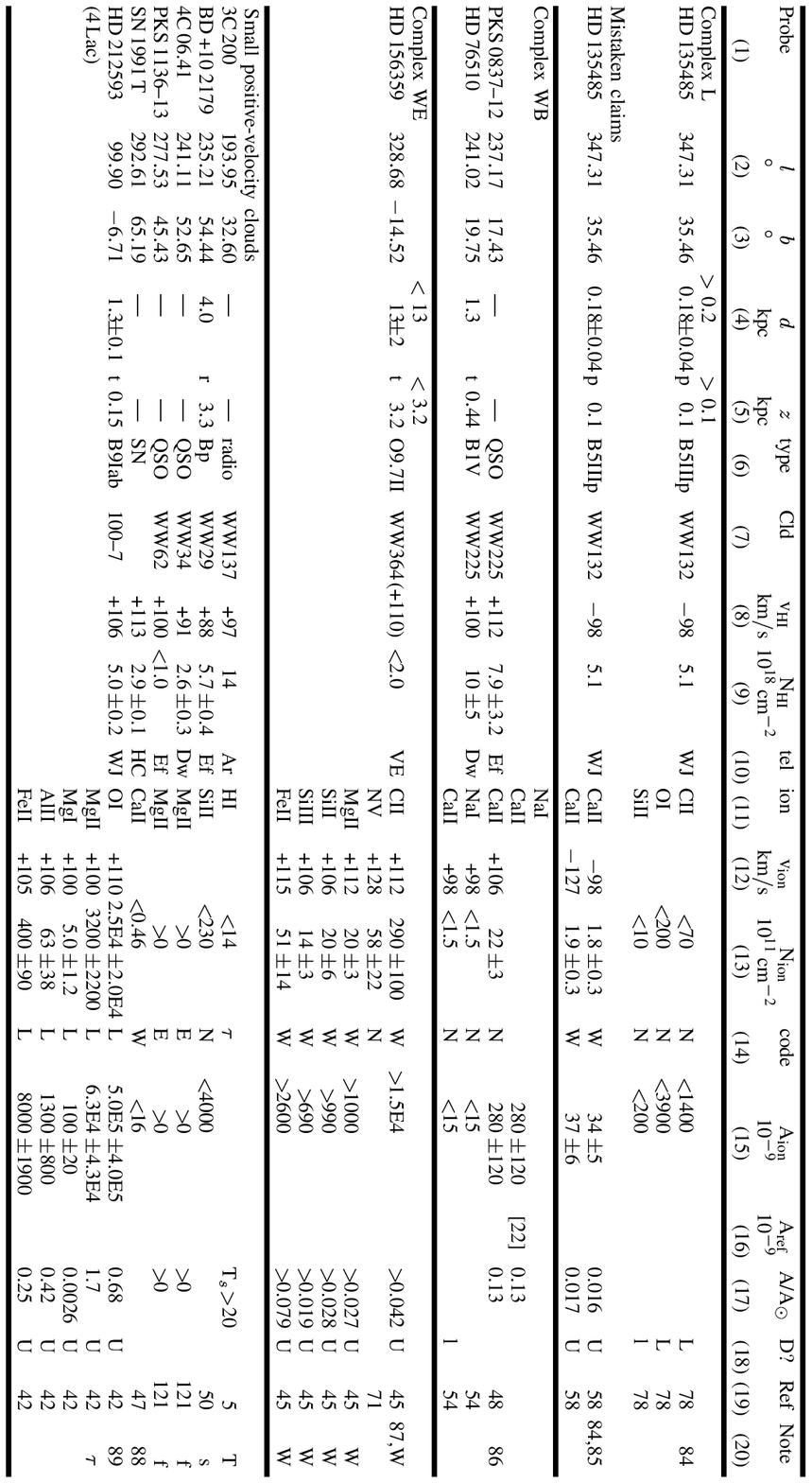}{-25}{-60}{\null} %AASPP
\InsertPage{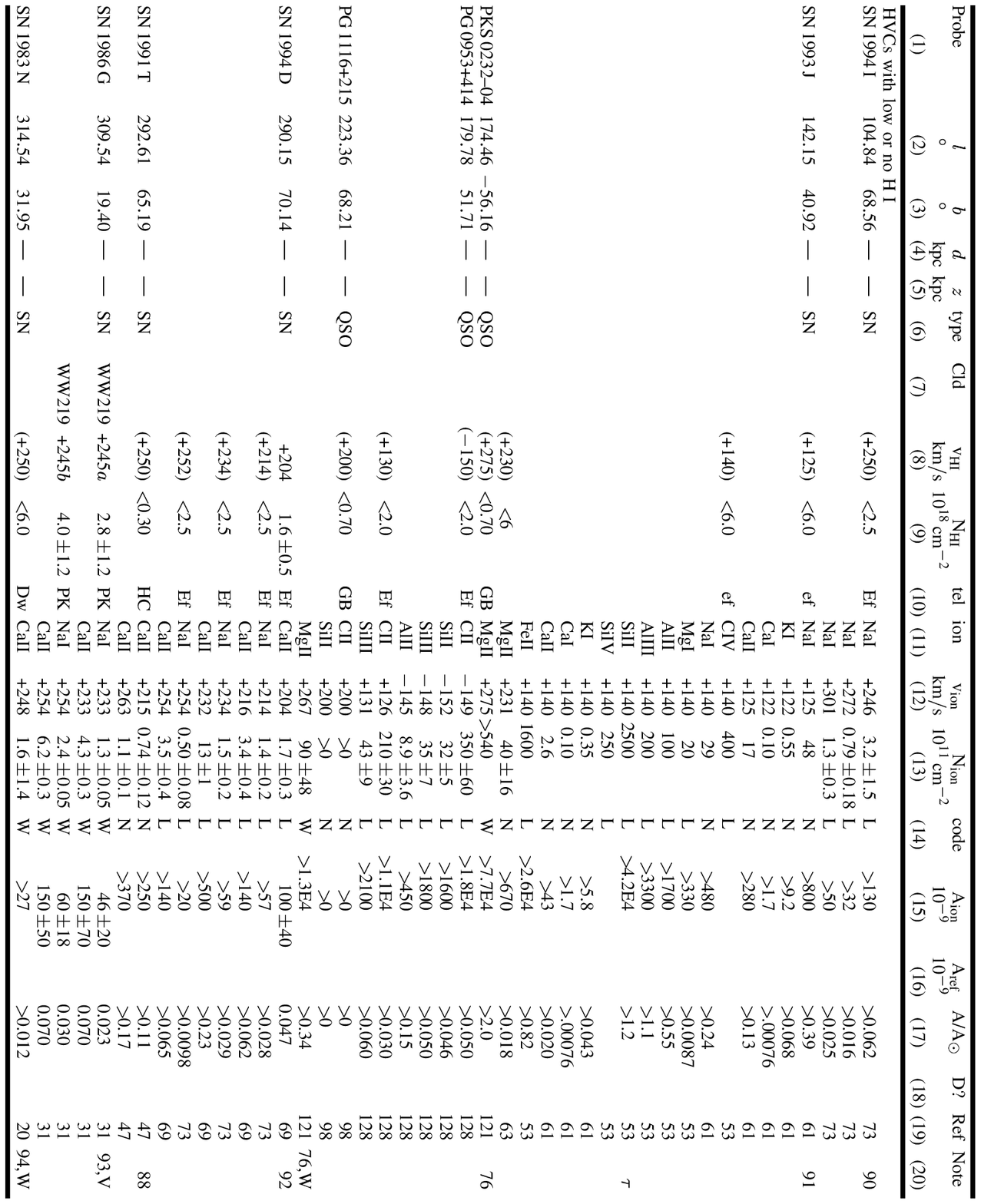}{-25}{-60}{\null} %AASPP
\InsertPage{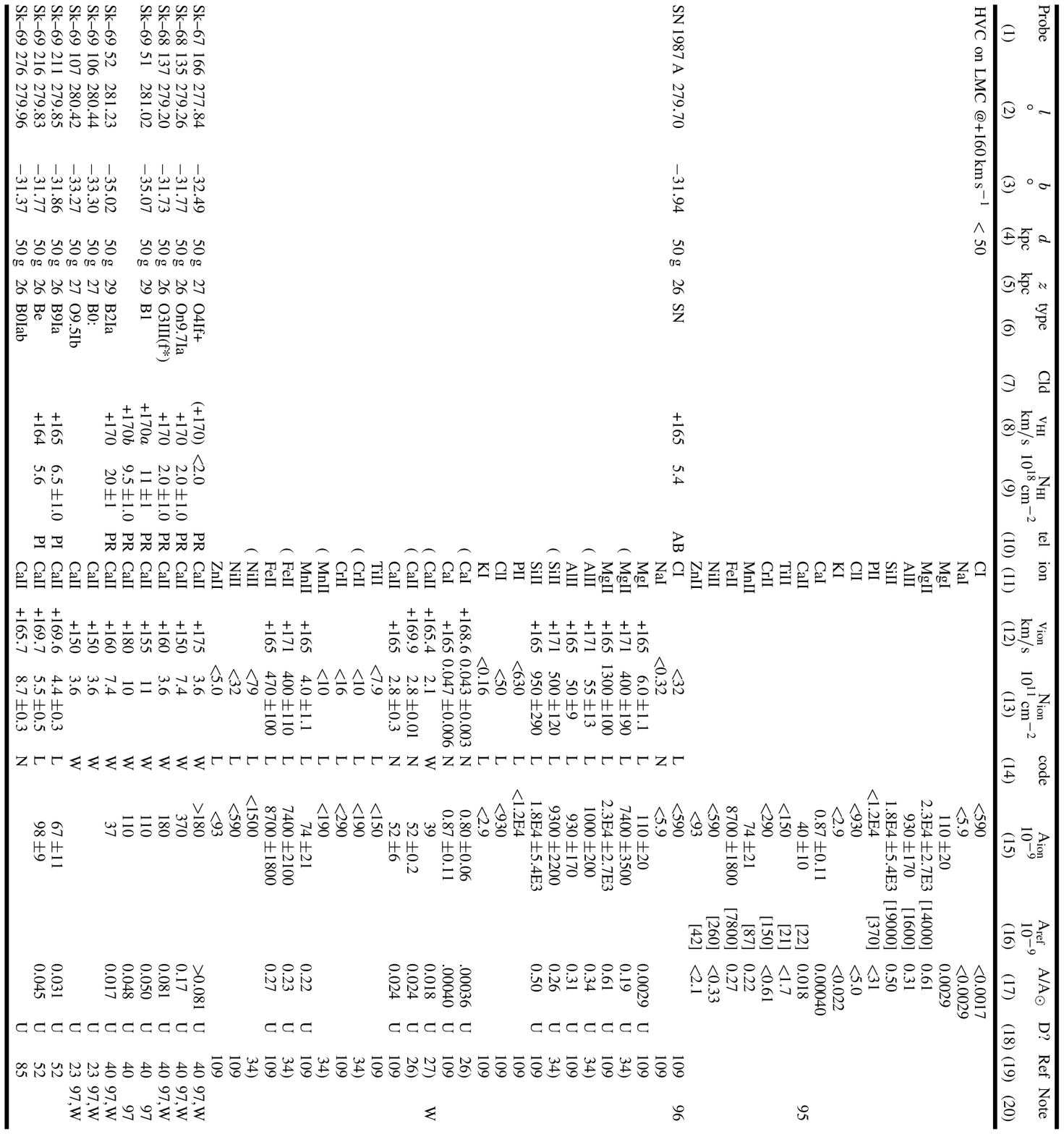}{-25}{-60}{\null} %AASPP
\InsertPage{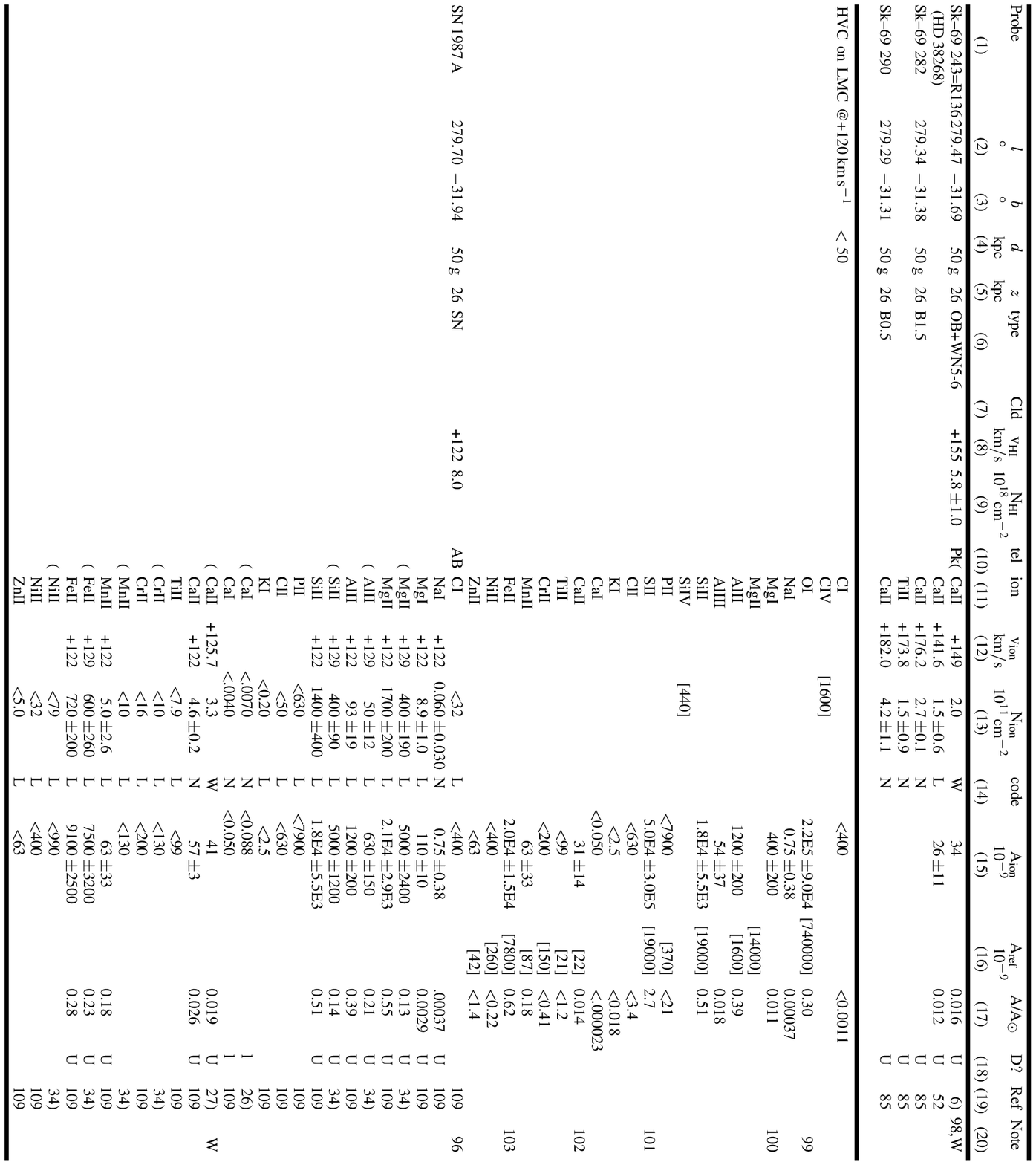}{-25}{-60}{\null} %AASPP
\InsertPage{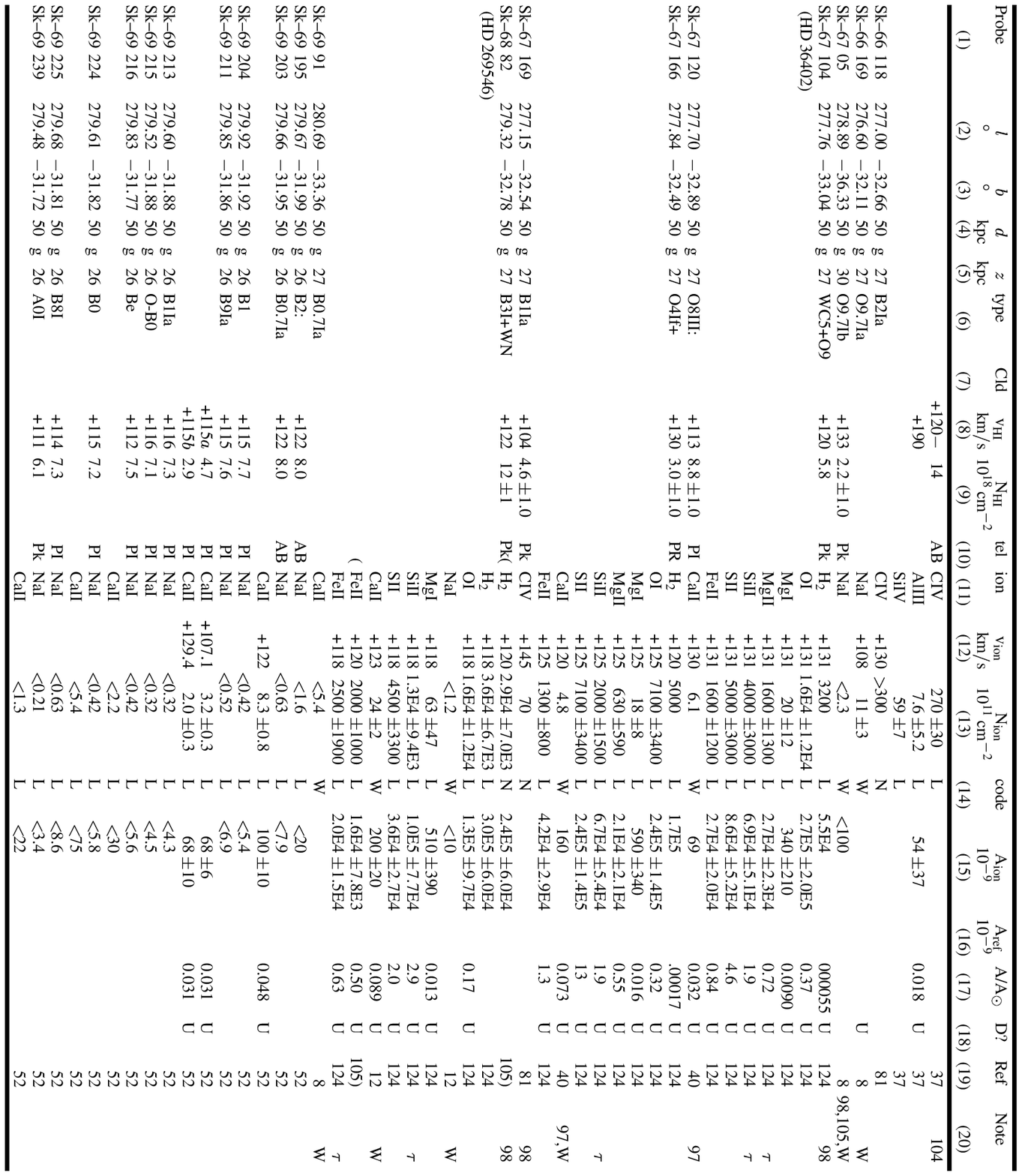}{-25}{-60}{\null} %AASPP
\InsertPage{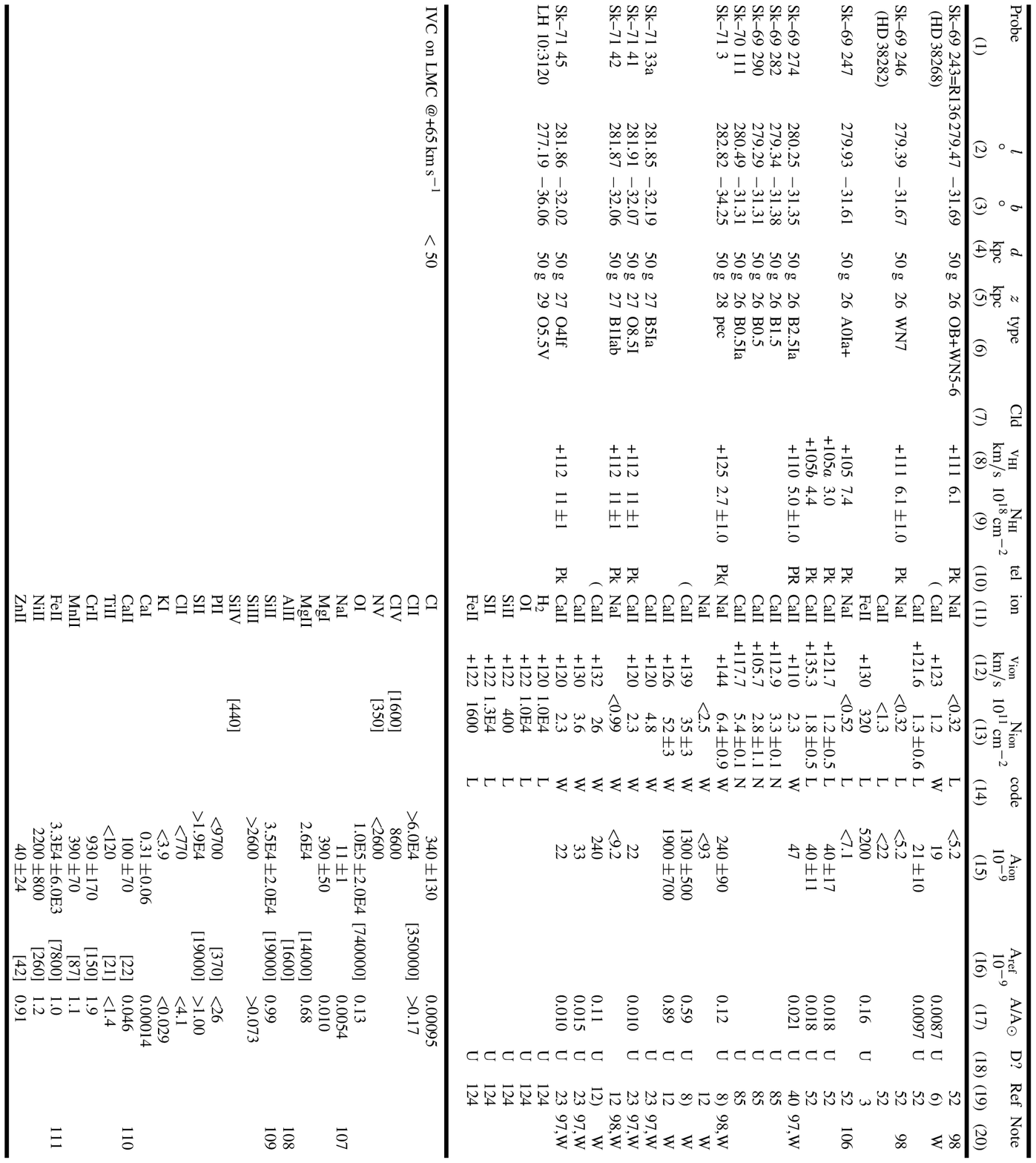}{-25}{-60}{\null} %AASPP
\InsertPage{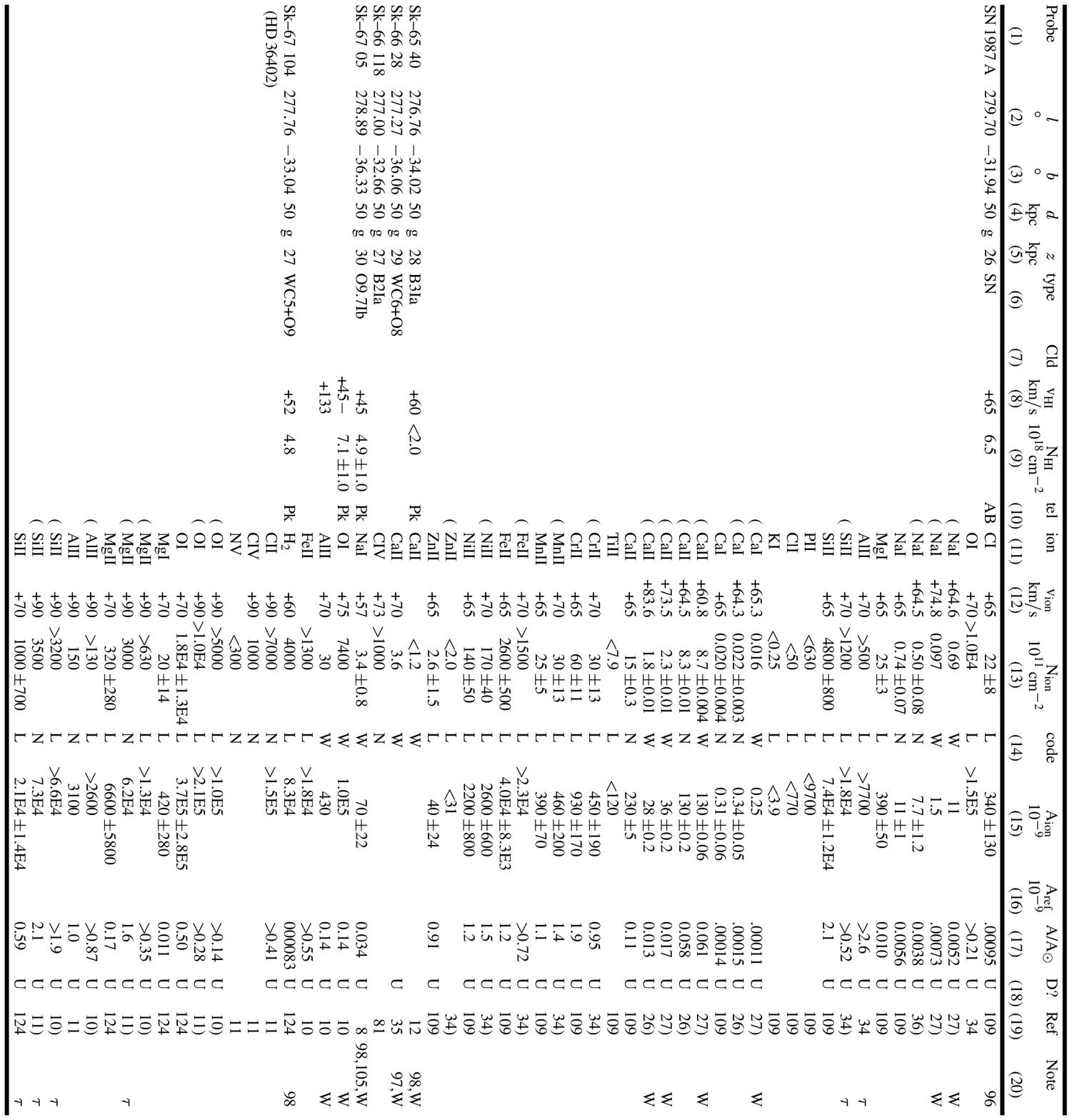}{-25}{-60}{\null} %AASPP
\InsertPage{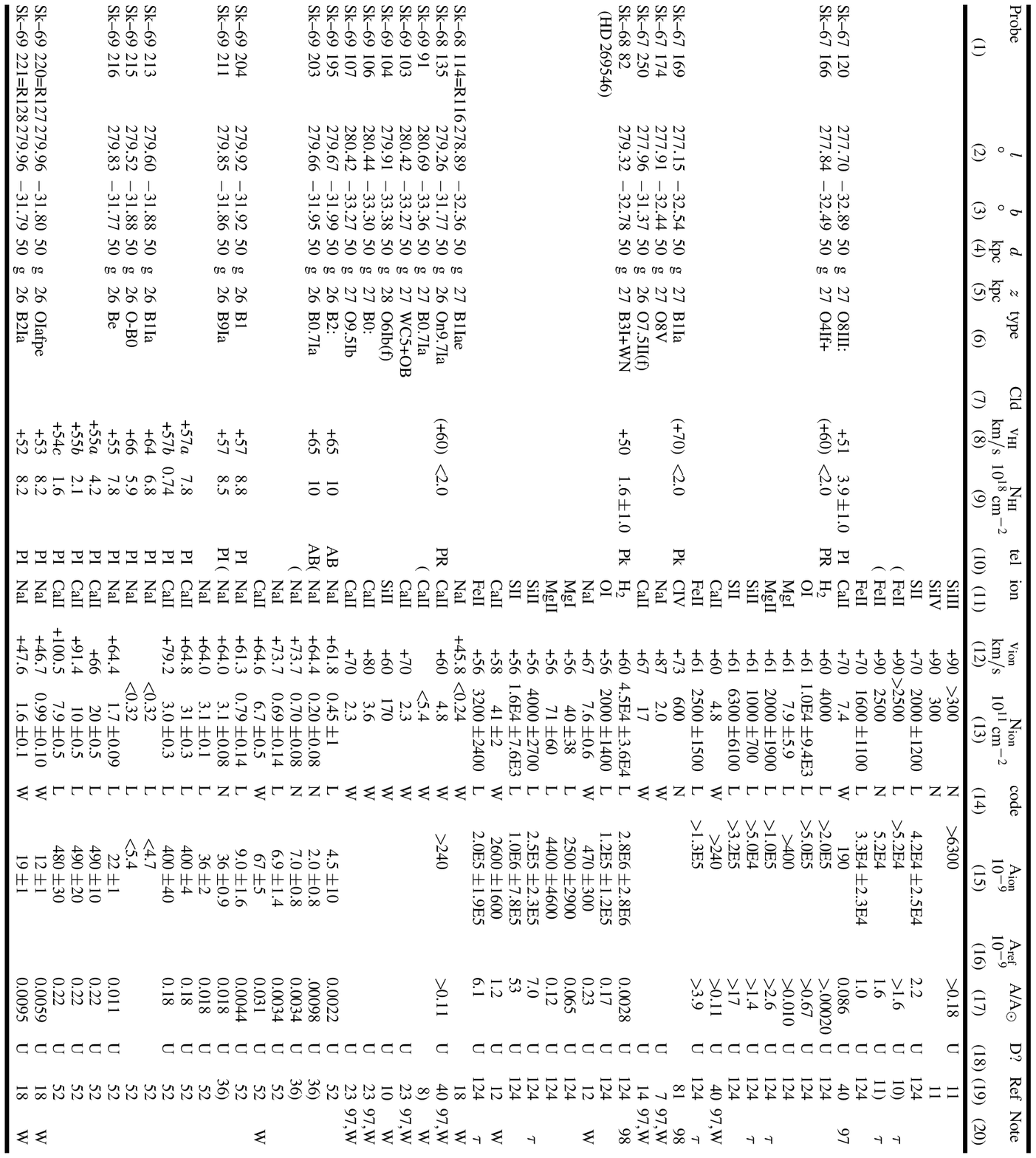}{-25}{-60}{\null} %AASPP
\InsertPage{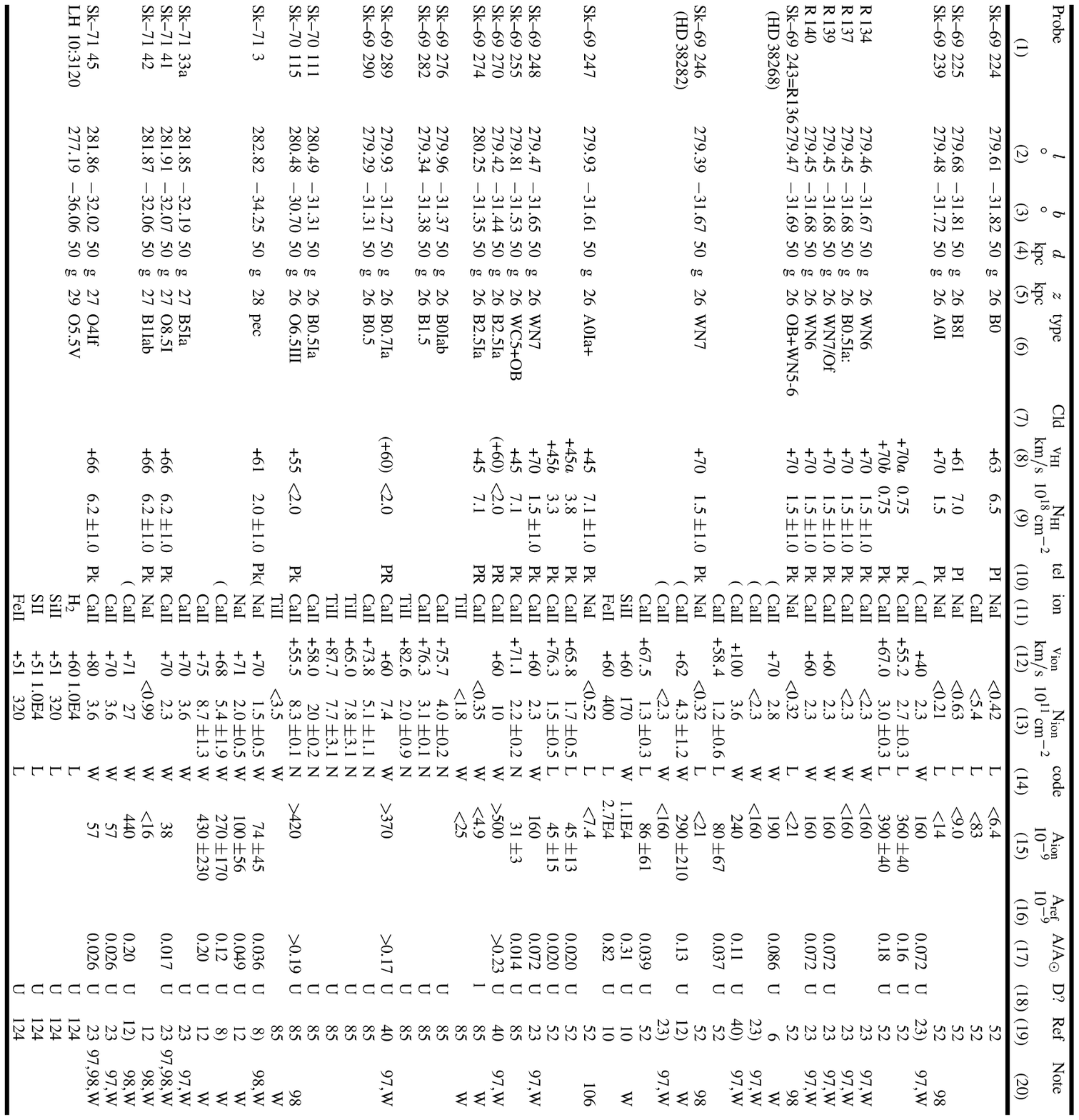}{-25}{-60}{\null} %AASPP
\InsertPage{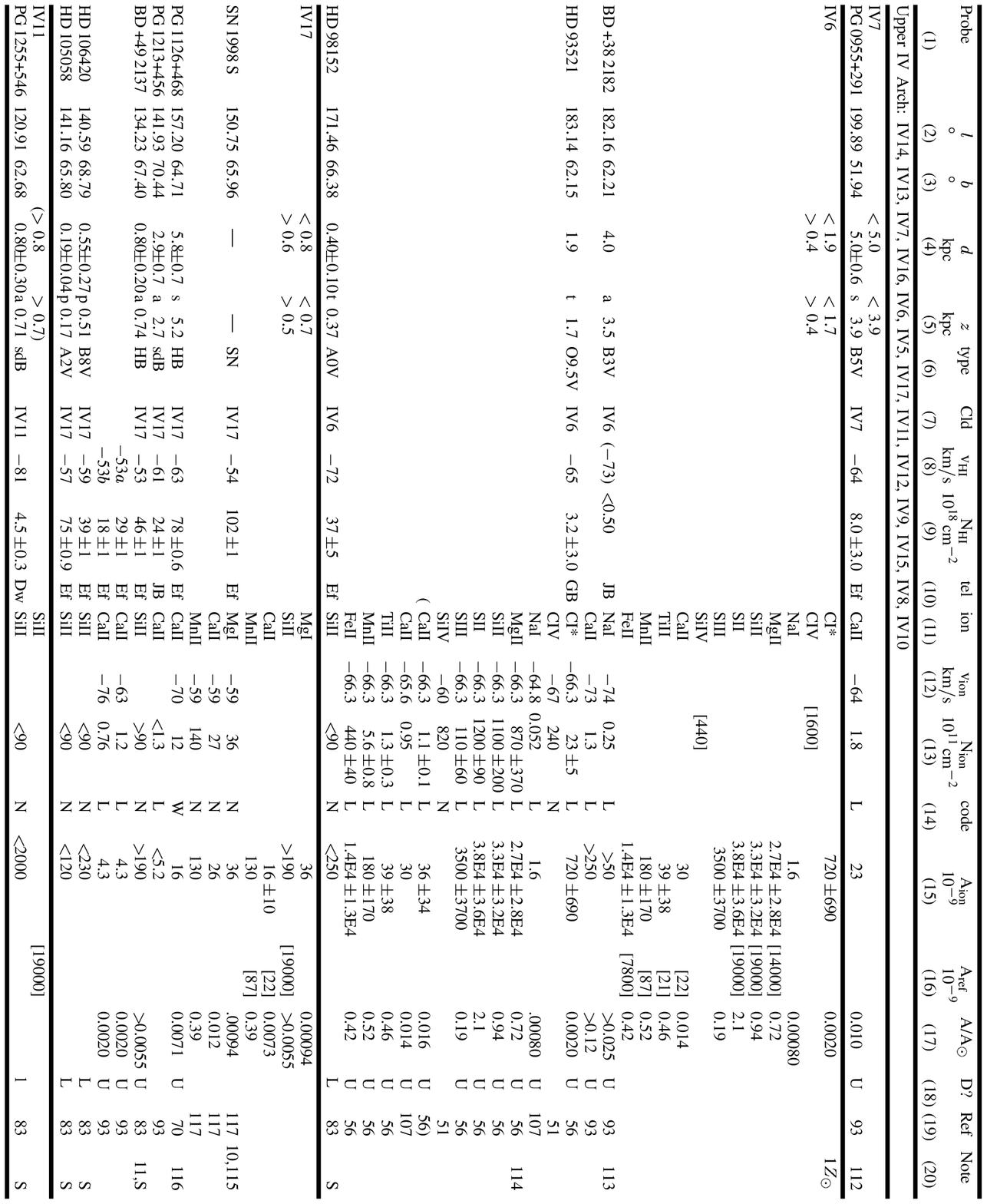}{-25}{-60}{\null} %AASPP
\InsertPage{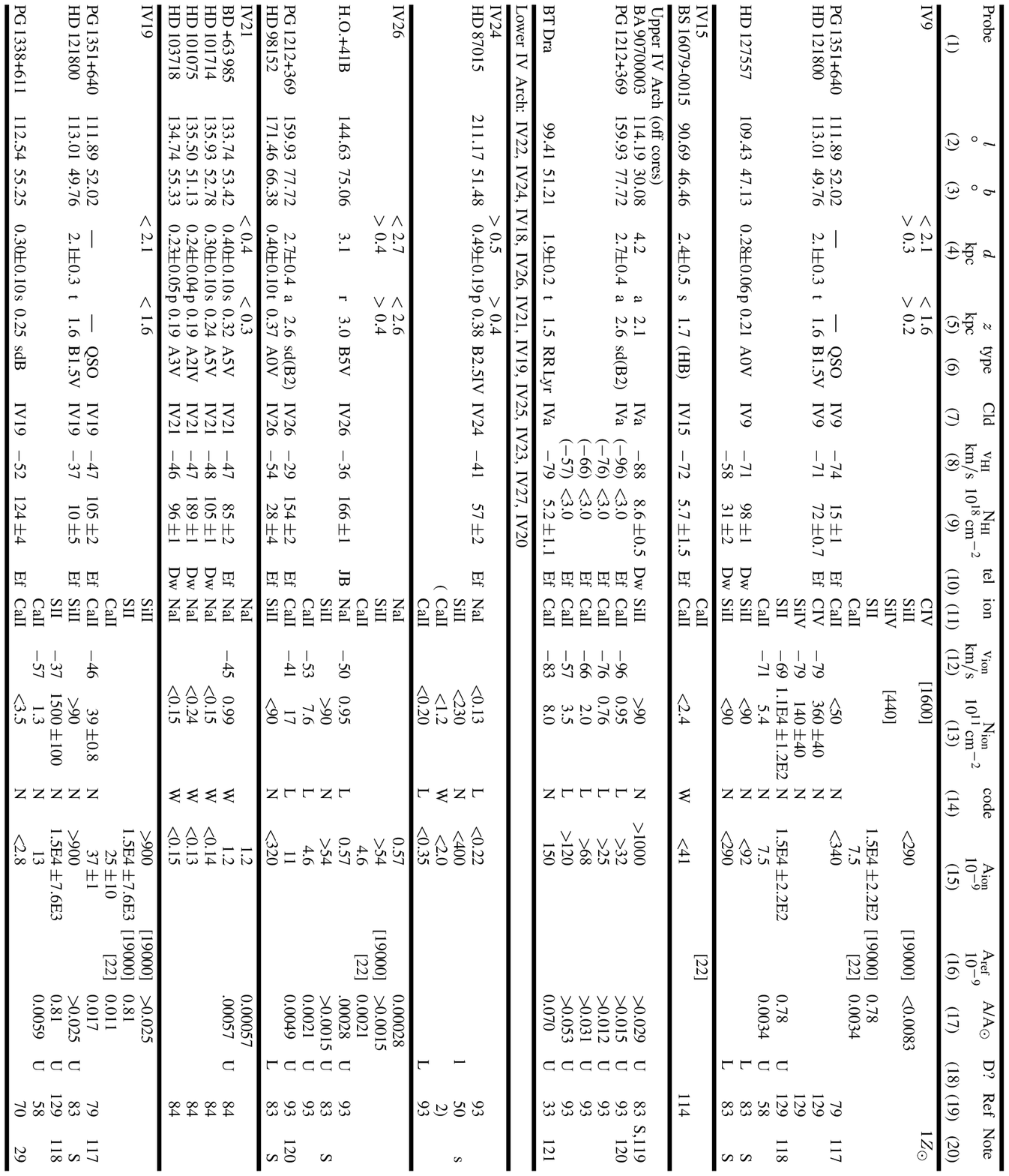}{-25}{-60}{\null} %AASPP
\InsertPage{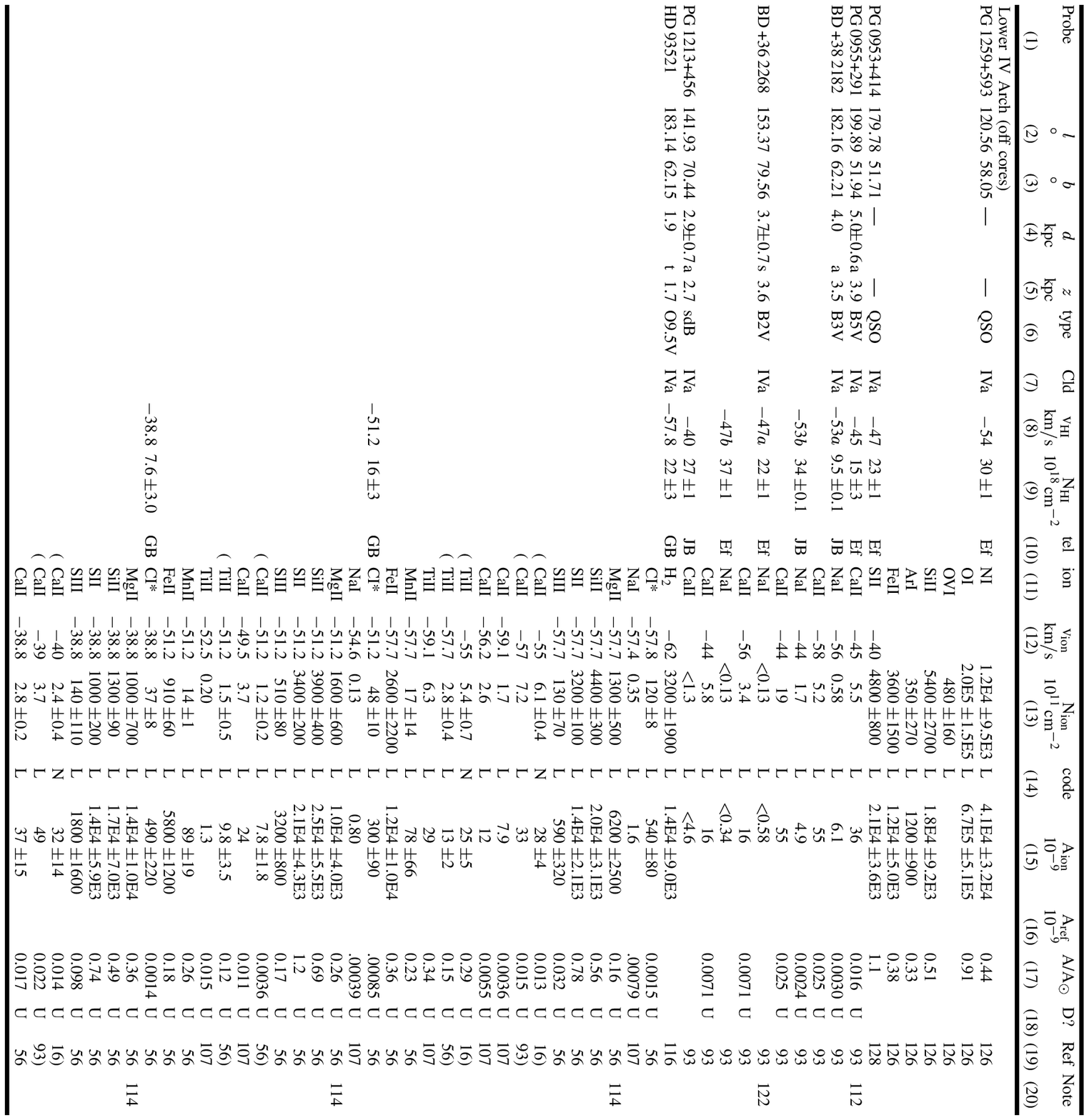}{-25}{-60}{\null} %AASPP
\InsertPage{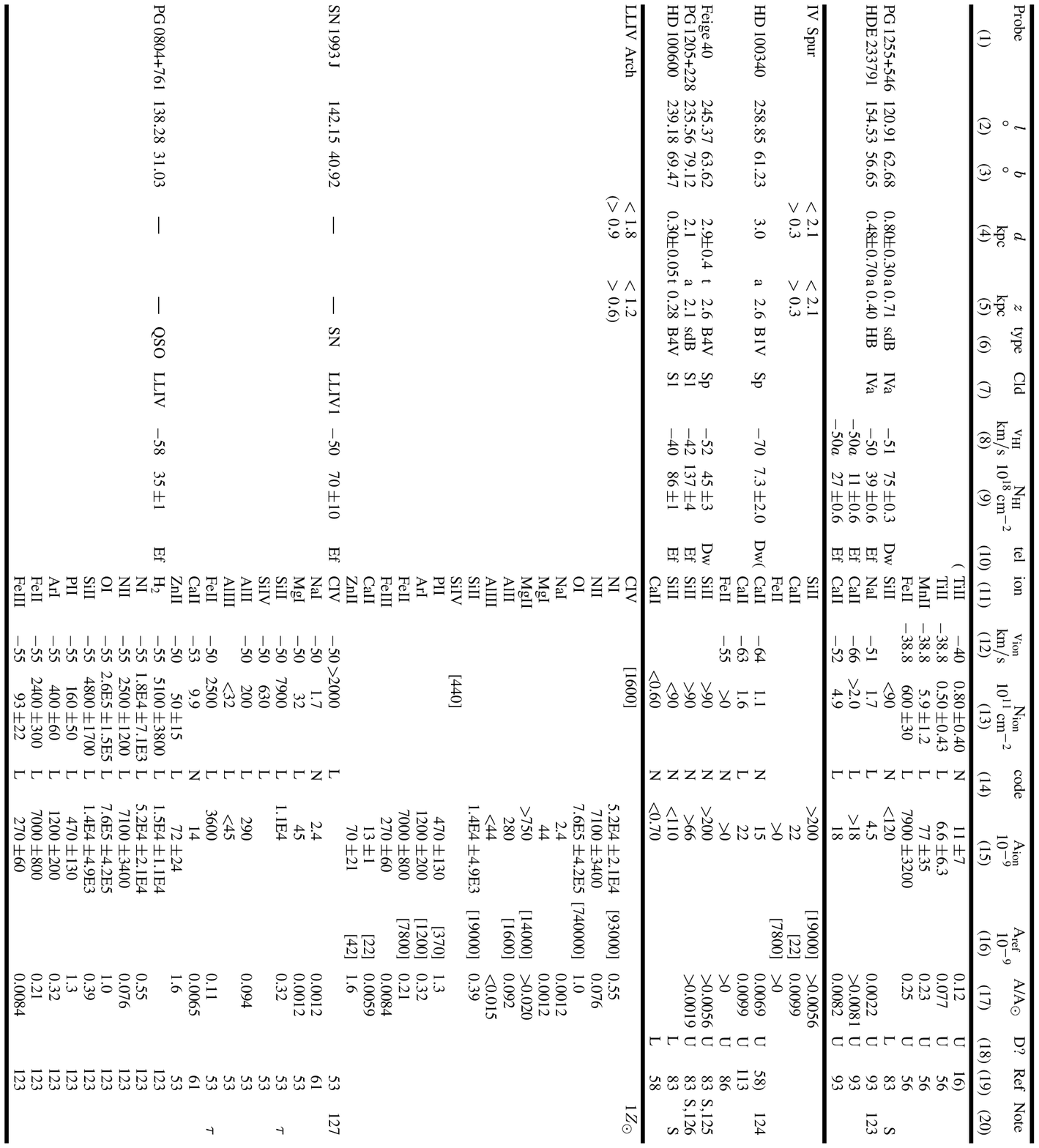}{-25}{-60}{\null} %AASPP
\InsertPage{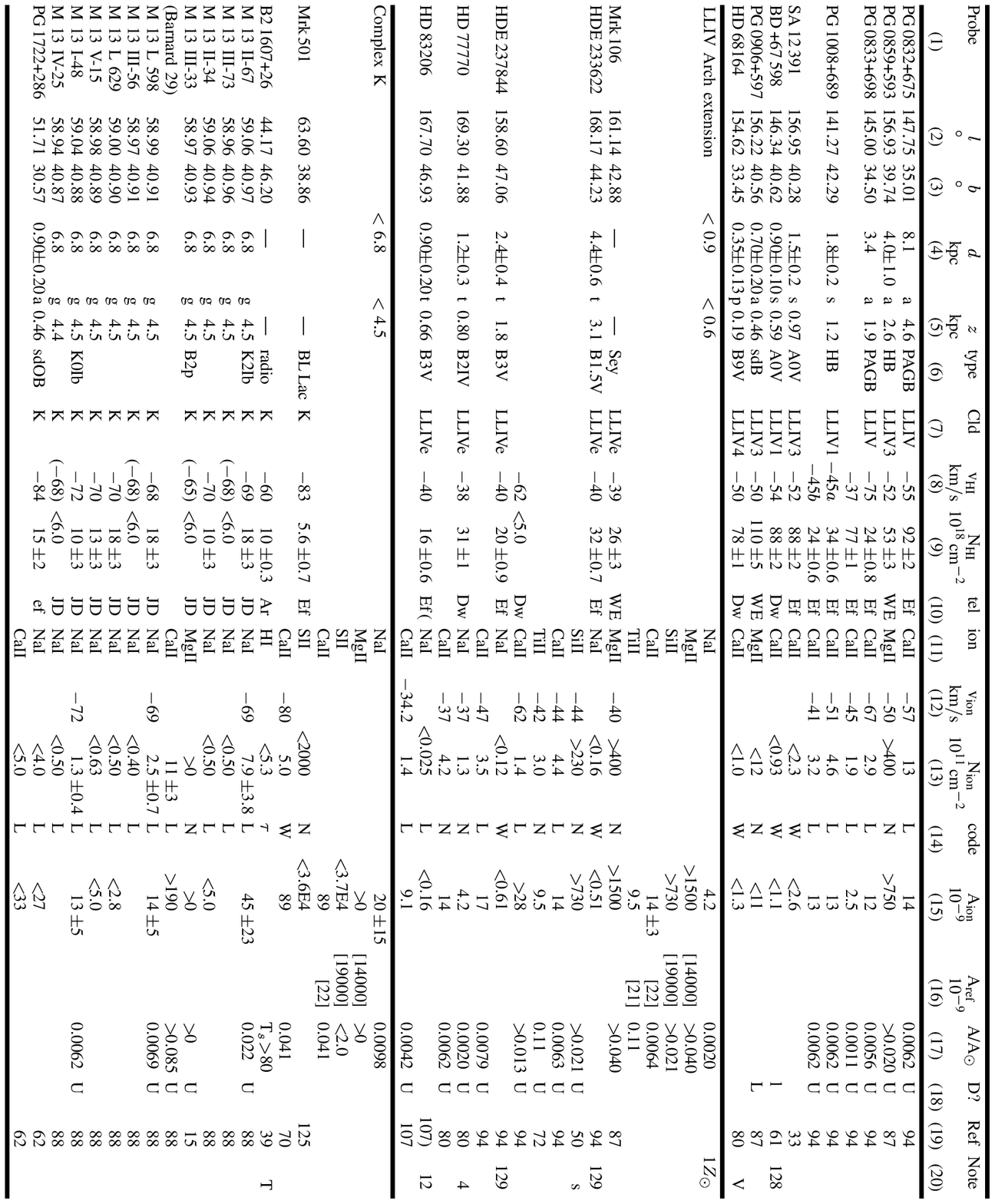}{-25}{-60}{\null} %AASPP
\InsertPage{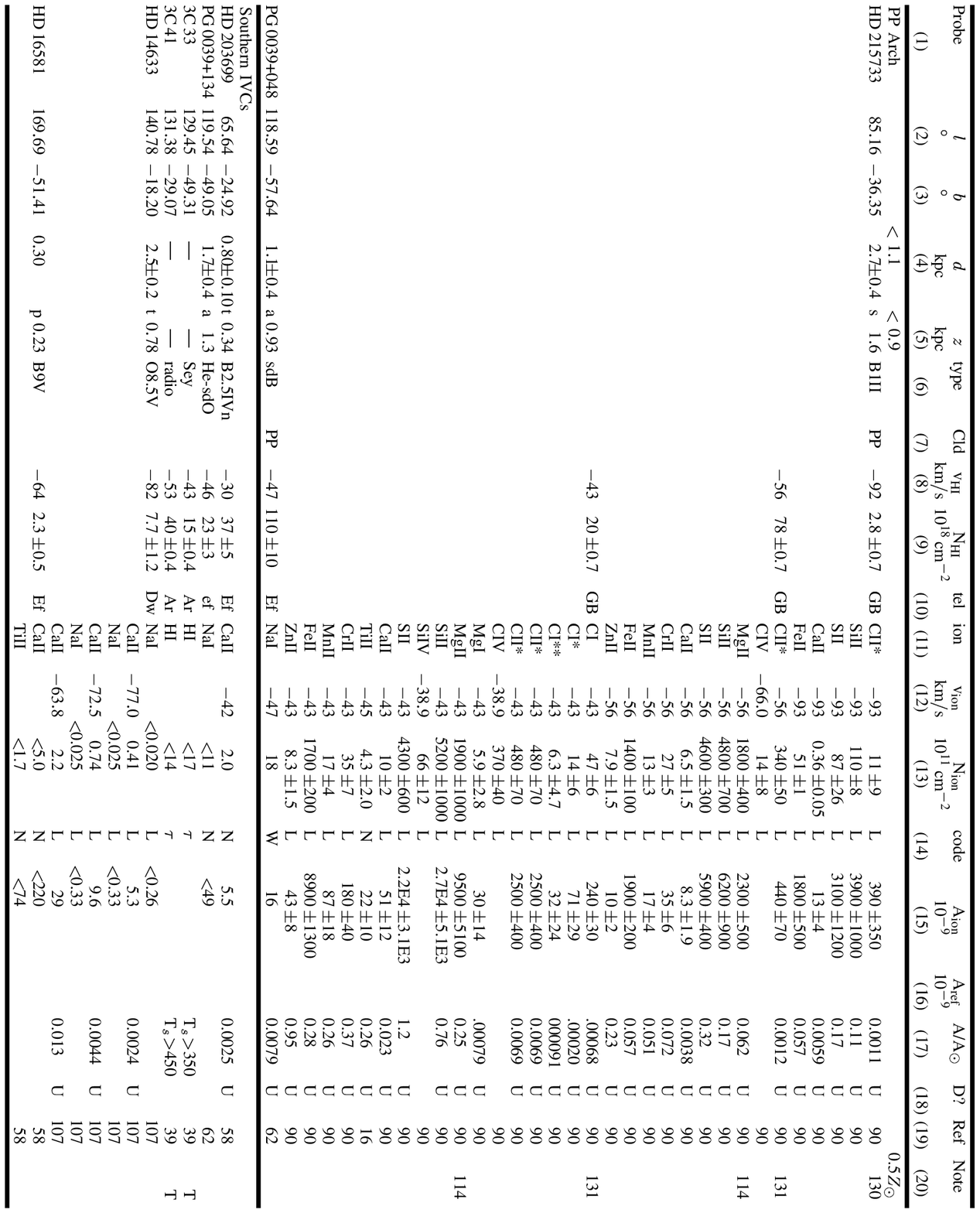}{-25}{-60}{\null} %AASPP
\InsertPage{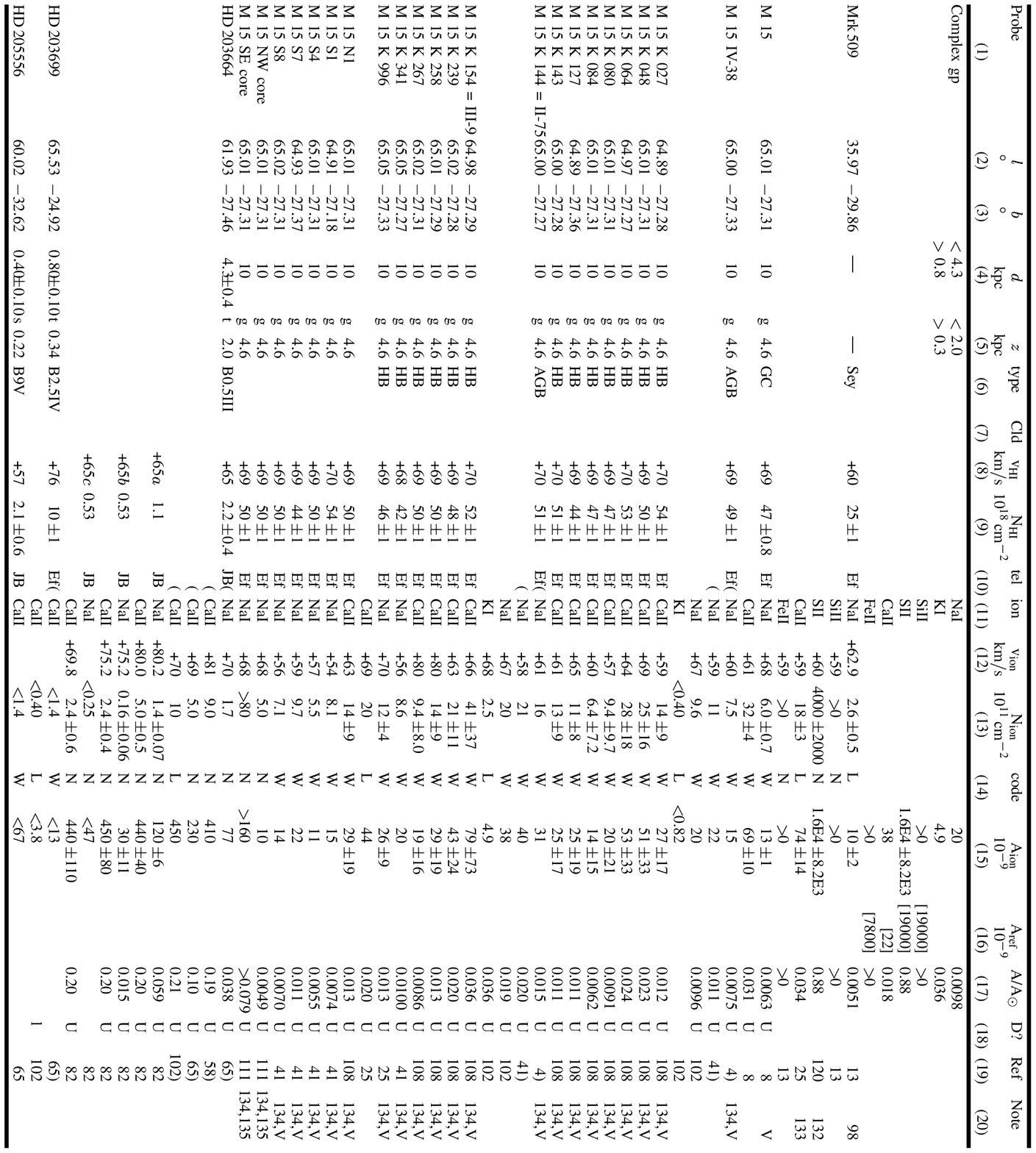}{-25}{-60}{\null} %AASPP
\InsertPage{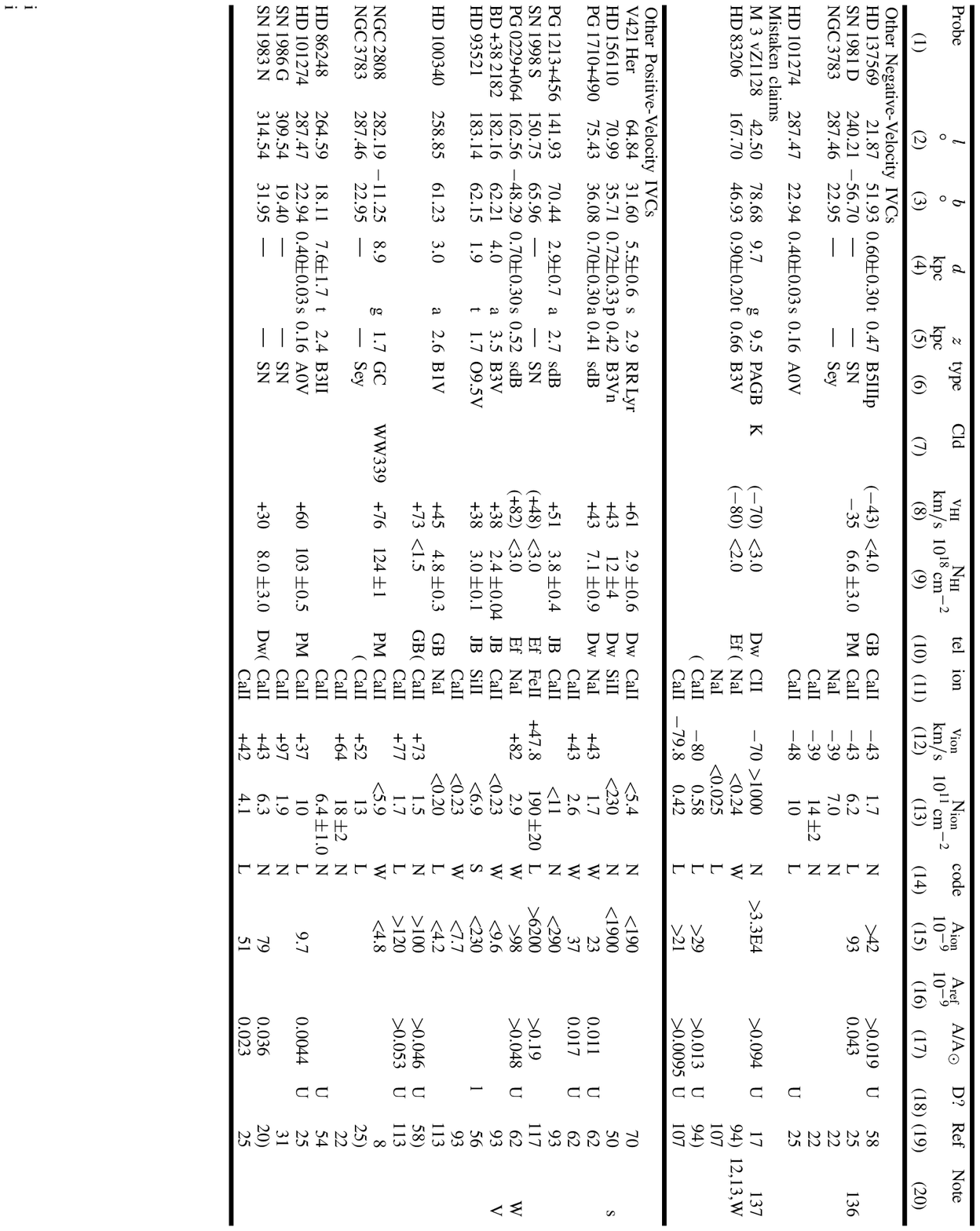}{-25}{-60}{\null} %AASPP
\InsertPage{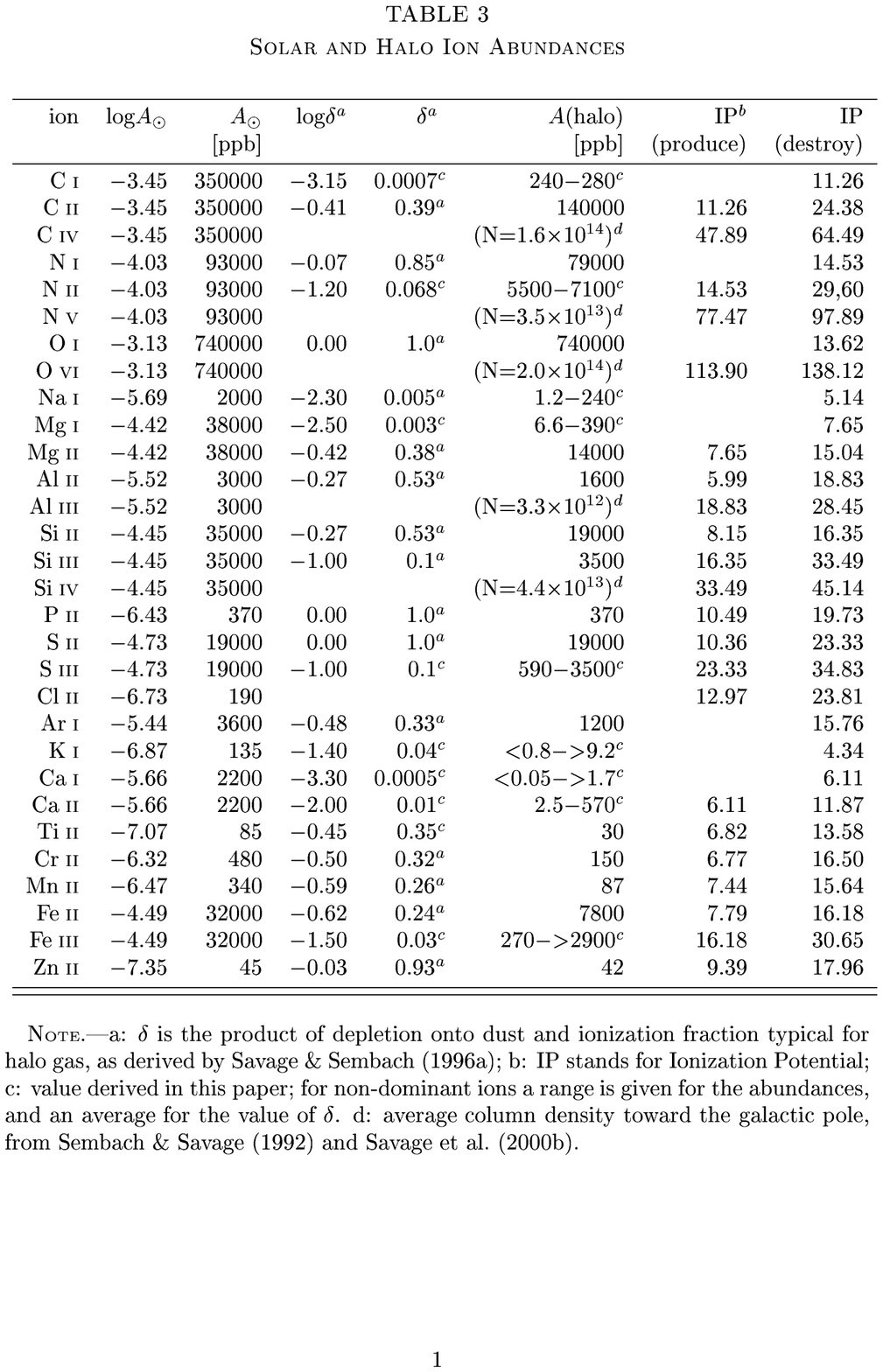}{-20}{-75}{\null} %AASPP
\InsertPage{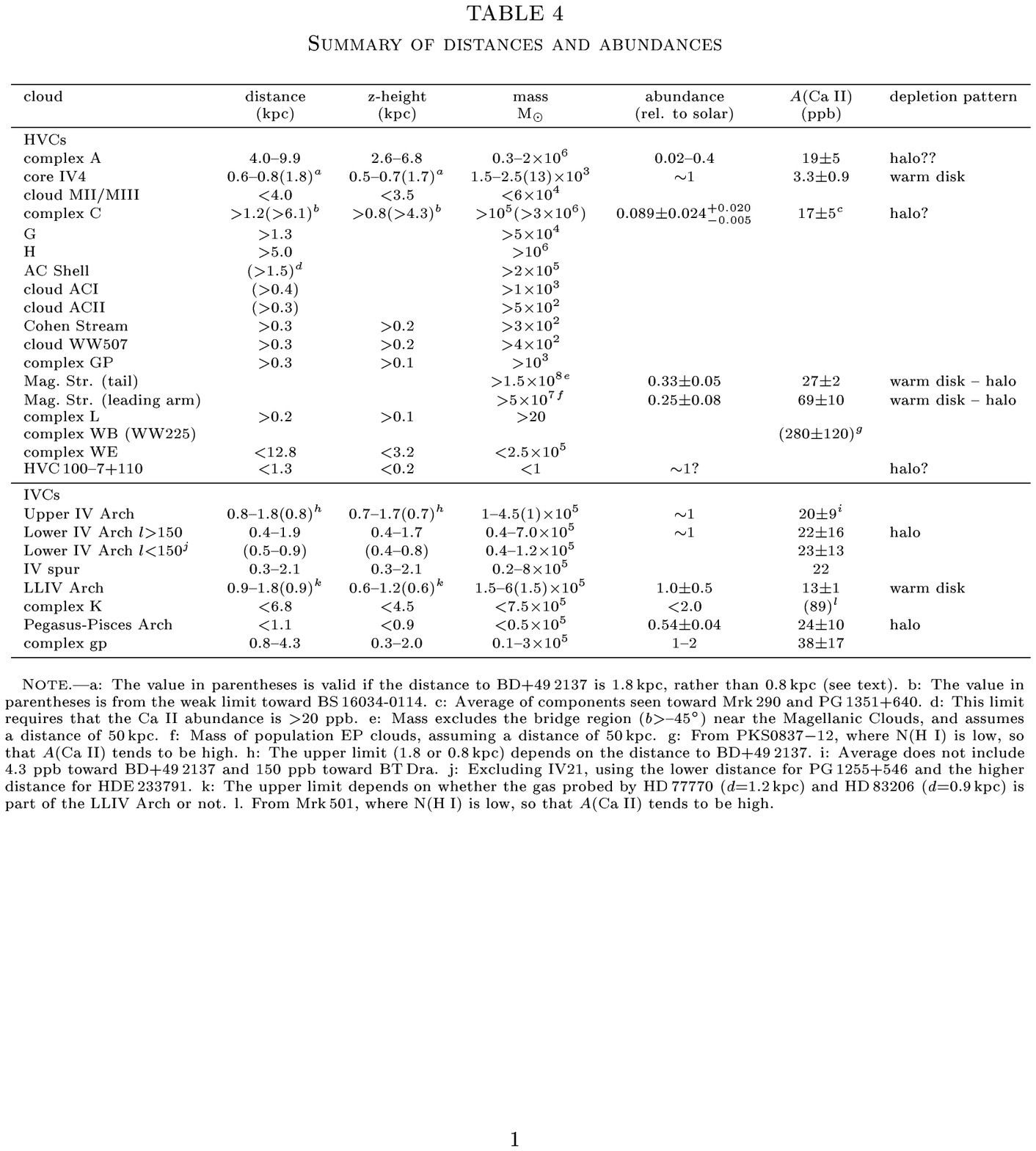}{-20}{-75}{\null} %AASPP

%%%%%%%%%%%%%%%%%%%%%%%%%%%%%%%%%%%%%%%%%%%%%%%%%%%%%%%%%%%%%%%%%%%%%%%%%%%%%%%%

\InsertPage{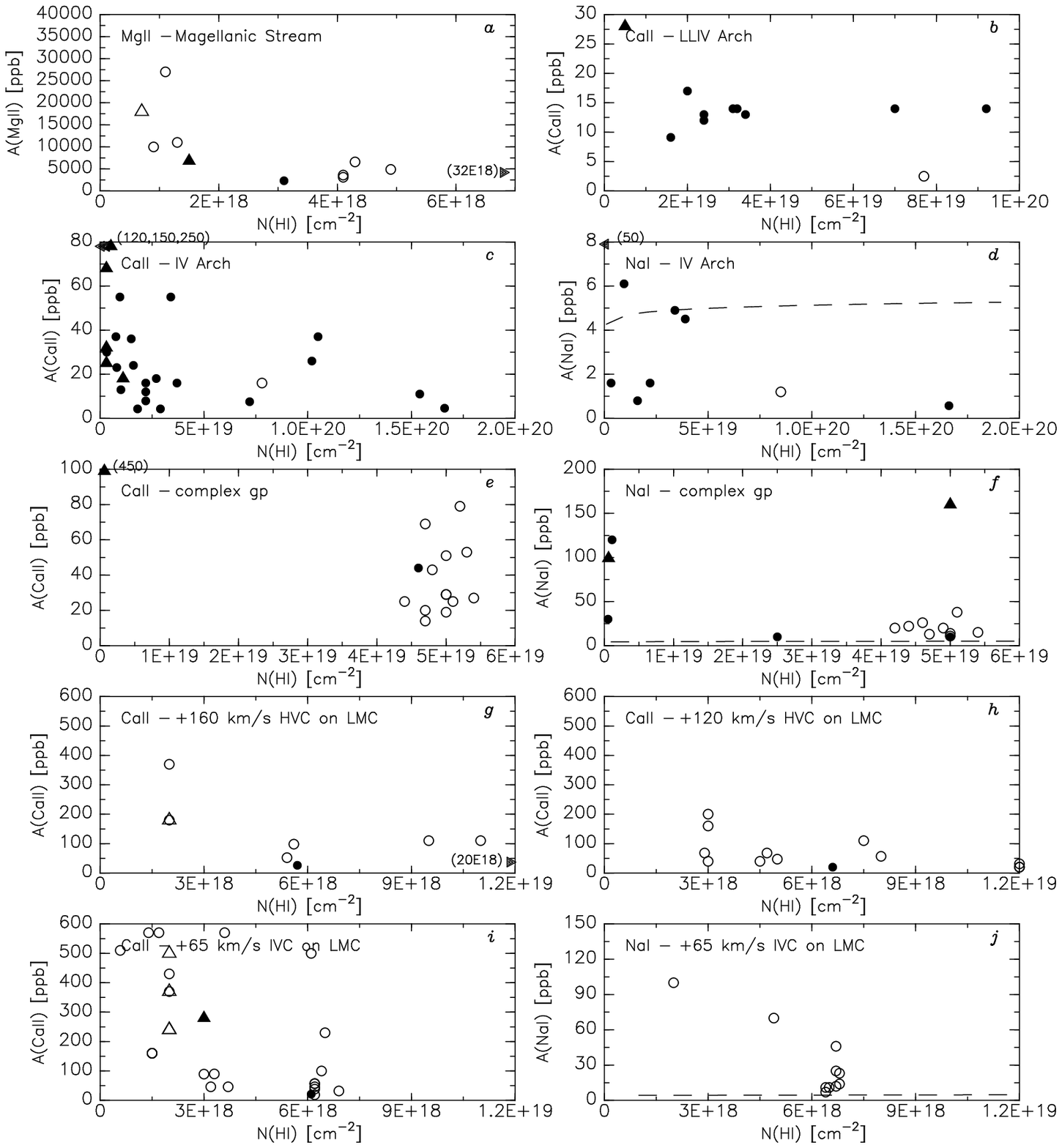}{-75}{-45}{% %AASPP
\fgnumber{\Fmultiple} Scatter plots of [N(ion)/N(\HI)] vs N(\HI) for the ions
for which multiple determinations exist in a single cloud. Circles indicate
actual measurements, triangles show lower limits. Closed symbols show
measurements where both the \HI\ and the ion column density were carefully
measured. Open symbols refer to measurements where only the equivalent width was
given, or where the \HI\ data were interpolated (in the case of LMC stars). Note
that three \CaII\ limits in the IV Arch and one in complex~gp are off-scale.
Most of the scatter at high N(\HI) may be due to fine structure in N(\HI), or to
variations in depletion/ionization fraction. The scatter at low N(\HI) is
probably due to the ionization of hydrogen in the case of \MgII, or to low dust
content in the case of \CaII\ (see Wakker \& Mathis 2000).
}

\InsertPage{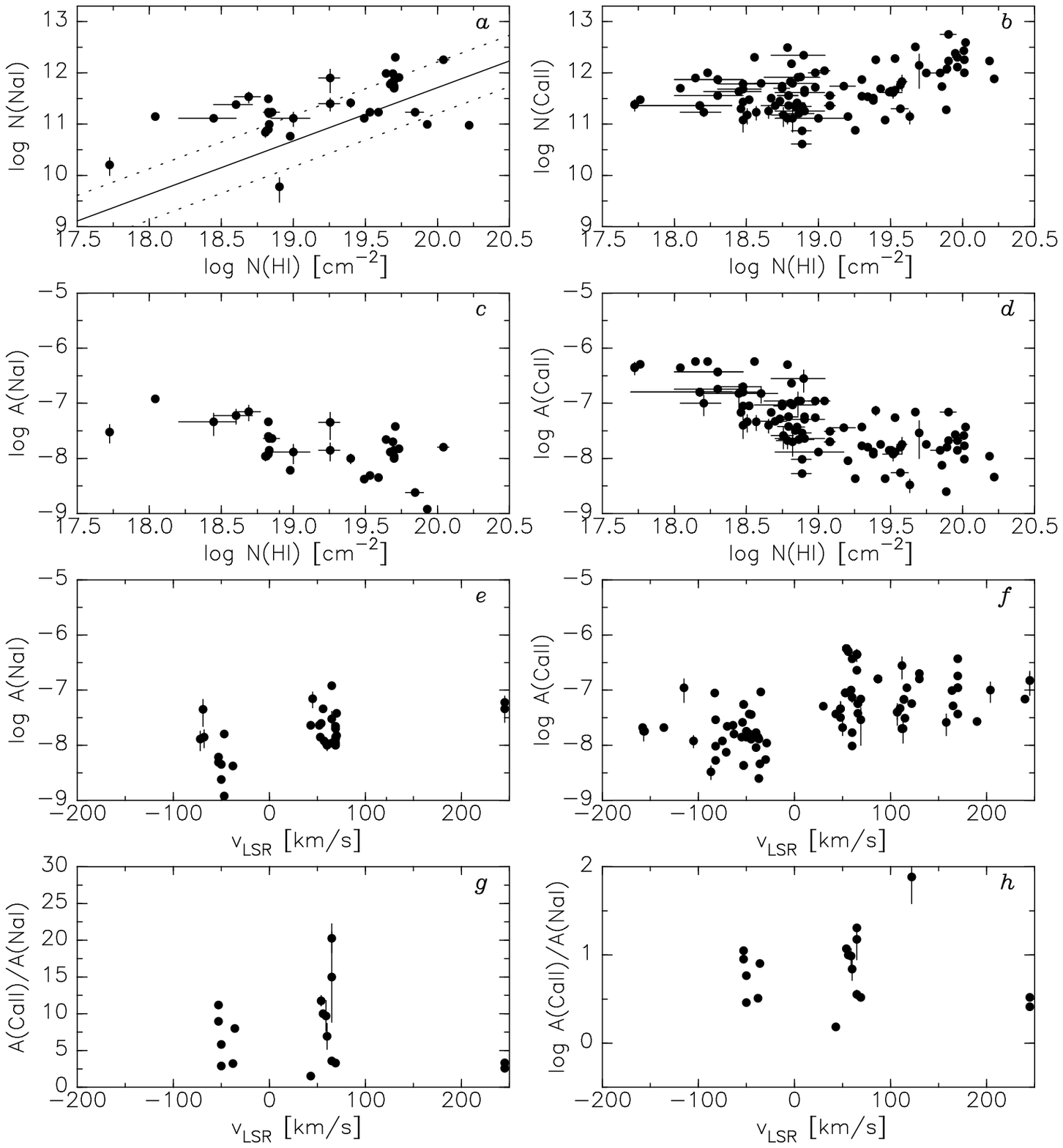}{-60}{-45}{% %AASPP
\fgnumber{\FCaNa} Correlations for \NaI\ and \CaII. a), b): log N(\NaI) and log
N(\CaII) vs log N(\HI). The straight line in panel a shows the relation claimed
by Ferlet et al.\ (1985). The dotted line indicates the 1$\sigma$ spread in that
relation. c), d) log $A$(\NaI) and log $A$(\CaII) vs \vlsr; $A$(\CaII) appears
to be higher in positive-velocity clouds, but this may be biased by the fact
that many of these directions have low N(\HI). e, f) ratio of $A$(\CaII) and
$A$(\NaI) vs \vlsr, showing that above $\vert$\vlsr$\vert$=50\,\kms\ there is no
obvious Routly-Spitzer effect.
}

\InsertPage{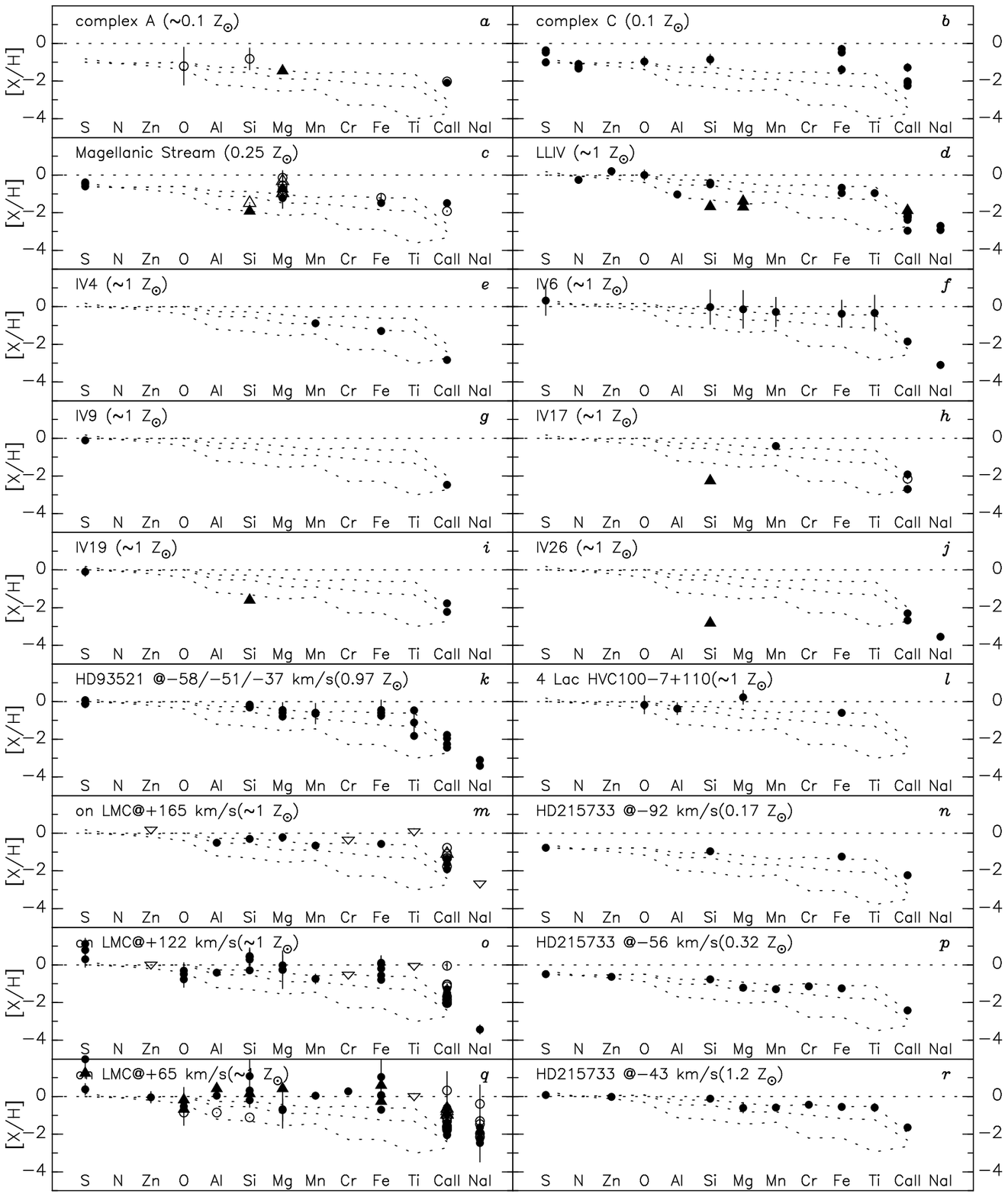}{-55}{-45}{% %AASPP
\fgnumber{\Fpattern} Abundance (i.e.\ N(ion)/N(\HI)) patterns in 18 selected
HVCs/IVCs where multiple elements have been measured. The elements are ordered
in terms of decreasing reference abundance in halo gas. The dotted lines shows
the patterns for cool disk, warm disk and halo gas (Savage \& Sembach 1996a),
shifted by the overall metallicity. The latter is given in parentheses after the
cloud name and is derived from the results for undepleted elements. Note that
the halo pattern was {\it derived} from the IVCs in the direction of HD\,93521
and HD\,215733. Error bars are usually smaller than the size of the dots. Closed
circles show good measurements. Open circles indicate only the equivalent width
was published. Downward pointing open triangles indicate an upper limit. Upward
pointing closed triangles indicate lower limits.
}

\InsertPage{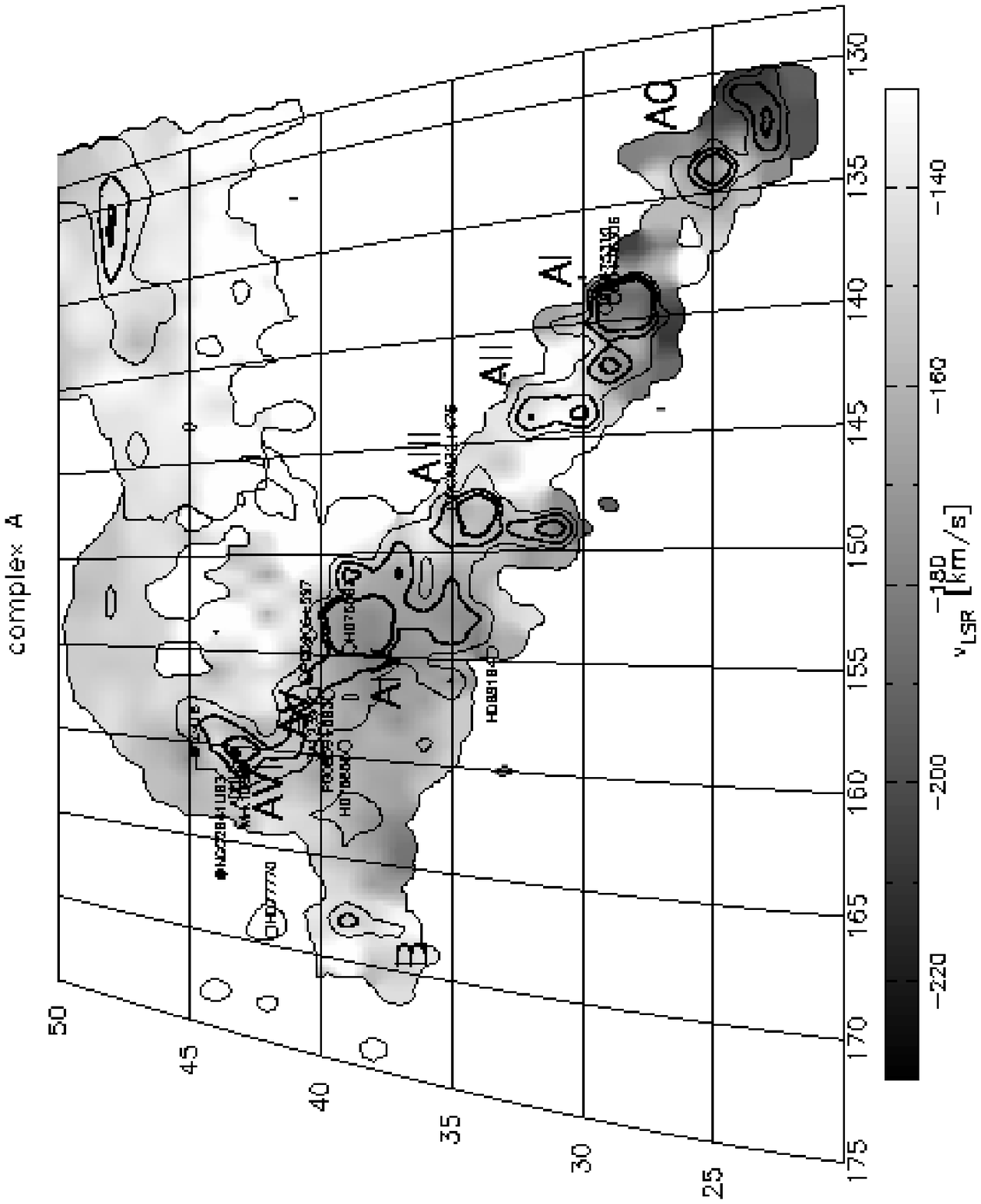}{-45}{-90}{% %AASPP
\fgnumber{\FgA} Map of HVC complex~A, from Hulsbosch \& Wakker (1988). Contours
indicate brightness temperature levels of 0.05, 0.3, 0.6 and 1\,K. The greyscale
shows velocities, as identified by the wedge. The core names are shown, as are
the positions of probe stars. Closed symbols refer to detections, open symbols
to non-detections. The structure in the upper right corner is HVC complex~C.
}

\InsertPage{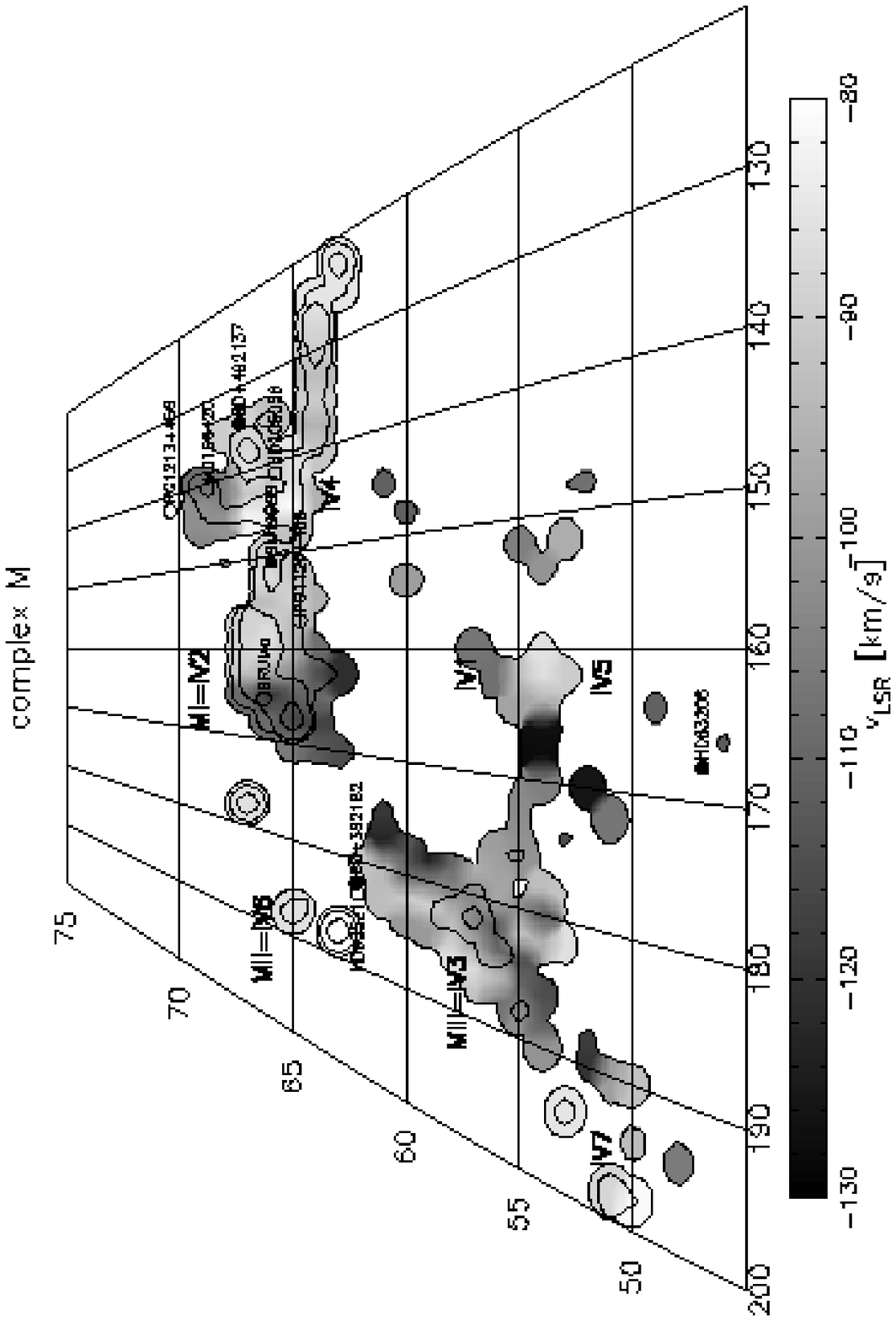}{-45}{-90}{% %AASPP
\fgnumber{\FgM} Map of HVC complex~M, from Hulsbosch \& Wakker (1988). Contours
indicate brightness temperature levels of 0.05, 0.3, 0.8 and 2.5\,K. The
greyscale shows velocities, as identified by the wedge. The core names are
shown, as defined by Wakker \& van Woerden (1991) and Kuntz \& Danly (1996).
Closed symbols refer to detections, open symbols to non-detections.
}

\InsertPage{fig06_C.cps}{-20}{-90}{% %AASPP
\fgnumber{\FgC} Map of HVC complex~C, from Hulsbosch \& Wakker (1988). Contours
indicate brightness temperature levels of 0.05, 0.4 and 1.0\,K. Colors indicate
velocities, as identified by the wedge. At many positions along the central
ridge emission is seen at two velocities. In these directions a half circle is
shown for the second component. Core names are also given, as are the positions
of probe stars. CI-A,B and CIII-A,B were defined by Giovanelli et al.\ (1973),
as was CeI, which they named C-extension. CeII-V are defined here by analogy.
Giovanelli et al.\ (1973) also define a core CIIIC, which is named CIC here,
while CIIIC is used for a core not observed by Giovanelli et al.\ (1973).
Complex~D was defined by Wakker \& van Woerden (1991), core ``CD'' is defined
here, as is core ``C/K'' (see text). Closed symbols refer to detections, open
symbols to non-detections.
}

\InsertPage{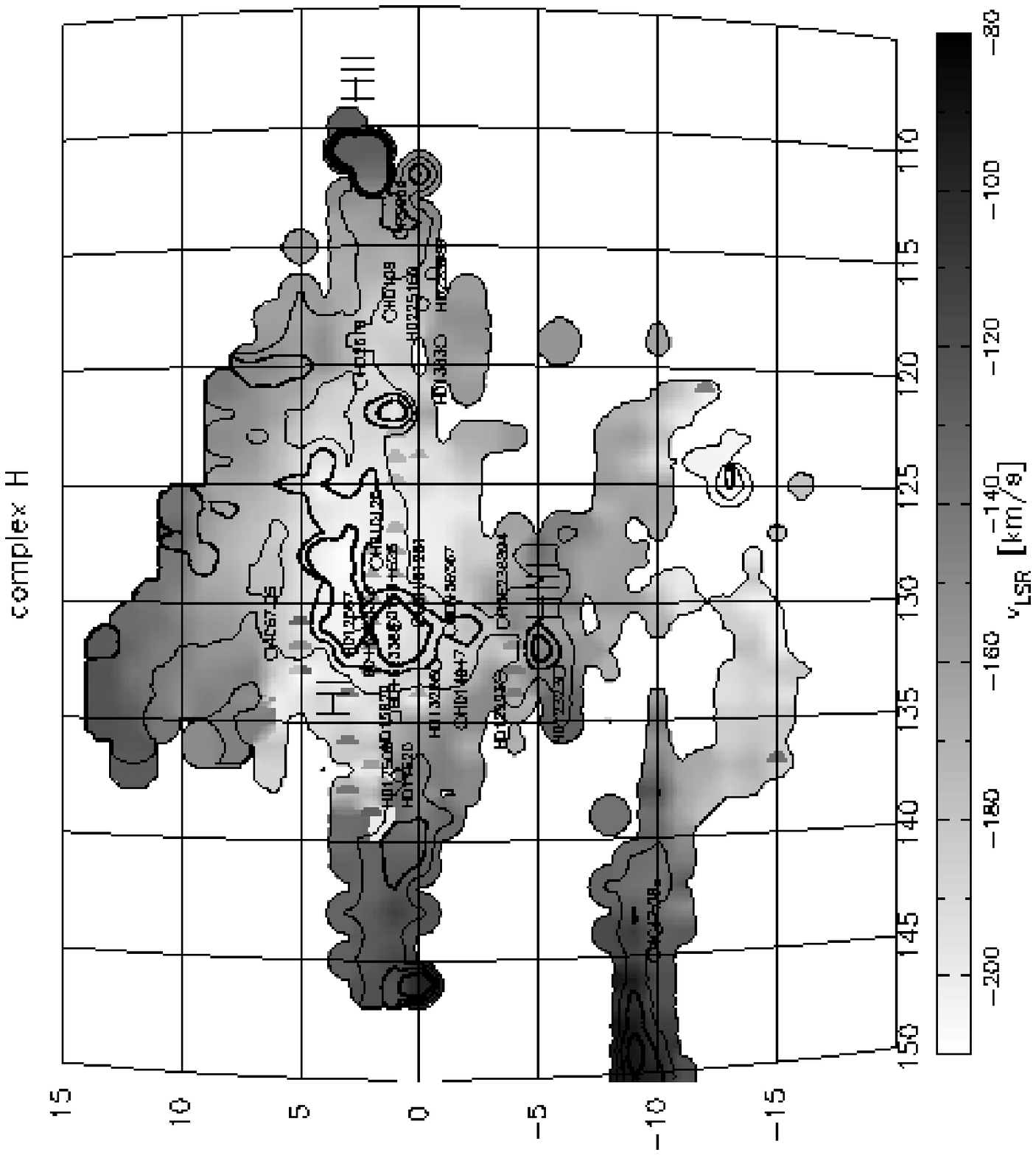}{-45}{-90}{% %AASPP
\fgnumber{\FgH} Map of HVC complex~H, from Hulsbosch \& Wakker (1988). Contours
indicate brightness temperature levels of 0.05, 0.3, 0.6 and 1\,K. The greyscale
shows velocities, as identified by the wedge. The core names are shown, as are
the positions of probe stars.
}

\InsertPage{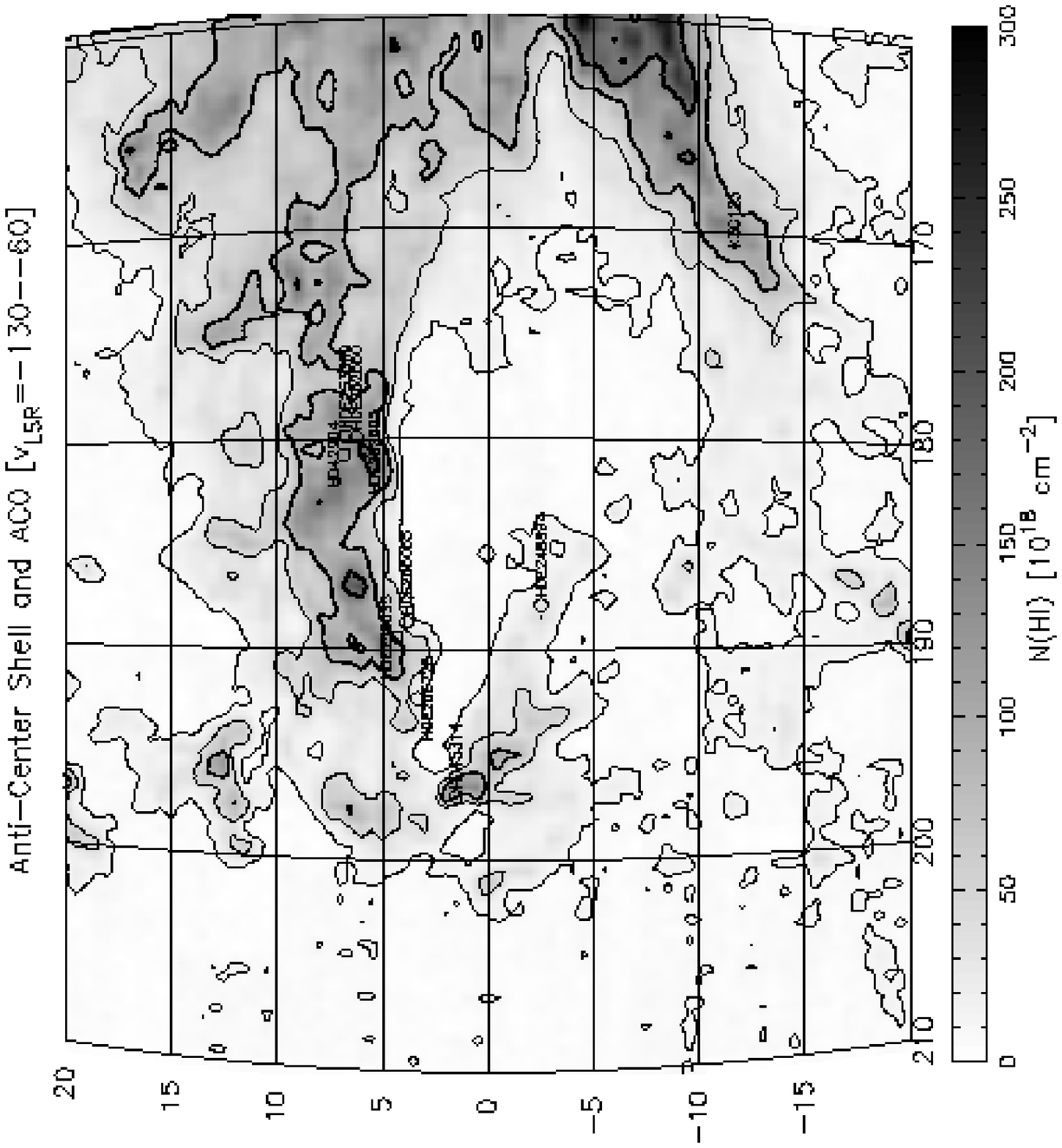}{-45}{-90}{% %AASPP
\fgnumber{\FgACS} Map of the Anti-Center shell, from Hartmann \& Burton (1997).
Contours indicate column densities of 10, 40, 80 and 160\tdex{18}\,\cmm2\ for
gas in the velocity range $-$130 to $-$60\,\kms. The open symbols show the
positions of the probes with non-detections. The structure at l=180\deg,
b=+6\deg\ is cloud AC0.
}

\InsertPage{fig09_ACc.cps}{-45}{-90}{% %AASPP
\fgnumber{\FgACc} Map of the Anti-Center HVCs, from Hulsbosch \& Wakker (1988).
Contours indicate brightness temperature levels of 0.05, 0.3, 0.8 and 1.5\,K.
Colors indicate velocities, as identified by the wedge. At many positions along
the central ridge emission is seen at two velocities. In these directions two
half circles are shown, with separate colors for each component. Core names are
also shown, as are the positions of probe stars.
}

\InsertPage{fig10_MS.cps}{-15}{-90}{% %AASPP
\fgnumber{\FgMS} Map of the Magellanic Stream, using data from Hulsbosch \&
Wakker (1988) and Morras et al.\ (2000). Added are the clouds in populations EN
and EP of Wakker \& van Woerden (1991), as well as scattered small
positive-velocity clouds in the northern Galactic hemisphere. Contour levels are
at brightness temperatures of 0.05 and 1\,K. Colors indicate velocities as
identified by the wedge. The projection has galactic longitude 270\deg\ along
the middle and the LMC near the center. The thick solid line crossing the poles
represents the Galactic equator. The curved lines are the orbits of the LMC and
SMC in the model of Gardiner \& Noguchi (1996); the small dots are the
present-day positions of the particles in that model. Labeled probes are those
of the Magellanic Stream, clouds WW187, WW211 and WW487, as well as the
extra-galactic objects in which \NaI\ or \CaII\ absorption is seen without
associated \HI\ (see Sect.~\Scloud.\SSnoHI). Filled symbols indicate detections,
open symbols non-detections.
}

\InsertPage{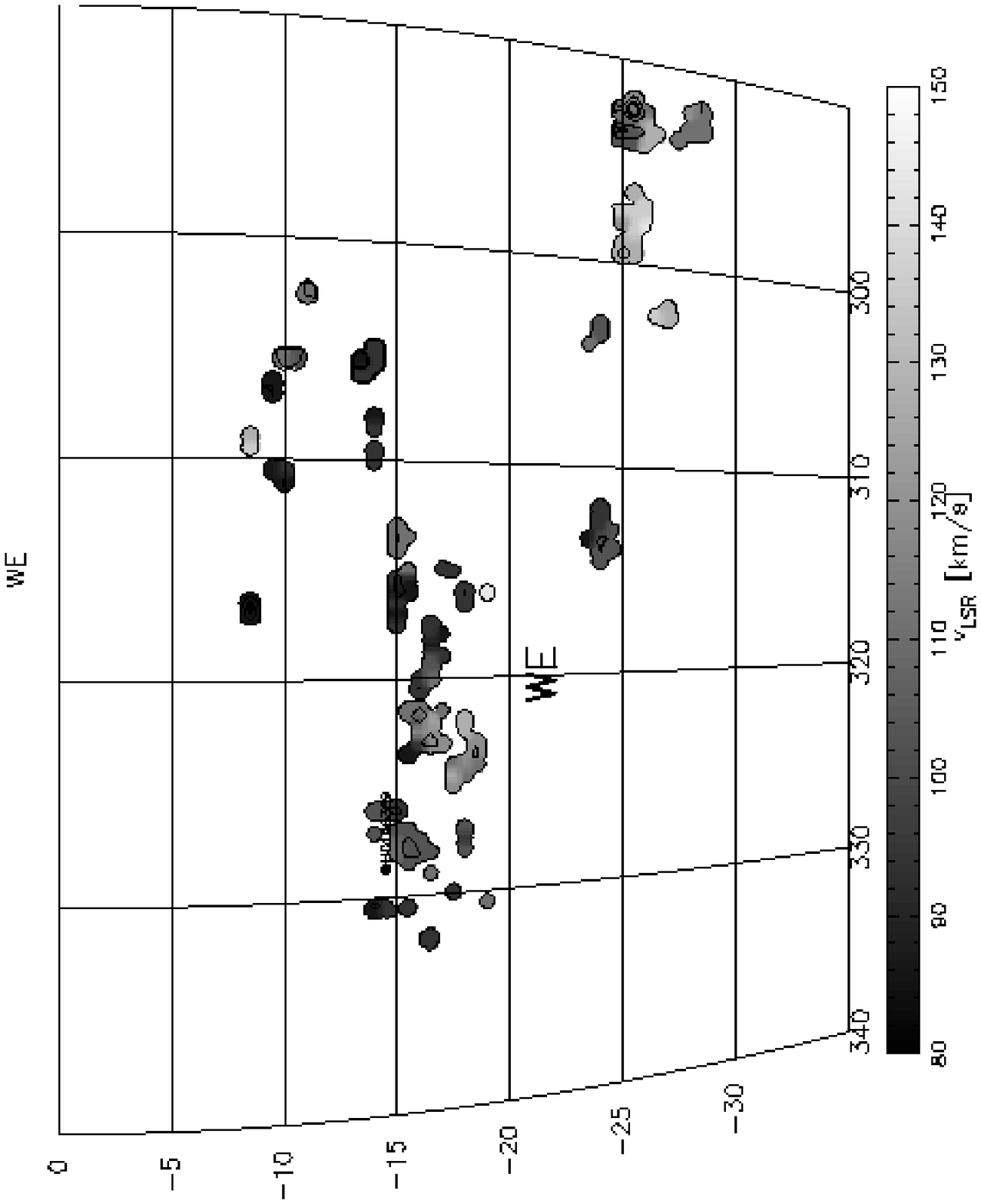}{-45}{-90}{% %AASPP
\fgnumber{\FgWE} Map of complex~WE, based on the data of Morras et al.\ (2000).
The greyscale indicates velocities as identified by the wedge.. Contours are at
brightness temperatures of 0.05 and 0.3\,K. The position of HD\,156359 is
indicated.
}

\InsertPage{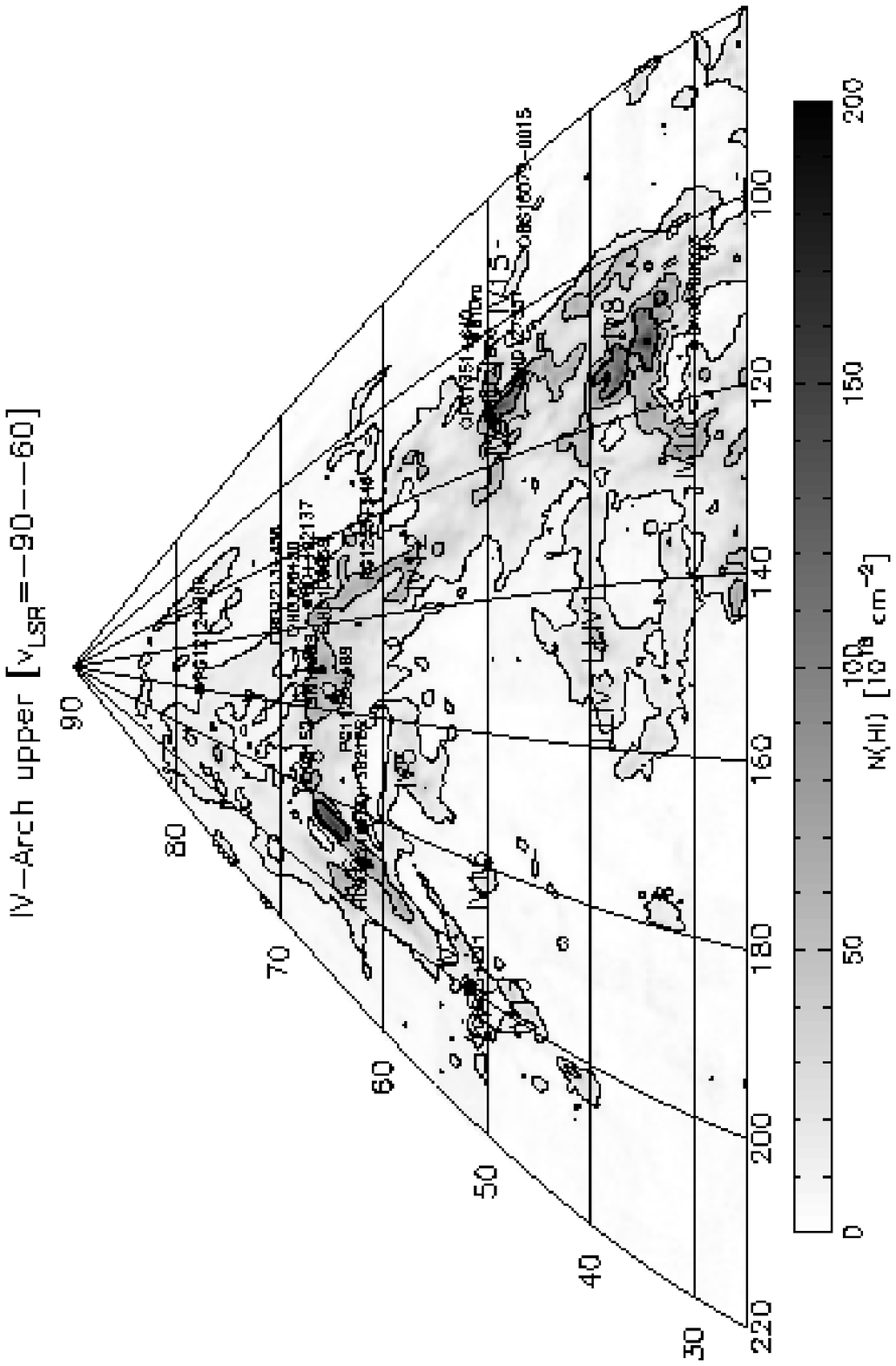}{-45}{-90}{% %AASPP
\fgnumber{\FgIVupp} Map of the Intermediate-Velocity Arch at velocities between
$-$90 and $-$60\,\kms, from Hartmann \& Burton (1997). The greyscale and
contours indicate the column density in this velocity range, with contour levels
at 10, 40, and 80\tdex{18}\,\cmm2. The names of the cores defined by Kuntz \&
Danly (1996) are indicated. The positions of the probes are shown, with closed
circles representing detections; open circles are non-detections.
} %AASPP  %note that there should not be an empty line now
\InsertPage{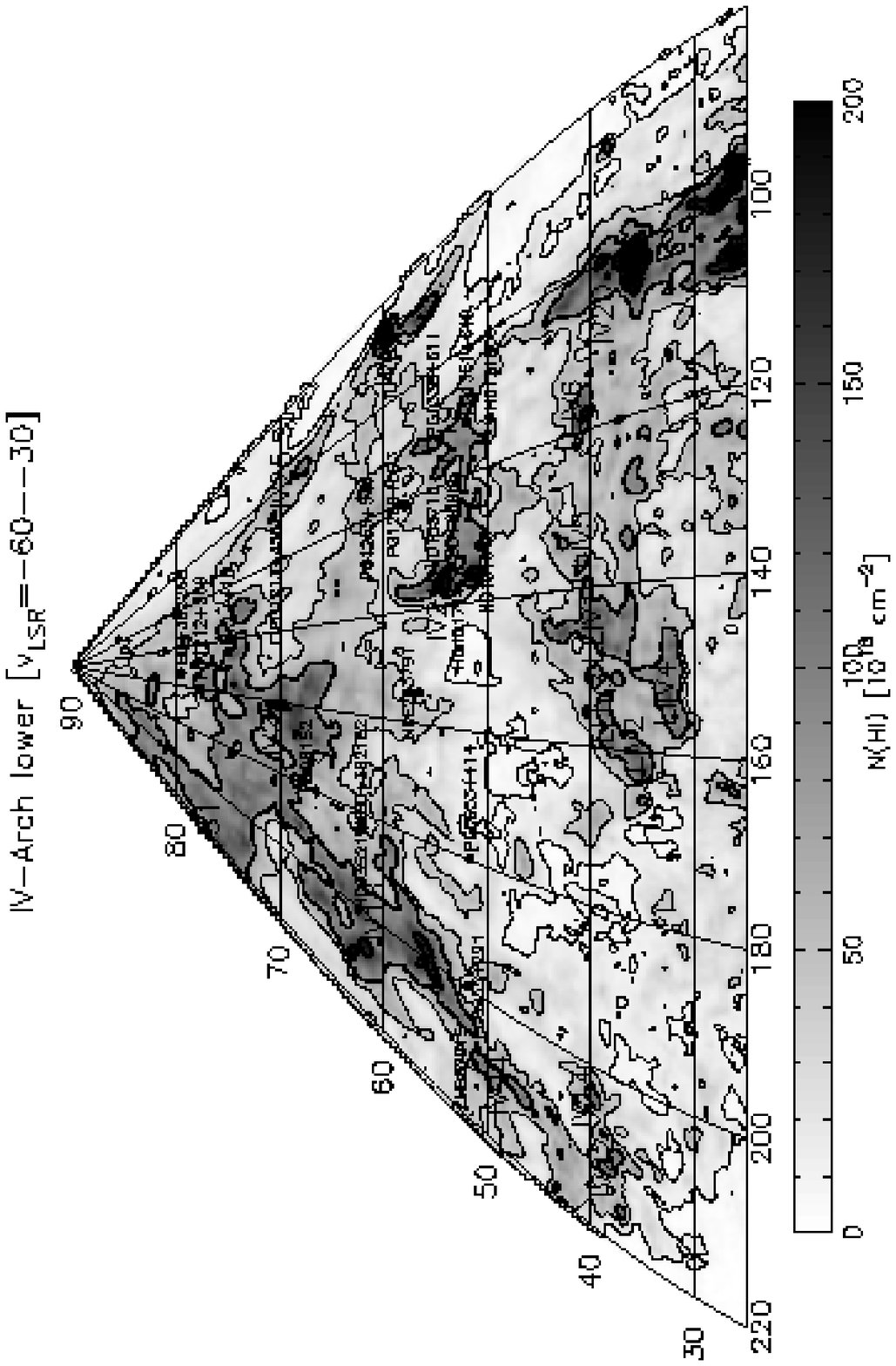}{-45}{-90}{% %AASPP
\fgnumber{\FgIVlow} Map of the Intermediate-Velocity Arch at velocities between
$-$60 and $-$30\,\kms, from Hartmann \& Burton (1997). See further
Fig.~\FgIVupp, except that there is an additional contour at 160\tdex{18}\,cmm2.
}

\InsertPage{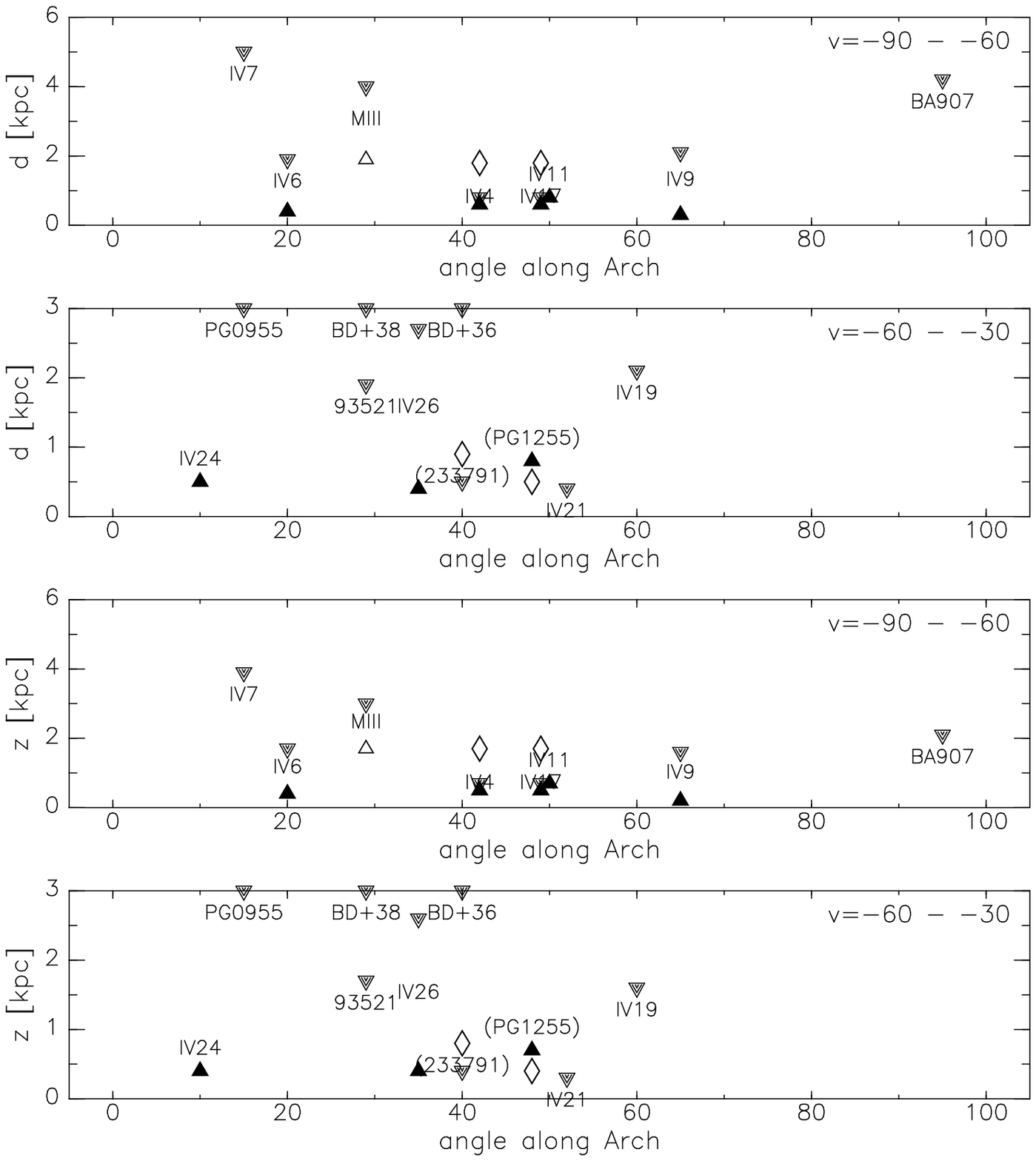}{-70}{-45}{% %AASPP
\fgnumber{\Fivdist} Summary of upper and lower distance limits for the IV Arch.
The x-axis shows an approximate ``angle along the Arch'', with the origin at
$l$=210\deg, $b$=30\deg. The top two panels show the distances, the bottom two
the z-heights. Core names are indicated, as are the stars outside cores, for
which just the beginning part of the name is shown. Three stars in the second
and fourth panel are plotted at y-max, but they are more distant than that.
Downward-pointing triangles are upper limits, upward-pointing triangles are
lower limits. The open triangle for MIII refers to the possible lower limit from
HD\,93521. Diamond shapes are for the possible alternative distance for three
stars with uncertain distances (BD+49\,2137 toward IV4 and IV17, HDE\,233791 and
PG\,1255+546).
}

\InsertPage{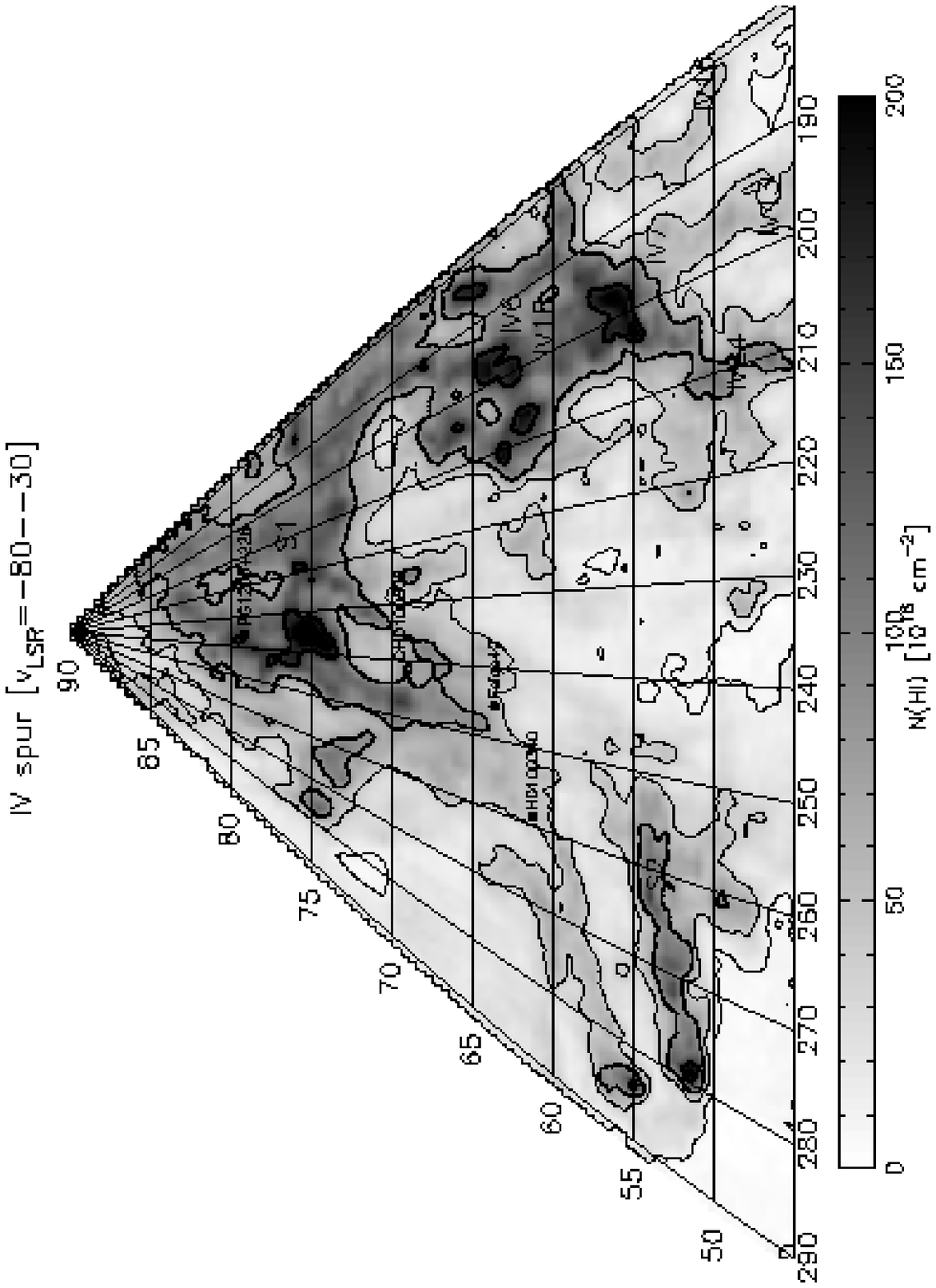}{-45}{-90}{% %AASPP
\fgnumber{\Fgspur} Map of the IV spur, from Hartmann \& Burton (1997). Contours
are at 10, 40 and 80\tdex{18}\,\cmm2. The core names are shown, as are the
positions of probe stars. Closed symbols refer to detections, open symbols to
non-detections.
}

\InsertPage{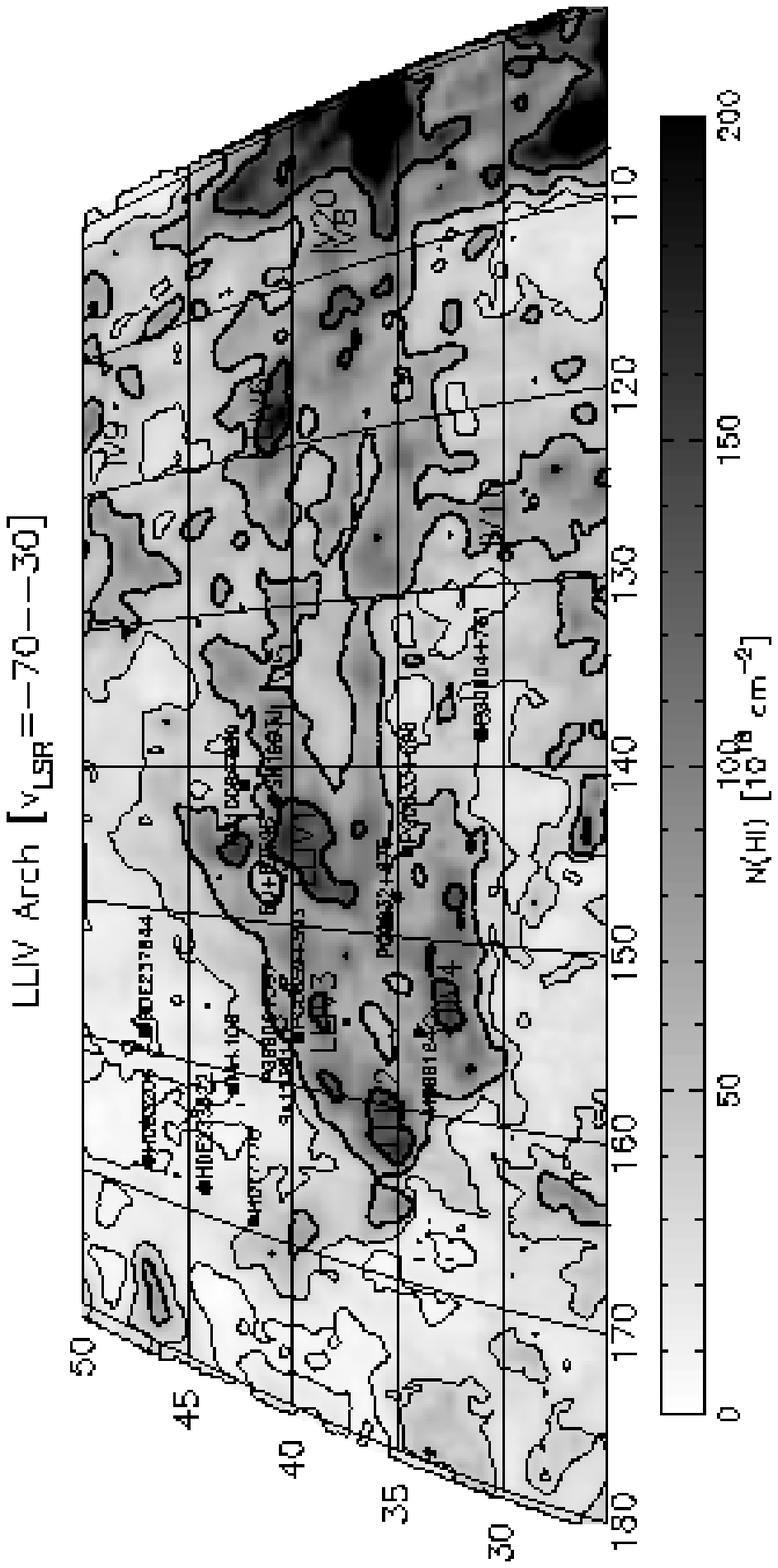}{-45}{-90}{% %AASPP
\fgnumber{\FgLLIV} Map of the Low-Latitude Intermediate-Velocity Arch, from
Hartmann \& Burton (1997). Contours are at 10, 40 and 80\tdex{18}\,\cmm2. The
core names are shown, as are the positions of probe stars. Closed symbols refer
to detections, open symbols to non-detections. Note that HVC complex~A lies
between the probes Mrk\,106 and PG\,0804+761.
}

\InsertPage{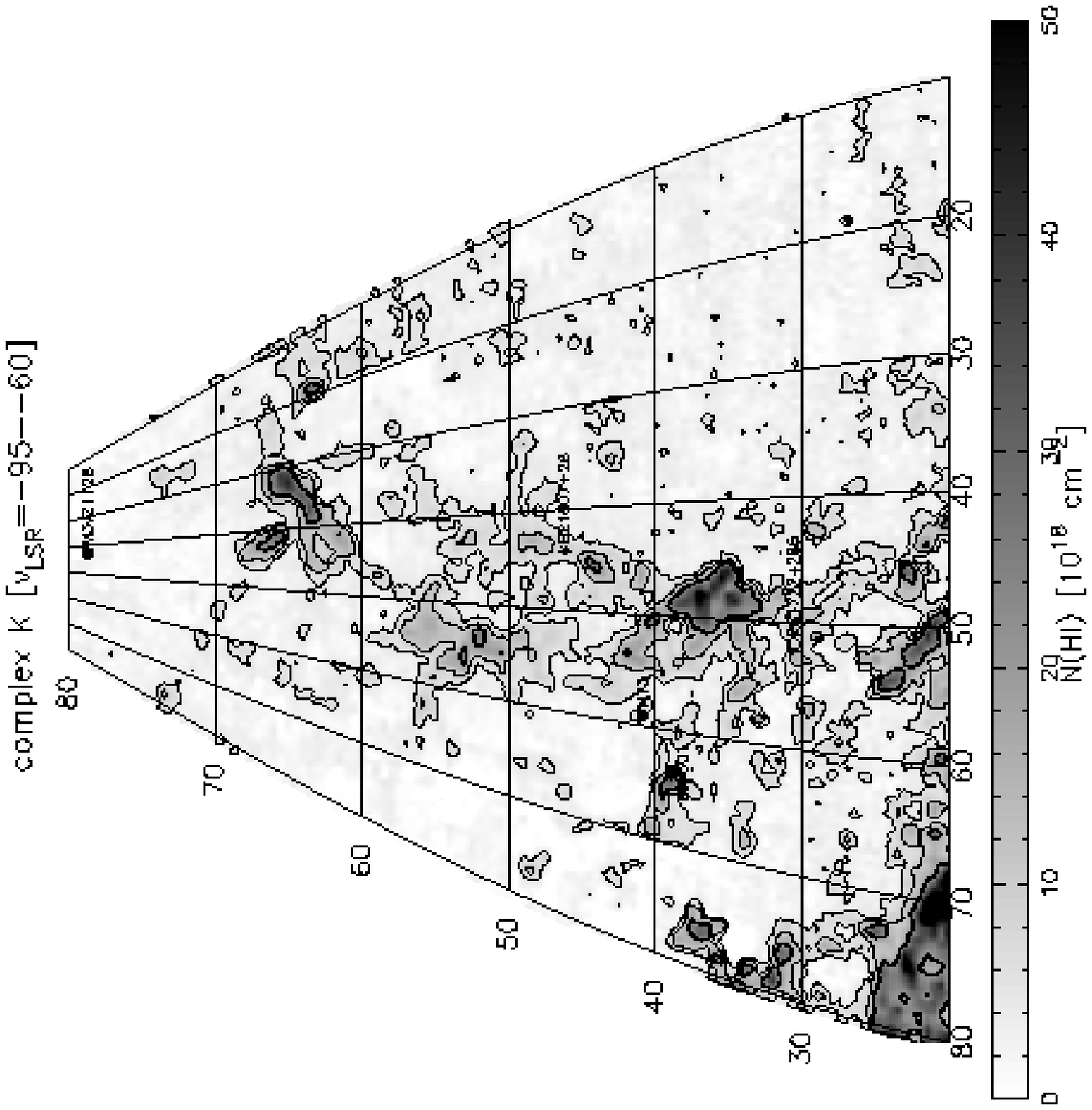}{-45}{-90}{% %AASPP
\fgnumber{\FgK} Map of complex~K, from Hartmann \& Burton (1997). Contours
indicate column densities of 5, 10, and 20\tdex{18}\,\cmm2\ for gas in the
velocity range $-$95 to $-$60\,\kms. The open symbols show the positions of the
probes with non-detections, closed symbols refer to detections.
}

\InsertPage{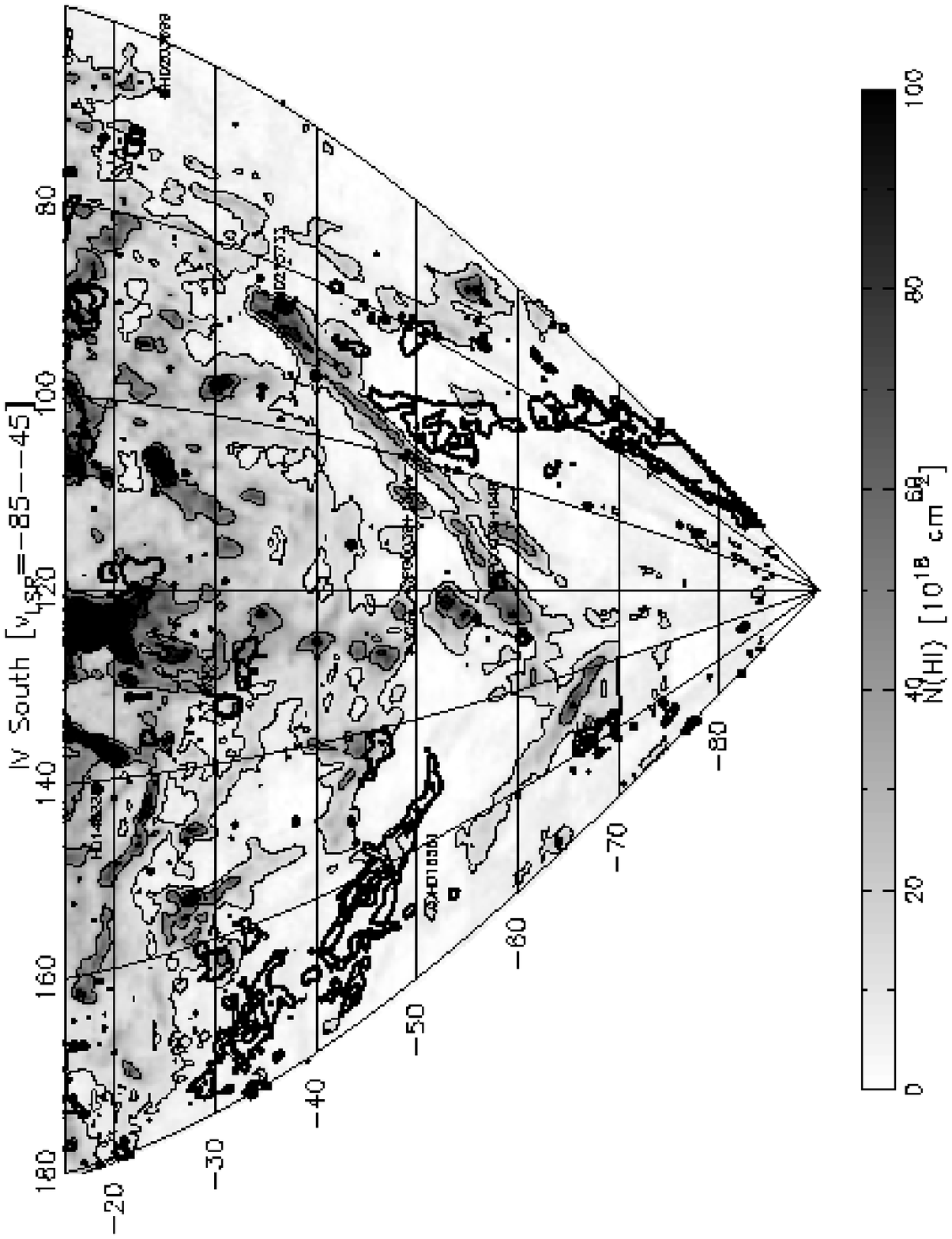}{-20}{-90}{% %AASPP
\fgnumber{\Fgsouth} Map of southern IVCs, from Hartmann \& Burton (1997).
Contour levels are at 10, 40 and 80\tdex{18}\,\cmm2\ for the gas with velocities
relative to the LSR between $-$85 and $-$45\,\kms. Closed symbols refer to
detections, open symbols to non-detections. Thick solid lines outline the gas
with \vlsr$<$$-$85\,\kms\ and are contour levels at 15 and 40\tdex{18}\,\cmm2.
The following HVCs can be seen: the Cohen Stream and WW507 at $l$=160\deg,
$b$=$-$45\deg, the Magellanic Stream ($l$$\sim$80\deg, $b$$<$$-$50\deg), the
VHVC near M\,33 ($l$=125\deg, $b$=$-$30\deg), and complex~G ($l$=90\deg,
$b$=$-$15\deg). The features at $l$=120\deg, $b$=$-$20\deg\ and $l$=134\deg,
$b$=$-$31\deg\ are M\,31 and M\,33.
}

\InsertPage{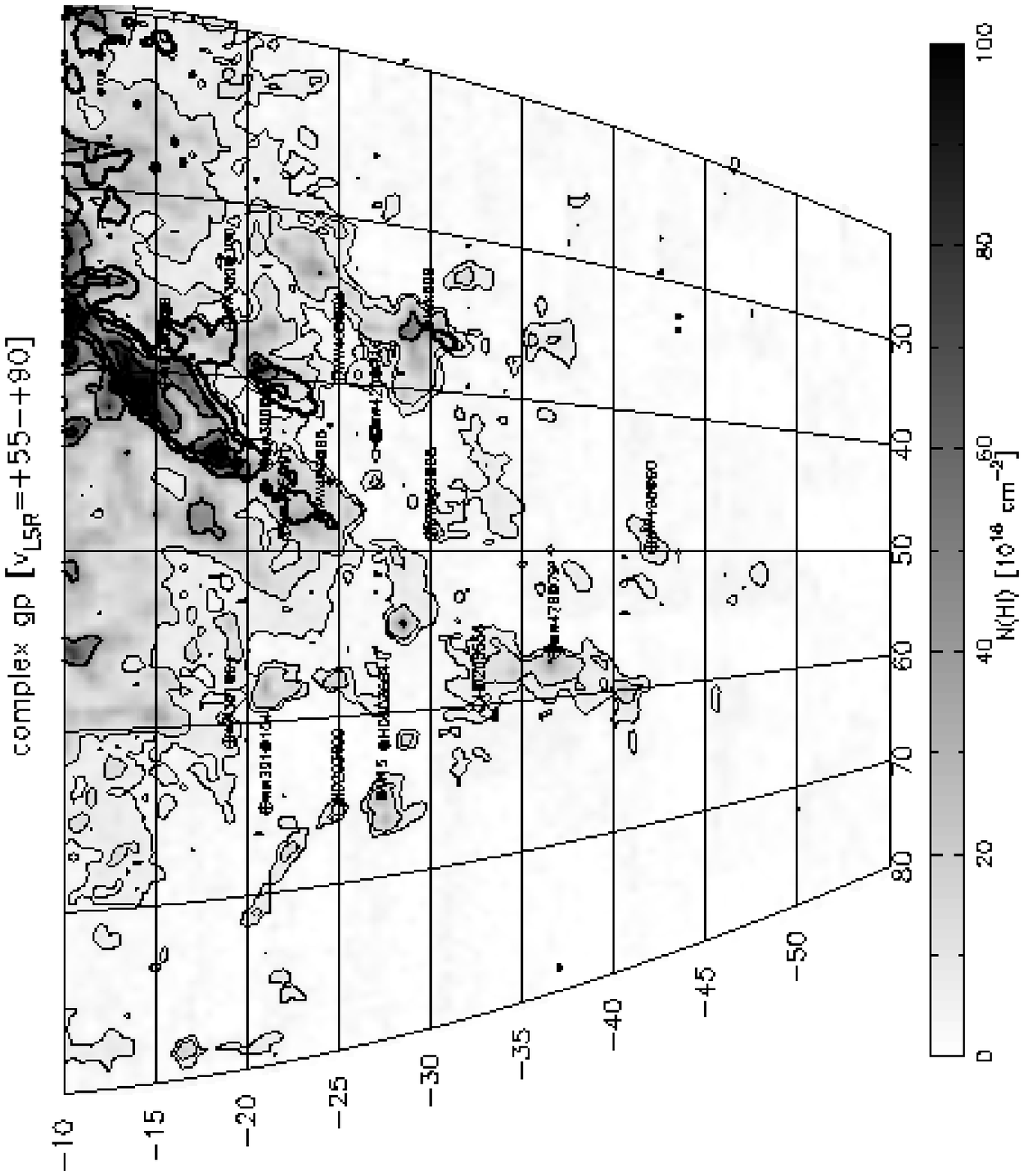}{-20}{-90}{% %AASPP
\fgnumber{\Fggp} Map of the positive-intermediate-velocity gas in the region
near HVC complex~GP, from Hartmann \& Burton (1997). The greyscale and contours
show the gas with velocities between +50 and +200\,\kms, with contour levels at
5, 10, 40, and 80\tdex{18}\,\cmm2. The positions of probes are show. Closed
symbols refer to detections, open symbols to non-detections. The thick contours
show the gas with \vlsr$>$+90\,\kms, i.e.\ HVC~40$-$15+100 (also known as the
``Smith cloud''). The positions of the other WW clouds within complex~GP are
also shown, with a label of the form WW\#@v, where WW\# is the catalogue number
of Wakker \& van Woerden (1991) and v the velocity in that catalogue.
}
\end{document}